\documentclass[dvipsnames,11pt,reqno,twoside,final]{article}
\usepackage[a4paper,left=3cm,right=3cm,top=3cm,bottom=3cm]{geometry}
\usepackage[T1]{fontenc}
\usepackage[utf8]{inputenc}
\usepackage{xcolor}
\usepackage{amssymb,mathtools}  % Also loads "amsmath"
\usepackage{dsfont}
\usepackage{enumitem}
\setlist{nosep} % no separation for lists
\setlist[enumerate]{label=(\roman*)} % romain numerals for enumerate by default
\usepackage{subcaption}
\usepackage{booktabs}
\usepackage{preamble/mhequ}

\usepackage{amsthm,bm}

\newcounter{counterEnvMain}
\newcounter{counterEnvDefault}
\numberwithin{counterEnvDefault}{section}

\theoremstyle{plain}

\newtheorem{lemma}[counterEnvDefault]{Lemma}
\newtheorem*{lemma*}{Lemma}
\newtheorem{conjecture}[counterEnvMain]{Conjecture}

\newtheorem{theorem}[counterEnvDefault]{Theorem}
\newtheorem*{theorem*}{Theorem}

\newtheorem{proposition}[counterEnvDefault]{Proposition}
\newtheorem{corollary}[counterEnvDefault]{Corollary}

\theoremstyle{definition}

\newtheorem{definition}[counterEnvDefault]{Definition}
\newtheorem*{definition*}{Definition}

\newtheorem{remark}[counterEnvDefault]{Remark}
\newtheorem*{example*}{Example}

\newtheorem*{claim*}{Claim}

\newtheorem*{assertion*}{Assertion}

\newtheorem*{proposition*}{Proposition}

\usepackage{microtype}
\renewcommand\epsilon\varepsilon

\usepackage{hyperref}
\definecolor{colorlinks}{RGB}{0, 24, 168}
\definecolor{colorcites}{RGB}{124, 10, 2}
\hypersetup{
    colorlinks=true,
    linkcolor=colorlinks,
    citecolor=colorcites,
    urlcolor=colorlinks,
    pdfborder={0 0 0}
}

\usepackage{xargs}
\usepackage[colorinlistoftodos,prependcaption,textsize=tiny]{todonotes}
\newcommandx\work[2][1=]{\todo[linecolor=RoyalBlue,backgroundcolor=RoyalBlue!25,bordercolor=RoyalBlue,#1]{\textsc{todo} #2}}
\newcommandx\commentPiet[2][1=]{\todo[linecolor=OliveGreen,backgroundcolor=OliveGreen!25,bordercolor=OliveGreen,#1]{\textsc{comment} Piet: #2}}
\newcommandx\commentIoan[2][1=]{\todo[linecolor=OliveGreen,backgroundcolor=OliveGreen!25,bordercolor=OliveGreen,#1]{\textsc{comment} Ioan: #2}}
\newcommandx\commentHugo[2][1=]{\todo[linecolor=OliveGreen,backgroundcolor=OliveGreen!25,bordercolor=OliveGreen,#1]{\textsc{comment} Hugo: #2}}
\newcommandx\commentKarol[2][1=]{\todo[linecolor=OliveGreen,backgroundcolor=OliveGreen!25,bordercolor=OliveGreen,#1]{\textsc{comment} Karol: #2}}
\newcommandx\noticePiet[2][1=]{\todo[linecolor=black,backgroundcolor=white,bordercolor=black,#1]{\textsc{notice} Piet: #2}}
\newcommandx\noticeIoan[2][1=]{\todo[linecolor=black,backgroundcolor=white,bordercolor=black,#1]{\textsc{notice} Ioan: #2}}
\newcommandx\noticeHugo[2][1=]{\todo[linecolor=black,backgroundcolor=white,bordercolor=black,#1]{\textsc{notice} Hugo: #2}}
\newcommandx\noticeKarol[2][1=]{\todo[linecolor=black,backgroundcolor=white,bordercolor=black,#1]{\textsc{notice} Karol: #2}}
\newcommandx\todoPiet[2][1=]{\todo{\textsc{todo} Piet: #2}}
\newcommandx\todoIoan[2][1=]{\todo{\textsc{todo} Ioan: #2}}
\newcommandx\todoHugo[2][1=]{\todo{\textsc{todo} Hugo: #2}}
\newcommandx\todoKarol[2][1=]{\todo{\textsc{todo} Karol: #2}}
\newcommandx\mistake[2][1=]{\todo[linecolor=red,backgroundcolor=red!25,bordercolor=red,#1]{\textsc{mistake} #2}}
\newcommandx\improve[2][1=]{\todo[linecolor=orange,backgroundcolor=orange!25,bordercolor=orange,#1]{\textsc{improve} #2}}
\newcommandx\change[2][1=]{\todo[linecolor=yellow,backgroundcolor=yellow!25,bordercolor=yellow,#1]{\textsc{change} #2}}
\newcommandx\mem[2][1=]{\todo[linecolor=orange,backgroundcolor=orange!25,bordercolor=orange,#1]{\textsc{mem} #2}}
\newcommandx\status[2][1=]{\todo[linecolor=Blue,backgroundcolor=Blue!25,bordercolor=Blue,#1]{\textsc{Status} #2}}

\newcommand\printtodolist{
    \newpage

    \listoftodos
}
\newcommand\hidetodos{
    \renewcommandx\todo[2][1=]{}
    \renewcommandx\work[2][1=]{}
    \renewcommandx\comment[2][1=]{}
    \renewcommandx\mistake[2][1=]{}
    \renewcommandx\improve[2][1=]{}
    \renewcommandx\change[2][1=]{}
    \renewcommandx\mem[2][1=]{}
    \renewcommandx\status[2][1=]{}
    \renewcommand\printtodolist{}
}

\newcommand\blank{\,\cdot\,}

\newcommand\Var{\operatorname{Var}}
\newcommand\Cov{\operatorname{Cov}}

\newcommand\indi[1]{\mathds{1}_{#1}}
\newcommand\true[1]{\mathds{1}[#1]}
\newcommand\diffi{{\,\mathrm{d}}}
\newcommand\diff{{\mathrm{d}}}

\newcommand\C{\mathbb C}

\newcommand\E{\mathbb E}

\renewcommand\H{\mathbb H}
\newcommand\I{\mathbb I}

\renewcommand\P{\mathbb P}

\newcommand\R{\mathbb R}

\newcommand\T{\mathbb T}

\newcommand\Z{\mathbb Z}

\newcommand\bbR{\mathbb R}

\newcommand\calA{\mathcal A}
\newcommand\calB{\mathcal B}
\newcommand\calC{\mathcal C}
\newcommand\calD{\mathcal D}
\newcommand\calE{\mathcal E}

\newcommand\calG{\mathcal G}
\newcommand\calH{\mathcal H}
\newcommand\calI{\mathcal I}

\newcommand\calK{\mathcal K}
\newcommand\calL{\mathcal L}
\newcommand\calM{\mathcal M}
\newcommand\calN{\mathcal N}
\newcommand\calO{\mathcal O}

\newcommand\calR{\mathcal R}
\newcommand\calS{\mathcal S}
\newcommand\calT{\mathcal T}
\newcommand\calU{\mathcal U}

\newcommand\calX{\mathcal X}

\newcommand\calZ{\mathcal Z}

\newcommand\frakA{\mathfrak A}
\newcommand\frakB{\mathfrak B}
\newcommand\frakC{\mathfrak C}

\newcommand\frakF{\mathfrak F}

\newcommand\frakO{\mathfrak O}

\newcommand\frakR{\mathfrak R}

\newcommand\fraka{\mathfrak a}

\newcommand\frake{\mathfrak e}

\newcommand\frako{\mathfrak o}

\makeatletter
\@namedef{subjclassname@2020}{\textup{2020} Mathematics Subject Classification}
\makeatother
\renewcommand{\Im}{\mathrm{Im}}
\renewcommand{\Re}{\mathrm{Re}}
\newcommand\GFF{\operatorname{GFF}}
\renewcommand\a{\mathbf{a}}
\renewcommand\b{\mathbf{b}}
\renewcommand\i{\mathrm{i}}
\renewcommand\u{{\boldsymbol{u}}}
\newcommand\n{{\boldsymbol{n}}}
\renewcommand\v{{\boldsymbol{v}}}
\newcommand\w{{\boldsymbol{w}}}
\newcommand\s{{\boldsymbol{s}}}
\renewcommand\t{{\boldsymbol{t}}}

\newcommand\sixv{6\mathrm{V}}
\newcommand\CYL[1]{\operatorname{Cyl}_{#1}}
\usepackage{rotating}

\usepackage{xcolor}
\usepackage{framed}

\usepackage{pifont} %flowers

\usepackage{stmaryrd}
\usepackage{nicefrac}

\usepackage{comment}
\usepackage{accents}

\usepackage{longtable,geometry}

\usepackage{hyperref}
\hypersetup{
	colorlinks=true, %set true if you want colored links
	linktoc=all,     %set to all if you want both sections and subsections linked
	linkcolor=blue  %choose some color if you want links to stand out
}

\usepackage{centernot}

\usepackage{mdframed}
\mdfdefinestyle{MyFrame}{%
    linecolor=black,
    outerlinewidth=3pt,
    innertopmargin=\baselineskip,
    innerbottommargin=\baselineskip,
    innerrightmargin=2pt,
    innerleftmargin=2pt,
    backgroundcolor=gray!10!white}

\newcommand{\eps}{\varepsilon}

\renewcommand{\Im}{\mathrm{Im}}
\renewcommand{\Re}{\mathrm{Re}}

\newcommand{\rk}[1]{\bgroup\color{red}%
	\par\medskip\hrule\smallskip%
	\noindent\textbf{#1}%
	\par\smallskip\hrule\medskip\egroup}

    \newcommand\sleven{\circ}
    \newcommand\slodd{\bullet}
    \newcommand\muSpin{\mu}

    \newcommand\connect[3]{\{#1\xleftrightarrow{#3}#2\}}

    \newcommand\HORI[1]{\operatorname{Hor}(#1)}
    \newcommand\VERTI[1]{\operatorname{Ver}(#1)}
    \newcommand\CIRCUIT[1]{\operatorname{Circuit}(#1)}
    \newcommand\ARM[1]{\operatorname{Arm}(#1)}

    \newcommand\pHORI[2]{\operatorname{Hor}^{#1}(#2)}
    \newcommand\pVERTI[2]{\operatorname{Ver}^{#1}(#2)}
    \newcommand\pCIRCUIT[2]{\operatorname{Circuit}^{#1}(#2)}

    \newcommand\ALTHORI[2]{\operatorname{AltHor}_{#1}(#2)}
    \newcommand\ALTVERTI[2]{\operatorname{AltVer}_{#1}(#2)}
    \newcommand\ALTCIRCUIT[2]{\operatorname{AltCircuit}_{#1}(#2)}
    \newcommand\ALTARM[2]{\operatorname{AltArm}_{#1}(#2)}

    \newcommand\pALTRIDGE[3]{\operatorname{Ridge}^{#1}_{#2}(#3)}
    \newcommand\pALTHORI[3]{\operatorname{AltHor}^{#1}_{#2}(#3)}
    \newcommand\pALTVERTI[3]{\operatorname{AltVer}^{#1}_{#2}(#3)}
    \newcommand\pALTCIRCUIT[3]{\operatorname{AltCircuit}^{#1}_{#2}(#3)}
    \newcommand\pALTARM[3]{\operatorname{AltArm}^{#1}_{#2}(#3)}

    \newcommand\COUNTHORI[2]{\operatorname{\#AltHor}^{#1}(#2)}
    \newcommand\COUNTVERTI[2]{\operatorname{\#AltVer}^{#1}(#2)}
    
    \newcommand\COUNTARM[2]{\operatorname{\#AltArm}^{#1}(#2)}

    \newcommand\rectleft[1]{\operatorname{Left}_{#1}}
    \newcommand\rectright[1]{\operatorname{Right}_{#1}}
    \newcommand\recttop[1]{\operatorname{Top}_{#1}}
    \newcommand\rectbottom[1]{\operatorname{Bottom}_{#1}}

    \newcommand\bbox[1]{\operatorname{Ball}_{#1}}
    \newcommand\ann[2]{\operatorname{Annulus}_{#1,#2}}

    \newcommand\ccircuit{c_{\mathrm{circuit}}}
    \newcommand\carm{c_{\mathrm{arm}}}

    \newcommand\symInt[1]{[\![#1]\!]}
    \newcommand\symRect[2]{[\![#1 \times #2]\!]}

\usetikzlibrary{shapes.geometric, arrows}

\usepackage{showlabels}

 \title{\vspace{-2em}Gaussian free field convergence of the six-vertex model\\with $-1\le \Delta\le -\frac12$}
\author{
    Hugo Duminil-Copin%
    \footnote{Institut des Hautes Études Scientifiques, \texttt{duminil@ihes.fr}}
    \footnote{Université de Genève, \texttt{hugo.duminil@unige.ch}},
    Karol Kajetan Kozlowski%
    \footnote{ENSL, CNRS, Laboratoire de Physique, F-69342 Lyon, France, \texttt{karol.kozlowski@ens-lyon.fr}},
    Piet Lammers%
    \footnote{CNRS, Sorbonne Université, LPSM, \texttt{piet.lammers@cnrs.fr}},
    and
    Ioan Manolescu%
    \footnote{Université de Fribourg, \texttt{ioan.manolescu@unifr.ch}}%
}

 \date{6 March 2026}

\usepackage[
    style=alphabetic,
    date=year,
    maxnames=3,
    maxbibnames=99
]{biblatex}
\AtEveryBibitem{%
  \clearfield{url}%
  \clearfield{urldate}%
  \clearfield{urlday}%
  \clearfield{urlmonth}%
  \clearfield{urlyear}%
  \clearfield{doi}%
  \clearfield{isbn}%
  \clearfield{issn}%
}

\usepackage{tocloft,lipsum,pgffor}

\begin{document}
\hidetodos%
{\renewcommand\c{\mathbf{c}}

\maketitle

\vfill
\vspace{-10em}

\begin{abstract}
We study the isotropic six-vertex model on $\mathbb{Z}^2$ with spectral parameter $\Delta\in[-1,-1/2]$,
that is, with weights $\a=\b=1$ and $\c\in[\sqrt{3},2]$.
We show that the associated height function converges, in the scaling limit, to a properly scaled full-plane Gaussian free field.
The result extends to anisotropic weights $\a\neq \b$ upon using a suitable embedding of the lattice.
\end{abstract}

\renewcommand{\contentsname}{}

\vspace{-10em}
\vfill

\tableofcontents
\vspace{1em}

\newpage

% <input src="sections/PART_A/main.tex" root="." version="0.0.1">
\part{Introduction}

% <input src="sections/PART_A/1_0_preface.tex" root="." version="0.0.1">
\section{Motivation}

\subsection{Phase transitions in a nutshell}\label{sec:1.1}

Physical systems undergoing a continuous phase transition can often be understood mathematically through lattice models, in which the microscopic degrees of freedom are encoded by variables attached to the sites, edges  or faces of a graph.
These models serve as effective descriptions of the underlying physical interactions while remaining amenable to rigorous analysis.
 In this framework, $k$-point correlation functions correspond to limits of expectations of products of \emph{local} operators
$\langle \prod_i O^{(i)}_{u_i}\rangle_\delta$.
In this context, $\langle\blank\rangle_\delta$ encodes the correlations of the model
on a lattice of mesh size $\delta$,
$u_1,\dots,u_k$ are points in space,
the $O^{(i)}$ are (potentially different) observables measuring local properties of the system near the origin $0$,
and $O^{(i)}_{u_i}$ are the translates of these observables to the points $u_i$
so that $O^{(i)}_{u_i}$ is measurable in terms of the behaviour in a small neighbourhood of $u_i$.
In the sequel, we suppose each observable to be centred.

Away from the phase transition, correlations are extremely weak: for any collection of distinct points, the correlation functions decay exponentially fast in the separation distance (here of order $1/\delta$) between the points $u_1,\dots,u_k$. In other words, individual observables decorrelate fast and become asymptotically independent.

At a continuous phase transition, however, asymptotic independence still occurs, but correlations are expected to be much stronger.
The $k$-point functions exhibiting a power-law decay with the distance:
\begin{equation}\label{ecriture scaling regime k pt fcts continuous PT}
\lim_{\delta\rightarrow 0}\delta^{-(\alpha_1+\dots+\alpha_k)}\big\langle \prod_{i=1}^k O^{(i)}_{u_i} \big\rangle_\delta=
\mathcal{C}_{O^{(1)},\dots,O^{(k)}}(u_1,\dots,u_k)
\end{equation}
 for certain non-trivial functions $\mathcal{C}_{O^{(1)},\dots,O^{(k)}}$ and {\em critical exponents} $\alpha_i=\alpha_i(O^{(i)})$ determining the rate of algebraic decay of the correlators at large distances.
%  Although the latter depend on the observables, we suppress this dependence to lighten notation.

The functions $\mathcal{C}_{O^{(1)},\dots,O^{(k)}}$ are predicted to be {\em invariant under dilations} and {\em rotations}, as well as \emph{universal} in the sense that they do not depend on the specific local interaction potential appearing in the definition of the lattice model.
One purpose of statistical mechanics is to
group models undergoing continuous phase transitions into \emph{universality classes} with matching critical exponents $\alpha_i$
and limiting functions $\mathcal{C}_{O^{(1)},\dots,O^{(k)}}$.

The very ideas on the behaviour of systems at continuous phase transitions date back to the foundational works in theoretical physics from the mid-1960s to the early 1970s, notably those of Fisher
\cite{Fisher_1967_TheoryCondensationCritical,Fisher_1966_QuantumCorrectionsCriticalPoint,Fisher_1964_CorrelationFunctionsCritical},
Kadanoff \cite{Kadanoff_1966_ScalingLawsIsing},
and Widom
\cite{Widom_1965_SurfaceTensionMolecular,Widom_1965_EquationStateNeighborhood},
culminating in Wilson's renormalisation-group theory
\cite{Wilson_1971_RenormalizationGroupCriticala,Wilson_1971_RenormalizationGroupCritical}.
The predictions have been tested extensively against perturbative calculations  in a wide range of settings, effective models, and experimental data, as well as through numerical simulations
\cite{KadanoffGotzeHamblen_1967_StaticPhenomenaCritical,EssamFisher_1963_PadeApproximantStudies,DombSykes_1957_SusceptibilityFerromagneticCurie,Baker_1961_ApplicationPadeApproximant,El-ShowkPaulosPoland_2012_Solving3DIsing,KosPolandSimmons-Duffin_2014_BootstrappingMixedCorrelators}.

\subsection{The CFT conjectural limit of lattice models}

It is not a priori clear what the hypothetical universality classes should be.
One line of thought, originating in the works of Patashinskii and Pokrovskii
\cite{PatashinskiiPokrovskii_1964_SecondOrderPhase} and later Polyakov
\cite{Polyakov_1968_MicroscopicDescriptionCritical,Polyakov_1970_ConformalSymmetryCritical,Polyakov_1970_PropertiesLongShort},
is that universality classes should be captured by quantum field theories that are invariant under not only scaling and rotations, but also {\em conformal transformations}.
Such quantum fields theories are called Conformal Field Theories (CFTs).
This led to the following loosely stated conjecture.

\begin{conjecture}
\label{Conjecture universalite PT}
The scaling limits in the sense of Equation~\eqref{ecriture scaling regime k pt fcts continuous PT}
of two-dimensional
statistical mechanics models undergoing a continuous phase transition, are given by the correlation functions of a CFT.
\end{conjecture}
 In general dimension, the additional requirement of being invariant under conformal maps is not so restrictive, since the local conformal group is finite-dimensional and thus imposes only limited constraints. However, as observed in the groundbreaking work of Belavin, Polyakov, and Zamolodchikov \cite{BelavinPolyakovZamolodchikov_1984_InfiniteConformalSymmetry}, the situation improves dramatically in two dimensions: there, the conformal group is infinite-dimensional and the resulting local conformal symmetry leads to the integrability of numerous CFTs in the plane; we refer to the mathematically-oriented textbook \cite{GawedzkiBook} for details.

Two-dimensional CFTs are classified by their central charge $c \in \mathbb R_{>0}$.
\commentPiet{Is this rigorously known? If so, we could maybe delay its statement, and then mentioned immediately after what is known for each central charge $c$?}
After extensive investigations in the 1980s and 1990s, these theories turned out to be amenable to a substantial degree of analysis on physical grounds \cite{TeschnerCFTLectureNotes,DiFrancescoMathieuSenechal_1997_ConformalFieldTheory}. On the mathematical side, apart from the numerous early developments (cf.~\cite{GawedzkiBook,Segal_1988_DefinitionConformalField}), there has recently been a renewed interest in these CFTs within the probability community, driven by the development of the Schramm--Loewner Evolution \cite{Schramm_2000_ScalingLimitsLooperased,Lawler_2014_ConformallyInvariantProcesses} (and the associated models constructed from it \cite{SheffieldWerner_2012_ConformalLoopEnsembles,MillerSheffield_2016_ImaginaryGeometryInteracting,Sheffield_2009_ExplorationTreesConformal,MillerSheffield_2020_LiouvilleQuantumGravity,GarbanRhodesVargas_2016_LiouvilleBrownianMotion}) as well as by rigorous techniques based on \cite{DavidKupiainenRhodes_2016_LiouvilleQuantumGravity,RhodesVargas_2014_GaussianMultiplicativeChaos,GuillarmouRhodesVargas_2019_PolyakovsFormulation2d}.

In many cases, one can provide a rather detailed description of the CFTs correlation functions. The understanding of the CFT with $c=1$ -- the free boson model, known to mathematicians as the Gaussian free field -- is straightforward thanks to its Gaussian nature. In particular, Wick's rule expresses $k$-point correlations in terms of products of $2$-point correlations, which significantly streamlines their analysis. We refer to \cite{Sheffield_2007_GaussianFreeFields,BerestyckiPowell_2025_GaussianFreeField} for details on this specific case. The rational CFTs, notably the minimal models corresponding to
$c =1-6\frac{(p-q)^2}{pq}$ with $p,q\ge 2$ coprime integers, are also very well-understood thanks to algebraic structures specific to these central charges
\cite{DamioliniGibneyTarasca_2024_FactorizationVectorBundles,DamioliniWoike_2025_ModularFunctorsConformal,
GuiZhang_2026_AnalyticConformalBlocks,FjelstadFuchsRunkel_2006_TFTConstructionRCFT,FuchsRunkelSchweigert_2002_TFTConstructionRCFT,
FuchsRunkelSchweigert_2004_TFTConstructionRCFTa,FuchsRunkelSchweigert_2004_TFTConstructionRCFT,FuchsRunkelSchweigert_2005_TFTConstructionRCFT}. For $c\ge 25$, the theory is well developed, as in
\cite{Teschner_1995_LiouvilleThreepointFunction,Teschner_2001_LiouvilleTheoryRevisited}, and more recently in the mathematics literature in
\cite{DuplantierSheffield_2011_LiouvilleQuantumGravity,BerestyckiPowell_2025_GaussianFreeField,DuplantierSheffield_2011_LiouvilleQuantumGravity,DuplantierSheffield_2011_LiouvilleQuantumGravity,GuillarmouKupiainenRhodes_2024_ConformalBootstrapLiouville}. Fully determining the correlations of CFTs still is the a subject of
intense study both on the physical and mathematical sides, see
e.g.~\cite{AngCaiSun_2021_IntegrabilityConformalLoop,KupiainenRhodesVargas_2020_IntegrabilityLiouvilleTheory,GuillarmouKupiainenRhodes_2024_ConformalBootstrapLiouville,CercleRhodesVargas_2023_ProbabilisticConstructionToda,
GuiZhang_2026_AnalyticConformalBlocks,DamioliniWoike_2025_ModularFunctorsConformal}.

Today's understanding of CFTs, even though not fully exhaustive, is very impressive. Still, even if one were given a complete and rigorous description of all two-dimensional CFTs, a thorough understanding of the continuous phase transition exhibited by a given lattice model still requires the following two challenging steps:

\begin{enumerate}
\item identifying which CFT (that is, which value of the central charge $c$ and which representation of the corresponding CFT algebraic structure), if any, captures the relevant limit at criticality;
\item identifying which correlation functions of that CFT describe the  limits of the $k$-point functions of the chosen local observables. Concretely, in the setting of~Equation~\eqref{ecriture scaling regime k pt fcts continuous PT}, this amounts to identifying the functions $\mathcal{C}_{O^{(1)},\dots,O^{(k)}}$ with appropriate correlation functions of the CFT in question.
\end{enumerate}

\subsection{From discrete to continuum}

In order to go in this direction, two principal strategies have been developed over the years.
\commentPiet{Reading the two strategies, it feels like: for transferMatrix, it is more about development of general techniques, while for discreteHolo it feels like it's more about solving one model at a time. Is this what we want to convey? HDC: I think it is right, and fine to say it.}

\subsubsection{Extracting information from the transfer matrix formalism}

For models that admit a transfer-matrix formulation, analysing the leading eigenvalues of the transfer matrix
$T_L$ of a system of size
$L$ with periodic boundary conditions, yields valuable information about the limiting behaviour. It is well known that the exponential growth rate of the largest eigenvalue of $T_L$ determines the free energy of the model. Moreover, as argued heuristically in \cite{Affleck_1986_UniversalTermFree,BloteCardyNightingale_1986_ConformalInvarianceCentral}, the finite-size correction to the free energy -- equivalently, the subleading correction to the principal eigenvalue -- produces a constant that is interpreted, conjecturally, as the central charge of the conformal field theory expected to govern the model's behaviour at criticality.

Additional information can be extracted from the large-$L$ asymptotics of the subleading eigenvalues of  the transfer matrix and the so-called translation operator (both of which will appear in this paper). In particular, conjectural equations relate the spacing of these eigenvalues to the possible values of critical exponents associated with local observables.
%$\lambda_0(L)>\lambda_1(L)>\lambda_2(L)>\dots$ of~$T_L$ and the associated eigenvalues $\widetilde\lambda_0(L),\widetilde\lambda_1(L),\widetilde\lambda_2(L),\dots$ of the so-called {\em translation operator} (recall that the two operators commute and are therefore codiagonalizable). Namely,
%\begin{align}
% \ln\Big(\frac{\lambda_k(L)}{\lambda_0(L)}\Big)
% &=
% -\,\frac{2\pi v_F}{2L+1}\,(\Delta_{k;+} + \Delta_{k;-})
% \;+\; O(L^{-2}),
%\label{eq:intro1} \\[1ex]
% \ln\Big(\frac{\widetilde\lambda_k(L)}{\widetilde\lambda_0(L)}\Big)
% &=
% -\,\frac{2\pi v_F}{2L+1}\,(\Delta_{k;+} - \Delta_{k;-})
% \;+\; O(L^{-2}),
%\label{ecriture DA grd L vp sous dominantes}
%\end{align}
%where the quantities $\Delta_{k;\pm}$ correspond to the possible critical exponents associated with local observables appearing in the lattice model.
The above physical picture is thus rather comprehensive, although it does not specify which critical exponent corresponds to a given lattice correlation function. This last step is typically achieved, at a physical level of rigour, through symmetry considerations. We refer to \cite{Cardy_1986_OperatorContentTwodimensional,Cardy_2008_ConformalFieldTheory} for reviews.

While constructing a transfer matrix from a local Hamiltonian is often straightforward, extracting the large-volume asymptotics
%above
is an arduous task. For a generic model, this problem is hopeless -- even at a heuristic level. However, the situation improves dramatically for integrable models of two-dimensional statistical mechanics \cite{Baxter_1982_ExactlySolvedModels}.

The simplest example in this direction is the two-dimensional Ising model in vanishing external field, originally solved by Onsager \cite{Onsager_1944_CrystalStatisticsTwoDimensional} and later revisited in many different ways (see \cite{McCoyWu_2014_TwoDimensionalIsingModel,Duminil-Copin_2023_100YearsCritical} for historical accounts). Kaufman \cite{Kaufman_1949_CrystalStatisticsII} and later Lieb-Schultz-Mattis \cite{LiebSchultzMattis_1961_TwoSolubleModels} observed that the Ising model becomes equivalent, through a simple algebraic transformation, to a model of non-interacting (or \emph{free}) fermions. This free-fermionic structure explains, to a large extent, the particularly simple exact solvability of the model. In particular, it leads to a closed and fully explicit formula for the eigenvalues of its transfer matrix, see for example \cite{Baxter_1982_ExactlySolvedModels} for a modern treatment. This analysis provides evidence
that the scaling limit of the model at criticality is governed by the conformal field theory with central charge $c=1/2$.

The situation is considerably more involved for other integrable models of two-dimensional statistical mechanics undergoing continuous phase transitions. These models are genuinely interacting, meaning that no simple mapping reduces them to free fermions. They remain solvable in the sense that their transfer matrices can be diagonalised in a relatively explicit manner, but doing so requires solving a system of Bethe Ansatz equations, determining which solutions correspond to the dominant and and to the tower of sub-dominant eigenvalues, and analysing their behaviour as $L$ tends to infinity.

In the mid-1980s and 1990s, substantial effort produced, on heuristic grounds, expansions of the top eigenvalues of the transfer matrix
for a wide range of integrable two-dimensional models at criticality
\cite{DestrideVega_1995_UnifiedApproachThermodynamic,
deVegaWoynarovich_1985_MethodCalculatingFinite,
IzerginKorepinReshetikhin_1989_ConformalDimensionsBethe,
KlumperBatchelor_1990_AnalyticTreatmentFinitesize,
KlumperBatchelorPearce_1991_CentralCharges6,
KlumperWehnerZittartz_1993_ConformalSpectrumSixvertex}.
Although non-rigorous, this body of work yielded extensive conjectures for the central charges and critical exponents governing the scaling limits of many integrable lattice models.

A main obstacle to full mathematical rigour lies in the difficulty of proving the \emph{condensation} of Bethe roots, a property originally conjectured by H\"{u}lten \cite{Hulthen_1938_UeberAustauschproblemKristalles}. Significant progress was made in
\cite{DorlasSamsonov_2009_ThermodynamicLimit6vertex,
Duminil-CopinGagnebinHarel_2021_DiscontinuityPhaseTransition,
Duminil-CopinKozlowskiKrachun_2022_SixVertexModelsFree,
Gusev_1980_WeakConvergenceWave,
Gusev_1985_ThermodynamicsExcitedStates,
Kozlowski_2018_CondensationPropertiesBethe},
culminating in the first rigorous derivations of expansions
of top eigenvalues for the transfer matrix of the six-vertex model \cite{Kozlowski_2018_CondensationPropertiesBethe} and for the staggered six-vertex model
\cite{FaulmannGohmannKozlowski_2025_LowtemperatureSpectrumQuantum}, which arises in the analysis of the XXZ spin-$1/2$ chain at finite temperature.

%Finally, in the non-integrable setting, it is worth mentioning the only available rigorous result of this type: the derivation of an expansion of the form \eqref{ecriture DA grd L energie libre} for a perturbation of the two-dimensional Ising model, obtained by constructive renormalisation methods in \cite{GiulianiMastropietro_2013_UniversalFiniteSize}.

\subsubsection{Discrete holomorphicity and conformal invariant scaling limits}

A second strategy roots in the development of \emph{discrete holomorphicity}.
The development of boundary CFTs, which incorporate the effect of boundary conditions, has significantly deepened the physical analysis of conformal field theories. The intuition that studying physical systems in planar domains can reveal additional structure has led, within the mathematical literature, to major breakthroughs in our understanding of the scaling limits of two-dimensional models. The idea is to harvest the fact that conformally invariant (or rather covariant) families of correlations in the continuum are often harmonic or holomorphic solutions of certain boundary value problems. It is therefore natural to expect that in a certain sense, discrete ancestors to these correlations are discrete harmonic or holomorphic solutions to the corresponding discrete boundary value problem.
This observation proved successful for a few models during the first decade of the millennium.

One of the early rigorous proofs of conformal invariance concerns domino tilings, which possess an underlying free-fermionic structure.
Kenyon \cite{Kenyon_2000_ConformalInvarianceDomino,CohnKenyonPropp_2001_VariationalPrincipleDomino} established conformal invariance in the scaling limit for the height-function distribution of domino tilings on bipartite Temperleyan planar graphs by showing that certain observables are discrete holomorphic and satisfy a Dirichlet boundary value problem.
He further proved the convergence of the height function (seen as a distribution) to the Gaussian free field, corresponding to the CFT with central charge $c=1$
(the free boson). This analysis was extended to periodic isoradial graphs in \cite{deTiliere_2007_ScalingLimitIsoradial},
a class preserving exact solvability and discrete holomorphic structure.

In 2001, Smirnov \cite{Smirnov_2001_CriticalPercolationPlane} proved  the celebrated Cardy formula \cite{Cardy_1992_CriticalPercolationFinite} for Bernoulli site percolation on the triangular lattice. Smirnov's argument relies on a certain approximately discrete holomorphic observables satisfying certain boundary conditions, which can be shown to converge to the solution of the continuum analogue of this boundary value problem. This major achievement led to a very precise description of the critical regime, including the determination of critical exponents and links to $c=0$ CFT \cite{LawlerSchrammWerner_2002_ScalingLimitPlanar,Smirnov_2001_CriticalPercolationPlane,CamiaNewman_2007_CriticalPercolationExploration,CamiaNewman_2006_TwoDimensionalCriticalPercolation,CamiaNivesvivat_2026_BoundaryOperatorsBrownian,CamiaFeng_2024_LogarithmicCorrelationFunctions}.

In 2004, Lawler, Schramm, and Werner \cite{LawlerSchrammWerner_2004_ConformalInvariancePlanar} derived the conformal invariance of the loop-erased random walk and the associated Uniform Spanning Tree (UST) by exploiting properties of the discrete Green function, making the approach close in spirit to strategies above. Note that the UST is closely connected to the dimer model and also enjoys a free-fermionic structure.

In 2008, Smirnov \cite{Smirnov_2010_ConformalInvarianceRandom} and Chelkak-Smirnov \cite{ChelkakSmirnov_2012_Universality2DIsing} developed a discrete-holomorphic framework for fermionic observables of the two-dimensional Ising model, showing that they converge in the scaling limit
to holomorphic solutions of appropriate Riemann-Hilbert boundary-value problems.
 Extensions of this method to isoradial graphs yielded robust convergence results
for fermionic observables and demonstrated universality and conformal symmetry in the scaling limit.
Further developments led to a multitude of scaling-limit results for interfaces \cite{ChelkakDuminil-CopinHongler_2014_ConvergenceIsingInterfaces}, the energy-density observable \cite{HonglerSmirnov_2013_EnergyDensityPlanar},
its $n$-point generalisations \cite{Hongler__ConformalInvarianceIsing},
spin correlators \cite{ChelkakHonglerIzyurov_2015_ConformalInvarianceSpin},
and eventually correlation functions corresponding to primary CFT operators
\cite{ChelkakHonglerIzyurov_2022_CorrelationsPrimaryFields}.
These works collectively yield the  link to the $c=\tfrac12$ CFT.

\subsection{Beyond the free fermion point}

Except for Bernoulli site percolation \cite{Smirnov_2001_CriticalPercolationPlane} and a reverse-engineered model known as the harmonic explorer \cite{SchrammSheffield_2005_HarmonicExplorerIts},  all rigorous progress on scaling limits achieved so far concern models that reduce to \emph{non-interacting} (free) fermions. Yet the main appeal of the CFT universality conjecture (Conjecture~\ref{Conjecture universalite PT}) for continuous phase transitions lies precisely in its breadth: it is intended to encompass genuinely \emph{interacting} models, far removed from any structure reminiscent of free theories.
However, once interactions are present, the analysis becomes considerably more intricate.

Rigorous progress has been achieved for small perturbations of free-fermionic models, such as weakly interacting dimers.
These results rely on \emph{constructive renormalisation}, the rigorous embodiment of Wilson's renormalisation-group program.
Initiated in \cite{PinsonSpencerFirstApproachToUniversalityInPertOf2DIsing}, this method has been further developed to establish
scaling properties for a variety of perturbative models
\cite{Mastropietro_2004_IsingModelsFour,GiulianiGreenblattMastropietro_2012_ScalingLimitEnergy,
BenfattoFalcoMastropietro_2014_UniversalityOneDimensionalFermia,
BenfattoFalcoMastropietro_2014_UniversalityOneDimensionalFermi,GiulianiMastropietroToninelli_2017_HeightFluctuationsInteracting,
GiulianiToninelli_2019_NonintegrableDimerModels}.
While these works represent a significant advance, they remain confined to perturbative regimes: the very nature of constructive renormalisation
makes it extremely difficult to reach non-perturbative settings or scaling limits far from free fermionic ones. %A notable exception is the remarkable proof of conformal invariance for Bernoulli site percolation on the triangular lattice \cite{Smirnov_2001_CriticalPercolationPlane}, which remains a singular and isolated achievement.

%In fact, on the physics side, even the renormalisation-group picture changes qualitatively far from the free fermionic point: interactions require summing infinite series of diagrams in nontrivial ways, effectively ``dressing'' free-theory quantities with interaction effects. This passage is notoriously non-smooth and highlights the need for methods conceptually distinct from those used in free fermionic or perturbative settings.

In this light, the genuinely interacting \emph{integrable} models of two-dimensional statistical mechanics form a natural testing ground for Conjecture~\ref{Conjecture universalite PT}.
These models are not equivalent to free fermions except, possibly, at isolated parameter values, and they possess a rich algebraic structure arising from the representation theory of quantum groups. This structure manifests in key identities such as the star--triangle relation \cite{Kennelly_1899_EquivalenceTrianglesThreepointed,Duminil-CopinLiManolescu_2018_UniversalityRandomclusterModel} and in the commutativity of families of transfer matrices \cite{Baxter_1972_PartitionFunctionEightVertex,SklyaninTakhtadzhyanFaddeev_1979_QuantumInverseProblem}, providing powerful tools that are unavailable for generic models.

A particularly prominent example of a genuinely interacting integrable model is the \emph{six-vertex model}; see, for instance,
\cite{LiebWu_1980_TwodimensionalFerroelectricModels,Baxter_1982_ExactlySolvedModels}.
\bigbreak
{\em  This paper provides the first scaling-limit result for the six-vertex model across a substantial range of parameters.    Specifically, for $\Delta \in [-1,-\tfrac12]$, we prove that the height function of the full-plane model converges to the Gaussian free field as the mesh size tends to zero. Although our result is currently restricted to the full-plane setting, it constitutes an important first step towards a more general framework for establishing scaling limits of planar models, as it applies to a broad class of genuinely interacting models. As such, it belongs to a larger program aimed at determining
    the behaviour of critical lattice models in two dimensions deep in the interaction regime.}
    \bigbreak
It is of course premature to discuss the proof (especially since we did not yet introduce the model nor the result properly) in full detail, but let us briefly indicate that the argument draws on ideas inspired by both the transfer-matrix formalism and discrete holomorphicity. In this sense, it synthesises elements of the two historical approaches outlined above. Roughly speaking, we exploit the properties of a certain spectral measure that encodes the averaged behaviour of the eigenvalues of the transfer matrix and shift operator, and use this to show that the
$k$-point correlations of the six-vertex model are harmonic in the limit. To achieve this, we combine the rotational invariance of the model obtained in \cite{magicformula} with a novel analysis of spectral properties. This constitutes the main innovation of the paper. The harmonicity, combined with an analysis of the behaviour near singular points, allows us to identify these limiting
$k$-point correlations, which can then be bootstrapped to obtain stronger modes of convergence.
To be more precise, our proof begins by establishing convergence only along certain suitable sub-sequences. It ends up being sufficient to deduce the full result, which in turn retroactively ensures that passing to sub-sequences was unnecessary.

The ability to extract such sub-sequential limits is in fact another main innovations of the paper. The framework in which we work permits the use of qualitative estimates for the model -- reminiscent of the RSW-type theory familiar to percolation specialists (we shall discuss RSW theory in depth) -- to obtain compactness and extract convergent sub-sequences of the spectral measure mentioned above. This strategy resonates with what made the proof of Cardy's formula for Bernoulli site percolation on the triangular lattice possible. In \cite{Smirnov_2001_CriticalPercolationPlane}, Smirnov goes around the problem of having observables that are only approximately holomorphic by showing that RSW-theory enables one to extract sub-sequential limits for these discrete observables, which end up being holomorphic. A leitmotif emerges: in order to move beyond the `ultra-integrable' cases of models possessing a free-fermion structure in which exactly discrete harmonic or holomorphic observables can be found, one likely needs to work in a setting where the absence of exact discrete harmonicity or holomorphicity can be compensated by \emph{a priori} estimates that ensure the existence of sub-sequential limits. The objects introduced in this paper illustrates the advantage and the potential of such a perspective.
\commentIoan{I find the intro great. The last two paragraphs do seem to suggest that our main innovation is that we use RSW. I do like the idea that regularity allows one to go beyond perfect discrete harmonicity or holomorphicity. However, I do think we should also mention that the original reason for harmonicity (in our setting it's the harmonicity of the two point function), is different from all previous arguments: it does not use exact identities (such as in the works of Smirnov) nor BA computations. We could say that we will get back to this later on. \\
Alternatively, we could mention already here rotation + scaling -> conformal.\\
None of this is urgent. HDC: I added one sentence for now and we can optimize later.}

% </input src="sections/PART_A/1_0_preface.tex" root="." version="0.0.1">
% <input src="sections/PART_A/1_1_definitions_main.tex" root="." version="0.0.1">
\section{Statement of our main result}

This section formally introduces the model and states the main results.

\subsection{Definition of the six-vertex model}

 In this paper, the six-vertex model is defined on graphs which locally look like the square lattice graph.
    We start with a definition on tori.
    For $M,L\in\Z_{\geq 4}$, let $\T_{M,L} =
    (V(\T_{M,L}), E(\T_{M,L}))$ denote the toroidal square grid on the vertex set
    $V(\T_{M,L}):=(\Z/M\Z)\times (\Z/L\Z)$, with edges placed between vertices at
    Euclidean distance $1$ from each other.

    An \emph{arrow configuration} $\omega$  is an assignment of an orientation to each edge.
    An arrow configuration is said to satisfy the \emph{ice rule} (or be a \emph{six-vertex configuration}) if every vertex has exactly two incoming and two outgoing edges. As a result, there are six possible arrangements of incoming and outgoing edges around each vertex,
    labelled according to Figure~\ref{fig:the_six_vertices}.

    \begin{figure}[b]
                \centering
            \includegraphics{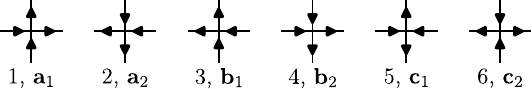}
            \caption{The six vertex configurations with labels and weights.}
            \label{fig:the_six_vertices}
    \end{figure}

    \begin{definition}[Six-vertex model on the torus]\label{def:6V_torus}
    For parameters $\a_1,\a_2,\b_1,\b_2,\c_1,\c_2 > 0$, the \emph{weight} of an arrow configuration $\omega$ on $\T_{M,L}$ is given by
    \begin{equation}
    \label{eq:sixvertexweights}
    W_{\sixv}(\omega) = \true{\text{$\omega$ satisfies the ice rule}}\cdot \a_1^{n_1} \a_2^{n_2} \b_1^{n_3} \b_2^{n_4} \c_1^{n_5} \c_2^{n_6}
       ,
    \end{equation}
    where $n_i$ denotes the number of vertices in $V(\T_{M,L})$ of type $i$ in $\omega$.
    The Gibbs measure $\P_{\T_{M,L}}$ on arrow configurations $\omega$ is given by
    \begin{equation}
        \P_{\T_{M,L}}[\{\omega\}]
    	:=\frac1{Z_{\T_{M,L}}}\cdot
         W_{\sixv}(\omega),
    \end{equation}
    where $Z_{\T_{M,L}}$ is the unique constant,  called the \emph{partition function}, rendering $\P_{\T_{M,L}}$ a probability measure.
\end{definition}

The weights were taken to be positive reals for a probabilistic interpretation.
In this work, we further specialise to the case where
\begin{equation}
    \a_1 = \a_2 =: \a;\qquad
    \b_1 = \b_2 =: \b;\qquad
    \c_1 = \c_2 =: \c,
\end{equation}
which renders $W_{\sixv}(\omega)$ invariant under flipping all orientations in $\omega$.
The parameters $\a$, $\b$, and $\c$ are always fixed in this article,
which is why they do not appear in notations.
It is standard to introduce the \emph{spectral parameter} defined by the formula
\commentKarol{Maybe we should add a comment somwhere about continuity of the phase transition?
I couldn't find it in the text... HDC: I think continuity is mentioned already, maybe not explicitly enough but we can fix this in a second time...}
\begin{align}\label{eq:Delta(a,b,c)}
\Delta = \Delta(\a,\b,\c) = \frac{\a^2 + \b^2 - \c^2}{2\a\b}.
\end{align}

The model may be extended to infinite volume in the following fashion.
Any configuration on $\T_{M,L}$ that obeys the ice rule has the same number of left-arrows on each vertical column -- we call this \emph{preservation of horizontal arrows}.
A configuration is called \emph{balanced} if, for every vertical column of horizontal arrows, the number of left-arrows equals the number of right-arrows.
    Write $\{\operatorname{balanced}\}$ for the collection of balanced arrow configurations satisfying the ice rule.
    From now on, $L$ is always even, so that balanced configurations exist.

For $\Delta<-1$,
the six-vertex model is known to be in a localised regime, implying trivial limiting behaviour~\cite{Duminil-CopinGagnebinHarel_2021_DiscontinuityPhaseTransition,RaySpinka_2022_FinitaryCodingsGradient,GlazmanPeled_2023_TransitionDisorderedAntiferroelectric}. Therefore, we focus  in the whole paper on the case $\Delta\ge-1$, which corresponds, when $\a=\b=1$, to $\c\le 2$.
 We shall  derive the following know result \emph{en passant} in Lemma~\ref{lemma:TorusLimits}).

\begin{theorem}[Infinite-volume six-vertex model]
    \label{thm:infinite_volume_6V}
    Fix $\a = \b=1$ and $\c \in [1,2]$.
The weak limit of the measures $\P_{\T_{M,L}}[\blank|\{\operatorname{balanced}\}]$ exists when the limits are taken in the following order: first $M$ tends to infinity, and then $L$ tends to infinity.
    We denote it $\P_{\Z^2}$ and call it the \emph{six-vertex measure in the plane with slope zero}.
    It is invariant under the automorphism group of $\Z^2$.
\end{theorem}

This result asserts that $\P_{\Z^2}$ is the unique probability measure on arrow configurations
of the square lattice graph
$\Z^2$ such that
\begin{equation}
    \lim_{L\to\infty}\lim_{M\to\infty}\P_{\T_{M,L}}[A|\{\operatorname{balanced}\}]=\P_{\Z^2}[A]
\end{equation}
for any event $A$ that is measurable in terms of the orientation of finitely many
edges of the square lattice graph $\Z^2$.
The measure $\P_{\Z^2}$ may be characterised in several other ways.
In addition to the above description, it is also the weak limit of $\P_{\T_{M,L}}$ as $M,L$ tend to infinity in arbitrary fashion, it is the unique ergodic Gibbs measure which is invariant under a global arrow flip, and it is the unique minimiser of a free energy functional related to the six-vertex model.
The last two equivalent statements do not play a role in this work.

Theorem~\ref{thm:infinite_volume_6V} and the above equivalences
follow from the general analysis of height functions in~\cite{Sheffield_2005_RandomSurfaces}
combined with \emph{delocalisation} of the height function.
Delocalisation was first derived at $\c=1$~\cite{ChandgotiaPeledSheffield_2021_DelocalizationUniformGraph}, $\c=2$~\cite{GlazmanPeled_2023_TransitionDisorderedAntiferroelectric},
and $\c\in[(2+2^{1/2})^{1/2},2]$~\cite{Lis_2021_DelocalizationSixVertexModel}, before the full range $\c\in[1,2]$ was
covered in~\cite{Duminil-CopinKarrilaManolescu_2024_DelocalizationHeightFunction} via a Bethe Ansatz argument,
and later in~\cite{GlazmanLammers_2025_DelocalisationContinuity2D} using a percolation approach.

\subsection{Height function of the six-vertex model}

A \emph{height function} is a function $h:F(\Z^2) \to \Z$ on the faces $F(\Z^2)$ of the square lattice
which differs by exactly $\pm1$ between any two adjacent faces.
We also require the face on the north-east of the origin to have an \emph{even} height.

We will consider height functions up to addition of an even constant. Formally, consider two height functions $h'$ and $h$ \emph{equivalent} if there exists some constant $a \in 2\Z$ such that $h'(u) = h(u) + a$ for all $u\in F(\Z^2)$. Gradients of height functions are simply the equivalence classes of height functions for this equivalence relation.
For all practical purposes, we  identify gradients of height functions with any representative of the equivalence class.
Finally, we shall write also $h$ for the piecewise constant function
\begin{equation}
\label{eq:simpleheightfunctionextensiontoplane}
h:\R^2\to\Z,\,(x,y)\mapsto h(\text{the face whose bottom-left corner is $(\lfloor x\rfloor,\lfloor y\rfloor)$}),
\end{equation}
and, for any $\delta>0$,
we define the scaled height function
\(
    h^{(\delta)}:\R^2\to\Z,\,u\mapsto h(u/\delta)\).

\begin{definition}[Height function of a six-vertex configuration]
    Full-plane six-vertex configurations are in bijection with gradient height
    functions.  More precisely, we associate any six-vertex configuration
    $\omega$ with the height functions for which the height of the face on the
    \emph{left} of each arrow is one unit higher than the height of that on its
    \emph{right}; see Figure~\ref{fig:six_vertex2}.
\end{definition}

    \begin{figure}
                \centering
         \includegraphics{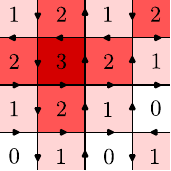}
            \caption{The six-vertex configuration and gradient height function are related such that the higher height is on the left of each arrow.}
            \label{fig:six_vertex2}
      \end{figure}

The multi-point correlation functions of gradient height-functions will be the core observables in our study of the six-vertex model's height function.

\begin{definition}[Six-vertex multi-point correlation functions]
    \label{def:Six-vertex multi-point correlation functions}
    The \emph{$k$-point correlation function} of the six-vertex model
    assigns to the vector $\u = ( u_1,u_1',\dots,u_{k},u_{k}' ) \in (\R^2)^{2k}$ whose $2k$ coordinates are built out of points
    $u_1,u_1',\dots,u_{k},u_{k}'\in\R^2$, the quantity
    \begin{align}
        \label{eq:multi-point_cor_function}
       \Phi_{k}(\u)
        :=\E_{\Z^2}\left[\prod_{i=1}^{k}\left(h(u_i')-h(u_i)\right)\right],
    \end{align}
    where $h$ denotes any height function associated with the six-vertex configuration sampled according to $\P_{\Z^2}$.
    Moreover, for $\delta >0$, the \emph{scaled $k$-point correlation function} is defined as
    \begin{equation}
        \textstyle
        \Phi_{k}^{(\delta)}(\u)
        :=\Phi_{k}(u_1/\delta,u_1'/\delta,\dots,u_{k}/\delta,u_{k}'/\delta).
    \end{equation}
\end{definition}

The $k$-point correlation function is well-defined,
as the integrand in the expectation is depends only on the gradient of the height function.
We immediately recognise a few basic properties of $\Phi_k$, which we often use
without further mention:
\begin{itemize}
    \item  $\Phi_k\equiv\Phi_k^{(\delta)}\equiv 0$ for $k$ odd since $h$ and $-h$ have the same distribution,
    \item $\Phi_k$ and $\Phi_k^{(\delta)}$ are antisymmetric under swapping $u_i$ and $u_i'$,
    \item $\Phi_k$ and $\Phi_k^{(\delta)}$ are invariant under permuting the pairs $\{u_i,u'_i\}$,
    \item $\Phi_k$ and $\Phi_k^{(\delta)}$ satisfy the following \emph{additivity property}
    for any $u_1$, $u'_1$, $u''_1$ and for any fixed $\v =(u_2,u_2',\ldots,u_k,u_k')$:
\begin{equation}
    \label{eq:Phi-additive}
    \Phi_k(u_1,u_1',\v )+\Phi_k(u_1',u_1'',\v )=\Phi_k(u_1,u_1'',\v ).
\end{equation}
\end{itemize}

Although one primary object of interest is the
$k$-point correlation, we will also treat the height function
$h$ as a random distribution. To that end, we introduce the distribution defined by integrating
$h$ against test functions.

\begin{definition}[Six-vertex test functions]
    \label{def:Six-vertex test functions}
        A \emph{generalised test function} is a finite, compactly supported, signed measure $\varphi$ on
    $\R^2$ with $\varphi(\R^2)=0$.
    For any generalised test function $\varphi$, define
    \begin{equation}
        \langle h^{(\delta)} , \varphi \rangle
        :=
        \int h^{(\delta)}(x) \diff\varphi(x)
        =
        \int h(x/\delta) \diff\varphi(x)
        .
    \end{equation}
    Observe that $\langle h^{(\delta)} , \varphi \rangle$ is a gradient measurable random variable
    because $\varphi(\R^2)=0$.
    We may therefore interpret it as a random variable.
\end{definition}

\subsection{Background on the Gaussian free field}\label{Sec:2.3}

With the model in place, we next describe the scaling limit that will ultimately arise. Write $|\cdot|$ for the Euclidean norm on $\R^2$.
Define the full-plane Green function $G_{\R^2}$ as
\begin{equation}
    \label{eq:GreenFunction}
   G_{\R^2}:\R^2\times\R^2\to(-\infty,\infty],\,(x,y)\mapsto -\tfrac1{2\pi}\log|y-x|.
\end{equation}
Our main result below says that as $\delta$ tends to zero,
the random height function $h^{(\delta)}$ obtained from $\P_{\Z^2}$
converges (up to scaling by a constant $\sigma=\sigma(\Delta)$) to the \emph{Gaussian free field} (GFF), which we denote $\Gamma$.
There are various mathematical ways to view the GFF depending on the desired regularity of the object;
here, we need the following three  (which resonate with Definitions~\ref{def:Six-vertex multi-point correlation functions} and~\ref{def:Six-vertex test functions}).
For more details on GFF, see \cite{BerestyckiPowell_2025_GaussianFreeField}.

\begin{definition}[GFF definitions]
    We consider three perspectives on the GFF.
\begin{enumerate}
\item \textbf{Multi-point correlation functions.}
                    For any $k \geq 1$, define
                    	\begin{equation}
    \calD_{k}:=\{(u_i,u_i')_{i=1,\dots,k}\in (\R^2\times\R^2)^{k}:\forall i\neq j,\,\{u_i,u_i'\}\cap\{u_j,u_j'\} = \emptyset\},
	\end{equation}
    and
 define the %analytic
 functions
                    \begin{equation}
                        \label{eq:GFF_multi-point_cor_function}
                        \Psi_{k}^{\GFF}:\calD_k\to\R,\,
                        \u \mapsto
                        \sum_{a,\,\pi}
                        (-1)^{\epsilon(a)}
                        \prod_{ij\in\pi}
                        G_{\R^2}(a_i,a_j)
                        ,
                    \end{equation}
                                where:
            \begin{itemize}
                \item The sum over $a$ runs over maps $\{1,\dots,k\}\to\{u_1,u'_1,\dots,u_k,u_k'\}$ with
                $a_i\in\{u_i,u_i'\}$,
                \item $\epsilon(a)\in\Z/2\Z$ is the parity of the number of indices $i$ such that $u_i$ is chosen,
                \item The sum over $\pi$ runs over all \emph{pairings} of $\{1,\dots,k\}$,
                that is, all partitions of $\{1,\dots,k\}$
                into \emph{pairs} (i.e.~sets containing two points).
            \end{itemize}
                        These are precisely the correlation functions corresponding to a Gaussian process
            with covariance $G_{\R^2}$.
            Notice that $\Psi_k^{\GFF}\equiv 0$ when $k$ is odd since $\pi$ is then empty.
\item
\textbf{Finite-dimensional marginals.}
            We say that a generalised test function $\varphi$ has \emph{finite Dirichlet energy} if $\int G_{\R^2}(u,v) \diff\varphi(u) \diff\varphi(v)<\infty$.
            For any finite family $\bm{\varphi}=(\varphi_1,\ldots,\varphi_n)$ of finite Dirichlet energy generalised test functions,
            introduce the associated $n\times n$ covariance matrix $\Sigma(\bm\varphi)$ defined via
            \begin{equation}
                \Sigma(\bm\varphi)_{ij}:=\int G_{\R^2}(u,v) \diff\varphi_i(u) \diff\varphi_j(v).
            \end{equation}
            We then think of $(\langle \Gamma,\varphi_i\rangle)_i$ as a random variable
            having the law $\calN(0,\Sigma(\bm\varphi))$.
            \commentIoan{I find this last sentence a bit confusing. I would have given the definition as follows:
            \\
                        Then $\Gamma = (\langle \Gamma,\varphi\rangle)_\varphi$ may be defined as a centred Gaussian process indexed by the
            generalised test functions  of finite Dirichlet energy, with covariance matrix given by
            $$ {\rm Cov}  ( \langle \Gamma,\varphi_1\rangle, \langle \Gamma,\varphi_2\rangle):=\int G_{\R^2}(u,v) \diff\varphi_1(u) \diff\varphi_2(v).$$
	But maybe I am missing something}
\item
\textbf{Random element of a negative regularity Hölder space.}
Fix $\alpha\in(-1,0)$ and a bounded open set $U\subset\R^2$.
Recall that the \emph{Hölder space $\calC^\alpha(U)$ of regularity $\alpha$ on $U$} is the completion of $C^\infty_c(\R^2)$ with respect to the semi-norm
 \begin{equation}
        \|\cdot\|_{\calC^{\alpha}(U)} : C^\infty_c(\R^2) \to \R
        ,\,
        f\mapsto
        \sup_{\substack{(\epsilon,\varphi)\in(0,1]\times\calT_1((-1,1)^2),\\\operatorname{support}(\varphi)\subset U/\epsilon}}
        \epsilon^{-\alpha}
        \int f(x/\epsilon)\diff \varphi(x),
    \end{equation}
where $\calT(U)$ denotes the set of generalised test functions
    whose support is included in $U$ and $\calT_1(U)\subset\calT(U)$ the set of such test functions whose density is $1$-Lipschitz.
We view $\Gamma$ as a random element in $\calC^\alpha(U)$
such that, for any finite $\bm\varphi=(\varphi_1,\dots,\varphi_k)$ with $\varphi_i\in \calT(U)$ for every $i$, the law of $(\langle \Gamma,\varphi_i\rangle)_i$ is $\calN(0,\Sigma(\bm\varphi))$.
 \end{enumerate}
\end{definition}

\subsection{Statement of the main result}

Three modes of convergence are considered.

\begin{definition}[GFF convergence]\label{def:GFF_convergence}
    Consider some fixed value $\sigma\in\R_{\geq 0}$
    as well as a random gradient height function $h:\R^2\to\R$ in some probability measure $\P$.
    We say that the \emph{scaling limit} of $h$ is $\sigma\Gamma$ or $\sigma\cdot\GFF$ if \emph{all} of the statements below hold true.
    \begin{enumerate}
            \item {\bf Convergence of multi-point correlation functions.}
            For any $k\geq 1$, $\Phi_{k}^{(\delta)}$ converges uniformly on compact subsets of $\calD_k$ to  $\sigma^k\Psi_{k}^{\GFF}$ as $\delta$ tends to zero.
    \item {\bf Convergence of finite-dimensional marginals.}
            For any family $\bm\varphi=(\varphi_1,\ldots,\varphi_n)$ of finite Dirichlet energy generalised test functions,
            the law of $(\langle h^{(\delta)},\varphi_i\rangle)_i$ converges weakly to $\calN(0,\sigma^2\Sigma(\bm\varphi))$ as $\delta$ tends to zero.
    \item {\bf Convergence in law in a negative regularity Hölder space.}
        For any $\alpha\in(-1,0)$ and any open bounded $U\subset\R^2$, the law of
        $h^{(\delta)}$ converges to that of $\sigma \Gamma$ in $\calC^{\alpha}(U)$ as $\delta$ tends to zero.
        Finally, we also require convergence in the classical Besov spaces $\calB_{p,q}^\alpha(U)$
        and Sobolev spaces $W^{\alpha,p}(U)$ for any $p\in[1,\infty)$ and $q\in[1,\infty]$ (see Section~\ref{sec:GFF_convergence_criteria} for details).
        \end{enumerate}

\end{definition}

We are now in a position to state our main result.

\begin{theorem}[Scaling limit of the six vertex model with isotropic weights]\label{thm:GFF_convergence}
	The scaling limit of the six-vertex model's height function on $\Z^2$ with $\a=\b=1$ and $\sqrt 3\le \c \le 2$ is $\sigma\cdot\GFF$, where
	\begin{align}\label{eq:sigma_expression}
		\sigma^2
        =  \frac{2}{\arccos \Delta}
		 = \frac{1}{\arcsin \tfrac{\c}{2}}.
	\end{align}
\end{theorem}

Let us comment on our requirement on the parameters.  In terms of $\Delta$, the previous theorem covers the regime $-1\leq \Delta \leq -\frac12$. The restriction on the isotropic case (i.e.,~with weights $\a=\b=1$) is lifted in Theorem~\ref{thm:GFF_convergence_isoradial} below. Recall that the isotropic six-vertex model with $\c>2$ ($\Delta<-1$) is known to be in a localised regime.
Theorem~\ref{thm:GFF_convergence} is expected to hold true for all $\c\in(0,2]$ (i.e., $\Delta\in[-1,1)$) and was previously obtained at the free fermion point $\c=\sqrt 2$ ($\Delta=0$) in~\cite{Kenyon_2000_ConformalInvarianceDomino} and for $\c$ close to $\sqrt 2$ in~\cite{GiulianiMastropietroToninelli_2017_HeightFluctuationsInteracting}.

Two aspects of our proof require lower bounds on $\c$.
First, the Fortuin--Kasteleyn--Ginibre (FKG) property for several representations of the six-vertex model
requires that $\c\geq 1$.
This property is used to obtain regularity of our objects at various stages of the proof.
Second, our proof relies on the asymptotic
rotational invariance of the multi-point correlation functions $\Phi_k$, which was obtained in \cite{magicformula}
using the corresponding random-cluster model \cite{Duminil-CopinKozlowskiKrachun_2020_RotationalInvarianceCritical}.
For that result to apply, the random-cluster model must exhibit the FKG property,
which requires $\c\geq \sqrt3$.

An extension of this asymptotic rotational invariance result to $\c\geq 1$ is conceivable, for instance by using the above-mentioned representations of the six-vertex model.
This would immediately allow the extension of Theorem~\ref{thm:GFF_convergence} to the interval $1\leq \c\leq 2$ (corresponding to $-1\leq\Delta\leq 1/2$).
However, circumventing the absence of the FKG inequality for the representations of the six-vertex model appears to be a formidable challenge, placing the case
$0<\c<1$ beyond the reach of current methods.

\section{First applications of our result}\label{sec:applications}

The six-vertex model in the regime $-1 \le \Delta < 1$ lies at the crossroads of a large family of two-dimensional lattice models.
It is closely related to the dimer model, the Ising and Potts models, the critical random-cluster model, loop $O(n)$ models,
Ashkin--Teller models, random permutations, and quantum spin chains \cite{FortuinKasteleyn_1972_RandomclusterModelIntroduction,Nienhuis_1982_ExactCriticalPoint,GlazmanPeled_2023_TransitionDisorderedAntiferroelectric,Lis_2022_SpinsPercolationHeight,McCoyWu_1968_HydrogenbondedCrystalsAnisotropic}.
While it is not yet clear how much information can ultimately be extracted from the GFF convergence established here, several significant
applications are already available.

\subsection{Critical exponents of other two-dimensional lattice models}

Thanks to the Baxter--Kelland--Wu (BKW) correspondence \cite{BaxterKellandWu_1976_EquivalencePottsModel}, the six-vertex model is intimately connected, at its continuous
phase transition, to the critical {\em random-cluster model} -- also called the {\em Fortuin-Kasteleyn (FK) percolation} -- introduced in \cite{FortuinKasteleyn_1972_RandomclusterModelIntroduction}.
Our main result enables the following consequences for random-cluster models.

\begin{itemize}
\item \textbf{One-arm exponent $\alpha_1$.}
For the random-cluster model with cluster-weight $q\in[1,4]$, \cite{alpha1}  obtains the existence and value of the one-arm critical exponent $\alpha_1$, describing the decay of the probability that a vertex connects to distance~$n$.
In turn, this derives the classical exponents $\eta$, $\zeta$, and $\delta$ governing the behaviour of the two-point function, the cluster-size tail at criticality, and the ghost-field connectivity; see \cite{alpha1} for details.

\item \textbf{Two-arm exponent $\alpha_2$.}
For the random-cluster model with $q\in[1,4]$, \cite{alpha2} obtains the existence and value of the two-arm exponent $\alpha_2$, describing the probability that a vertex lies on a primal/dual interface extending to distance~$n$.
As a consequence, one deduces the fractal dimension of any sub-sequential scaling limit of critical interfaces, in agreement with the predicted $\mathrm{CLE}(\kappa)$; see \cite{alpha2} for details.

\item \textbf{Energy exponent $\iota$.}
For the random-cluster model with $q\in[4-\epsilon,4]$ (for some small value of $\epsilon$), the articles \cite{qequal4,qcloseto4} determine the so-called \emph{influence exponent}~$\iota$, which controls the covariance of observables at criticality.
Combined with the scaling relations established in \cite{Duminil-CopinManolescu_2022_PlanarRandomclusterModel}, this yields the thermodynamic critical exponents $\alpha$, $\beta$, $\gamma$, and $\nu$, governing respectively the behaviour of the free energy, the spontaneous magnetisation, the susceptibility, and the correlation length.
\end{itemize}

Finally, since the random-cluster model is coupled to the Potts model, all the above critical exponents transfer directly to the two, three and four state Potts models.
These results were already known for two-state Potts case, better known as the Ising model, but are new for the three- and four-state Potts models.

We expect that more results can be obtained in this direction, both for the random-cluster model and other models.

\begin{remark}
We stress that the critical exponents obtained here do not rely on evaluating the top eigenvalues of the transfer matrix. Instead, they harvest Theorem~\ref{thm:GFF_convergence} which is based on an analysis of the average behaviour of certain eigenvalues, well-separated from the spectral edge. In this sense, our approach circumvents the major difficulty of providing a rigorous justification for computations of the leading eigenvalues.
\end{remark}

\subsection{Scaling limit of specific random-cluster observables}

Harvesting the BKW correspondence \cite{BaxterKellandWu_1976_EquivalencePottsModel} allows one to derive the scaling limit of certain random-cluster observables at criticality. In particular, the characteristic function of six-vertex test functions $\langle h^{(\delta)},\varphi\rangle$ can be expressed as the expectation -- under the random-cluster model -- of a product of suitably twisted weights associated with the loops of a percolation configuration. More precisely, if $\mu = \tfrac1{2\pi}\arccos(\sqrt q/2)$ and
$
\cos_\mu(\cdot) := \cos(\cdot + 2\pi\mu)/\cos(2\pi\mu),
$
then

\begin{equation}\label{eq:magicformuladiscrete}
\mathbb E^{\rm 6V}_{\mathbb Z^2}\Big[e^{\i\langle h^{(\delta)},\varphi\rangle}\Big]
= \phi_{\delta\mathbb Z^2,q}\!\Big[\prod_{\ell\in \mathcal L}
\cos_\mu\!\big(\varphi(\mathrm{int}(\ell)) \big)\Big],
%\cos_\mu\!\Big(\int_{\mathrm{int}(\ell)} \varphi(x)\, dx \Big)\Big],
\end{equation}
where $\phi_{\delta\mathbb Z^2,q}$ denotes the law of the ensemble $\mathcal L$ of loops on $\delta\mathbb Z^2$ arising from the loop representation of the critical random-cluster measure with cluster-weight $q$, and where $\mathrm{int}(\ell)$ is the interior of the loop $\ell$, i.e., the bounded connected component of $\mathbb R^2 \setminus \ell$ when $\ell$ is viewed as a continuous simple path. Variants of this identity have appeared repeatedly in the literature; see, for instance, \cite{Dubedat_2011_TopicsAbelianSpin}. We refer to \cite{magicformula} for more details on this formula.

The convergence to the Gaussian free field yields the following corollary, which underpins the derivation of the critical exponents presented in the previous section.

\begin{corollary}
Fix $q\in[1,4]$. For every finite Dirichlet energy  generalized test function $\varphi$,
\begin{equation}
\lim_{\delta\rightarrow 0}\phi_{\delta\mathbb Z^2,q}\Big[\prod_{\ell\in \calL}\cos_\mu\big(\varphi({\rm int}(\ell))\big)\Big]
% \Big[\prod_{\ell\in \calL}\cos_\mu\Big(\int_{{\rm int}(\ell)}\varphi(x)dx\Big)\Big]
=\exp\Big(-\tfrac12\sigma^2\iint G_{\mathbb R^2}(x,y)\varphi(x)\varphi(y)dxdy\Big),
\end{equation}
where $\sigma^2=2/\arccos(-\tfrac{\sqrt q}2)$.
\end{corollary}

\subsection{Applications to the anisotropic six-vertex model}

The universality of the the random-cluster model derived in \cite{Duminil-CopinKozlowskiKrachun_2020_RotationalInvarianceCritical} enables us to transfer our main result to the anisotropic six-vertex model (weights $\a\neq\b$).
It is customary to parametrize $(\a, \b, \c)$ in the following way: let $\zeta  = \arccos(- \Delta)$ and $\theta \in (0,\pi)$ be the unique angle such that, if $1>\Delta>-1$
\begin{align}
\a \sin \tfrac{\zeta}{2} &=  \sin((1 - \tfrac{\theta}{\pi}) \zeta), \qquad
\b \sin \tfrac{\zeta}{2} =  \sin \tfrac{\theta \zeta}{\pi}, \qquad
\c = 2\cos \tfrac{\zeta}{2}.
\label{eq:parametrization}
\end{align}
and if $\Delta=-1$,
\begin{align}
\a  &=  2\tfrac{\pi-\theta}{\pi}, \qquad
\b  =  2\tfrac{\theta}{\pi}, \qquad
\c = 2.
\label{eq:parametrization-1}
\end{align}
% Here the parameter $\zeta$ is chosen such that $\Delta=-\cos(\zeta)$.
The parameter $\theta$ encodes the natural embedding of the square lattice;
$\theta=\pi/2$ encodes the isotropic case.
More precisely, let
\begin{equation}
\mathbf L_\theta:\R^2\rightarrow \R^2,\, (x,y)\mapsto (x+\cos(\theta)y,\sin(\theta)y).
\end{equation}
The combination of the universality result of \cite{Duminil-CopinKozlowskiKrachun_2020_RotationalInvarianceCritical}, the consequences of the Baxter--Kelland--Wu coupling obtained in \cite{magicformula}, and Theorem~\ref{thm:GFF_convergence} implies the following result.

\begin{theorem}[Scaling limit of the six vertex model with general weights]\label{thm:GFF_convergence_isoradial}
	The height function of the six-vertex model on $\mathbf  L_\theta\Z^2$ with $\a, \b, \c > 0$ such that $\Delta \in [-1, -1/2]$  is converging in the sense of Definition~\ref{def:GFF_convergence},
    Items (i) and (ii) to $\sigma\cdot\GFF$, where
	\begin{align}\label{eq:sigma_expression2}
		\sigma^2=  \frac{2}{\arccos \Delta}
		 = \frac{1}{\arcsin \tfrac{\c}{2}}
		= \frac{2}{( \pi - \zeta)}.
	\end{align}
 \end{theorem}

Note that we do not claim convergence in the sense of
Definition~\ref{def:GFF_convergence}(iii), i.e.,~in H\"older spaces. This omission is purely technical. In order to keep the
paper to a reasonable length, we chose to rely as much as possible on the so-called spin representation of the
six-vertex height function (discussed at length below). Establishing the necessary RSW theory of this spin
representation is relatively direct in the isotropic case but is not available the anisotropic setting.
While \cite{magicformula} works directly in the anisotropic regime, it does
not provide the regularity estimate required here (see Remark~\ref{rmk:anisotropic} for further
discussion).
\commentIoan{I don't think we currently have RSW in the anisotropic case, unless we're willing to extract it from FK. HDC I agree but I think what you said is coherent with what is written no?}

% </input src="sections/PART_A/1_1_definitions_main.tex" root="." version="0.0.1">
% <input src="sections/PART_A/1_2_ingredients.tex" root="." version="0.0.1">
\section{Overview of the proof and ingredients}
\label{sec:overview_steps_ingredients}

\newcommand\INGREDROTATION{Ingredient~1}
\newcommand\INGREDSCALE{Ingredient~2}
\newcommand\INGREDREGULARITY{Ingredient~3}
\newcommand\INGREDSPECTRAL{Ingredient~4}

We give below a roadmap to the proof of our main result, Theorem~\ref{thm:GFF_convergence}.
Theorem~\ref{thm:GFF_convergence_isoradial} will be derived from Theorem~\ref{thm:GFF_convergence} in Section~\ref{sec:anisotropic};
outside of that section, we only consider the isotropic case $\a=\b=1$.

\subsection{Overview of the proof structure}

The proof of the convergence result in the isotropic case (Theorem~\ref{thm:GFF_convergence}) consists of four steps performed in Part~\ref{part:proofs},
informally described as follows.  We say that $h$ converges to $\sigma\cdot\GFF$
        \emph{along some sequence} $(\delta_n)_n$ tending to zero
        if each of the convergences in Definition~\ref{def:GFF_convergence} holds true along that sequence of scales.
\begin{enumerate}[label=\arabic*.]
\item \textbf{Theorem~\ref{thm:dicht_2p_final_form}}
    asserts that the two-point function $\Phi_2^{(\delta)}$ satisfies a dichotomy:
either $\Phi_2^{(\delta)}$ converges to $\sigma^2\cdot\Psi_2^{\GFF}$ for some $\sigma\geq 0$, or
  such a convergence holds true along {\em two sub-sequences}  with {\em two distinct} values of $\sigma$.
\item \textbf{Theorem~\ref{thm:k2impliesallk}}
   enables one to pass from two-point to multi-point correlation functions in the following sense: if along some sub-sequence $(\delta_n)_n$,
    $\Phi_2^{(\delta_n)}$ tends to $\sigma^2\cdot\Psi_2^{\GFF}$,
    then
    $\Phi_k^{(\delta_n)}$ tends to $\sigma^k\cdot\Psi_k^{\GFF}$ for {\em every} $k$.
    This extends the dichotomy of Theorem~\ref{thm:dicht_2p_final_form} to all multi-point correlation functions.

\item \textbf{Theorem~\ref{thm:criterion}} states that if along some sub-sequence $(\delta_n)_n$, all correlation functions converge to those of $\sigma\cdot\GFF$,
    then the limit of $h^{(\delta_n)}$ is $\sigma\cdot\GFF$ (in the sense of Definition~\ref{def:GFF_convergence}).
    This extends the dichotomy of Theorem~\ref{thm:dicht_2p_final_form} to all modes of GFF convergence.

\item {\bf Theorems~\ref{thm:glimpse_scale_invariance} and \ref{thm:computation_deriv_f}} jointly imply that if for some $(\delta_n)_n$,
    $h^{(\delta_n)}$ converges to $\sigma\cdot\GFF$,
    then $\sigma^2=2/\arccos\Delta$.
    This makes the dichotomy collapse to a single case,
    and completes the proof of the main result.
\end{enumerate}

\begin{table}[b]
\centering
\begin{tabular}{l|cccc}
\hline
 & \begin{tabular}{@{}c@{}}\INGREDROTATION:\\\emph{Rotation}\\\emph{invariance}\end{tabular} & \begin{tabular}{@{}c@{}}\INGREDSCALE:\\\emph{Scale}\\\emph{invariance}\end{tabular} & \begin{tabular}{@{}c@{}}\INGREDREGULARITY:\\\emph{Regularity of}\\\emph{correlations}\end{tabular} & \begin{tabular}{@{}c@{}}\INGREDSPECTRAL:\\\emph{Spectral}\\\emph{representation}\end{tabular} \\
\hline
Step 1: \emph{Two-point} & X & & X & X \\
Step 2: \emph{Multi-point} & X & & X & X \\
Step 3: \emph{Test functions} & & & X & \\
Step 4: \emph{Finding $\sigma$} & & X & X & \\
\hline
\end{tabular}
\caption{Use of the proof ingredients in the global proof steps}
\label{tab:proof_structure}
\end{table}

% \medbreak
To compactly state the proofs of these steps in Part~\ref{part:proofs},
we rely on four proof ingredients which are used as ``black boxes''
in Part~\ref{part:proofs} (see Table~\ref{tab:proof_structure}). These ``black boxes'' are developed in Parts~\ref{part:spectral}--\ref{part:glimpse}.
In short, they are described as follows.
\begin{enumerate}[label=\arabic*.]
    \item \textbf{Rotational invariance.}
    The correlation functions are asymptotically rotationally invariant.
    More precisely, any \emph{sub-sequential} scaling limit of the correlation functions,
    is rotationally invariant.
    \item \textbf{Glimpse of scale invariance.}
    A suitably chosen ``observable'' converges in the scaling limit,
    and we can calculate its limit explicitly.
    \item \textbf{Regularity estimates and qualitative behaviour.}
    We establish suitable bounds on the correlation functions
    which hold true at all scales.
    \item \textbf{Spectral representation of correlation functions.}
    Correlation functions may be expressed in terms of the spectra
    of two commuting transfer matrices.
\end{enumerate}

The first two ingredients (rotation invariance and a weak form of scale invariance) echo the discussion at the beginning of the paper and are consistent with physics predictions, especially those arising from the renormalization-group formalism. We emphasize, however, that these properties are \emph{not} obtained through a rigorous renormalization-group analysis. The third ingredient (regularity estimates) corresponds to qualitative bounds expected for generic continuous phase transitions.

The fourth ingredient is more mysterious. We interpret it as follows. Since the work of Polyakov
\cite{Polyakov_1968_MicroscopicDescriptionCritical,Polyakov_1970_ConformalSymmetryCritical,Polyakov_1970_PropertiesLongShort},
it has been predicted that conformal invariance should follow from rotation, scale, and translation invariance (the latter being trivial for our model), provided the theory also satisfies a suitable \emph{locality} principle.
While our spectral representation does not directly yield locality of the observables, it allows us to relate the effect of applying the Laplacian to correlation functions at different spatial positions. Even if \emph{a priori} of a different kind than locality, this remarkable feature provides the additional structure needed to carry out our analysis.

The subsections below formally describe these ingredients so that they can be used as black boxes in Part~\ref{part:proofs},
and proved in later parts.

%{\em Note that we only consider the isotropic case $\a=\b=1$, so that we omit to recall this fact in the next results.}

\subsection{\INGREDROTATION: Rotation invariance}

The following result serves as a key external input for the paper. It states that $k$-point correlations functions are invariant under rotations in the limit as $\delta$ tends to zero.

\begin{theorem}[Rotation invariance of $k$-point correlations \cite{magicformula}]
\label{thm:rotation_invariance}
   Fix  $\sqrt{3} \le\c \le 2$.
   Then, for any $k \in 2\mathbb{Z}_{\geq 1}$ and any compact set $K \subset \mathcal{D}_{k}$,
   \begin{equation}
       \lim_{\delta \to 0}
       \sup_{\u \in K}
       \sup_{I}
       \big|
         \Phi^{(\delta)}_k(\u)
         -
         \Phi^{(\delta)}_k(I\u)
       \big|
       = 0,
   \end{equation}
   where the second supremum is taken over all isometries
   $I : \mathbb{R}^2 \to \mathbb{R}^2$.
\end{theorem}

This is the origin of the restriction $\c \ge \sqrt{3}$.
Indeed, \cite{Duminil-CopinKozlowskiKrachun_2020_RotationalInvarianceCritical} proves asymptotic rotational invariance for the critical random-cluster model with cluster weight $q \in [1,4]$. Through the Baxter--Kelland--Wu (BKW) correspondence \cite{BaxterKellandWu_1976_EquivalencePottsModel}, this range of $q$ matches the regime  $\c \in [\sqrt{3}, 2]$ of the six-vertex model.
In \cite{magicformula}, this correspondence is used to transfer the asymptotic rotational invariance from the critical random-cluster model to the six-vertex height function.

We expect Theorem~\ref{thm:rotation_invariance} to hold for all $\c \in (0,2]$.
For $\c \in [1,2]$, we believe that there may exist a proof following~\cite{Duminil-CopinKozlowskiKrachun_2020_RotationalInvarianceCritical}, but working directly with the six-vertex model rather than its FK-percolation representation.
The arguments of the present paper would extend verbatim to $\c \in [1,2]$ if Theorem~\ref{thm:rotation_invariance} were available in that parameter range.

For $\c < 1$, although convergence to the GFF is still expected, the six-vertex model lacks positive association, and several steps of the proof
(related to Ingredient~3) fail (in their present form) without this positive association.

\subsection{\INGREDSCALE: A glimpse of scale invariance}

It may be natural to expect that, in addition to rotational invariance, scale invariance is an important ingredient in identifying the scaling limit.
While having such a property would simplify considerably our argument, it seems currently out of reach of direct techniques.
Still, a glimpse of scale invariance is provided by the fact
that the \emph{free energy} or \emph{surface tension} of the six-vertex model with a slope is twice differentiable at zero slope.
Indeed, its second derivative will be identified as the limit of a certain quantity as the scale $\delta$ tends to zero.
The convergence of said quantity will act as our indicator of scale invariance.

Below, we make the previous claim explicit. Let us start by recalling the definition of the free energy.

\begin{definition}[Free energy]
   Fix $\c>0$.
    Define the \emph{free energy at slope $s\in[-1,1]$} via
    \begin{equation}
        f(s)=\lim_{L\to\infty} \lim_{M\to\infty} \frac1{ML}\log\left( Z_{\T_{M,L}}
        \P_{\T_{M,L}}\left[\left\{\tfrac{
            I
        }{ML}=\tfrac{\lfloor Ls/2\rfloor}{L/2} \right\}\right]\right),
    \end{equation}
    where $I$ is the number of left arrows minus the number of right arrows on any given vertical column of horizontal edges.
\end{definition}

While the height function may not be defined for unbalanced six-vertex configurations on the torus,
$I/ML$ should be interpreted as its average slope in the vertical direction.
\commentIoan{Maybe add a few words about infinite-volume measures with slope as limits of the above. HDC: not sure it is needed here}
Note that $f:[-1,1]\rightarrow \mathbb R$ is an even function thanks to the symmetry by flipping all arrows.

Twice differentiability of the free energy was proved in \cite{Duminil-CopinKozlowskiKrachun_2022_SixVertexModelsFree} for the six-vertex model with $\c\in(0,2]$ using Bethe Ansatz techniques.
 It was used in \cite{Duminil-CopinKarrilaManolescu_2024_DelocalizationHeightFunction} to prove the delocalisation of the zero-slope six-vertex model for $\c\in[1,2]$.
By further harnessing~\cite{Duminil-CopinKozlowskiKrachun_2022_SixVertexModelsFree}, we explicitly compute the second derivative of $f$ at $0$ -- see Section~\ref{sec:computation_deriv_f} for $\c\in[1,2]$.
%That article also provides a characterisation of this second derivative, but does not compute it explicitly.

We use this result as a starting point for a two-step analysis, summarised in the following two results.
The first step shows that the second derivative is indeed related to scale invariance of our actual six-vertex model: it determines the amplitude of the GFF limit (assuming such a limit exists).
In the second step, we turn the characterisation of \cite{Duminil-CopinKozlowskiKrachun_2022_SixVertexModelsFree} into an explicit computation.

\begin{theorem}[GFF-LDP correspondence]
    \label{thm:glimpse_scale_invariance}
	Fix $\c \in [1,2]$. Assume that the six-vertex model has a sub-sequential scaling limit of the form $\sigma \cdot \GFF$.
	Then,
	\begin{equation}
	\sigma^2=-\frac{1}{f''(0)}.
	\end{equation}
\end{theorem}

\begin{theorem}[Computation of $f''(0)$]\label{thm:computation_deriv_f}
Fix $\c \in [1,2]$. Then, $f$ is symmetric and twice differentiable at $s=0$, with
\begin{equation}\label{eq:surfacetensiondefinition}
  	f''(0)=  -\tfrac12\arccos \Delta
	    	= - \arcsin (\c/2)
            ,
\end{equation}
where the relation between $\c$ and $\Delta$ is given in Equation~\eqref{eq:Delta(a,b,c)}.
\end{theorem}

Jointly, the two theorems prove that any sub-sequential scaling limit that is a multiple of the GFF
must have the explicitly computed variance.

Let us briefly comment on Theorem~\ref{thm:glimpse_scale_invariance}.
The formula $\sigma^2=-1/f''(0)$ would boil down to a ``back of the envelope''
calculation if the topology of the
sub-sequential convergence towards $\sigma\GFF$ were strong enough
to include convergence of probabilities of large deviation events.
Unfortunately, the topology is not compatible
with events whose probability decays exponentially fast to zero, and we must therefore obtain the formula
by different means.
We shall derive the formula by essentially expressing the probability of
a large deviation event as the product of many probabilities of GFF events that \emph{are}
compatible with the topology of the sub-sequential convergence.

\subsection{\INGREDREGULARITY: Regularity estimates and qualitative behaviour}

The recently developed Russo--Seymour--Welsh (RSW) theory for the six-vertex model
implies a \emph{circuit estimate} for a suitable percolation representation of the model.
%Informally, this theory asserts that if we take an annulus of aspect radius two, then with a uniformly positive probability, we observe a ``maximal circuit'' winding around the annulus when imposing ``minimal boundary conditions'' outside the annulus.
This RSW theory was first developed in~\cite{Duminil-CopinKarrilaManolescu_2024_DelocalizationHeightFunction} using the Bethe Ansatz.
The representation employed here first appeared in~\cite{Lis_2021_DelocalizationSixVertexModel,Lis_2022_SpinsPercolationHeight}.
The corresponding circuit estimate follows from~\cite{Duminil-CopinKarrilaManolescu_2024_DelocalizationHeightFunction} and was later obtained independently in~\cite{GlazmanLammers_2025_DelocalisationContinuity2D} through a different approach not relying on the Bethe Ansatz.
We will later return to these aspects in more detail.

For the purpose of this introduction, we adopt the following principle: this paragraph records the consequences of the representation and the circuit estimate without describing the representation of the circuit estimate explicitly, so that they may be treated as black boxes in Part~\ref{part:proofs}. All statements below will be proved in Part~\ref{part:qualitative}.

First,
define the \emph{scale separation functions} for any $\{a,a'\},\{b,b'\}\subset\R^2$:
\begin{align}
    \label{eq:scale_separation_functions_cts}
      &  S_{\R^2}(\{a,a'\},\{b,b'\}):=
    \log\frac{\operatorname{dist}(\{a,a'\},\{b,b'\})}{\min\{|a'-a|,|b'-b|\}};
    \\
    \label{eq:scale_separation_functions_discrete}
    &S_{\R^2}'(\{a,a'\},\{b,b'\}):=
    \log\frac{1\vee\operatorname{dist}(\{a,a'\},\{b,b'\})}{\min\{|a'-a|,|b'-b|\}}.
\end{align}
Recall that $|\cdot|$ denotes Euclidean distance in these formulas;
$\operatorname{dist}$ denotes the Euclidean distance between the two sets (that is, the minimum distance between any point in the first set and any point in the second set).
The first function is truly invariant under scaling;
the second function is more adapted to the discrete setting as it allows the pairs of points to overlap.

The following estimate bears a resemblance to Equation~\eqref{eq:GFF_multi-point_cor_function} and forms the basis of our qualitative analysis of correlation functions.
\begin{theorem}[Regularity estimate]\label{thm:Regularity}
    For any $k\in 2\Z_{\geq 1}$,
    there exist constants $\alpha_k>0$ and $C_k<\infty$     such that for any  $\c\in[1,2]$ and $\u \in (\Z^2)^{2k}$,
    \begin{align}\label{eq:Regularity}
       |\Phi_{k}(\u )|\leq C_k \sum_{\pi} \prod_{ij\in\pi}
       \begin{cases}
            e^{-\alpha_k S_{\R^2}(\{u_i,u_i'\},\{u_j,u_j'\})} & \text{if $S_{\R^2}(\{u_i,u_i'\},\{u_j,u_j'\})\geq 20k^2$,} \\
            1\vee -S_{\R^2}'(\{u_i,u_i'\},\{u_j,u_j'\}) &\text{if $S_{\R^2}(\{u_i,u_i'\},\{u_j,u_j'\})< 20k^2$,}
       \end{cases}
    \end{align}
    where $\pi$ runs over pairings of $\{1,\ldots,k\}$.
   % This bound holds uniformly in the interval $\c\in[1,2]$.
    % We extend the definition of $\operatorname{RegularityBound}$ to any $u\in\calD_{k}$ satisfying Equation~\eqref{eq:regularityCondition}.
\end{theorem}

The previous theorem has the following important corollary, which follows immediately from the additivity property in Equation~\eqref{eq:Phi-additive}.

\begin{corollary}[Precompactness of correlation functions]\label{cor:precompactness_phi}
    Fix $\c\in[1,2]$ and $k\in2\Z_{\geq 1}$.
    Let $(\delta_n)$ denote any sequence tending to zero.
    Then the following two are equivalent:
    \begin{enumerate}
        \item $\Phi_k^{(\delta_n)}$ converges to $\Psi_k$ pointwise on a countable dense subset of $\calD_k$;
        \item $\Phi_k^{(\delta_n)}$ converges to $\Psi_k$ uniformly on compact subsets of $\calD_k$.
    \end{enumerate}
    In this case, we say simply that $\Phi_k^{(\delta_n)}$ converges to $\Psi_k$ on $\calD_k$.

    Moreover, $\sup_\delta|\Phi_k^{(\delta)}|$ is finite on $\calD_k$,
    and therefore the family $(\Phi_k^{(\delta)})_{\delta}$ is precompact in this topology.
    Finally, any sub-sequential limit is continuous on $\calD_k$.
    \noticePiet{Changed to continuous for now, is easier to prove.
    The proof is added below.}
    % Finally, any sub-sequential limit is $\alpha_k$-H\"older continuous on compact subsets of $\calD_k$.
 \end{corollary}

\begin{proof}
    The proof follows by carefully manipulating the regularity estimate (Theorem~\ref{thm:Regularity}),
    the definition of the scale separation functions (Equations~\eqref{eq:scale_separation_functions_cts} and~\eqref{eq:scale_separation_functions_discrete}),
    and additivity (Equation~\eqref{eq:Phi-additive}).

    We first claim that,
    for any neighbourhood $\calN\subset\calD_k$ of some compact set $\calK\subset\calD_k$,
    we may find some $\delta_0>0$ such that
    \begin{equation}
      \label{eq:uniform_bound}
      \sup_{\v\in\calK,\,\delta\in(0,\delta_0)} |\Phi_k^{(\delta)}(\v)|
      \leq \sup_{\u\in\calN} B(\u)
    \end{equation}
    where $B(\u)$ is defined as
    \begin{equation}
        C_k \sum_{\pi} \prod_{ij\in\pi}
       \begin{cases}
            e^{-\alpha_k S_{\R^2}(\{u_i,u_i'\},\{u_j,u_j'\})} & \text{if $S_{\R^2}(\{u_i,u_i'\},\{u_j,u_j'\})\geq 20k^2$,} \\
            1\vee -S_{\R^2}(\{u_i,u_i'\},\{u_j,u_j'\}) &\text{if $S_{\R^2}(\{u_i,u_i'\},\{u_j,u_j'\})< 20k^2$.}
       \end{cases}
    \end{equation}
    This claim is convenient: it does not take into account the integer restriction
    in Theorem~\ref{thm:Regularity}, and it does not involve the function $S_{\R^2}'$.

    To prove the claim, we first want to find a $\delta_0>0$ such that:
    \begin{itemize}
        \item For any $\v\in\calK$ and $\delta\in(0,\delta_0)$, the point $\lfloor\v/\delta\rfloor\in(\Z^2)^{2k}$ \emph{with all coordinates rounded down} still lies in $(\delta^{-1})\calN\subset\calD_k$,
        \item For any $\delta\in(0,\delta_0)$ and $\u\in(\delta^{-1})\calN$, Equations \eqref{eq:scale_separation_functions_cts} and~\eqref{eq:scale_separation_functions_discrete}
        coincide for any $i\neq j$.
    \end{itemize}
    The first item holds true by basic topological considerations.
    For the second item, it is easy to see that \eqref{eq:scale_separation_functions_cts} and~\eqref{eq:scale_separation_functions_discrete}
    coincide on $(\delta^{-1})\calN$ for small enough $\delta$ by compactness of $\calN$,
    simply because $\operatorname{dist}(\{u_i/\delta,u_i'/\delta\},\{u_j/\delta,u_j'/\delta\})\geq 1$.

    It is easy to derive the claim from the two above items.
    Notice that $B$ inherits scale-invariance from $S_{\R^2}$,
    and therefore the left side of \eqref{eq:uniform_bound}
    is bounded by $\sup_{\u\in\calN,\,\delta\in(0,\delta_0)}B(\u/\delta)=\sup_{\u\in\calN}B(\u)$.
    This proves the claim (Equation~\eqref{eq:uniform_bound}).
    We will now derive the statements in the corollary from this claim.

    First, it follows immediately that $\sup_\delta|\Phi_k^{(\delta)}|$ is bounded on $\calD_k$.
    For the other two statements (locally uniform convergence towards a continuous function), it suffices to prove that for any point $\u\in\calD_k$
    and $\epsilon>0$,
    there exists a neighbourhood $\calN\subset\calD_k$ of $\u$ and some $\delta_0>0$ such that
    \begin{equation}
        \sup_{\v\in\calN,\,\delta\in(0,\delta_0)} |\Phi_k^{(\delta)}(\u)-\Phi_k^{(\delta)}(\v)|\leq \epsilon.
    \end{equation}
    This difference can be written as a telescopic sum of $2k$ terms,
    where each term is of the form $\Phi_k^{(\delta)}(\v')-\Phi_k^{(\delta)}(\v'')$ for some $\v'$ and $\v''$ that differ in only one entry.
    In that case, we can apply additivity (Equation~\eqref{eq:Phi-additive}) to write
    this difference as a single correlation function $\Phi_k^{(\delta)}(\w)$ where $\w$ has the property that two points are very close
    (since $\calN$ is a tiny neighbourhood of $\u$) and the other points are fixed.
    Equation~\eqref{eq:uniform_bound} then tells us that $\Phi_k^{(\delta)}(\w)$ can be made as small as desired by shrinking $\calN$.
    This implies the desired statement.
\end{proof}

The following cylinder estimates are straightforward adaptations of Theorem~\ref{thm:Regularity}. Below, $\Phi_{\CYL{L},2}$ is the two-point correlation function on the cylinder $\Z\times(\Z/L\Z)$, see
Subsection~\ref{subsection:ingredient_spectral} below for details.

\begin{corollary}[Regularity estimate for the cylinder]\label{cor:Regularity_cylinder}
There exist constants $C,c\in(0,\infty)$ such that for every $\c\in[1,2]$, $L\in 2\Z_{\geq 1}$ and $k\in\Z_{\geq 1}$,
    \begin{align}
        \label{eq1:cor:Regularity_cylinder}
            |\Phi_{\CYL{L},2}((0,0),(k,0),(2k,0),(3k,0))|&\leq C,\\
            \label{eq2:cor:Regularity_cylinder}
            |\Phi_{\CYL{L},2}((0,0),(0,\ell),(k,0),(k,\ell))|&\leq C(\ell/k)^c \quad\text{ for $0<\ell\leq \min\{8k,L/2\}$.}
   \end{align}
   \commentIoan{I found $4\pi k$ instead of $8k$. Piet: I'm not sure, see the proof... changed it back to $8k$ just to be safe...}
\end{corollary}

A more subtle manifestation of the RSW theory takes the form of a mixing estimate.
It is standard that RSW estimates induce polynomial mixing estimates between scales.
Here, we state a non-optimised version in terms of the multi-point correlation functions.

\begin{theorem}[Mixing estimate]\label{thm:mixing}
    Fix $\c\in[1,2]$,
    $k\in2\Z_{\geq 1}$.
    In this theorem, we consider $\u=(\u^{(1)},\u^{(2)})\in\calD_k$ with $\u^{(1)}=( u_1,u_1',u_2,u_2')$ and $\u^{(2)}=( u_3,\ldots u_k')$;
    we consider each of the $2k$ points fixed, except for the first point $u_1$
    which is variable.
    Then there exists some constant $C=C(u_1',u_2,u_2',\ldots)<\infty$ such that
	\begin{align}
    \limsup_{u_1\to u_2}
    \limsup_{\delta\to 0}
    \left|
	\Phi_{k}^{(\delta)}(\u)-\Phi_{2}^{(\delta)}(\u^{(1)})\Phi_{k-2}^{(\delta)}(\u^{(2)})
    \right|\leq C.
	\end{align}
\end{theorem}

Next, we state two intermediate results which
are useful in Part~\ref{part:glimpse},
where we identify the variance of the limiting GFF.
These intermediate results are stated in terms of (random)
subsets of $F(\Z^2)$.
We endow such subsets with nearest-neighbour connectivity: faces are neighbours if and only if they share an edge.
We identify a path of such faces with the union of the line segments connecting the centres
of the faces, so that we may view such paths as subsets of $\R^2$.

\begin{theorem}[Arm exponents]
    \label{thm:intro:arm_exponents}
   Fix $\c\in[1,2]$. There exists a constant $c>0$ such that for any $k\in\Z_{\geq 0}$ and $R,r\in\Z_{\ge 4}$ satisfying $R\geq 2r$,
    \begin{equation}
        \P_{\Z^2}\left[\left\{
            \parbox{24em}{$\exists a\in2\Z$
            such that the two sets of faces  $\{h+a\leq 0\}$ and\\
          $\{h+a\geq k\}$ both contain paths from $[-r,r]^2$
            to $\partial[-R,R]^2$}
            \right\}
        \right]
        \leq (r/R)^{ck^2}.
    \end{equation}
\end{theorem}

The last intermediate results follows directly from the representation (and not the RSW theory).
Informally, \emph{flip domination} says that if $h$ is below some fixed $m\in2\Z$
on a closed circuit of faces, then, on the faces surrounded by this circuit,
$h-m$
is stochastically dominated by $m-h$.
The formal statement is slightly more involved,
owing to the gradient nature of the six-vertex height function.

\begin{theorem}[Flip domination]
    \label{thm:intro:flip_domination}
    Fix $\c\in[1,2]$. Consider the following setup:
    \begin{itemize}
        \item $\gamma$ is an arbitrary self-avoiding $F(\Z^2)$-circuit,
        \item $F_\gamma\subset F(\Z^2)$ denotes the faces strictly surrounded by $\gamma$ (not those visited by $\gamma$),
        \item $m\in2\Z$ denotes the maximum of $2\lceil\frac12h|_{F_\gamma}\rceil$, so that $h-m$ is gradient measurable,
        \item $\calE$ is any gradient event measurable in terms of $h|_{F(\Z^2)\setminus F_\gamma}$ with positive probability.
    \end{itemize}
    Then, for the measure $\P_{\Z^2}[\blank|\calE]$,
    the height function $(h-m)|_{F_\gamma}$ is stochastically dominated by
    $(m-h)|_{F_\gamma}$.
    More precisely, for any bounded increasing function $X$, we have
    \begin{equation}
        \E_{\Z^2}\big[X((h-m)|_{F_\gamma})\big|\calE\big]
        \leq \E_{\Z^2}\big[X((m-h)|_{F_\gamma})\big|\calE\big].
    \end{equation}
\end{theorem}

% Finally, the RSW theory is used to prove one more intermediate result,
% namely Proposition~\ref{prop:split_into_ridge_events}.
% We do not state Proposition~\ref{prop:split_into_ridge_events} here because
% it would require us to state more definitions that are not very illuminating at this point.
% It is proved in Section~\ref{sec:ldp_split} (Part~\ref{part:qualitative}).

\subsection{\INGREDSPECTRAL: Spectral representation of correlation functions}
\label{subsection:ingredient_spectral}

The purpose of the next few paragraphs is to state a spectral representation
(which is a consequence of the transfer matrix formalism)
as it is used in the core of the proof, without going into detail on how it is obtained.
The spectral representation does not rely on the Yang--Baxter equations or the Bethe Ansatz.
We start with a definition of the six-vertex model on the cylinder.

\subsubsection{Six-vertex model on the cylinder}

The transfer matrix can be used to derive identities in the six-vertex model on a cylinder.
Recall that $\P_{\T_{M,L}}$ is the six-vertex measure on the torus $\T_{M,L}$.
For $L$ even, let $\CYL{L}$ denote the graph on the vertex set $\Z\times(\Z/L\Z)$
with nearest-neighbour connectivity.
The definition of a balanced six-vertex configuration extends to the cylinder:
a six-vertex configuration is called \emph{balanced} if, in each column of horizontal edges, there are exactly $L/2$ arrows pointing to the right and $L/2$ arrows pointing to the left.
The event of balanced six-vertex configurations is denoted $\{\operatorname{balanced}\}$ like before.
The following lemma forms the starting point of our spectral representation
(it is proved later on in Equation~\eqref{eq:cylop}).

\begin{lemma}[Cylinder measure]
    \label{lemma:Cylinder measure}
        Fix $\c>0$.
        The weak limit of $\P_{\T_{M,L}}[\blank|\{\operatorname{balanced}\}]$ as $M\to\infty$ exists and is denoted $\P_{\CYL{L}}$.
\end{lemma}

Recall that Theorem~\ref{thm:infinite_volume_6V} implies that when $\c\in[1,2]$, $\P_{\CYL{L}}$ converges weakly to $\P_{\Z^2}$ as $L$ tends to infinity.

Balanced six-vertex configurations $\omega$ on $\CYL{L}$ are in bijection with gradient height functions $h:F(\CYL{L})\to\Z$.
To see that this is true, we remark that any oriented loop on the dual graph of $\CYL{L}$ intersects the same number of left- and right-pointing arrows (relative to the orientation of the loop),
thanks to the ice rule and the balanced condition.
We may extend the domain of such gradient height functions to $\R\times (\R/L\Z)$
(via an analogue of Equation~\eqref{eq:simpleheightfunctionextensiontoplane})
and to $\R^2$ by a simple lift.

Recall Definition~\ref{def:Six-vertex multi-point correlation functions}.
For any $\u =(u_1,u_1',\ldots,u_k,u_k')\in (\R^2)^{2k}$, we shall write
\begin{align}\label{eq:multi-point_cor_function_cyl}
    \Phi_{\CYL{L},k}(\u )=
%    \Phi_{\CYL{L},k}(u_1,u_1',\dots,u_{k},u_{k}')
	:=\E_{\CYL{L}}\left[\prod_{i=1}^{k}\left(h(u_i')-h(u_i)\right)\right].
\end{align}

\subsubsection{Spectral representation of the two-point function}
Let us introduce some more notation.
    For $\u=(u_1,u_1',\ldots,u_{2k},u_{2k}')\in (\R^2)^{2k}$, define $x_i$, $y_i$, $x_i'$, and $y_i'$ such that
    \begin{equation}
        u_i'-u_i = (x_i,y_i)
        \qquad\text{and}\qquad
        u_{i+1}-u_i' = (x_i',y_i').
\label{definition xiyietc via uiuprimei}
    \end{equation}
    We say that a sequence $(u_1,u_1',\ldots,u_{2k},u_{2k}')$
    is \emph{horizontally ordered} whenever $x_i,x_i'\geq 0$
    for all $i$,
    and \emph{horizontally strictly ordered} whenever $x_i\geq 0$ and $x_i'> 0$
    for all $i$.

\begin{theorem}[Spectral representation of the two-point function]
    \label{thm:spectral_representation_four_point}
	Fix $\c>0$ and $L\in 2\mathbb Z_{>0}$. Then, there exists a finite positive measure $\mu_L$ on $\R_{>0}\times \R$
    such that
    \begin{equation}
        \label{eq:thm:spectral_representation_four_point}
      \Phi_{\CYL{L},2}(\u ) =
      \int   \left((1-a)^{x_2}e^{-\i by_2}-1\right)
        (1-a)^{x_1'}e^{-\i by_1'}
        \left(1-(1-a)^{x_1}e^{-\i by_1}\right)\diff\mu_L(a,b)
    \end{equation}
    for any horizontally ordered $\u =(u_1,u_1',u_2,u_2')\subset\Z^2$,
    and which is supported on the set $(0,2]\times [-\pi,\pi]$ and invariant
    under the map $(a,b) \mapsto (a,-b)$. Furthermore, $\mu_L(\{|b|\in (0,2\pi/L)\})=0$.
    % \commentIoan{The measure on $b$ should be supported on $[-\pi,\pi]$ if we want it to be symmetric under $(a,b) \mapsto (a,-b)$}
\end{theorem}

The measure $\mu_L$ is called \emph{positive} to distinguish it from \emph{signed}
or \emph{complex} measures.
We shall later see that $\mu_L$ is a finite sum of Dirac masses induced by the spectrum of the transfer matrix of the six-vertex model.
The measure is therefore referred to as a \emph{spectral measure}.
The heart of the proof of our main theorem will be to study the full-plane two-point function $\Phi_2$
via an appropriate limit (as $L$ tends to infinity) of the measures $\mu_L$.

\subsubsection{Spectral representation of general observables}
\label{SoussousSectionSpectralRepGnlObs}
We shall derive an expression for general observables similar to~\eqref{eq:thm:spectral_representation_four_point}, but less explicit.
At the heart of this expression is a crucial symmetry under \emph{reflection} and the associated \emph{reflection positivity}.

To describe it, let $\calR:=\{\frac12\}\times \R\subset \R^2$ denote the reflection line.
The definition is chosen such that $\calR$ traverses face centres
(and not vertices).
Let $\dagger:\R^2\to\R^2$ denote the reflection with respect to $\calR$.
Let $F^-(\CYL{L})$ denote the cylinder faces on the left of $\calR$,
and let $F^+(\CYL{L})$ denote the cylinder faces on the right of $\calR$;
the faces whose centres lie on $\calR$ are included in both sets.
The function $\dagger$ is also interpreted as an involution on $F(\CYL{L})$,
and acts on height functions via $h^\dagger:=h\circ\dagger$.

\begin{remark}
    It is important that $\dagger$ is applied to the height function,
    not the arrows.
\end{remark}

Let $\frakA$ denote the set of \emph{local observables},
that is, real-valued random variables ${\bf X}$ which are measurable
in terms of the restriction of the gradient of $h$ to \emph{finitely many} faces (such observables
are necessarily bounded).
Let
\begin{equation}
    \frakA^\pm:=\{{\bf X}\in\frakA:\text{${\bf X}$ is measurable in terms of $h|_{F^\pm(\CYL{L})}$}\}.
\label{definitionfrakApm}
\end{equation}
Notice that the reflection $\dagger$ may be interpreted as a bijection
from $\frakA^-$ to $\frakA^+$, via
\begin{equation}
    {\bf X}^\dagger: h \mapsto {\bf X}(h^\dagger).
\end{equation}
%Since $\dagger$ is an involution, its inverse is also $\dagger$.
Finally, for any $v\in\Z_{\geq 0}\times\Z$,
define the (translation) map $\tau_v:\Z^2\to\Z^2,\, u\mapsto u+v$,
which is extended to $\frakA^+$ via
\begin{equation}
    \tau_v:\frakA^+\to\frakA^+,\,
    {\bf X}\mapsto (h\mapsto {\bf X}(h\circ\tau_v)).
\end{equation}

\begin{remark}
To better grasp the definition, consider the following example.
\commentKarol{ I agree it works if we understand the writing periodically in the second coordinate or take L large enough.
Do you think it is worthy to frame that better?}
Let $(u_1,u_1',\ldots, u_k,u_k')\subset (\Z_{\geq 0}+\frac12)\times(\Z+\frac12)$
denote a family of centres of faces in $F^+(\CYL{L})$.
Then:
\begin{itemize}
    \item The observable ${\bf X}$ defined via ${\bf X}:h\mapsto \prod_i (h(u_i')-h(u_i))$
    belongs to $\frakA^+$,
    \item For any $v\in\Z_{\geq 0}\times\Z$, we have $\tau_v({\bf X})\in\frakA^+$ with $\tau_v({\bf X})(h)=\prod_i (h(u_i'+v)-h(u_i+v))$,
    \item We have ${\bf X}^\dagger\in\frakA^-$ with ${\bf X}^\dagger(h)=\prod_i (h((u_i')^\dagger)-h(u_i^\dagger))$.
\end{itemize}
\end{remark}

We now state the main theorem about the spectral representation of general observables.
\begin{theorem}[Spectral representation of general observables]\label{thm:spectral_representation_multi_point}
    Fix $\c>0$. For any triple
    $({\bf X},{\bf Y},L)\in \frakA^-\times\frakA^+ \times 2\Z_{\geq 1}$,  there exists a finite complex-valued measure
    $\mu_{{\bf X},{\bf Y},L}$ supported on $[0,2)$ such that:
    \begin{enumerate}
        \item  For any $k\ge0$,
        \begin{equation}\label{eq:spe}
            \E_{\CYL{L}}[{\bf X}\cdot\tau_{(k,0)}({\bf Y})]
            = \int  (1-a)^{k} \diffi\mu_{{\bf X},{\bf Y},L}(a),
        \end{equation}
        \item   (\textbf{Cauchy--Schwarz inequality}) $\mu_{{\bf X},{\bf X}^\dagger,L}$ and $\mu_{{\bf Y}^\dagger,{\bf Y},L}$ are positive measures and
        \begin{equation}\label{eq:Cauchy--Schwarz}
            |\E_{\CYL{L}}[{\bf X}{\bf Y}]|^2\leq\|\mu_{{\bf X},{\bf Y},L} \|^2
            \leq
            \|\mu_{{\bf X},{\bf X}^\dagger,L}\|\cdot\|\mu_{{\bf Y}^\dagger,{\bf Y},L}\|
            =\E_{\CYL{L}}[{\bf X}{\bf X}^\dagger]\E_{\CYL{L}}[{\bf Y}^\dagger{\bf Y}]
        \end{equation}
	   where $\|\cdot\|$ denotes the total variation metric.
    \end{enumerate}
\end{theorem}
Note that \eqref{eq:spe} holds true and is non-trivial for $k=0$. The second property can be understood as a reflection positivity property, see e.g.~\cite{Biskup_2009_ReflectionPositivityPhase} and references therein.
%We shall see that the Cauchy--Schwarz inequality, which may appear to be a soft tool,
%enables us to translate properties of the spectral measure $\mu_\infty$ (introduced above for the two-point case)
%into the desired properties of the spectral measures $\mu_{{\bf X},{\bf Y},\infty}$.

% </input src="sections/PART_A/1_2_ingredients.tex" root="." version="0.0.1">
% <input src="sections/PART_A/1_4_organisation.tex" root="." version="0.0.1">

\section{Organisation of the paper}

The paper is organised into four parts,
according to Table~\ref{tab:organisation}.
Recall from Section~\ref{sec:overview_steps_ingredients} that
Part~\ref{part:proofs} contains the main proofs and depends on all of the four ingredients.
The other parts are independent of one another, except that \INGREDSCALE\ (Part~\ref{part:glimpse})
relies on some ideas developed in \INGREDREGULARITY\ (Part~\ref{part:qualitative}).
\INGREDROTATION\ (the rotational invariance of Theorem~\ref{thm:rotation_invariance}) was proved in prior work, and no part is dedicated to it.

\begin{table}[ht]
    \centering
    \begin{tabular}{@{}llll@{}}
        \toprule
        \textbf{Part} & \textbf{Content} & \textbf{Techniques} & \textbf{External inputs} \\
        \midrule
        \ref{part:proofs} & Main proofs & measure theory; & Ingredients 1--4 \\
        &(Theorems~\ref{thm:GFF_convergence} and \ref{thm:GFF_convergence_isoradial})&complex analysis&\\[0.4em]
        \ref{part:spectral} & \INGREDSPECTRAL & elementary; & \\
        &Spectral representation& linear algebra  &\\[0.4em]
        \ref{part:qualitative} & \INGREDREGULARITY & percolation & \cite{Duminil-CopinKarrilaManolescu_2024_DelocalizationHeightFunction} (or \cite{GlazmanLammers_2025_DelocalisationContinuity2D})\\
        &Regularity estimates & & \\[0.4em]
          %& \INGREDROTATION &  & \cite{Duminil-CopinKozlowskiKrachun_2020_RotationalInvarianceCritical}  \\[0.4em]
      %    &Rotational invariance&  &and \INGREDREGULARITY\\
        \ref{part:glimpse} & \INGREDSCALE & percolation; & \cite{Duminil-CopinKozlowskiKrachun_2022_SixVertexModelsFree}  \\
        & Glimpse of scale invariance & complex analysis&and \INGREDREGULARITY\\
        \bottomrule
    \end{tabular}
    \caption{Organisation of the paper: content, techniques, and dependencies of each part.}
    \label{tab:organisation}
\end{table}
% \commentIoan{Shouldn't part D also make reference to papers by AG and PL? Is  \cite{Duminil-CopinKarrilaManolescu_2024_DelocalizationHeightFunction} really used as an input? }
% \noticePiet{Both can be used equivalently. Maybe add my paper with Sasha}

% </input src="sections/PART_A/1_4_organisation.tex" root="." version="0.0.1">

% </input src="sections/PART_A/main.tex" root="." version="0.0.1">
% <input src="sections/PART_B/main.tex" root="." version="0.0.1">
\part{Proof of the main results}
\label{part:proofs}

This part implements the main proofs as outlined in Section~\ref{sec:overview_steps_ingredients}.
The ingredients stated formally in the introduction are used as external inputs.
The main results of Sections~\ref{sec:twopoint}, \ref{sec:multi-point}, and~\ref{sec:GFF_convergence_criteria}
are Theorems~\ref{thm:dicht_2p_final_form},~\ref{thm:k2impliesallk}, and~\ref{thm:criterion}, respectively,
and they correspond to Steps~1--3 outlined in Section~\ref{sec:overview_steps_ingredients}.
Sections~\ref{sec:twopoint}--\ref{sec:GFF_convergence_criteria} may be read independently of one another.
Section~\ref{sec:main_result} then combines these results to prove
the main result in the isotropic case (Theorem~\ref{thm:GFF_convergence}).
Section~\ref{sec:anisotropic} presents the proof of the anisotropic case (Theorem~\ref{thm:GFF_convergence_isoradial}).
The value of $\c\in[\sqrt{3},2]$ does not play a role in the proofs and is omitted from notations.

% <input src="sections/PART_B/2_0_intro.tex" root="." version="0.0.1">
\section{Sub-sequential GFF limits for the two-point function}
\label{sec:twopoint}

\subsection{Statement of the theorem and outline of the proof}

Section~\ref{sec:twopoint} is dedicated to proving the following result.

\begin{theorem}[Dichotomy for the two-point function]
    \label{thm:dicht_2p_final_form}
    One of the following two properties holds true:
    \begin{itemize}
        \item There exists some $\sigma\in\R_{\geq 0}$ such that $\Phi_2^{(\delta)}$ converges to $\sigma^2\Psi_2^{\GFF}$ uniformly on every compact subset of $\mathcal D_2$ as $\delta$ tends to zero,
        \item There exist two distinct $\sigma,\sigma'\in\R_{\geq 0}$ and two sequences $(\delta_n)_n,(\delta'_n)_n$ tending to zero
        such that $\Phi_2^{(\delta_n)}$ and $\Phi_2^{(\delta_n')}$ converge to $\sigma^2\Psi_2^{\GFF}$ and $(\sigma')^2\Psi_2^{\GFF}$ respectively, uniformly on every compact subset of $\mathcal D_2$ as $n$ tends to infinity.
    \end{itemize}
\end{theorem}

\begin{remark}\label{rmk:2.1}
    The theorem does not assert that all sub-sequential limits are multiples of $\Psi_2^{\GFF}$.
    We believe that its proof allows for the possibility of other sub-sequential scaling limits,
    such as, for example, $\sigma^2\Psi_2^{\GFF}+(\sigma')^2\Psi_2'$,
    where $\Psi'$ is the two-point correlation function of a \emph{massive} GFF.
    \commentIoan{Are we sure that the other possible limits are massive GFFs? I would limit this remark to the first sentence and would provide a second remark, later on, where we explain that the measures satisfying the bounds and the rotation invariance property form a (positive cone in a)  vector space and give the general expression for the measures not concentrated on $a^2 = b^2$. Karol, do you remember the general form of such measures? }
    \noticePiet{I am definitely not sure, but I find it a very useful example, because it is clear that it has ``two scaling limits''. The above text was not meant to be misleading, but we can change it.}
\end{remark}

Let us sketch the proof. The first step (Subsection~\ref{subsec:2.2}) consists in taking the limit as $L$ tends to infinity and taking a scaling limit along a sub-sequence in
the spectral representation formula from Theorem~\ref{thm:spectral_representation_four_point} for the two-point correlation function of the six-vertex model on the cylinder. By compactness arguments, this yields a limiting two-point function $\Psi_2$ and a positive measure $\mu$ on $\R_{>0}\times\R$, invariant under $(a,b) \mapsto (a,-b)$, such that for every
$\u =(u_1,u_1',u_2,u_2')\in (\R^2)^4$
satisfying some simple geometric constraints,
\begin{equation}\label{eq:spec rep scaling limit 2 point introduction}
    \Psi_2(\u )
    =   \int
        \left(e^{-ax_2-\i by_2}-1\right)
        e^{-ax_1'-\i by_1'}
        \left(1-e^{-ax_1-\i by_1}\right)\diff\mu(a,b)
\end{equation}
(recall Equation~\eqref{definition xiyietc via uiuprimei} for a definition of $(x_1,y_1,x'_1,y'_1,x_2,y_2)$ associated with such a $\u $). The measure $\mu$ fully encodes the two-point function $\Psi_2$, and our aim is to analyse its structure.

Assuming we may differentiate under the integral,
applying the Laplacian at {\em any} argument of $\Psi_2$ produces a factor $a^2+(\i b)^2$ in the integrand, which hints that harmonicity of $\Psi_2$ in each argument on $\mathcal D_2$ is equivalent to the concentration property $\mu[\{b^2 \ne a^2\}]=0$. In fact, it is a simple exercise (see also the proof of Theorem~\ref{thm:main measure}) to check from the expression above that
$\Psi_2=\sigma^{2}\cdot\Psi_{2}^{\GFF}$ for some $\sigma>0$ if and only if $\mu[\{b^2 \ne a^2\}]=0$ and the density of the first marginal of $\mu$ is $\frac{\sigma^2}{2\pi a}\diffi a$.

We will not be able to establish these two properties of $\mu$ directly. The key ingredient we do exploit, however, is the rotational invariance of $\Psi_2$ (Theorem~\ref{thm:rotation_invariance}),
which provides a collection of identities relating values of $\Psi_2$ at different points.
These identities translate into constraints on the joint distribution of $a$ and $b$ under $\mu$ -- see \eqref{eq:Psi_in_terms_of_I} and \eqref{eq:rotation_of_F}.
By combining them with certain analyticity properties (of the function $I_F$ defined in Subsection~\ref{sec:2.3}), we will derive in Subsection~\ref{sec:2.4} that
\begin{align}\label{eq:a^2>b^2}
	\mu[\{a^2 < b^2\}] = 0.
\end{align}

One might hope that Theorem~\ref{thm:rotation_invariance} would also yield the opposite bound $\mu[\{a^2 > b^2\}] = 0$, thereby implying $\mu[\{b^2 \ne a^2\}]=0$. The density of the first marginal would then follow readily from the rotational invariance constraints, thus determining $\mu$ up to multiplicative constant.
Unfortunately this is not the case: as pointed out in Remark~\ref{rmk:2.1}, one may construct an entire family of scaling limits consistent with rotational invariance and all regularity assumptions,
but for which $\mu[\{b^2 \ne a^2\}]\ne0$.
\commentIoan{This may be a good place to give the actual expressions for these measures. If so, turn this paragraph into a remark and refer to it in Remark~\ref{rmk:2.1}. Also specify the rotational invariance contraints: to me they are manifested entirely in \eqref{eq:Psi_in_terms_of_I} and the ``regularity assumptions'' of Definition~\ref{definition compact space of measures} }

Although \eqref{eq:a^2>b^2} is insufficient to uniquely determine $\Psi_2$, it is nonetheless enough to show (see Subsection~\ref{sec:2.5}) that the large-scale and small-scale behaviour of $\Psi_2$ is compatible with that of two-point correlation functions of the GFF.
In particular, if $\Psi_2$ is not a multiple of $\Psi_{2}^{\GFF}$, then its sub-sequential scaling limits at large and small scales must be distinct multiples of $\Psi_{2}^{\GFF}$. This yields the dichotomy of Theorem~\ref{thm:dicht_2p_final_form}.

% </input src="sections/PART_B/2_0_intro.tex" root="." version="0.0.1">
% <input src="sections/PART_B/2_1_spec_rep_2p.tex" root="." version="0.0.1">
\subsection{Compactness of the spectral representation}
\label{subsec:2.2}

Here and below, we keep the convention from Equation~\eqref{definition xiyietc via uiuprimei} for $(x_1,y_1,x'_1,y'_1,x_2,y_2)$ introduced in the introduction for every horizontally ordered $\u $.
Recall from Theorem~\ref{thm:spectral_representation_four_point}
that
\begin{equation}\label{eq:2.2.1}
      \Phi_{\CYL{L},2}(\u ) =
      \int  \chi^{\rm discr}_\u (a,b) \diff\mu_L(a,b)
\end{equation}
where
\begin{equation}
    \label{eq:def_chi_tilde}
    \chi^{\rm discr}_\u (a,b):=\left((1-a)^{x_2}e^{-\i by_2}-1\right)
        (1-a)^{x_1'}e^{-\i by_1'}
        \left(1-(1-a)^{x_1}e^{-\i by_1}\right)
\end{equation}
for any horizontally ordered $\u =(u_1,u_1',u_2,u_2')\in (\Z^2)^4$.
This subsection proves that the family $(\mu_L)_L$ belongs to
a compact space of measures, which allows us to derive a spectral representation
for the sub-sequential scaling limit of the two-point function.

\subsubsection{A compact space of measures}

In this section, we introduce a compact space $\calM$ of measures satisfying some scale-invariant qualitative bounds, and then show that the cylinder measures $\mu_L$ belong to this space.
\begin{definition}[The compact space $\calM_{c,C}$]
\label{definition compact space of measures}
    For $c,C\in(0,\infty)$, let $\calM_{c,C}$ denote the set of positive measures $\nu$ on $\R_{>0}\times\R$,
    invariant under $(a,b) \mapsto (a,-b)$ and
    satisfying the following bounds:
        \begin{enumerate}
            \item\label{def:allbounds:1}
            \(\nu[\{a\in(\alpha,2\alpha)\}]\leq C,\) for any $\alpha>0$,
            \item\label{def:allbounds:2}
            \(\nu[\{(a,|b|)\in(0,\alpha)\times(\beta,2\beta)\}]\leq C(\alpha/\beta)^{c},\) for any $\beta\geq\alpha>0$.
        \end{enumerate}
    We endow $\calM_{c,C}$ with the \emph{weak} (or \emph{vague}) topology, that is, the topology
    making the map
    \(
        \nu\mapsto\nu[f]
    \)
    continuous for any continuous function $f:\R_{>0}\times\R\to\C$ whose support is a compact subset
    of $\R_{>0}\times\R$.
\end{definition}

We now show that the cylinder measures $\mu_L$ belong to $\calM_{c,C}$ for suitable constants $c,C\in(0,\infty)$.
\begin{lemma}
    \label{lemma:allbounds}
    There exist constants $c,C\in(0,\infty)$ such that, for every $L$, the measure $\mu_L$ belongs to $\calM_{c,C}$.
\end{lemma}

From this point on, we fix such a pair $c,C\in(0,\infty)$ once and for all, and we write $\calM:=\calM_{c,C}$ in what follows.

\begin{proof}
Reflection symmetry suggests that some $\u\in (\Z^2)^4$ are special:
namely those $\u$ where the first two points are chosen on the left of some vertical reflection line,
and where the last two points are the reflections of the first two points.
This leads to $\chi^{\rm discr}_\u $ being real and of constant sign, which is crucial for the proof.
We shall fix $L$ throughout.

\paragraph{Step 1: Checking (i) in the definition of $\calM$.}
    For $k\in\Z_{\geq 0}$ and $\ell\in 2\Z_{\geq 0}$, set
    \begin{equation}
         \u =\begin{pmatrix}
            u_1\\
            u_1'\\
            u_2\\
            u_2'\\
            \end{pmatrix}:=\begin{pmatrix}
            0 &0\\
            k &0\\
            k+\ell&0 \\
            2k+\ell&0
            \end{pmatrix}
               \end{equation}
               so that
               \begin{equation}
            \chi^{\rm discr}_\u (a,b)=-
            (1-a)^\ell  (1-(1-a)^{k})^2
            \leq 0.\end{equation}

We now split the argument in two cases depending on the value of $\alpha$. First, consider the case $\alpha\leq 1/16$.
    Let $k=\ell\in [\frac1{8\alpha},\frac1{4\alpha}]\cap 2\Z$.
    Then, for any \( (a,b)\in[\alpha,2\alpha] \times \R\), a short computation gives that    \(
            |\chi^{\rm discr}_\u (a,b)|
            \geq
            1/1000\). Also,  Corollary~\ref{cor:Regularity_cylinder}  implies that $|\Phi_{\CYL{L},2}(\u )|\le C$ uniformly in $L$ and $\alpha$.
    Combining the claims of the two last sentences and $\chi^{\rm discr}_\u (a,b)<0$ leads to   \begin{equation}
        \mu_L[\{a\in[\alpha,2\alpha]\}]
        \leq
        -1000 \int\chi^{\rm discr}_\u (a,b) \diff\mu_L(a,b)= 1000|\Phi_{\CYL{L},2}(\u )|
        \leq 1000 C.
    \end{equation}
    For the case $\alpha\geq 1/16$, simply set $k=1$ and $\ell=0$ and use a similar strategy
    to get
    \begin{equation}
        \mu_L[\{a\in[1/16,\infty)\}]
        \leq
        256|\Phi_{\CYL{L},2}(\u )|
        \leq 256.
    \end{equation}
    On the right we used that height differences are bounded by $1$,
    and thus $|\Phi_{\CYL{L},2}(\u )|\leq 1$.

 \paragraph{Step 2: Checking (ii) in the definition of $\calM$.}
 	We will prove the existence of constants $c,C > 0$ such that
	\begin{align}\label{eq:allbounds:21}
		\mu_L[\{(a,|b|)\in(0,\alpha)\times(\beta,2\beta)\}]\leq C\big(\tfrac\alpha\beta\big)^{c}\qquad
		\text{ for $\alpha < \tfrac1{16}$ and $\beta  = \tfrac{\pi}{2\ell}$}
	\end{align}
           with $1 \leq \ell \leq L/2$ integer.
           Condition (ii) may be deduced from~\eqref{eq:allbounds:21},
           together with point (i) above and the fact that $\mu_L$ is supported on $\{(a,b): 0 < a \leq 2 \text{ and } 2\pi/L \leq |b|\leq  \pi\}$ (see Theorem~\ref{thm:spectral_representation_four_point}),
           by simple algebraic manipulations.

           Fix values of $\alpha$ and $\beta$ as in \eqref{eq:allbounds:21} and choose some even integer $k \in [\frac1{8\alpha},\frac1{4\alpha}]$.
	Set
        \begin{equation}
            \u =\begin{pmatrix}
                u_1\\
                u_1'\\
                u_2\\
                u_2'\\
                \end{pmatrix}:=\begin{pmatrix}
                0&0 \\
                0&\ell\\
                k  &0\\
                k &\ell
                \end{pmatrix}
                  \end{equation}
                  so that $x_1 = x_2 = 0$ and $x_1 ' = k $ and
           \begin{equation}
            \chi^{\rm discr}_\u (a,b)=
            2(1-a)^k   (1-\cos \ell b)
   	    \ge0.
        \end{equation}
  Then,  for any
                \(
                     (a,b)\in[0,\alpha]\times [\beta,2\beta]\), a short computation gives that
         \(
                \chi^{\rm discr}_\u (a,b)\geq 1.
                \)
                Also, Corollary~\ref{cor:Regularity_cylinder} implies that $|\Phi_{\CYL{L},2}(\u )|\leq C'(\alpha/\beta)^{c'}$ for some constants $C',c'\in(0,\infty)$ that are uniform in $L$, $\alpha$, and $\beta$.
The conclusion follows by the same argument as in point~(i).
    \end{proof}

We now start manipulating the measures in $\calM$.
By construction, these measures have the property that
\begin{equation}\sup_{\nu\in\calM}\int (a \wedge \tfrac1a)\diff\nu(a,b)<\infty.\end{equation}
This means that the function $a\wedge \frac1a$ is a good domination function for
applying the dominated convergence theorem.
In the following lemma, we collect a few more properties of $\calM$.
We leave it as a straightforward exercise to the reader.

\begin{lemma}[Properties of $\calM$]
    We have the following properties:
    \label{lemma:propertiesOfM}
    \begin{enumerate}
        \item $\calM$ is a compact topological space,
        \item Suppose that $f:\R_{>0}\times\R\to\C$ is a continuous function
        such that
        \begin{equation}
            \sup_{a,b} (a\vee \tfrac1{a}) |f(a,b)|<\infty,
        \end{equation}
        then the function $\calM\to\C,\,\nu\mapsto\nu[f]$ is continuous,
        \item Let $(\nu_n)_n\subset\calM$ denote a sequence of measures,
        and let $(f_n)_n$ denote a sequence of continuous functions $f_n:\R_{>0}\times\R\to\C$ such that $\sup_{n,a,b} (a\vee \tfrac1{a}) |f_n(a,b)|<\infty$. If $(\nu_n)_n$ converges to $\nu$ and $(f_n)_n$
        converges uniformly on every compact subset of $\R_{>0}\times\R$ to a function $f$,
            then
            \begin{equation}
            \lim_{n\rightarrow\infty}\nu_n[f_n]\to\nu[f].\end{equation}
    \end{enumerate}
\end{lemma}

\subsubsection{Passage to the full-plane limit of cylinder measures}

In this section, we analyse the limit of the cylinder measures as $L$ tends to infinity.
Since the measures $\mu_L$ belong to $\calM$, we may fix, once and for all, a sub-sequential limit $\mu_\infty \in \calM$ of the sequence $(\mu_L)_L$.
The full sequence $(\mu_L)_L$ may actually be shown to converge --- for instance using~\eqref{eq:phi_chi_infty} and the uniqueness of
of the $L\to \infty$ limit of the two-point function stemming from the existence of the full-plane measure ---  but this fact will not play a role in what follows.
Therefore we simply work with a fixed sub-sequential limit.

The full-plane two-point correlation function may then be expressed in terms of $\mu_\infty$ as in the case of the cylinder.

\begin{lemma}
    \label{lemma:LToInfty}
    For any horizontally ordered sequence $\u  \subset \Z^2$ with $\{y_1,y_2\} \ni 0$,
    we have
    \begin{equation}\label{eq:phi_chi_infty}
        \Phi_2(\u)=\int \chi^{\rm discr}_\u (a,b)\diffi\mu_\infty(a,b).
    \end{equation}
\end{lemma}

\begin{proof}
Suppose that $\u $ satisfies the assumptions of the lemma.
Using the full-plane limit of the six-vertex model (Theorem~\ref{thm:infinite_volume_6V}), we obtain
\begin{equation}
    \Phi_2(\u)
    =
    \lim_{L\to\infty}\Phi_{\CYL{L},2}(\u)
    \stackrel{\eqref{eq:2.2.1}}{=}
    \lim_{L\to\infty}\mu_L[\chi^{\rm discr}_\u ].
\end{equation}
Since $\mu_L$ is supported on $(0,2]\times\R$, we may insert the indicator $\true{a\le 2}$ and write
\begin{equation}
    \Phi_2(\u)
    =
    \lim_{L\to\infty}\mu_L[\true{a\le 2}\cdot\chi^{\rm discr}_\u ].
\end{equation}
Because $y_1=0$ or $y_2=0$, we have $\true{a\le 2}\cdot\chi^{\rm discr}_\u =O(a\wedge \frac{1}{a})$
(see Equation~\eqref{eq:def_chi_tilde}). Lemma~\ref{lemma:propertiesOfM} therefore implies that
\begin{equation}
    \Phi_2(\u)
    =
    \mu_\infty[\true{a\le 2}\cdot\chi^{\rm discr}_\u ].
\end{equation}
The proof is completed by discarding the indicator $\true{a\le 2}$, which is justified by the fact that $\mu_\infty[|a|>2]=0$.
This property is inherited from the corresponding bound for the measures $\mu_L$, by the definition of $\calM$ and a second application of Lemma~\ref{lemma:propertiesOfM}.
\end{proof}

\begin{figure}[h]
    \centering
    \includegraphics{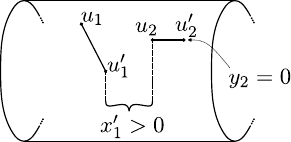}
    \caption{The spectral representation passes to the scaling limit
    when the points $\u\in(\R^2)^4$ satisfies the geometric constraints of the figure:
    the pairs of points cannot overlap ``horizontally'' (in the figure, $x_1\geq 0$ and $x_1'>0$),
    and one pair must line on a horizontal line (in the figure, $y_2=0$).}
    \label{fig:scaling_measures}
\end{figure}

\subsubsection{Passage to the scaling limit}

We now pass to the scaling limit by letting the mesh size of the lattice tend to zero.
For any $\nu \in \calM$ and $\delta>0$, denote by $\nu^{(\delta)} \in \calM$ the measure defined by
\begin{equation}
    \nu^{(\delta)}(U) := \nu(\delta U)
\end{equation}
for any measurable subset $U \subset \R_{>0} \times \R$.
By construction, for every integrable function $f$,
\begin{equation}
    \int f(\delta a,\delta b)\diffi\nu^{(\delta)}(a,b)
    =
    \int f(a,b)\diffi\nu(a,b).
\end{equation}

For $u=(x,y) \in \R^2$, set $u/\delta := (\lfloor x/\delta\rfloor,\lfloor y/\delta\rfloor)\in\Z^2$,
and for any $\u =  (u_1, u_1', \ldots, u_k, u_k') \in (\R^2)^{2k}$ for some $k$, write
$\u /\delta := (u_1/\delta, u_1'/\delta,  \ldots, u_k/\delta, u_k'/\delta) \subset (\Z^2)^{2k}$.

The following notion will be used throughout.

\begin{definition}[Convergence sequence]
    A \emph{convergence sequence} is a sequence of positive reals $(\delta_n)_n$ tending to zero such that
    $\mu_\infty^{(\delta_n)}$ converges in the compact topological space $\calM$.
    We denote by $\mu$ its limit.
    The dependence on $(\delta_n)_n$ will always be clear from context.
\end{definition}

Convergence sequences exist by compactness of $\calM$. Moreover, from any sequence of scales $(\delta_n')_n$ tending to zero,
one may extract a convergence sub-sequence $(\delta_n)_n$.

To state the next result, we introduce a variant $\chi$ of the function $\chi^{\rm discr}$ from
Equation~\eqref{eq:def_chi_tilde}:
for any $\u =(u_1,u_1',u_2,u_2')\subset\R^2$, define
\begin{equation}
    \label{eq:chi_u_def}
    \chi_\u (a,b):=
    (e^{-ax_2-\i b y_2}-1)
    e^{-ax_1'-\i b y_1'}
    (1-e^{-a x_1 -\i b y_1}).
\end{equation}

\begin{lemma}[Spectral representation of the sub-sequential scaling limit]
    \label{lemma:initial_scaling}
  Let $(\delta_n)_n$ be a convergence sequence.
Then, for any horizontally strictly ordered $\u  = (u_1,u_1',u_2,u_2') \in (\R^2)^4$ satisfying $\{y_1,y_2\}\ni 0$
    \begin{equation}\label{eq:initial_scaling}
        \Psi_2(\u ) := \lim_{n\to\infty} \Phi_2^{(\delta_n)}(\u )
        = \int \chi_\u (a,b) \diffi\mu(a,b).
    \end{equation}
\end{lemma}

\begin{remark}
% We required that the first coordinates of $u_1$ and $u'_1$ be \emph{strictly} smaller than those of
%$u_2$ and $u'_2$, rather than merely less or equal.
Compared to \eqref{eq:phi_chi_infty}, we require above that the points be horizontally {\em strictly} ordered.
This additional strictness is necessary because, under scaling, one must also control the behaviour of
$\chi_\u $ for \emph{large} values of $a$,
and this geometric requirement is precisely what ensures that such behaviour is properly handled.
\end{remark}

\begin{proof}%Since both $\Phi_2$ and $\chi_\u $ are antisymmetric under interchanging $u_i$ and $u_i'$, we may assume without loss of generality that  $\u \subset \R^2$ is horizontally strictly ordered.
    For notational simplicity we assume
    that the points $\u $ belong to $(\Z^2)^{2k}\subset(\R^2)^{2k}$,
    that $1/\delta_n\in\Z$ for all $n$,
    and that $y_2=0$.
    Since $\u$ is horizontally strictly ordered and due to the smoothness of $\u \mapsto \chi_\u(a,b)$, the bounds in the definition of $\calM$ and the Hölder nature of the correlators,
    these assumptions are harmless.

    The proof would be easy if $\mu_\infty$ was supported on $\{a\leq 1\}$
    (meaning that all eigenvalues of the transfer matrix were non-negative).
    The negative eigenvalues, corresponding to $a\in(1,2]$,
    make the proof slightly more technical.

    For any $n$, split $\Phi_2^{(\delta_n)}(\u )
        =
        P_n+N_n$ depending on the contributions of positive and negative eigenvalues respectively:
    \begin{align}
        P_n&:=\int_{\{a\leq 1\}}\chi^{\rm discr}_{\u /\delta_n}(a,b)\diffi\mu_\infty(a,b)
        ;\\
        N_n&:=
                \int_{\{1<a\leq 2\}}\chi^{\rm discr}_{\u /\delta_n}(a,b)\diffi\mu_\infty(a,b).
    \end{align}
    To conclude, it suffices to show that $P_n$ tends  to $\int \chi_\u (a,b)\diff\mu(a,b)$ (as $n\to\infty$), and $N_n$ to zero.

    \paragraph{Step 1: Limit of $N_n$.}
    Recall from the definition of $\calM$ that $\mu_\infty[\{1<a\leq 2\}]<\infty$.
    By writing out $\chi^{\rm discr}_{\u /\delta_n}$ explicitly, it is easy to see that
    \begin{equation}
        |\chi^{\rm discr}_{\u /\delta_n}(a,b)|
        \leq
        4(1-a)^{x_1'/\delta_n}.
    \end{equation}
    For any fixed $a\in(1,2)$, the bound on the right converges to zero as $n$ tends to infinity.
    Thus, using the dominated convergence theorem,
    the only contribution to the limit can come from
    an atom at $a=2$.
    However, $\mu_\infty[\{a=2\}]=0$. Indeed, writing $e_1=(1,0)$, one has, as $k$ tends to infinity over the odd integers,
    \begin{equation}
        \Phi_2(0,e_1, (k+1)e_1, (k+2)e_1)=
        -\int a^2(1-a)^{k}\diff \mu_\infty(a,b) \to 4 \mu_\infty[\{a=2\}].
    \end{equation}
    But we know that the correlation functions on the left tend to zero,
    thanks to our regularity estimate (Theorem~\ref{thm:Regularity}). This concludes Step~1.

    \paragraph{Step 2: Limit of $P_n$.}
    We may write
    \begin{align}
         P_n    =
        \int_{\{a\leq 1/\delta_n\}}\chi^{\rm discr}_{\u /\delta_n}(\delta_n a,\delta_n b)\diffi\mu_\infty^{(\delta_n)}(a,b)
        =
        \int f_n(a,b)\diffi\mu_\infty^{(\delta_n)}(a,b)
        ,
    \end{align}
    where
    \(
        f_n(a,b):=
        \true{a\leq 1/\delta_n}\cdot
        \chi^{\rm discr}_{\u /\delta_n}(\delta_n a,\delta_n b)\).
    By working out an explicit expression for $f_n$,
    it is straightforward to see that $f_n$ converges on every compact subset of $\mathbb R_{>0}\times \mathbb R$ to $\chi_\u $ as $n$ tends to infinity. Also, the fact that $\{y_1,y_2\}\ni 0$ implies that $\sup_{n,a,b} (a\vee \tfrac1{a}) |f_n(a,b)|<\infty$.
    Since $\mu_\infty^{(\delta_n)}$ tends to $\mu$
    (by hypothesis), the convergence of $P_n$ now follows from Lemma~\ref{lemma:propertiesOfM}.
\end{proof}

% </input src="sections/PART_B/2_1_spec_rep_2p.tex" root="." version="0.0.1">
% <input src="sections/PART_B/2_1b_reg.tex" root="." version="0.0.1">

%\subsection{Regularity of the sub-sequential scaling limit}
\subsection{Consequences of rotational invariance}
\label{sec:2.3}
\commentIoan{I would insist more on the contribution of rotational invariance. This is the essential (and new) ingredient that makes the whole analysis work}

Let us briefly recap what we did so far.
We introduced the compact set of measures $\calM$ described in Definition~\ref{definition compact space of measures}
    and studied some of its properties in Lemma~\ref{lemma:propertiesOfM}.
This enabled us to prove that any sequence of mesh sizes tending to zero has a ``convergence sub-sequence'' $(\delta_n)_n$ along which, for some $\mu\in\calM$,
    \begin{equation}
        \label{eq:Phi2_limit_recap}
        \Psi_2(\u ):=\lim_{n\to\infty}
        \Phi_2^{(\delta_n)}(\u )
        =
        \int\chi_\u (a,b)\diffi\mu(a,b)
    \end{equation} for any horizontally strictly ordered $\u =(u_1,u'_1,u_2,u'_2)\in \calD_2$ with
    $\{y_1,y_2\}\ni0$
    % and whose first coordinates of $u_1$ and $u'_1$ is strictly smaller than those of $u_2$ and $u'_2$
    (see Equation~\eqref{eq:chi_u_def}, Lemma~\ref{lemma:initial_scaling}, and Figure~\ref{fig:scaling_measures}).

 % In this section, we introduce an analytic function $I_F$ that will be useful to encode the regularity of our two-point correlation functions $\Psi_2$. Set $\C_+:=\{z\in\C:\Re(z)>0\}$.

 In this section we study the properties of the sub-sequential scaling limits $\mu$ of the spectral measures $\mu_\infty$,
 in particular the consequences of the rotational invariance of Theorem~\ref{thm:rotation_invariance}.
These are specifically manifested in~\eqref{eq:Psi_in_terms_of_I}, which will eventually lead to Theorem~\ref{thm:dicht_2p_final_form}.
 Rotational invariance will be used again in the proof of Theorem~\ref{thm:k2impliesallk}, but we consider~\eqref{eq:Psi_in_terms_of_I} to be its main consequence.

 Set $\C_+:=\{z\in\C:\Re(z)>0\}$.

\begin{definition}
   To a convergence sequence $(\delta_n)_n$,
associate the following functions:
\begin{align}
    &F:\C_+\times\R\to\C,\,(x,y)\mapsto \int a e^{-ax-\i by}\diffi\mu(a,b),\label{eq:h20}\\
    &I_F:\R_{>0}\to\R,\,s\mapsto -\int_1^s F(x,0)\diff x.
\end{align}
%(we omit the dependency on $(\delta_n)_n$ -- which manifests itself via the definition of $\mu$ -- in the notation as it will be clear from context).
\end{definition}

The bounds in the definition of $\calM$ ensure that the above functions are indeed well-defined.
In the definition of $I_F$, the integral is taken with a sign, so that $I_F'(s) = - F(s,0)$ for all $s > 0$.

We gather a few properties of $F$ first.
\begin{lemma}
    \label{lemma:F_properties}
     For every convergence sequence $(\delta_n)_n$, the function $F$ satisfies the following properties:
    \begin{enumerate}
        \item For any horizontally strictly ordered $(u_1,u_1',u_2,u_2')\in (\R^2)^4$,
        we have
        \begin{equation}
            \label{eq:horizontal_derivative_of_Psi}
            \lim_{\epsilon\to 0} \tfrac1\epsilon
            \Psi_2(u_1,u_1',u_2,u_2+(\epsilon,0))
            =
            F(x_1+x_1',y_1+y_1')-F(x_1',y_1').
        \end{equation}
        \item The function $F$ is continuous on $\C_+\times\R$.
        \item For any fixed $y\in\R$, the function $F(\blank,y)$ is holomorphic on $\C_+$,
        \item For any $(x,y)$, we have $|F(x,y)|\leq F(\Re(x),0)$, and $F(x,0)$ is decreasing in $x\in\R_{>0}$,
        \item We have $\lim_{x\to\infty} F(x,0) = 0$.
    \end{enumerate}
\end{lemma}

\begin{proof}
	All properties follow from \eqref{eq:initial_scaling}, the bounds in the definition of $\calM$ and the dominated convergence theorem
    with the dominating function $a\wedge \frac1a$.
\end{proof}

We now turn to properties of $I_F$. Recall that $|\cdot|$ is the Euclidean norm on $\mathbb R^2$.

\begin{lemma}
    \label{lemma:Psi_in_terms_of_I}
    For every convergence sequence $(\delta_n)_n$, the functions $F$ and $I_F$ satisfy the following properties.
    \begin{enumerate}
    \item The function $I_F$ is analytic.
        \item For {\em every} $\u =(u_1,u'_1,u_2,u'_2)\in\calD_2$,
        the limit
        $\Psi_2(\u ):=\lim_{n\to\infty}
            \Phi_2^{(\delta_n)}(\u )$
        is well-defined and equals
        \begin{equation}
            \label{eq:Psi_in_terms_of_I}
            \Psi_2(\u )
            =
            I_F(|u_2-u_1|)+I_F(|u_2'-u_1'|)-I_F(|u_2-u_1'|)-I_F(|u_2'-u_1|).
        \end{equation}
        \item For any $(x,y)\in\R_{>0}\times\R$, we have
        \begin{equation}
            \label{eq:F_as_derivative_of_I}
            F(x,y)=-\partial_x (I_F(|(x,y)|)).
        \end{equation}
    \end{enumerate}
\end{lemma}

\begin{remark}
	Note the similarity between \eqref{eq:Psi_in_terms_of_I} and the expression of the GFF correlation function \eqref{eq:GFF_multi-point_cor_function}.
	That $\Psi_2(\u)$ may be decomposed as a sum of four terms depending on the pairwise differences of $\u$
	 is a simple consequence of \eqref{eq:initial_scaling} -- at least for $\u$ to which \eqref{eq:initial_scaling} applies.
	That the terms only depend on the Euclidean norm of the differences is a crucial fact, which encodes the rotational invariance of the scaling limits (Theorem~\ref{thm:rotation_invariance}).
\end{remark}
\commentIoan{Added this remark on the consequence of rotational invariance. \\I find \eqref{eq:Psi_in_terms_of_I} very elegant. But it's a very important property and should be stressed more. I wonder whether changing the lemma to a prop or theorem makes sense. HDC: yes a proposition sounds good}

\commentKarol{I'm fine with it, but can also keep things like they are}

\begin{proof} Analyticity of $I_F$ follows from the properties of $F$. We turn to proofs of (ii) and (iii).

    Define $\R^2_\pm:=\{(x,y)\in\R^2:\pm x>0\}$.
    Recall that $\Psi_2$ is well-defined and expressed in terms of $\mu$ for all horizontally strictly ordered $\u$ with $y_1 =0$ or $y_2 = 0$ -- see Lemma~\ref{lemma:initial_scaling}
    The proof has four steps.
 First, we prove that for fixed $(u_1,u_1',u_2)\in(\R^2_-)^2\times\R_+^2$, the function $\Psi_2(u_1,u_1',u_2,\blank)$ is well-defined and $C^1$ on $\R^2_+$.
Second, we prove Equation~\eqref{eq:F_as_derivative_of_I}.
  Third, we prove Equation~\eqref{eq:Psi_in_terms_of_I} on $\R^2_-\times\R^2_-\times\R^2_+\times\R^2_+$ when $y_2=0$.
Finally, we extend Equation~\eqref{eq:Psi_in_terms_of_I} to all of $\calD_2$.

    \paragraph{Step~1: Definition of $\Psi_2(u_1,u_1',u_2,\blank)$.} Readers may help themselves with Figure~\ref{fig:rotation_trick}.
    Fix $(u_1,u_1')\in(\R_-^2)^2$.
    Let us first prove well-definedness of $\Psi_2(u_1,u_1',\blank,\blank)$
    on $(\R_+^2)^2$.
    Fix two points $u_2,u_2'\in(\R_+^2)^2$.
    Let $U$ denote an open set containing the line segment between $u_2$ and $u_2'$,
    and such that its closure is a compact subset of $\R_+^2$
    (see Figure~\ref{fig:rotation_trick}).
    Let $\theta:\R^2\to\R^2$ denote the rotation by a sufficiently small angle
    so that $\theta u_1,\theta u_1'\in\R^2_-$
    and $\theta U\subset\R_+^2$.
    We may now find a path from $u_2$ to $u_2'$ consisting of finitely many straight line segments $S\subset U$,
    which each have the property that
    either $S$ is horizontal or $\theta S$ is horizontal (see Figure~\ref{fig:rotation_trick}).
    Write $S_-$ and $S_+$ for the starting point and endpoint of $S$, respectively.
    By additivity of $\Phi_2^{(\delta_n)}$, we get
    \begin{equation}
            \Phi^{(\delta_n)}_2(\u )
            =
            \sum_{S}
            \Phi_2(\tfrac{u_1}{\delta_n},\tfrac{u_1'}{\delta_n},\tfrac{S_-}{\delta_n},\tfrac{S_+}{\delta_n})
    \end{equation}
    As $n\to\infty$, all terms in the finite sum converge:
    if $S$ is horizontal, then the term converges by Lemma~\ref{lemma:initial_scaling};
    if $\theta S$ is horizontal,
    then the term converges by applying Lemma~\ref{lemma:initial_scaling} to the rotated system and
    using Theorem~\ref{thm:rotation_invariance}.

    We now prove that $\Psi_2(u_1,u_1',u_2,\blank)$ is $C^1$ on $U$.
    By Lemma~\ref{lemma:F_properties}(i), the horizontal partial derivative exists and is continuous
    (we also use the additivity property of $\Psi_2$).
    By applying the same reasoning to the rotated system,
    we find that the partial derivative in the direction $\theta^{-1}e_1$ (with $e_1=(1,0)$)
    also exists and is continuous.
    Since the vectors $\{e_1,\theta^{-1}e_1\}$ span the tangent space of $\R^2$,
    we conclude that $\Psi_2(u_1,u_1',u_2,\blank)$ is $C^1$ on $U$.

    \begin{figure}
                \centering
        \includegraphics{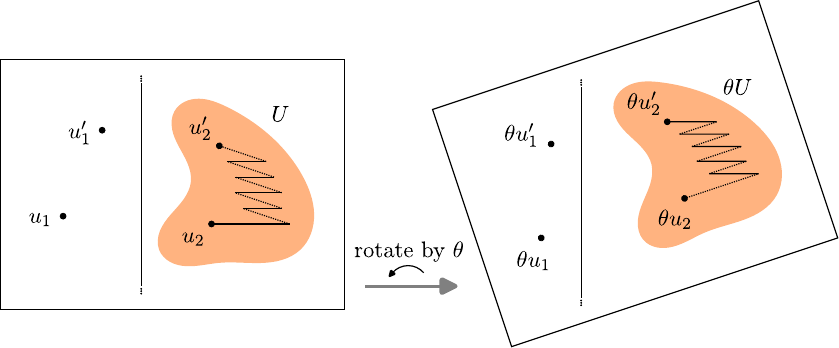}
        \caption{In Step~1 in the proof of Lemma~\ref{lemma:Psi_in_terms_of_I}, we prove that $\Psi_2(u_1,u_1',\blank,\blank)$ is well-defined and $C^1$ on $(\R^2_+)^2$
        by using Lemma~\ref{lemma:F_properties} and rotation invariance of $\Psi_2$.}
        \label{fig:rotation_trick}
    \end{figure}

    \paragraph{Step~2: Proof of Equation~\eqref{eq:F_as_derivative_of_I}.} Readers may help themselves with Figure~\ref{fig:hor_of_rot}.
    Set $u_1'=(0,0)$, $u_2\in\R^2_+$, and $u_1=-(n-1)\cdot u_2$ for some large $n>1$.
    As before, let $(x_1', y_1') = u_2 - u_1'$.
   Recall Equation~\eqref{eq:horizontal_derivative_of_Psi} from Lemma~\ref{lemma:F_properties}.
    There are two observations to make.

First, the dominated convergence theorem implies that $\lim_{n\to\infty}F(nx_1',ny_1')=0$, and therefore
 \begin{equation}\label{eq:h1}
        \lim_{n\to\infty}
        \lim_{\epsilon\to 0} \tfrac1\epsilon
        \Psi_2(u_1,u_1',u_2,u_2+(\epsilon,0))
        =
        \lim_{n\to\infty}
        F(nx_1',ny_1')-F(x_1',y_1')
        =-F(x_1',y_1').
    \end{equation}

	Second, we proved above that $u_2' \mapsto \Psi_2(u_1,u_1',u_2,u_2')$ is totally differentiable on $\bbR^2_+$.
	Furthermore, by Theorem~\ref{thm:rotation_invariance}, it is invariant under the reflection orthogonal to the line containing $u_1$, $u_1'$, and $u_2$.
%   Second,  we know that $u_2' \mapsto \Psi_2($ is rotation invariant and totally differentiable.
%        Moreover, the points $u_1$, $u_1'$, and $u_2$ lie on a line.
        We conclude that the partial derivative at $u_2$ \emph{orthogonal} to this line  is zero.
        It is therefore natural to decompose the partial derivative we analysed above into two:
        one along the line and one orthogonal to the line.
        The derivative along the line can be rewritten in terms of the function $F(\blank,0)$,
        using the rotation invariance of Theorem~\ref{thm:rotation_invariance}.
        By doing so, we get
        \begin{equation}
        \lim_{\epsilon\to 0} \tfrac1\epsilon
        \Psi_2(u_1,u_1',u_2,u_2+(\epsilon,0))
        =
        \left(
            \partial_{x_1'} |(x_1',y_1')|
        \right)\big(
        F(n\cdot |u_2|,0)-F(|u_2|,0)\big).
        \end{equation}
        By sending $n$ to infinity, the first term disappears by dominated convergence,
        so that
        \begin{align}
            \lim_{n\to\infty}
        \lim_{\epsilon\to 0} \tfrac1\epsilon
        \Psi_2(u_1,u_1',u_2,u_2+(\epsilon,0))
        &= -\left(
            \partial_{x_1'} |(x_1',y_1')|
        \right)F(|u_2|,0)\\
        &=\partial_{x_1'} \big[ I_F(|(x_1',y_1')|) \big],\label{eq:h2}
        \end{align}
   where the second line follows from the definition of $I_F$.

       Combining \eqref{eq:h1} and \eqref{eq:h2} yields $-F(x_1',y_1')=\partial_{x_1'} (I_F(|(x_1',y_1')|))$
    as desired.

    \begin{figure}
        \centering
        \includegraphics{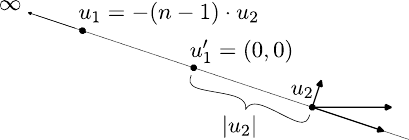}
        \caption{Step~2 in the proof of Lemma~\ref{lemma:Psi_in_terms_of_I}}
        \label{fig:hor_of_rot}
    \end{figure}

    \paragraph{Step~3: Proof of Equation~\eqref{eq:Psi_in_terms_of_I} on $\R^2_-\times\R^2_-\times\R^2_+\times\R^2_+$ when $y_2=0$.}
    Fix $(u_1,u_1',u_2)\in(\R^2_-)^2\times\R^2_+$.
    If $u_2'=u_2$,
    then both sides of Equation~\eqref{eq:Psi_in_terms_of_I} equal zero.
    To extend Equation~\eqref{eq:Psi_in_terms_of_I}  to every $u_2'\in\R^2_+$ on the same horizontal line as $u_2$, it  suffices to prove that the horizontal partial derivatives of both sides
    with respect to $u_2'$ are equal.
    This follows from Equations~\eqref{eq:horizontal_derivative_of_Psi} and~\eqref{eq:F_as_derivative_of_I}.

   \paragraph{Step~4: Proof of Equation~\eqref{eq:Psi_in_terms_of_I} to all of $\calD_2$.}
    The left and right of Equation~\eqref{eq:Psi_in_terms_of_I} are
    compatible with respect to the additivity property first introduced
    in Equation~\eqref{eq:Phi-additive}, and under applying a fixed isometry of the
    plane to all four points.    One may therefore extend Equation~\eqref{eq:Psi_in_terms_of_I}
    from $(\R^2_-)^2\times(\R^2_+)^2$ to all of $\calD_2$.
\end{proof}

% </input src="sections/PART_B/2_1b_reg.tex" root="." version="0.0.1">
% <input src="sections/PART_B/2_2_concentration.tex" root="." version="0.0.1">
\subsection{Concentration on the sub-diagonal}\label{sec:2.4}

For every convergence sequence, the function $\Psi_2$ is entirely determined by the function $I_F$
introduced above. The problem therefore reduces to identifying $I_F$.
If $I_F$ happens to be a multiple of $\log$, then $\Psi_2$ is a scaled version of
$\Psi_2^{\GFF}$, the two-point function of the GFF.

The remainder of this section is devoted to analysing $I_F$.
The analysis consists of two steps.
First, we show that for any convergence sequence, the corresponding measure $\mu$ is supported on
$\{|b|\le a\}$.
In the second step, we use this information to derive further properties of $I_F$,
which ultimately yield the dichotomy stated in
Theorem~\ref{thm:dicht_2p_final_form}.
The first step is essential and constitutes a central component of the article.
We establish it now.

\begin{theorem}
    \label{thm:upperboundonb}
    For every convergence sub-sequence $(\delta_n)_n$, we have
    $\mu[\{|b|>a\}]=0$.
\end{theorem}

We start with two remarks motivating the proof. In the argument, we leverage these two perspectives
and the definition of $F$ in terms of $\mu$
to derive Theorem~\ref{thm:upperboundonb}.
\begin{remark}
 The fact that $F$ is the horizontal partial derivative of a
    radially symmetric function imposes strong constraints on $F$.
    \end{remark}

    \begin{remark} Consider the definition of $F$
and suppose for a second that $|b|=a$ almost everywhere.
Then for any $(x,y)\in\R_+^2$, we get
\begin{equation}
    F(x,y)=\int a e^{-ax-\i by}\diffi\mu(a,b)
    =
    \Re\left(
    \int a e^{-ax-\i |b|y}\diffi\mu(a,b)
    \right)
    =
    \Re(F(x+\i y,0)).
\end{equation}
Although we will not really justify that $|b|=a$,
the above perspective suggests that the expression $F(x+\i y,0)$
provides an interesting, alternative way to pass two-dimensional data to $F$.
\end{remark}

Below, we shall use basic complex analysis tools
such as holomorphic extensions and contour integrals.
Issues such as singularities (poles)
and branch cuts then become important.
In this context, we shall write $\sqrt{\blank}$
for the unique branch cut
\begin{equation}
    \sqrt{\blank}:(\C\setminus\R_{\leq 0})\to\C_+
\end{equation}
which maps positive real numbers to positive real numbers.
This means that we explicitly discard real nonpositive function values for the function $\sqrt{\blank}$.

\begin{lemma}
        \label{lemma:initial_rot_inv}
    For any $(x,y)\in\C_+\times\R$, we have
    \begin{equation}
        \label{eq:rotation_of_F}
        F(x,y) =
        \frac{x}{\sqrt{x^2+y^2}}
        F(\sqrt{x^2+y^2},0).
    \end{equation}
\end{lemma}

\begin{proof}
    If $x$ is real, then this follows from Equation~\eqref{eq:F_as_derivative_of_I} in Lemma~\ref{lemma:Psi_in_terms_of_I}.
    For fixed $y$, both sides in Equation~\eqref{eq:rotation_of_F}
    are holomorphic functions in $x\in\C_+$, thanks to Lemma~\ref{lemma:F_properties}(iii). Since they coincide on the positive real axis, they must be equal for all $x\in\C_+$.
\end{proof}

The following auxiliary lemma is a standard computation of a Fourier transform. We omit the proof.

\begin{lemma}\label{lem:h}
	For fixed $\sigma,\alpha\in\R_{>0}$, the Fourier transform of $\frac{\sigma^2}{t^2+\sigma^2} \frac{\sin(\alpha t)}{t}$
	satisfies
	\begin{equation}
        \label{eq:FFsin}
		\int_\R \frac{\sigma^2}{t^2+\sigma^2} \frac{\sin(\alpha t)}{t} e^{-\i bt}\diffi t
		= \pi\int_{b-\alpha}^{b+\alpha} \tfrac\sigma2e^{-\sigma |y|} \diffi y
		= \pi\cdot(\true{|\cdot| \leq \alpha} * \tfrac\sigma2 e^{-\sigma |\cdot|})(b)
	\end{equation}
    for any $b\in\R$.
\end{lemma}

\begin{proof}[Proof of Theorem~\ref{thm:upperboundonb}]
    Since $ae^{-a}$ is a strictly positive function on $\R_{>0}$,
    it suffices to prove that, for any $\lambda>1$,
    we have
    \begin{equation}
        \label{eq:targetequationcutoff}
        F(1,0)=\int ae^{-a}\diff\mu(a,b)
        =
        \int \true{|b|\leq \lambda a} ae^{-a}\diff\mu(a,b).
    \end{equation}
    Fix $\lambda > 1$.
    The idea is to mollify the indicator and apply a double Fourier transform,
    so that extra factors appear that are compatible with the general definition of $F$
    and Equation~\eqref{eq:rotation_of_F}.
    First, for $\sigma > 0$, define $J_\sigma$ as the integral on the right
    in Equation~\eqref{eq:targetequationcutoff} but with a mollified indicator:
    \begin{equation}
       J_\sigma:=\int \Big(\big(\true{|\cdot|\leq \lambda a}*\tfrac\sigma2e^{-\sigma|\cdot|}\big)(b)\Big) ae^{-a}\diff\mu(a,b).
    \end{equation}
    By dominated convergence, $J_\sigma$ converges to the integral on the right in Equation~\eqref{eq:targetequationcutoff} as $\sigma$ tends to infinity.
    Thus, to prove \eqref{eq:targetequationcutoff}, it suffices to show that
    \begin{equation}
    J_\sigma \xrightarrow[\sigma \to\infty]{}F(1,0). \label{eq:final_limit_to_prove2}
    \end{equation}

    Plugging in the formula of Lemma~\ref{lem:h} above yields
    \begin{equation}
       J_\sigma =
        \int \Big(\int_\R \frac{\sigma^2}{t^2+\sigma^2} \frac{\sin(\lambda a t)}{\pi t} e^{-\i bt}\diffi t\Big) ae^{-a}\diff\mu(a,b).
    \end{equation}
    Since the integrand is of order $O(ae^{-a}/(1+t^2))$,
    which is integrable in this product measure, we can apply Fubini's theorem to the product measure $\diff t\otimes \mu$.
    Rearranging yields
    \begin{equation}
       J_\sigma=
        \frac1\pi
        \int_\R
        \frac1{t}
        \frac{\sigma^2}{t^2+\sigma^2}
       \Big( \int
        \sin(\lambda a t)
        ae^{-a-\i b t}
        \diff\mu(a,b)\Big)
        \diff t.
    \end{equation}
    Using that $\sin(\lambda a t) = \frac1{2\i} (e^{\i \lambda a t} - e^{-\i \lambda a t})$,
    the definition of $F$ gives
     \begin{equation}
        2\i \int
        \sin(\lambda a t)
        ae^{-a-\i b t}
        \diff\mu(a,b)
        =F(1-\i\lambda t,t) - F(1+\i\lambda t,t) =:F_-(t)-F_+(t)
    \end{equation}
   Plugging this into our previous expression for $J_\sigma$ enables to write
    \begin{equation}
        \label{eq:J_sigma_intermediate}
       J_\sigma=
        \frac{1}{2\pi \i}
        \int_\R
        \frac1{ t}
        \frac{\sigma^2}{t^2+\sigma^2}
        \big(F_-(t)-F_+(t)\big)
        \diff t.
    \end{equation}
     We are going to use complex analysis to study this integral,
    see the Figure~\ref{fig:contours} for the different poles and integration paths.

    \begin{figure}
        \centering
        \includegraphics{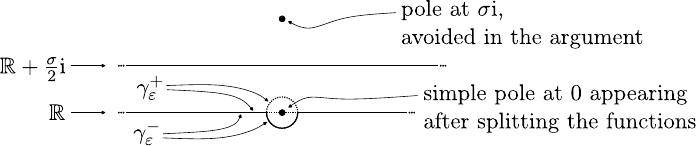}
        \caption{The paths $\gamma^+_\eps$ and $\gamma^-_\eps$ run along $\bbR$ and surround $0$ using a small half-circle in the upper and lower half-plane, respectively. The function $F_- - F_+$ is analytic on the strip $|\Im(z)| < \sigma$, hence $J_\sigma$ may be written as its integral along $\gamma^-_\eps$. When replacing the integral of $F_-$ along $\gamma^-_\eps$ with that along $\gamma^+_\eps$, one should take into account the pole of $F_-$ at $0$. }
        \label{fig:contours}
    \end{figure}

    Let us make some remarks regarding Equation~\eqref{eq:J_sigma_intermediate}.
    \begin{itemize}
    \item Equation~\eqref{eq:rotation_of_F} implies that
 \begin{equation}\label{eq:F_pm}
        F_\pm(t)
        =
        \frac{1\pm\i\lambda t}{\sqrt{(1\pm \i\lambda t)^2+t^2}}
        F(\sqrt{(1\pm \i\lambda t)^2+t^2},0).
    \end{equation}

        \item The function $F_-$ is \emph{a priori} defined on $\R$.
        We may extend $F_-$ and \eqref{eq:F_pm} as equal holomorphic functions on the open set $\calN_-\supset\R$,
        where $\calN_-\subset\C$ is the largest subset of $\C$ such that the square root
        in the definition of $F_-$ is well-defined. Indeed, a soon as the square root is well defined, $F(\sqrt{(1\pm \i\lambda t)^2+t^2},0)$ is also defined, as the square root takes values in $\C_+$.
        Similar considerations apply to $F_+$, which extends to a holomorphic function on the similarly defined $\calN_+$.
        \item The integrand in Equation~\eqref{eq:J_sigma_intermediate} is a holomorphic function on $(\calN_- \cap \calN_+)\setminus\{\pm\sigma\i\}$.
        In particular, there is no simple pole at $t=0$, because the factor $1/t$ is cancelled
        by the vanishing difference of holomorphic functions $F_-(t)-F_+(t)$.
    \end{itemize}
    Let $\gamma^-_\epsilon$ and $\gamma^+_\epsilon$ denote the following paths:
    \begin{itemize}
        \item $\gamma^-_\epsilon$ first runs from $-\infty$ to $-\epsilon$ along $\R$,
        then follows a half-circle of radius $\epsilon$ in the lower half-plane
        from $-\epsilon$ to $\epsilon$,
        and finally runs from $\epsilon$ to $+\infty$ along $\R$,
        \item $\gamma^+_\epsilon$ is defined similarly, except that the half-circle
        runs in the upper half-plane.
    \end{itemize}
    If $\epsilon\in(0,\sigma)$ is so small that a ball of radius $2\epsilon$ around $0$ is contained
    in $\calN_-\cap\calN_+$,
    then the Cauchy integral theorem yields
        \begin{equation}
        \label{eq:J_sigma_intermediate2}
       J_\sigma=
        \frac{1}{2\pi \i}
        \int_{\gamma^-_\epsilon}
        \frac1{ t}
        \frac{\sigma^2}{t^2+\sigma^2}
        \big(F_-(t)-F_+(t)\big)
        \diff t.
    \end{equation}
    We now argue that the functions $F_\pm$ are uniformly bounded on $\gamma^-_\epsilon$.
    First, since they are continuous, it suffices to bound these functions on $\R$.
    But that is easy, since for real $t$, we get
    $|F_\pm(t)|=|F(1\pm \i\lambda t,t)|\leq F(1,0)<\infty$ (see the bound in Lemma~\ref{lemma:F_properties}(iv)).
    We may now split the integral in Equation~\eqref{eq:J_sigma_intermediate2}, which yields
    \begin{equation}
       J_\sigma=
        \frac{1}{2\pi \i}
        \left(
        \int_{\gamma^-_\epsilon}
        \frac1{ t}
        \frac{\sigma^2}{t^2+\sigma^2}
        F_-(t)
        \diff t
        -
        \int_{\gamma^-_\epsilon}
        \frac1{ t}
        \frac{\sigma^2}{t^2+\sigma^2}
        F_+(t)
        \diff t
        \right).
    \end{equation}
    For the integral on the left,
    we would like to replace the path $\gamma^-_\epsilon$ by $\gamma^+_\epsilon$.
    The integrand is holomorphic in the region enclosed by the two paths,
    except for a simple pole at $t=0$.
    By the residue theorem, we get
    \begin{equation}
        \int_{\gamma^-_\epsilon}
        \frac1{ t}
        \frac{\sigma^2}{t^2+\sigma^2}
        F_-(t)
        \diff t
        =
        2\pi\i F_-(0)+
        \int_{\gamma^+_\epsilon}
        \frac1{ t}
        \frac{\sigma^2}{t^2+\sigma^2}
        F_-(t)
        \diff t.
    \end{equation}
    Since $F_-(0)=F(1,0)$,
    \begin{align}
       J_\sigma
       & =
        F(1,0)+
        \frac{1}{2\pi \i}
        \left(
        \int_{\gamma^+_\epsilon}
        \frac1{ t}
        \frac{\sigma^2}{t^2+\sigma^2}
        F_-(t)
        \diff t
        -
        \int_{\gamma^-_\epsilon}
        \frac1{ t}
        \frac{\sigma^2}{t^2+\sigma^2}
        F_+(t)
        \diff t
        \right) \\
      &  =     F(1,0)+
        \frac{1}{\pi}
        \Im \left(
        \int_{\gamma^+_\epsilon}
        \frac1{ t}
        \frac{\sigma^2}{t^2+\sigma^2}
        F_-(t)
        \diff t
              \right),
    \end{align}
    since the two integrals in the first line are equal up to complex conjugation.
%    Since our goal was to derive that $J_\sigma$ tends to $F(1,0)$ as $\sigma$ tends to infinity, to complete the proof it suffices to demonstrate that
We will now prove that
    \begin{equation}
        \label{eq:final_limit_to_prove}
        \lim_{\sigma\to\infty}
        \int_{\gamma^+_\epsilon}
        \frac1{ t}
        \frac{\sigma^2}{t^2+\sigma^2}
        F_-(t)
        \diff t
        =0.
    \end{equation}

    Recall that $F_-$ is a holomorphic function on $\calN_-\supset\R$,
%    where $\calN_-$ is the largest subset of $\C$ such that the square root
%    in the definition of $F_-$ remains well-defined.
%    Recall also
and that $F_-$ is uniformly bounded on $\R$ by $F(1,0)$.
    The following claim asserts an even better control on the function $F_-$.

    \begin{claim*}
       {\em
        All of the following hold true:
            \begin{enumerate}
                \item The set $\calN_-$ contains $\H^+:=\{z\in\C:\Im(z)>0\}$,
                \item The function $|F_-|$ is uniformly bounded on $\H^+$,
                \item For any $s\in\H^+$, we have
            $\lim_{\sigma\to\infty}F_-(\sigma s)=0$.
            \end{enumerate}
        }
    \end{claim*}

    We shall first see that the Claim implies Equation~\eqref{eq:final_limit_to_prove}.
    By another application of the Cauchy integral theorem,
    and a change of variables $t=\sigma s$,
    \begin{equation}
        \int_{\gamma^+_\epsilon}
        \frac1{ t}
        \frac{\sigma^2}{t^2+\sigma^2}
        F_-(t)
        \diff t
        =
        \int_{\R+\tfrac\sigma2\i}
        \frac1{ t}
        \frac{\sigma^2}{t^2+\sigma^2}
        F_-(t)
        \diff t
        =
        \int_{\R+\tfrac12\i}
        \frac1{s}
        \frac{1}{s^2+1}
        F_-(\sigma s)
        \diff s.
    \end{equation}
    By the dominated convergence theorem,
    the right-hand side tends to zero as $\sigma\to\infty$,
    which implies Equation~\eqref{eq:final_limit_to_prove} and Equation~\eqref{eq:final_limit_to_prove2} follows.
   Thus, the proof is complete, modulo the Claim which is proved below.\end{proof}

 To conclude this section, we prove the claim invoked in the argument above.
 Notice that we have not used the property that $\lambda>1$;
    we shall do so now (we use that $\lambda^2-1>0$).

  \begin{proof}[Proof of the Claim]
    Introduce $\alpha:\C\to\C,\,t\mapsto(1- \i\lambda t)^2+t^2$ so that
    \begin{equation}
        \label{eq:F_minus_in_terms_of_alpha}
        F_-(t)
        =
        \frac{1-\i\lambda t}{\sqrt{\alpha(t)}}
        F(\sqrt{\alpha(t)},0).
    \end{equation}
    We now make some simple observations (see also Figure~\ref{fig:parabola}).
    \begin{itemize}
        \item For any $t\in\C \setminus \R_{<0}$, we have $\Re(\sqrt{t})\geq 1$ if and only if $t$ is on or to the right of the parabolic curve
        \begin{equation}
            \{(1+\i t)^2:t\in\R\}
            =
            \{1-t^2+2\i t:t\in\R\}
            \subset\C.
        \end{equation}
        \item From the definition of $\alpha(t)$, it is immediate that $\alpha(t)$ lies on or to the right of this parabola
        for real $t$.
        Thus, we get $\Re(\sqrt{\alpha(t)})\geq 1$ for any $t\in\R$.
        \item For $x\in\R$ and $y\in\R_{\geq 0}$, we may write out $\alpha(x+\i y)$ explicitly:
        \begin{equation}
            \label{eq:explicit_alpha}
            \alpha(x+\i y)=
            1+2\lambda y + (\lambda^2-1)(y^2 - x^2)
            -2\i x(\lambda + (\lambda^2-1) y).
        \end{equation}
        From this, it is immediate that:
        \begin{gather}
            \Re(\alpha(x+\i y))\geq \Re(\alpha(x));\qquad
            |\Im(\alpha(x+\i y))|\geq |\Im(\alpha(x))|.
        \end{gather}
        In particular, since $\alpha(x)$ lies on or to the right of the parabola,
        $\alpha(x+\i y)$ also lies on or to the right of the parabola.
        We conclude that $\Re(\sqrt{\alpha(t)})\geq 1$ for any $t\in\H^+$.
    \end{itemize}

    \begin{figure}
        \centering
        \includegraphics{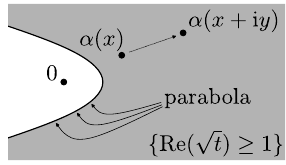}
        \caption{}
        \label{fig:parabola}
    \end{figure}

    We have now proved $\calN_-\supset\H^+$, i.e.~Property (i) of the claim.
    Also, with the help of Lemma~\ref{lemma:F_properties}(iv) we get that, on $\H^+$,
    \begin{equation}
        |F(\sqrt{\alpha(t)},0)|\leq F(\Re(\sqrt{\alpha(t)}),0) \leq F(1,0).
    \end{equation}
    This yields Property~(ii)
    since the factor $(1-\i\lambda t)/\sqrt{\alpha(t)}$ is also bounded on $\H^+$.
    It remains to establish Property~(iii). Fix $s\in\H^+$
    and set $t=\sigma s$.
    The factor $(1-\i\lambda t)/\sqrt{\alpha(t)}$  stays bounded like before;
    it suffices to show that $\Re(\alpha(\sigma s))$  converges to infinity as $\sigma$ tends to infinity,
    so that Lemma~\ref{lemma:F_properties}(iv) implies that
    \begin{equation}
        |F(\sqrt{\alpha(\sigma s)},0)|\leq F(\Re(\sqrt{\alpha(\sigma s))},0) \to 0.
    \end{equation}

    The explicit formula for $\alpha(\sigma s)$ above enables us to choose
     $\epsilon=\epsilon_s\in(0,\pi)$ such that,
    for any $R>0$ and sufficiently large $\sigma$,
    \begin{equation}
        \alpha(\sigma s)\in \{r e^{\i \theta} : \theta \in (-(\pi-\epsilon), \pi-\epsilon),\,r>R^2\}.
    \end{equation}
    By the standard way of viewing the square root (half the angle, square root of the modulus),
    this implies that
    \begin{equation}
        \liminf_{\sigma\to\infty} \Re(\sqrt{\alpha(\sigma s)}) \geq \sin(\tfrac{\pi-\epsilon}{2})\cdot R.
    \end{equation}
    Since $R$ was arbitrary, the limit is infinite.
\end{proof}

\begin{remark}[The role of $\lambda>1$]
    Since our main result implies that $\mu[\{a\neq b\}]=0$,
    we may wonder how the argument fails when $\lambda<1$.
    In that case, $\lambda^2-1<0$,
    and Equation~\eqref{eq:explicit_alpha} implies
    that $\alpha(t)$ has a zero in the upper half-plane
    at some point $t=\i y$ with $y>0$.
    The function $F_-$ (Equation~\eqref{eq:F_minus_in_terms_of_alpha})
    exhibits singular behaviour around this point, which breaks the above argument.
\end{remark}

% </input src="sections/PART_B/2_2_concentration.tex" root="." version="0.0.1">
% <input src="sections/PART_B/2_3_subseq.tex" root="." version="0.0.1">
\subsection{Derivation of the dichotomy}\label{sec:2.5}

For any convergence sequence,
we know that the associated analytic function $I_F$ is decreasing,
since $I_F'(x)=-F(x,0)\leq 0$.
If $I_F(s)\propto -\log s$,
then $\Psi_2\propto\Psi_2^{\GFF}$
by Equation~\eqref{eq:Psi_in_terms_of_I} of Lemma~\ref{lemma:Psi_in_terms_of_I}.
To derive the dichotomy of Theorem~\ref{thm:dicht_2p_final_form},
we must understand what happens when the analytic function $I_F$ is not of this form.
It turns out to be beneficial to encode the property that $I_F(s)\propto -\log s$
in terms of a new function $\Xi$, defined as
\begin{equation}
    \Xi:\R_{>0}\to\R_{\geq 0},\,s\mapsto -sI_F'(s)=sF(s,0).
\end{equation}
Note that $I_F(s)\propto-\log s$ if and only if $\Xi$ is constant.
We now prove the following result on $\Xi$.
\begin{theorem}
    \label{thm:Xi_derivative}
    For any convergence sequence $(\delta_n)_n$,
   the function $\Xi$ is bounded and
    \begin{equation}\label{eq:non-increasing}
        -\Xi'(s)
    =
    s \Delta I_F(|u|)|_{u=se_1}
    =
    s \int (a^2-b^2) e^{-as} \diffi\mu(a,b)
    \geq 0.
    \end{equation}
    In particular, $\Xi$ is a non-increasing function.
\end{theorem}

Before diving into the proof, let us extend the applicability of Lemma~\ref{lemma:initial_scaling}. %Below, let $x(u)$ and $y(u)$ be the first and second coordinates of the point $u$.
\begin{lemma}[Extension of Lemma~\ref{lemma:initial_scaling}]
    \label{lemma:dropcondition}
    Let $(\delta_n)_n$ denote a convergence sequence.
    For every $\u =(u_1,u_1',u_2,u_2')\in (\R^2)^4$  whose first coordinates of $u_1$ and $u_1'$ are strictly smaller than those of $u_2$ and $u_2'$,
      \begin{equation}
        \label{eq:dropcondition_recap}
        \Psi_2(\u ):=\lim_{n\to\infty}
        \Phi_2^{(\delta_n)}(\u )
        =
        \int\chi_\u (a,b)\diffi\mu(a,b).
    \end{equation}
\end{lemma}

\begin{proof}
   Fix $\u $ satisfying the assumptions of the lemma. Above, we proved already that $\Psi_2(\u )$ exists,
    and that the function $\chi_\u $ is $\mu$-integrable.
    Indeed, any term of the form $1-e^{-ax-\i b y}$ is of order $O(a)$
    thanks to Theorem~\ref{thm:upperboundonb}.
    Thus, the left and right of Equation~\eqref{eq:dropcondition_recap}
    are well-defined; it remains to prove equality.

    Set $e_1=(1,0)$. We may use the additivity property to write
    the height difference $h(u_2'/\delta_n)-h(u_2/\delta_n)$ as the sum of height differences
    along the following path:
    \begin{equation}
       \frac{u_2}{\delta_n} \to \frac{u_2+ne_1}{\delta_n}  \to \frac{u_2'+ne_1}{\delta_n} \to \frac{u_2'}{\delta_n} .
    \end{equation}
    Set ${\s ^n}:=(u_1,u_1',u_2,u_2+ne_1)$ and ${\t ^n}:=(u_1,u_1',u_2'+ne_1,u_2')$.
    We claim that
    \begin{equation}
        \Psi_2(\u )=\lim_{n\to\infty}(\Psi_2({\s ^n})+\Psi_2({\t ^n})).\end{equation}
      For this to be true, we must justify that the correlation function of the middle step in the path
    does not appear. Yet, it tends to zero thanks to the regularity estimate (Theorem~\ref{thm:Regularity}).

  We may therefore apply (twice) the case where the two last points are on the same horizontal line to get
   \begin{equation}
        \Psi_2(\u )=\lim_{n\to\infty}\int(\chi_{{\s ^n}}+\chi_{{\t ^n}})(a,b)\diffi\mu(a,b).\end{equation}
  Finally, the dominated convergence   theorem,
    where we use Theorem~\ref{thm:upperboundonb}
    to find the desired dominating function, enables us to insert  the missing part of the function $\chi_\u $ to get
    \begin{equation}
          \Psi_2(\u )
    =\int\chi_\u (a,b)\diffi\mu(a,b).
    \end{equation}
 This concludes the proof.   \end{proof}

\begin{proof}[Proof of Theorem~\ref{thm:Xi_derivative}]
    To see that $\Xi$ is bounded, recall from the definitions that
    \begin{equation}
        \Xi(s)=s\int a e^{-as} \diffi\mu(a,b).
    \end{equation}
    The bounds on $\mu\in\calM$ thus imply that $\Xi$ is bounded.

    We now focus on proving \eqref{eq:non-increasing}.
The equality
on the right of \eqref{eq:non-increasing} is Theorem~\ref{thm:upperboundonb}.
For the equality on the left, note that  the radial symmetry of $I_F(|\cdot|)$ enables us to replace $\partial_x$ by $s\partial_{yy}$,
to get that
\begin{align}
    -\Xi'(s)
    &=
    I_F'(s) + s I_F''(s)\\
    &=
    (\partial_x + s \partial_{xx} ) I_F(|(x,y)|)|_{(x,y)=se_1}
    \\
    &=
    s(\partial_{yy}+\partial_{xx}) I_F(|(x,y)|)|_{(x,y)=se_1}\\
    &=
    s \Delta I_F(|u|)|_{u=se_1}
    .
\end{align}
To get \eqref{eq:non-increasing}, it only remains to justify that   \begin{equation}
        \Delta I_F(|u|)|_{u=se_1}
        =
        \int (a^2-b^2) e^{-as} \diffi\mu(a,b).
    \end{equation}
    Set $u_1=-ne_1$, $u_1'=(0,0)$,
    $u_2=se_1$, and $u_2'=u_2+(x,y)$ (with $|(x,y)|$ tiny).
    Using Lemma~\ref{lemma:dropcondition} and letting $n$ tend to infinity gives       \begin{equation}
        \lim_{n\to\infty}
        \Psi_2(u_1,u_1',u_2,u_2')
        =
        \int
        e^{-a s}
        (e^{-a x - \i b y}-1)
        \diffi\mu(a,b).
    \end{equation}
Using Lemma~\ref{lemma:Psi_in_terms_of_I} and letting $n$ tend to infinity, we get another expression for the left-hand side of the previous equality:
    \begin{equation}
        \lim_{n\to\infty}
        \Psi_2(u_1,u_1',u_2,u_2')
        =
        I_F(|se_1+(x,y)|)-I_F(|se_1|).
    \end{equation}
    Equalling the two previous identities gives
       \begin{equation}
         I_F(|se_1+(x,y)|)-I_F(|se_1|)= \int
        e^{-a s}
        (e^{-a x - \i b y}-1)
        \diffi\mu(a,b)
      .
    \end{equation}
    Now, calculating the Laplacian in $(x,y)$ at the point $(0,0)$
    yields
    \begin{equation}
        \Delta I_F(|u|)|_{u=se_1}
        =
        \int (a^2-b^2) e^{-as} \diffi\mu(a,b).
    \end{equation}
   We used that we may differentiate under the integral
    thanks to the dominated convergence theorem,
    using that $a\wedge \frac1a$ is integrable,
    and that $|b|\leq a$ (Theorem~\ref{thm:upperboundonb}).
\end{proof}

Before concluding the proof of the dichotomy, we summarise what has been established so far.
    For any sequence $(\delta_n)_n$ tending to zero,
    we may extract a sub-sequence $(\delta'_n)_n$
    and find some analytic function $I_F:\R_{>0}\to\R$
    such that all of the following hold true:
    \begin{itemize}
        \item For any $\u \in\calD_2$,
        \begin{equation}
            \Psi_2(\u ):=\lim_{n\to\infty}
            \Phi_2(\u /\delta_n)
            =
            I_F(|u_2-u_1|)+I_F(|u_2'-u_1'|)-I_F(|u_2'-u_1|)-I_F(|u_2-u_1'|).
        \end{equation}
        \item The function $-sI_F'(s)$ is nonnegative, bounded, and non-increasing.
    \end{itemize}
We are now ready to prove the dichotomy result.

\begin{proof}[Proof of Theorem~\ref{thm:dicht_2p_final_form}]
%    In this proof, we will manipulate different sequences. We therefore add a superscript to refer to the sequences.
%
  We start by observing that trivially at least one of the following three statements
  must hold:
    \begin{enumerate}
        \item Either there exists $\sigma$ such that for {\em every} convergence sequence $(\delta_n)_n$, $I_F=-\frac{\sigma^2}{2\pi}\log,$
        \item Or there exist two different $\sigma,\sigma'$ and two convergence sequences $(\delta_n)_n$ and $(\delta'_n)_n$ such that
        $I_F$ is equal respectively to $-\frac{\sigma^2}{2\pi}\log$ and $-\frac{(\sigma')^2}{2\pi}\log$,        \item Or for some convergence sequence $(\delta_n)_n$, the function $I_F$ is not proportional to $-\log$.
    \end{enumerate}
    The first two cases clearly enter into the framework of the dichotomy of Theorem~\ref{thm:dicht_2p_final_form}. Indeed, Lemma~\ref{lemma:Psi_in_terms_of_I}(ii) then implies that $\Phi_2^{(\delta_n)}$ converges to its GFF counterpart.   As a consequence, (i) corresponds to the first case of the dichotomy, and (ii) to the second.  To conclude the proof, we only need prove that (iii) also implies the second case of the dichotomy.
    From now on, fix some convergence sequence $(\delta_n)_n$ such that $I_F\not\propto-\log$.

    Using Theorem~\ref{thm:Xi_derivative}, define $\sigma,\sigma'\in\R_{\geq 0}$ such that
    \begin{equation}
        \frac{\sigma^2}{2\pi}=\lim_{s\to 0}-sI_F'(s)
        >
        \lim_{s\to \infty}-sI_F'(s)
        = \frac{(\sigma')^2}{2\pi}.
    \end{equation}
    From the properties in the previous lemma we see that, for any $s\in\R$, we get
    \begin{equation}
        \lim_{\theta\to 0}I_F(\theta s)-I_F(\theta) = -\frac{\sigma^2}{2\pi}\log s;
        \qquad
        \lim_{\Theta\to\infty} I_F(\Theta s)-I_F(\Theta) = -\frac{(\sigma')^2}{2\pi}\log s.
    \end{equation}
    In particular, Lemma~\ref{lemma:Psi_in_terms_of_I}(ii) implies that for any $\u \in\calD_2$,
    \begin{equation}
        \lim_{\theta\to 0}\lim_{n\rightarrow\infty} \Phi_2^{(\delta_n/\theta)}(\u )
        =
        \sigma^2\Psi_2^{\GFF}(\u );
        \qquad
        \lim_{\Theta\to\infty}\lim_{n\rightarrow\infty}\Phi_2^{(\delta_n/\Theta)}(\u )
        =
        (\sigma')^2\Psi_2^{\GFF}(\u ).
    \end{equation}
    This means that we can take two ``scaling limits of the sub-sequential scaling limit'' :
    one obtained by ``zooming in'' and the other by ``zooming out''. We present the zooming-in construction; the zooming-out case is analogous.

Fix a dense countable family $(\u _k)_k$ of quadruplets in $\calD_2$.
Choose a sequence $(\theta_k)_k$ tending to zero such that, for every $k$,
\begin{equation}
    \bigl| \Psi_2(\theta_k \u _i) - \sigma^2 \Psi_2^{\GFF}(\u _i) \bigr|
    \le \tfrac{1}{k}
    \qquad \forall\, i \le k.
\end{equation}
Next, choose an increasing sequence $(n_k)_k$ such that, for every $k$, $\delta_{n_k}/\theta_k\le \tfrac1k$ and
\begin{equation}
     \bigl| \Phi_2^{(\delta_{n_k})}(\theta_k \u _i)
           - \Psi_2(\theta_k \u _i) \bigr|
    \le \tfrac{1}{k}
    \qquad \forall\, i \le k.
\end{equation}
With this choice the sequence $(\delta_{n_k}/\theta_k)_k$  satisfies
\begin{equation}
    \bigl| \Phi_2^{(\delta_{n_k}/\theta_k)}(\u _i)
           - \sigma^2 \Psi_2^{\GFF}(\u _i) \bigr|=\bigl| \Phi_2^{(\delta_{n_k})}(\theta_k\u _i)
           - \sigma^2 \Psi_2^{\GFF}(\u _i) \bigr|
    \le \tfrac{2}{k}
    \qquad \forall\, i \le k.
\end{equation}
Hence, for every $i$, the sequence
$(\Phi_2^{(\delta_{n_k}/\theta_k)}(\u _i))$ converges pointwise as $k$ tends to infinity
to $\sigma^2 \Psi_2^{\GFF}(\u _i)$.
By Corollary~\ref{cor:precompactness_phi},
this means that $\Phi_2^{(\delta_{n_k}/\theta_k )}$ converges to $\sigma^2 \Psi_2^{\GFF}$
uniformly on compact subsets of $\calD_2$.
The same construction yields a sequence converging to $(\sigma')^2 \Psi_2^{\GFF}$.
\end{proof}

% </input src="sections/PART_B/2_3_subseq.tex" root="." version="0.0.1">

% <input src="sections/PART_B/3_0_intro.tex" root="." version="0.0.1">
\section{Sub-sequential GFF limits for multi-point functions}
\label{sec:multi-point}

This section is dedicated to proving the following result.

\begin{theorem}
    \label{thm:k2impliesallk}
    Fix $\sigma\in\R_{\geq 0}$ and a sequence $(\delta_n)_n$ tending to zero.
    Consider the following statement for fixed $k$:
    \begin{equation}
        \Phi_{k}^{(\delta_n)}|_{\calD_k} \xrightarrow[n\to\infty]{}       \sigma^k  \Psi_{k}^{\GFF} \qquad\text{uniformly on compacts subsets of $\calD_k$.}
    \end{equation}
    If this statement holds true for $k=2$, then it holds true for all $k\in\Z_{\geq 1}$.
\end{theorem}

Recall Corollary \ref{cor:precompactness_phi}, which guarantees precompactness. Hence, after passing to a sub-sequence, we may assume that the functions converge.
Throughout this section, we therefore fix $\sigma$ and $(\delta_n)_n$ such that:
\begin{itemize}
\item   $\Phi_2^{(\delta_n)}|_{\calD_2}$ converges to $\sigma^2  \Psi_2^{\GFF}$,
\item
$  \Phi_{k}^{(\delta_{n})}|_{\calD_k}$ converges to some continuous function $\Psi_k$.
\end{itemize}
It suffices to prove that $\Psi_k=\sigma^k \Psi_{k}^{\GFF}$ for every $k$.

The proof goes in two steps.
First, we prove that the scaling limit $\Psi_k$ of the $ \Phi_{k}^{(\delta_n)}$ is harmonic in each coordinate, harvesting the concentration of the measure $\mu$ on $\{b = \pm a\}$.
Once harmonicity has been established, we identify $\Psi_k=\sigma^k\Psi_{k}^{\GFF}$
as the unique function that is harmonic and satisfies a few other properties (limit at infinity, behaviour when merging points)
that $\Psi_k$ possesses.
The second step uses induction on $k$,
with the base case $k=2$ being our hypothesis.

% </input src="sections/PART_B/3_0_intro.tex" root="." version="0.0.1">
% <input src="sections/PART_B/3_1_harmonicity.tex" root="." version="0.0.1">

\subsection{Harmonicity}

\begin{proposition}
    \label{proposition:harmonicity}
    The function $\Psi_k$ is harmonic in each coordinate.
\end{proposition}

In order to prove this proposition, we introduce the set
\begin{equation}
    \calH:=\{\u =(u_1,u'_1,\dots,u_k,u'_k)\in\calD_k: \text{$\Psi_k$ is harmonic in $u_1$ at $\u $}\}.
\end{equation}
Since $\Psi_k$ is naturally invariant or antisymmetric under various permutations
of its arguments, it suffices to show that $\calH = \calD_k$.
Observe also that, for harmonicity in $u_1$, the position of $u_1'$ plays no role.
We begin with the following lemma, which asserts that as long as $u_1$ does not lie in the
convex hull of the points $u_2, u_2', \ldots, u_k, u_k'$, harmonicity basically follows
from the arguments established so far.

\begin{lemma}
    \label{lemma:harmonic_outside_convex_hull}
    We have
    \begin{equation}
        \calZ_0:=\{\u =(u_1,u'_1,\ldots,u_k,u'_k)\in\calD_k:u_1\not\in\operatorname{ConvexHull}(\{u_2,u_2',\ldots\})\}\subset\calH.
    \end{equation}
\end{lemma}

\begin{proof}
    Fix $\u \in\calZ_0$.
    Let $C_r\subset \R^2$ denote the circle of radius $r>0$ centred
    at $u_1$.
    It suffices to prove that if $r$ is smaller than the distance from $u_1$
    to the convex hull,
    then
    \begin{equation}
        \Upsilon:=\int_{C_r}\Psi(u_1,s,u_2,u_2',u_3,u_3',\ldots)\diffi s=0
    \end{equation}
    (the integral is just the uniform probability measure on $C_r$).

    By applying an isometry of the plane, we may assume without loss of generality that $C_r$ is strictly on the left of the vertical axis $\{0\}\times\R$,
    while the convex hull is strictly on the right of this axis.

    To prove that $\Upsilon=0$ we are going to use the
    spectral representation of general observables, the Cauchy--Schwarz inequality
    (Theorem~\ref{thm:spectral_representation_multi_point}),
    and harmonicity of $\sigma^2\Psi_2^{\GFF}$.
    For fixed $n$, introduce the following observables belonging to
    $\frakA^-$ and $\frakA^+$ (cf. Equation~\eqref{definitionfrakApm})  respectively:
    \begin{align}
        {\bf X}_n (h)
        &:=
        \int_{C_r} \left(h(s/\delta_n)-h(u_1/\delta_n)\right)\diff s
        ;\\
        {\bf Y}_n(h)&:=
        \prod_{i=2}^k \left(h(u_i'/\delta_n)-h(u_i/\delta_n)\right).
    \end{align}
    By Theorem~\ref{thm:spectral_representation_multi_point},
    we have
    \begin{equation}
        |\E_{\CYL{L}}[{\bf X}_n {\bf Y}_n]|
        \leq
        \sqrt{
        \E_{\CYL{L}}[{\bf X}_n {\bf X}_n^\dagger ]\E_{\CYL{L}}[{\bf Y}_n^\dagger {\bf Y}_n ]
        }.
    \end{equation}
    The left-hand side converges to $|\Upsilon|$
    as we take $L$ and then $n$ to infinity.
    It suffices to prove that the right-hand side converges to $0$ in the same double limit.
    Yet,
    \begin{equation}
        ({\bf X}_n {\bf X}_n^\dagger)(h)=
        \left({\textstyle \int_{C_r} \left(h(s/\delta_n)-h(u_1/\delta_n)\right)\diffi s}\right)
        \left({\textstyle \int_{C_r} \left(h^\dagger(s/\delta_n)-h^\dagger(u_1/\delta_n)\right)\diffi s}\right).
    \end{equation}
 The $\E_{\CYL{L}}$-expectation of this random variable tends to zero in the double limit
    by uniform convergence and harmonicity of $\sigma^2\Psi_2^{\GFF}$.

    It remains to demonstrate that the expectation of ${\bf Y}_n^\dagger {\bf Y}_n$
  stays uniformly bounded in the double limit.
    But this is obvious since it tends to the bounded function
    \begin{equation}
        \Psi_{2k-2}\big(u_2,u_2',\ldots,u_k,u_k', (u_2)^{\dagger},(u_2')^{\dagger},\dots,(u_k)^{\dagger},(u_k')^{\dagger})<\infty.
    \end{equation}
    Here, we remind that $(u)^{\dagger}$ has been introduced in Subsection~\ref{SoussousSectionSpectralRepGnlObs} and corresponds to the reflection of $u$
    over $\{ \tfrac{1}{2}\} \times \R$.
    This concludes the proof.
\end{proof}

We now extend harmonicity from $\calZ_0$ to all of $\calD_k$ (see Figure~\ref{fig:split} for an illustration). We start by a technical lemma.
\begin{lemma}
    Suppose that $\tilde \u \in\calD_k$ may be obtained from $\u =(u_1,u'_1,\dots,u_k,u'_k)\in\calD_k$
    by choosing a line $L\subset\R^2$ such that
        $u_i$ and $u_i'$ lie strictly on the same side of $L$ for each $i$,
        and then moving all points on one side of the line by a distance of $\lambda_0\in\R_{\geq 0}$ perpendicularly away from $L$.

    Then, $\u \in\calH$ whenever $\calN_{\tilde \u }\subset\calH$ for some neighbourhood $\calN_{\tilde \u }$ of $\tilde \u $.
\end{lemma}

\begin{proof}
    By applying an isometry of the plane and permuting the different $\{u_i,u'_i\}$, we can assume without loss of generality that $L=\{0\}\times\R$,
    that the pairs $\{u_i,u_i'\}$ lie on the left of $L$
    for $i\leq \ell$ and on the right of $L$ for $i>\ell$,
    and that we move the points on the right of $L$ to the right by $\lambda_0\in\R_{\geq 0}$.

    Set $e_1=(1,0)$.
    For $s\in \R^2$ and $\lambda\in\R_{\geq 0}$, define
    \begin{equation}
        \u ^{-,s}:=(u_1,s,u_2,u_2',\ldots,u_\ell');
        \qquad
        \u ^{+,\lambda}:=(u_{\ell+1}+\lambda e_1,u_{\ell+1}'+\lambda e_1,\ldots,u_k'+\lambda e_1).
    \end{equation}
    For any $r>0$, we let $C_r$ denote the circle of radius $r$ around $u_1$.
    We claim that the function
    \begin{equation}
        \Upsilon_r(\lambda):=\int_{C_r}\Psi_k(\u ^{-,s},\u ^{+,\lambda})\diffi s
    \end{equation}
    satisfies $\Upsilon_r(0)=0$ for small enough $r$, which would establish the harmonicity and the lemma.

    Fix $r$ at least so small that the convex hull of $C_r$ does not intersect
    $L$ or $\cup_{i>1}\{u_i,u_i'\}$.
   Since $\Upsilon_r:\R_{\geq 0}\to\R$ is a continuous function
       and $\Upsilon_r(\lambda)=0$ for $\lambda\approx\lambda_0$ and $r$ small enough and $\calN_{\tilde \u }\subset\calH$,
       it suffices to demonstrate that $\Upsilon_r|_{(0,\infty)}$ is an analytic function.

    For fixed $n$, introduce the following observables belonging to
    $\frakA^-$ and $\frakA^+$ respectively:
    \begin{align}
        &{\bf X}_n (h)
        :=
        \left(\textstyle\int_{C_r} \left(h(s/\delta_n)-h(u_1/\delta_n)\right)\diffi s\right)
        \prod_{i=2}^\ell
        \left(h(u_i'/\delta_n)-h(u_i/\delta_n)\right)
        ;
        \\
        &{\bf Y}_n(h):=
        \prod_{i=\ell+1}^k \left(h(u_i'/\delta_n)-h(u_i/\delta_n)\right).
    \end{align}
    Theorem~\ref{thm:spectral_representation_multi_point}
    implies that  for any $\lambda>0$,
        \begin{equation}
            \Upsilon_r(\lambda)
            =
            \lim_{n\to\infty}\lim_{L\to\infty}\int (1-a)^{\lfloor \lambda/\delta_n\rfloor} \diffi\mu_{{\bf X}_n,{\bf Y}_n,L}
            ,
        \end{equation}
where $\mu_{{\bf X}_n,{\bf Y}_n,L}$ satisfies the following bound on its total variation
        \begin{equation}
            \|\mu_{{\bf X}_n,{\bf Y}_n,L}\|
            \leq \sqrt{
                \E_{\CYL{L}}[{\bf X}_n{\bf X}_n^\dagger]
                \E_{\CYL{L}}[{\bf Y}_n^\dagger{\bf Y}_n]
            }.
        \end{equation}
    Now, Theorem~\ref{thm:Regularity} gives the following.
     Since there is some constant $C<\infty$ such that $\E_{\Z^2}[{\bf X}_n{\bf X}_n^\dagger]
    \E_{\Z^2}[{\bf Y}_n^\dagger{\bf Y}_n]\leq C^2$ for any $n$,
    we may find a complex measure $\mu_{{\bf X},{\bf Y}}$ on $[0,\infty]$
    such that $\|\mu_{{\bf X},{\bf Y}}\|\leq C$ and
 $\mu_{{\bf X},{\bf Y}}$ is a sub-sequential limit in the weak topology on complex measures on the compact
        set $[0,\infty]$ of the family $(\mu_{{\bf X}_n,{\bf Y}_n,L,\delta_n})_{n,L}$ as $L$
        and then $n$ tend to infinity.

    In particular, these statements imply that for any $\lambda>0$, we have
    \begin{equation}
        \Upsilon_r(\lambda)=\int e^{-a\lambda} \diffi\mu_{{\bf X},{\bf Y}}(a).
    \end{equation}
    Then, $\Upsilon_r|_{(0,\infty)}$ is the Laplace transform of a finite complex measure on $[0,\infty]$, or equivalently the linear combination (with complex coefficients) of four Laplace transforms
    of finite positive measure on $[0,\infty]$. It is automatically analytic.
\end{proof}

\begin{figure}
    \centering
    \includegraphics{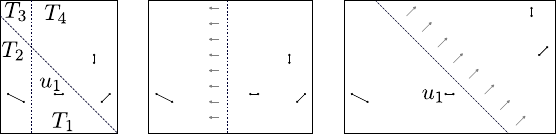}
    \caption{}
    \label{fig:split}
\end{figure}

We are now ready to prove full harmonicity.

\begin{proof}[Proof of Proposition~\ref{proposition:harmonicity}]
Consider the set (see Figure~\ref{fig:split})
\begin{equation}
    \label{eq:calD_k_prime}
    \calD_k':=\{\u \in\calD_k:\,\min_{i\neq j}|u_i-u_j|>1000k^2 \max_i |u_i'-u_i|\}.
\end{equation}
We proceed in two steps. We first show that $\calD'_k\subset \calH$, and then show that $\calD_k\subset \calH$, which is equivalent to the claim.

\paragraph{Step 1: proof that $\calD'_k \subset \calH$.}
Recall the definition of $\calZ_0$ from Lemma~\ref{lemma:harmonic_outside_convex_hull}.
Define inductively the sets $\calZ_{t+1}$ (for $t\in\mathbb{Z}_{\ge0}$) as the collection of elements
$\u \in\calD_k$ obtained from some $\tilde\u \in\calZ_t$ via the operation described in the previous lemma.
By construction, the previous lemma implies that $\cup_t\calZ_t\subset\calH$.

It suffices to prove that $\calD'_k \subset \calZ_2$.
Fix $\u \in\calD'_k$. We can choose two lines $L$ and $L'$ (see Figure~\ref{fig:split}) such that
\begin{itemize}
\item $\mathbb R^2\setminus (L\cup L')$ is composed of four infinite ``triangular'' connected components $T_1,\dots,T_4$  indexed in clockwise order around the intersection point $L\cap L'$,
    \item $u_1$ and $u_1'$ lie in $T_1$,
    \item each pair $\{u_i,u_i'\}$ (for $i\ge2$) is contained either in $T_2$ or in $T_4$.
\end{itemize}

Translate all points in $T_2$ by a distance $\lambda\gg1$ in the direction perpendicular to the line separating $T_1$ and $T_2$. Next, translate all points in $T_4$ by a distance $\lambda'\gg1$ in the direction perpendicular to the line separating $T_1$ and $T_4$.

After these translations, it is straightforward to verify that $u_1$ and $u_1'$ no longer lie in the convex hull of the remaining points.
The resulting $2k$-tuple therefore belongs to $\calZ_0$.
By definition of the sets $\calZ_t$, the configuration before the second translation lies in $\calZ_1$, and the original configuration $\u $ lies in $\calZ_2$.

\paragraph{Step 2: proof that $\calD_k\subset \calH$.}
     Fix $\u =(u_1,u'_1,\dots,u_k,u'_k)\in\calD_k$. We write $ \Psi_k(\u )$ as a sum of increments that are of the form  $\Psi_k(\v )$ for $\v \in\calD'_k$.

  Since harmonicity in $u_1$ does not depend on the position of $u'_1$, we may assume without loss of generality that $u_1'$ is extremely close to $u_1$.
    Let $(\gamma_i)_{i=2,\ldots,k}$ denote a family of disjoint smooth curves
    $[0,1]\to\R^2$
    of finite length
    where $\gamma_i$ starts at $u_i$ and ends at $u_i'$.
    Then, for any $N\in\Z_{\geq 1}$, linearity of expectation implies that
    \begin{equation}
        \Psi_k(\u )=\sum_{n_2=1}^{N}\cdots \sum_{n_k=1}^{N}
        \Psi_k(
            u_1,u_1',
            \gamma_2(\tfrac{n_2-1}N),\gamma_2(\tfrac{n_2}N),
            \ldots,
            \gamma_k(\tfrac{n_k-1}N),\gamma_k(\tfrac{n_k}N)
        ).
    \end{equation}
 By continuity, for $N$ sufficiently large,
all terms on the right correspond to $2k$-tuples in $\calD_k'$.
Since each term is harmonic in $u_1$, it follows that $\Psi(\u )$ is harmonic in $u_1$ as well.
\end{proof}

% </input src="sections/PART_B/3_1_harmonicity.tex" root="." version="0.0.1">
% <input src="sections/PART_B/3_2_full.tex" root="." version="0.0.1">

\subsection{Characterisation of GFF correlation functions}

We are now ready to prove Theorem~\ref{thm:k2impliesallk}.
The idea is to proceed inductively.

\begin{proof}[Proof of Theorem~\ref{thm:k2impliesallk}]
We wish to prove
that $\Psi_k=\sigma^k\Psi_k^{\GFF}$
by inducting on $k\in2\Z_{\geq 1}$.
The base case $k=2$ holds true by hypothesis.
Now, fix $k$ and suppose that the statement holds true for all $\ell<k$.

Let us gather gather the properties of $\Psi_k$ established so far.
  We view $\Psi_k$ as a function in $u_1\in \R^2\setminus\cup_{i>1}\{u_i,u_i'\}$ while all other arguments are fixed.
    This function satisfies the following properties.
    \begin{itemize}
        \item {\bf Harmonicity:} $\Psi_k$ is harmonic in $u_1$ by Proposition~\ref{proposition:harmonicity}.
        \item {\bf Full-plane asymptotics:} $\Psi_k$ converges to a constant in $\R$ as $|u_1|\to\infty$. Indeed,  the regularity estimate (Theorem~\ref{thm:Regularity}) implies that $\Psi_k(\u)\to 0$
    as $|u_1|,|u_1'|\to\infty$,
    which implies that $\Psi_k(\u)$ tends to a finite constant as $u_1'$ is fixed
    and $|u_1|\to\infty$.
        \item {\bf Fusion asymptotics:}
             for any $i>1$ and $v\in \{u_i,u_i'\}$, Theorem~\ref{thm:mixing} implies that as $u_1$ tends to $v$,
            \begin{equation}
                \Psi_k(\u)
                =
                  \Psi_2(u_1,u_1',u_i,u_i')
                \Psi_{k-2}(((u_j,u_j'))_{j\not\in\{1,i\}})
                  +O(1),\end{equation}
            where $\Psi_2=\sigma^2\Psi_2^{\GFF}$
            and $\Psi_{k-2}=\sigma^{k-2}\Psi_{k-2}^{\GFF}$ by the induction hypothesis.
        \item {\bf Value at one specific point:} $\Psi_k(\u)=0$ at $u_1=u_1'$.
\end{itemize}

By extension theorems around singularities for harmonic functions~\cite{Cesaroni_2014_RemovableSingularitiesHarmonic},
these four properties (harmonicity together with the analysis of the singularities) fully determine the function $\Psi_k$, and yield $\Psi_k=\sigma^k\Psi_k^{\GFF}$. This completes the proof.
\end{proof}

% </input src="sections/PART_B/3_2_full.tex" root="." version="0.0.1">

% <input src="sections/PART_B/4_criterion.tex" root="." version="0.0.1">
\section{Sub-sequential GFF limits}
\label{sec:GFF_convergence_criteria}

This section is dedicated to proving the following result.

\begin{theorem}[Convergence criterion]
	\label{thm:criterion}
	Fix $\sigma\in\R_{\geq 0}$ and let $(\delta_n)_n$ denote a sequence tending to zero.
    If for all $k \in\Z_{\geq 1}$,
    \begin{equation}
		\label{eq:assumptionconvergence}
        \Phi_{k}^{(\delta_n)}|_{\calD_k} \xrightarrow[n\to\infty]{}       \sigma^k  \Psi_{k}^{\GFF} \qquad\text{uniformly on compacts of $\calD_k$,}
	\end{equation}
    then the height function converges to $\sigma\cdot\GFF$ along the scaling sequence $(\delta_n)_n$
	in the sense of Definition~\ref{def:GFF_convergence}.
\end{theorem}

\begin{proof}
		The proof relies on a single
	input: the regularity estimate of Theorem~\ref{thm:Regularity}.
	The theorem is therefore valid for any model for which similar regularity estimates hold.

Fix $\sigma\in[0,\infty)$ and a sequence $(\delta_n)_n$ tending to zero.
Recall Definition~\ref{def:GFF_convergence} which describes GFF convergence.
Our objective is to prove convergence of finite-dimensional marginals,
and convergence in law.

\paragraph{Convergence of finite-dimensional marginals.}
The generalised test functions with finite Dirichlet energy form a Hilbert space
where the inner product is given by the Dirichlet form.
Therefore, it suffices to consider a single generalised
test function $\varphi$ with finite Dirichlet energy,
and prove that
\begin{equation}
	\textstyle
	\langle h^{(\delta_n)}, \varphi \rangle \xrightarrow[n\to\infty]{} \calN\left(0, \sigma^2 \int G_{\R^2}(u,v) \diffi\varphi(u) \diffi\varphi(v)\right)
\end{equation}
in law.
Since a normal distribution is determined by its moments,
it suffices to prove that all moments of the random variable $\langle h^{(\delta_n)}, \varphi \rangle$
converge to the desired limits~\cite[Example~30.1 and Theorem~30.2]{Billingsley_1995_ProbabilityMeasure},
which is precisely what we will do.

Without loss of generality, we may assume that $|\varphi|(\R^2)=2$
and $\varphi=\varphi_+-\varphi_-$ where the two measures
are probability measures.
We first observe that
\begin{align}
		\langle h^{(\delta_n)}, \varphi \rangle&=\int h^{(\delta_n)}(u')\diffi\varphi_+(u')-\int h^{(\delta_n)}(u)\diffi\varphi_-(u)\\
		&=\int\int  (h^{(\delta_n)}(u')-h^{(\delta_n)}(u))\diffi\varphi_+(u')\diffi\varphi_-(u).
\end{align}

Fix $k\in\Z_{\geq 1}$.
Nothing needs to be proven for odd values of $k$ since the corresponding  moment is zero.
We therefore restrict to $k$ even.
In the discrete setting, Fubini's theorem can be applied to get
\begin{align}
	\E_{\Z^2}[\langle h^{(\delta_n)}, \varphi \rangle^k]
	&=
	\E_{\Z^2}\left[\left(\iint \left(h^{(\delta_n)}(u')-h^{(\delta_n)}(u)\right) \diffi\varphi_+(u')\diffi\varphi_-(u)\right)^k\right]
	\\
	&=
		\int\Phi_k^{(\delta_n)}(\u ) \diffi\tilde\varphi^k(\u )
\end{align}
		where
	\begin{equation}
		\diffi\tilde\varphi^k(\u ):=\diffi\varphi_-(u_1)\diffi\varphi_+(u_1')\cdots\diffi\varphi_-(u_k)\diffi\varphi_+(u_k').
\end{equation}
The integrand converges to $\sigma^{k}\cdot \Psi_{k}^{\GFF}$ as $n\to\infty$ by our assumption (Equation~\eqref{eq:assumptionconvergence}).
Therefore, it suffices to justify an application of the dominated convergence theorem to ensure the convergence to the corresponding moments of the GFF.

By definition,
\begin{equation}
	\Phi_k^{(\delta_n)}(\u)
	=
	\Phi_k([\tfrac{u_1}{\delta_n}],\ldots,[\tfrac{u_k'}{\delta_n}])
\end{equation}
where $[z]\in F(\Z^2)$ denotes the face whose lower-left
corner is $(\lfloor z_1\rfloor,\lfloor z_2\rfloor)$ for any $z\in\R^2$.
We shall find a dominating function using the regularity estimate (Theorem~\ref{thm:Regularity}).
We may do so separately for each pairing $\pi$ of $\{1,\ldots,k\}$ involved in \eqref{eq:Regularity}.
We focus on the pairing $\pi=\{(1,2),\ldots,(k-1,k)\}$.
Since the integrals factorise over the pairs, it suffices to prove that the following integral converges:
\begin{equation}
	\int\diffi\tilde\varphi^2(\u )
	\sup_n
	       \begin{cases}
			e^{-\alpha_k S_n(\u )} & \text{if $S_n(\u )\geq 20k^2$;} \\
			1\vee -S'_n(\u ) &\text{if $S_n(\u )< 20k^2$;}
       \end{cases}
\end{equation}
where $ S^\#_n(\u ):=S_{\R^2}^\#(\{[\tfrac{u_1}{\delta_n}],[\tfrac{u_1'}{\delta_n}]\},\{[\tfrac{u_2}{\delta_n}],[\tfrac{u_2'}{\delta_n}]\}$, with $\#$ denoting either no superscript or the apostrophe $'$.

Since $\tilde\varphi^2(\u )$ is a probability measure,
it suffices to prove that
\begin{equation}
	\int\diffi\tilde\varphi^2(\u )
	\sup_n
	\max\{0,-S'_n(\u )\}<\infty.
\end{equation}
By going back to the definitions of $S_{\R^2}$ and $S_{\R^2}'$,
it is quite straightforward to find a constant $\eta$ such that
\begin{equation}
	\max\{0,-S'_n(\u)\} \leq \max\{0,-S_{\R^2}(\u)\} + \eta.
\end{equation}
Thus, it suffices to prove that
\begin{equation}
	\label{eq:integrability_S_R2}
	\int
	\max\{0,-S_{\R^2}(\u)\}
	\diffi\tilde\varphi^2(\u )
	<\infty.
\end{equation}
Now, recall the explicit formula for $S_{\R^2}$.
Since $\tilde\varphi^2$ is compactly supported, Equation~\eqref{eq:integrability_S_R2} follows from the fact that $\varphi$ has finite Dirichlet energy.

\paragraph{Convergence in the negative regularity Hölder space.}
Fix $U\subset\R^2$ bounded and open and $\alpha\in(-1,0)$.
It suffices to prove that the family of random distributions
$(h^{(\delta_n)}|_U)$ is tight in $\calC^\alpha(U)$,
since any sub-sequential limit must then coincide with $\sigma\cdot\GFF$ by the convergence of finite-dimensional marginals proved above.

Furlan and Mourrat established a general tightness criterion for random fields in negative regularity Hölder spaces~\cite[Theorem~1.1]{FurlanMourrat_2017_TightnessCriterionRandom}.
We apply this theorem (let us borrow notation from the paper) with $\beta=0$
and with some integer $p>-\frac{1}{\alpha}$.
Notice also that, since our random function $h^{(\delta_n)}$
is stationary, we do not need to take a supremum over $x$.
Recall that $\calT_1((-1,1)^2)$ denotes the set of generalised test functions
supported in $(-1,1)^2$ and whose density with respect to the Lebesgue measure
is Lipschitz.
By~\cite[Theorem~1.1]{FurlanMourrat_2017_TightnessCriterionRandom},
it suffices to prove that for any $\varphi\in \calT_1((-1,1)^2)$, we have
\begin{equation}
	\label{eq:FurlanMourrat_bound}
	\sup_{\delta\in(0,1)} \E_{\Z^2}[
		|\langle h^{(\delta)}, \varphi \rangle|^p
	]<\infty.
\end{equation}
Naturally, the supremum over $n$ in~\cite[Theorem~1.1]{FurlanMourrat_2017_TightnessCriterionRandom} is encoded in our scaling parameter $\delta$,
which is why it does not appear explicitly.
But Equation~\eqref{eq:FurlanMourrat_bound} was already established above:
the generalised test function $\varphi$ has finite Dirichlet energy since it is Lipschitz,
and the uniform bound in Equation~\eqref{eq:FurlanMourrat_bound}
comes from the existence of the dominating function discussed above.
To finish,~\cite[Theorem~1.1]{FurlanMourrat_2017_TightnessCriterionRandom} implies tightness
in the topological space induced by all semi-norms $(\|\cdot\|_{\calC^\alpha(V)})_V$,
which implies tightness in $\calC^\alpha(U)$ in particular.

For tightness in the Besov spaces $\calB_{p,q}^\alpha$ for $\alpha\in(-1,0)$,
$p\in[1,\infty)$, and $q\in[1,\infty]$,
notice that the bounds in Equation~\eqref{eq:FurlanMourrat_bound}
match the hypothesis in~\cite[Theorem~2.30]{FurlanMourrat_2017_TightnessCriterionRandom},
which implies the desired result.
By~\cite[Theorem~6.2.4]{BerghLofstrom_1976_InterpolationSpacesIntroduction},
this also implies the desired tightness in the Sobolev space $W^{\alpha,p}(U)$.
\end{proof}

% </input src="sections/PART_B/4_criterion.tex" root="." version="0.0.1">

% <input src="sections/PART_B/5_0_conclusions.tex" root="." version="0.0.1">
\section{Proof of the main result (Theorem~\ref{thm:GFF_convergence})}
\label{sec:main_result}

\begin{proof}[Proof of Theorem~\ref{thm:GFF_convergence}]
    Theorems~\ref{thm:dicht_2p_final_form},~\ref{thm:k2impliesallk},
    and~\ref{thm:criterion} jointly imply that, for $\a=\b=1$
    and for any fixed $\c\in[\sqrt{3},2]$,
    one of the following two statements holds true:
    \begin{itemize}
        \item There exists some $\sigma\in\R_{\geq 0}$ such that the height function converges to $\sigma\cdot\GFF$,
        \item There exist two distinct standard deviations $\sigma,\sigma'\in\R_{\geq 0}$ and
        two sequences $(\delta_n)_n$ and $(\delta'_n)_n$ tending to zero,
        such that the height function converges to $\sigma\cdot\GFF$ along the scaling sequence $(\delta_n)_n$,
        and to $\sigma'\cdot\GFF$ along the scaling sequence $(\delta'_n)_n$.
    \end{itemize}
  Yet, we may rule out the second case, since $\sigma$ and $\sigma'$ cannot be distinct.
  Indeed, the combination of Theorems~\ref{thm:glimpse_scale_invariance} and~\ref{thm:computation_deriv_f}
  yields that both $\sigma^2$ and $(\sigma')^2$ should be equal to $1/\arcsin \tfrac{\c}{2}$.
  Thus, we are in the first case with $\sigma^2=1/\arcsin \tfrac{\c}{2}$.
\end{proof}

We record an additional result that may be useful in future works. Although our proof of the main theorem required a somewhat indirect route, we are now in a position to establish the convergence of the spectral measures. In particular, the sub-sequential limiting measure $\mu$ introduced earlier is in fact unique, and is simply the limit of the spectral measures.

\begin{theorem}
  \label{thm:main measure}
Fix $\a=\b=1$ and $\c\in[\sqrt{3},2]$. Then, if $\mu_L$ denotes the measure given by Theorem~\ref{thm:spectral_representation_four_point}
(cf.~the explicit construction in Equation~\eqref{eq:def muL} that follows), then
\begin{equation}
\lim_{\delta\rightarrow0}\lim_{L\rightarrow\infty}\mu_L^{(\delta)}=\tfrac12(\delta_{b=a}+\delta_{b=-a})\cdot \frac{\sigma^2}{2\pi a}\diffi a.
\end{equation}
\end{theorem}

\begin{proof}
Since $(\mu_L)_{L\in 2\Z_{>0}}\subset\calM$, the family is precompact. It therefore suffices to show that any subsequential limit $\mu$ coincides with
$
\tfrac12 (\delta_{b=a}+\delta_{b=-a}) \cdot \frac{\sigma^2}{2\pi a}\diff a.
$

We invoke the convergence of $\Phi^{(\delta)}_2$ to $\sigma^2\cdot\Psi_2^{\text{GFF}}$. This yields
\begin{equation}
I_F(s)-I_F(1) = -\frac{\sigma^2}{2\pi}\log s,
\end{equation}
which, together with \eqref{eq:F_as_derivative_of_I}, implies
\begin{equation}
F(s,0)=\frac{\sigma^2}{2\pi s}.
\end{equation}
Let us turn to the measure $\mu$ to see what the previous formula implies.
We already know that $\mu[\{|b|>a\}]=0$. To prove that $\mu[\{|b|<a\}]=0$, observe that \eqref{eq:non-increasing} gives
\begin{equation}
\int (a^2-b^2)e^{-a}\diff \mu(a,b) = -\Xi'(1) = 0,
\end{equation}
since $\Xi(s)=sF(s,0)=\sigma^2/(2\pi)$ is constant. We therefore obtain that $\mu[\{|b|<a\}]=0$.

It remains to identify the marginal in the $a$-variable and show that
\begin{equation}
\diff \mu(a) = \frac{\sigma^2}{2\pi a}\diff a.
\end{equation}
By the definition and the formula for $F$, we get that for every $x>0$,
\begin{equation}
\int a e^{-ax}\diff \mu(a) = \frac{\sigma^2}{2\pi x}.
\end{equation}
The right-hand side is precisely the Laplace transform of the measure $
\frac{\sigma^2}{2\pi a}\diff a.
$
Since a $\sigma$-finite positive measure is uniquely determined by its Laplace transform, this identifies $\mu$ and completes the proof.
\end{proof}

% </input src="sections/PART_B/5_0_conclusions.tex" root="." version="0.0.1">

% <input src="sections/PART_B/6_0_ani.tex" root="." version="0.0.1">
\section{Proof for the anisotropic case (Theorem~\ref{thm:GFF_convergence_isoradial})}
\label{sec:anisotropic}

The convergence of multi-point correlation functions and finite-dimensional marginals is extended to the \emph{anisotropic} six-vertex models as follows: first express these quantities in terms of the
macroscopic loops of a random-cluster model on an associated isoradial graph, and then invoke the
universality result of \cite{Duminil-CopinKozlowskiKrachun_2020_RotationalInvarianceCritical}, which ensures that the scaling limit of the loop ensemble
is identical to that of the isotropic case. Consequently, the limiting correlations coincide with
those obtained in the isotropic setting.

\begin{proof}[Proof of Theorem~\ref{thm:GFF_convergence_isoradial}]
Fix $\Delta\in[-1,-1/2]$. Consider the six-vertex model with weights $\a,\b,\c$ given by
\eqref{eq:parametrization}. Throughout this section, we restrict to the regime where
$\Delta(\a,\b,\c)$ is set to be equal to $\Delta$. By definition of this parametrization, $\zeta$ is fixed and the only remaining degree of
freedom is the angle~$\theta$. We write $\mathbb{E}^{\rm 6V}_{\mathbf  L_\theta(\mathbb{Z}^2)}$ for the
corresponding anisotropic six-vertex measure on $\mathbf  L_\theta(\mathbb{Z}^2)$, and denote by $h$
its associated height function.

The Baxter--Kelland--Wu correspondence \cite{BaxterKellandWu_1976_EquivalencePottsModel} maps this six-vertex model on
$\mathbf  L_\theta(\mathbb{Z}^2)$ to the random-cluster model on an isoradial graph $\mathbb L(\theta)$ (a
rotated version of a rectangular lattice) with the corresponding isoradial weights; see
\cite{Duminil-CopinKozlowskiKrachun_2020_RotationalInvarianceCritical} for details. The explicit construction of this random-cluster model will not be needed here.
We will only use the fact that its loop representation is naturally supported on
$\mathbf  L_\theta(\mathbb{Z}^2)$. Following the notation of
\cite{Duminil-CopinKozlowskiKrachun_2020_RotationalInvarianceCritical,magicformula}, we denote by $\phi_{\delta\mathbb L(\theta),q}$ the law of the loop
ensemble of the random-cluster measure on $\delta\mathbb L(\theta)$ with the relevant isoradial
weights.

\paragraph{Convergence of multi-point correlation functions.}
Fix $k\in2\Z_{\geq 1}$ and $\u\in\calD_k$.
We assume without loss of generality that all $2k$ points in $\u$ are distinct.
In this section only, we shall write $\u=(u_1,u_2,\ldots,u_{2k})$.
Write $\frakF$ for the powerset of $\{1,\ldots,2k\}$
with subsets of cardinal $0$, $1$, and $2k$ removed.

For a given loop ensemble $\calL$ and $S\subset\{1,\ldots,2k\}$, let $\mathcal N_S$ denote the number of loops in $\calL$ that surround all points in $\{u_i: i\in S\}$ and none of the points in $\{u_j: j\notin S\}$. For $S=\{v\}$, we write simply $\mathcal N_v$.

% Let ${\rm Fin}$ be the set
% $\mathcal P(\{1,\dots,2N\})$ with the empty set and the full set $\{1,\dots,2N\}$ removed.
% Let ${\rm Fin}_2$ denote the subset of ${\rm Fin}$ obtained by further
% excluding the singletons $\{i\}$ for $1 \le i \le 2N$.
% Given a loop ensemble $\calL$ and $S \in {\rm Fin}$, define
% $\mathcal N_S$ to be the number of loops in $\calL$
% that surround all points $x_i$ with $i \in S$ and none of the points
% $x_i$ with $i \notin S$.
% For $S = \{i\}$, write simply $\mathcal N_i$.

Now \cite[Proposition~2.5]{magicformula} asserts that we may find
real polynomials
 $P_R \in \mathbb{R}\!\left[(X_S)_{S \in \frakF}\right]
$
associated with each $R\subset\{1,\ldots,k\}$ such that the following holds true:
for every $\u\in\calD_k$,
and every mesh size $\delta$ for which none
of the points lie on edges of $\delta\mathbf  L_\theta(\mathbb{Z}^2)$ (so that one may unambiguously decide whether the points $x_i$ are surrounded by the loops in $\calL$),
\begin{equation}
   \label{eq:magicformuladiscrete3}
\E^{\rm 6V}_{\mathbf  L_\theta(\mathbb{Z}^2)}
   \Bigg[
      \prod_{i=1}^k \big(h^{(\delta)}(u_{2i-1}) - h^{(\delta)}(u_{2i})\big)
   \Bigg]
   =
   \sum_{R \subset \{1,\dots,k\}}
   \phi_{\delta\mathbb L(\theta),q}
   \Bigg[
      \prod_{i \in R} (\mathcal N_{2i-1} - \mathcal N_{2i})
      \, P_R\big((\mathcal N_S)_{S \in \frakF}\big)
   \Bigg].
\end{equation}
Moreover, \cite[Lemma~4.5]{magicformula} asserts that we may restrict to ``large'' loops.
More precisely, fix $\epsilon>0$, and let $\mathcal N_j^{\rm big}$ be the number of loops contributing to
$\mathcal N_j$ that are \emph{not} contained in the ball of radius $\epsilon$
centred at $u_j$.
Then, there exists a universal constant
$c > 0$ such that for every $\epsilon > 0$, the terms on the right in~\eqref{eq:magicformuladiscrete3} satisfy
\begin{align}
\phi_{\delta\mathbb L(\theta),q}
\Bigg[
   \prod_{i \in R}(\mathcal N_{2i-1} - \mathcal N_{2i})&
   \, P_R\big((\mathcal N_S)_{S \in \frakF}\big)
\Bigg]\nonumber\\
&=
\phi_{\delta\mathbb L(\theta),q}
\Bigg[
   \prod_{i \in R}(\mathcal N_{2i-1}^{\rm big} - \mathcal N_{2i}^{\rm big})
   \, P_R\big((\mathcal N_S)_{S \in \frakF}\big)
\Bigg]
+ O(\epsilon^c).
\end{align}
Combining the previous two displayed equations, we obtain that the
$k$-point correlations of the anisotropic six-vertex model can be approximated (with an error tending to zero as $\epsilon$ tends to zero) in
terms of a functional depending only on the loops of radius at least $\epsilon$
in the loop ensemble obtained from the random-cluster model on $\delta\mathbb L(\theta)$.
The universality of the random-cluster model on isoradial graphs, proved in
\cite{Duminil-CopinKozlowskiKrachun_2020_RotationalInvarianceCritical}, implies that the law of the loop ensembles for different values of $\theta$ are equal. In particular, the one with angle $\theta$ is the same as the one with angle $\pi/2$, which is related to the {\em isotropic} six-vertex model. We deduce that for every $\eta > 0$ and every compact set
$K \subset \mathcal{D}_k$, there exists $\delta_0(k,K,\eta) > 0$ such that for every
$0 < \delta < \delta_0$ and $\u  \in K$,
\commentPiet{I totally believe this step but it doesn't yet feel rock-solid to me...}
\begin{equation}
\Bigg|
\E^{\rm 6V}_{\mathbf  L_\theta(\mathbb{Z}^2)}
   \Bigg[
      \prod_{i=1}^k \big(h^{(\delta)}(u_{2i}) - h^{(\delta)}(u_{2i-1})\big)
   \Bigg]
   - \Phi_k^{(\delta)}(\u )
\Bigg|
\leq \eta,
\end{equation}
where we recall that $\Phi_k^{(\delta)}(\u )$ denotes the correlation function of the isotropic case. In particular, uniform convergence of the $k$-point correlations on compact
subsets of $\mathcal{D}_k$ for the isotropic case $\theta = \pi/2$ (Theorem~\ref{thm:GFF_convergence})
immediately transfers to the anisotropic case with arbitrary $\theta$.

\paragraph{Convergence of finite-dimensional marginals.}

Having established convergence of the multi-point correlation functions, it is natural to attempt the strategy used in the isotropic setting.
The problem with this strategy is that it requires a dominating function (for applying the dominated convergence theorem), which we found in terms of uniform bounds on the correlation functions.
At this point, a technical difficulty arises: Theorem~\ref{thm:Regularity} relies on crossing estimates for the spin representation, and these are not readily available in the anisotropic case. Implementing such an approach would therefore require revisiting the entire analysis without the symmetry given by the $\pi/2$ rotation.

To bypass this issue, we work directly with Equation~\eqref{eq:magicformuladiscrete} instead.
Let $h^{(\delta)}_\theta$ be the height function of the anisotropic six-vertex model on $\mathbf  L_\theta(\mathbb Z^2)$, and fix a finite Dirichlet energy test function $\varphi$.
Recall that if the characteristic function of $\langle h^{(\delta)}_\theta,\varphi\rangle$ converges
pointwise to that of a Gaussian, then this sequence converges to that Gaussian in law in the weak topology.

For a loop ensemble $\mathcal L$, define (whenever the infinite product is well-defined)
\begin{equation}
A_\varphi(\mathcal L)
:= \prod_{\ell\in \mathcal L}
\cos_\mu\varphi(\mathrm{int}(\ell)).
\end{equation}

Let $\mathcal L^{(\delta)}$ and $\mathcal L^{(\delta)}_\theta$ denote the loop ensembles associated with the random-cluster model on $\delta\mathbb L(\tfrac\pi2)$ and $\delta\mathbb L(\theta)$, respectively. Extracting a sub-sequence of $(\delta_n)_n$ if necessary, we may assume that $\mathcal L^{(\delta_n)}$ and $\mathcal L^{(\delta_n)}_\theta$ converge in law to loop ensembles $\mathcal L^{\rm cont}$ and $\mathcal L^{\rm cont}_\theta$. Let $\phi^{\rm cont}$ and $\phi^{\rm cont}_\theta$ denote their laws. The result of \cite{magicformula} ensures that for every $t\in\mathbb R$, the quantities
$A_{t\varphi}(\mathcal L^{\rm cont})$ and $A_{t\varphi}(\mathcal L^{\rm cont}_\theta)$ are almost surely well-defined and integrable, and that
\noticePiet{here we can even remove the equation for $\theta=\pi/2$,
and say it's independent of $\theta$. I agree but I think for reference it is simpler. I would leave it as such.}
\begin{align}
\label{eq:magicformulacontinuum}
\phi^{\rm cont}\!\big[A_{t\varphi}(\mathcal L^{\rm cont})\big]
&= \lim_{n\to\infty} \phi_{\delta_n\mathbb L(\tfrac\pi2)}\!\big[A_{t\varphi}(\mathcal L^{(\delta_n)})\big]
= \lim_{n\to\infty} \mathbb E^{\rm 6V}_{\mathbb Z^2}\!\left[e^{\i t\langle h^{(\delta_n)},\varphi\rangle}\right],\\[1ex]
\label{eq:magicformulacontinuum twisted}
\phi^{\rm cont}_\theta\!\big[A_{t\varphi}(\mathcal L^{\rm cont}_\theta)\big]
&= \lim_{n\to\infty} \phi_{\delta_n\mathbb L(\theta)}\!\big[A_{t\varphi}(\mathcal L^{(\delta_n)}_\theta)\big]
= \lim_{n\to\infty} \mathbb E^{\rm 6V}_{\mathbf L_\theta(\mathbb Z^2)}\!\left[e^{\i t\langle h^{(\delta_n)}_\theta,\varphi\rangle}\right],
\end{align}
where the second equality in each line is simply \eqref{eq:magicformuladiscrete}.
The universality result in~\cite{Duminil-CopinKozlowskiKrachun_2020_RotationalInvarianceCritical}
asserts that the two limiting loop ensembles have the same law,
and therefore the quantities on the left are the same.
Since we already identified the characteristic function in the isotropic case,
this also implies the desired pointwise convergence of the characteristic function in the anistropic case.
 \end{proof}

 \begin{remark}\label{rmk:anisotropic}
The fact that we do not get regularity for the anisotropic case explains why we refrain from stating convergence in H\"older spaces, as establishing it would require substantial additional technical work. Nevertheless, we expect that such a result could be obtained without encountering any fundamental obstacles.
\end{remark}

% </input src="sections/PART_B/6_0_ani.tex" root="." version="0.0.1">

% </input src="sections/PART_B/main.tex" root="." version="0.0.1">
% <input src="sections/PART_C/new.tex" root="." version="0.0.1">
\part{\INGREDSPECTRAL: Spectral representation of correlation functions}
\label{part:spectral}

This part derives the results stated in \INGREDSPECTRAL,
namely Theorems~\ref{thm:spectral_representation_four_point} and~\ref{thm:spectral_representation_multi_point}.
The value of $L\in2\Z_{\geq 1}$ is fixed throughout this part.
The two theorems follow from the basic symmetries on the cylinder
(translation invariance, reflection invariance, and invariance under global arrow flip).
We stress that while the transfer matrix of the six-vertex model has an extremely rich structure,
the purpose of this part is underlining almost the \emph{opposite} fact:
demonstrating that the properties required for Part~\ref{part:proofs}
do not require this structure, and are very general.

\section{Transfer matrix formalism% (Theorems~\ref{thm:spectral_representation_four_point} and~\ref{thm:spectral_representation_multi_point}; Lemma~\ref{lemma:Cylinder measure})
}

%\subsection{Transfer matrix formalism}

Recall that $\CYL{L}=(V(\CYL{L}),E(\CYL{L}))$ is the nearest-neighbour graph
on the vertex set $V(\CYL{L}):=\Z\times(\Z/L\Z)$.
Write $E_i$ for the \emph{horizontal edges} between $\{i\}\times(\Z/L\Z)$
and $\{i+1\}\times(\Z/L\Z)$.
The set $E_i$ is called the \emph{$i$-th column of horizontal edges}.
We also write $E_{ii'}$ for the union of $E_i\cup E_{i+1}\cup\cdots\cup E_{i'}$
with the set of all \emph{vertical edges} between those columns (i.e.~with endpoints in $\{i+1,\dots,i'\}\times(\Z/L\Z)$).

A \emph{column configuration} is an element $\kappa\in\{\pm1\}^{\Z/L\Z}$
encoding arrow orientations in a column $E_i$ of \emph{horizontal arrows}.
There are $L$ vertical positions; $\kappa_j=+1$ means that the arrow in position $j$
is oriented to the right; $\kappa_j=-1$ means that the arrow is oriented to the left.
Write $\frakC$ for the set of \emph{balanced} column configurations,
meaning that $\sum_j\kappa_j=0$.
Similarly, the \emph{vertical column configuration} along the vertical line $\{k\} \times \R/L\Z$
is an element $\alpha^k\in\{\pm1\}^{\Z/L\Z}$
encoding arrow orientations of \emph{vertical arrows}.
There are $L$ vertical positions; $\alpha_j^k=+1$ means that the arrow between positions
$j$ and $j+1$ points up; $\alpha_j^k=-1$ means that the arrow points down.
An arrow configuration on $E_{ii'}$ may be written as a family
$((\kappa^{k})_{i\leq k\leq i'},(\alpha^k)_{i<k\leq i'})$,
where the $\kappa^k$ encode the horizontal arrows in column $E_k$,
and the $\alpha^k$ the vertical arrows on $\{k\} \times \R/L\Z$.

For $i\leq i'$ and $\zeta,\zeta'\in\frakC$, let
\(\P^{ii'}_{\zeta\zeta'}\) denote the probability measure
on $\{\pm 1\}^{E_{ii'}}$ with boundary conditions $\zeta$ and $\zeta'$,
defined by
\begin{equation}
    Z^{ii'}_{\zeta\zeta'}\P^{ii'}_{\zeta\zeta'}[((\kappa^{k})_{i\leq k\leq i'},(\alpha^k)_{i<k\leq i'})]
    :=\true{(\kappa^i,\kappa^{i'})=(\zeta,\zeta')}\cdot\true{\text{ice rule}}\cdot
    \c^{\#\{\text{$\c$-vertices}\}},
    \end{equation}
where   $Z^{ii'}_{\zeta\zeta'}$ is a normalisation factor given by
\begin{equation}
  Z^{ii'}_{\zeta\zeta'}:=
    \sum_{
        (\kappa^{k})_{i\leq k\leq i'}
        ,\,
        (\alpha^k)_{i<k\leq i'}
    }
    \true{(\kappa^i,\kappa^{i'})=(\zeta,\zeta')}\cdot\true{\text{ice rule}}\cdot
    \c^{\#\{\text{$\c$-vertices}\}}.
\end{equation}
We collect a few basic properties.

\begin{lemma}[Basic properties of the probability measures]~
    \begin{enumerate}
        \item \textbf{One-step law.}
        For any $i\in\Z$ and $\kappa,\kappa'\in\{\pm1\}^{\Z/L\Z}$,
        \begin{equation}
            Z^{i(i+1)}_{\kappa\kappa'}\P^{i(i+1)}_{\kappa\kappa'}[\alpha^{i+1}]
            =
            \true{\text{ice rule for $(\kappa,\alpha^{i+1},\kappa')$}}
            \cdot
            \c^{\#\{\text{$\c$-vertices in $(\kappa,\alpha^{i+1},\kappa')$}\}}.
        \end{equation}
        %The one-step partition function is obtained by summing $\alpha^{i+1}\in\{\pm1\}^L$.
        \item \textbf{Composition rule.}
       For any $i\le i'\le i''$ and $\kappa,\kappa''\in\{\pm1\}^{\Z/L\Z}$,
        \begin{equation}
            Z^{ii''}_{\kappa\kappa''}\P^{ii''}_{\kappa\kappa''}
            :=\sum_{\kappa'\in\frakC}
            Z^{ii'}_{\kappa\kappa'}Z^{i'i''}_{\kappa'\kappa''}
            (\P^{ii'}_{\kappa\kappa'}\times\P^{i'i''}_{\kappa'\kappa''}).
        \end{equation}
        \item \textbf{Torus measure.}
        We have
        \begin{equation}
            Z_{\T_{M,L}}
            \P_{\T_{M,L}}[\{\operatorname{balanced}\}]
            \P_{\T_{M,L}}[\blank|\{\operatorname{balanced}\}]
            =
            \sum_{\kappa\in\frakC}Z^{0M}_{\kappa\kappa}\P^{0M}_{\kappa\kappa}.
        \end{equation}
        \item \textbf{Reflection symetry.}
        Let $i\le i'$ and let $\mathbf X$ denote a random variable which is measurable in
        terms of the arrows in $E_{ii'}$.
        (Recall that the arrows encode the gradient of the height function.)
        Let $\mathbf X^\dagger$ denote the random variable measurable
        in terms of the arrows in $E_{(-i')(-i)}$
        obtained by composing $\mathbf X$ with the reflection of the gradient
        over $\calR$.
        Then,
        \begin{equation}
            \E^{ii'}_{\kappa\kappa'}[\mathbf X]=\E^{(-i')(-i)}_{\kappa'\kappa}[\mathbf X^\dagger].
        \end{equation}
        \item \textbf{Horizontal shift invariance.}
        Let $i\le i'$ and let $\mathbf X$ denote a random variable which is measurable in
        terms of the arrows in $E_{ii'}$.
        For $n\in\Z$, let $\tau_{(n,0)}$ denote the shift by $(n,0)$,
        and let $\tau_{(n,0)}(\mathbf X)$ denote the random variable
        measurable in terms of $E_{(i+n)(i'+n)}$ obtained by composing $\mathbf X$
        with the shift. Then,
        \begin{equation}
            \E^{ii'}_{\kappa\kappa'}[\mathbf X]
            =
            \E^{(i+n)(i'+n)}_{\kappa\kappa'}[\tau_{(n,0)}(\mathbf X)].
        \end{equation}
    \end{enumerate}
\end{lemma}

\begin{proof}
    The first three follow from bookkeeping manipulations of the sums.
    The final two are symmetries of the six-vertex model.
\end{proof}

The previous lemma reveals a Hilbert space structure.
Let $\Omega:=\ell^2(\frakC)$ denote the complex Hilbert space with orthonormal basis $(e_\kappa)_{\kappa\in\frakC}$.
Use the symbol $\dagger$ for the Hermitian conjugate of vectors and operators.
Associate a Hilbert space $\Omega_i:=\Omega$ to each column of horizontal edges $E_i$.
For each \emph{valid triple}, that is, a triple $(\mathbf X,i,i')$, where $i\leq i'$ and where $\mathbf X$ is a complex-valued operator
depending only on the edges in $E_{ii'}$,
we define the operator $\frako_{\mathbf X}^{ii'}:\Omega_i\to\Omega_{i'}$
via
\begin{equation}
    e_{\kappa'}^\dagger\; \frako_{\mathbf X}^{ii'}\; e_{\kappa} := Z^{ii'}_{\kappa\kappa'}\E^{ii'}_{\kappa\kappa'}[\mathbf X].
\end{equation}
Within this framework, the above lemma readily implies the following corollary.

\begin{corollary}[Basic properties of operators]~
    \begin{enumerate}
        \item \textbf{One-step partition function.}
        The operator $t(\pi/2):=\frako_{1}^{i(i+1)}$, which corresponds to the one-step
        partition function, is independent of the choice of $i$, and satisfies
        \begin{equation}
            e_{\kappa'}^\dagger t(\pi/2)e_{\kappa}:=
            \sum_{\alpha\in\{\pm1\}^L}\true{\text{ice rule for $(\kappa,\alpha,\kappa')$}}
            \cdot
            \c^{\#\{\text{$\c$-vertices in $(\kappa,\alpha,\kappa')$}\}}.
        \end{equation}
        \item \textbf{Composition rule.}
        For any two valid triples $(\mathbf X,i,i')$ and $(\mathbf Y,i',i'')$,
        the composition $(\mathbf{XY},i,i'')$ is a valid triple,
        and
        \begin{equation}
            \frako^{ii''}_{\mathbf{XY}}  = \frako^{i'i''}_\mathbf Y\circ\frako^{ii'}_\mathbf X.
        \end{equation}
        \item \textbf{Torus measure.}
      For any valid triple $(\mathbf X,0,M)$,
        \begin{equation}
            Z_{\T_{M,L}}
            \P_{\T_{M,L}}[\{\operatorname{balanced}\}]
            \E_{\T_{M,L}}[\mathbf X|\{\operatorname{balanced}\}]
            =
            \operatorname{Trace}\frako^{0M}_\mathbf X.
        \end{equation}
        \item \textbf{Reflection symmetry.}
        For any valid triple $(\mathbf X,i,i')$ where $\mathbf X$ is real-valued,
        \begin{equation}
            (\frako_{\mathbf X}^{ii'})^\dagger = \frako_{\mathbf X^\dagger}^{(-i')(-i)}.
        \end{equation}
        \item \textbf{Horizontal shift-invariance.}
     For any valid triple $(\mathbf X,i,i')$ and $n\in\Z_{\geq 0}$,
        \begin{equation}
            \frako_{\tau_{(n,0)}(\mathbf X)}^{(i+n)(i'+n)}=\frako_{\mathbf X}^{ii'}.
        \end{equation}
    \end{enumerate}
\end{corollary}

We now turn to the transfer matrix $t(\pi/2)$ itself.

\begin{lemma}[Basic properties of the transfer matrix]
    \label{lem:properties transfer matrix}
    The operator $t(\pi/2)$ is a Hermitian Perron-Frobenius matrix with a single block.
    Therefore, it is diagonal in some orthonormal basis $(v_k)_k$ with real eigenvalues,
    and the basis may be labelled such that the corresponding eigenvalues $(\lambda_k(\pi/2))_k$ satisfy
    \begin{equation}\lambda_{0}(\pi/2)>|\lambda_1(\pi/2)|\geq |\lambda_2(\pi/2)|\geq \lambda_3(\pi/2)\geq \cdots.\end{equation}
    Finally, $v_0$ may be chosen such that it has positive real entries in the basis $(e_\kappa)_\kappa$.
    \commentKarol{Here, I removed the absolute values}
    \commentPiet{But we can also have negative eigenvalues? HDC: agreed}
\end{lemma}

\begin{proof}
    The operator $t(\pi/2)$ is a real symmetric matrix in the basis $(e_\kappa)_\kappa$.
    Therefore, it is Hermitian.
    It may be checked directly that it is a Perron-Frobenius matrix.
    It is easy to see that $t(\pi/2)^k$ has positive entries for
    sufficiently large $k$.
    As a consequence, it has a single block.
    The rest follows.
\end{proof}

Define
\begin{equation}
    T(\pi/2):=\frac{t(\pi/2)}{\lambda_0(\pi/2)};\qquad
    \Lambda_k(\pi/2):=\frac{\lambda_k(\pi/2)}{\lambda_0(\pi/2)};
    \qquad
    \frakO_\mathbf X^{ii'}:=\frako_\mathbf X^{ii'}/\lambda_0(\pi/2)^{i'-i}.
\end{equation}

\begin{remark}[Lemma~\ref{lem:properties transfer matrix}]
The eigenspace of $T(\pi/2)$ associated with the top eigenvalue $1$ is one-dimensional.
The convention that $v_0$ has positive real entries simply fixes the complex phase of the normalised eigenvector corresponding to $\lambda_0(\pi/2)$.
The spectrum of $T(\pi/2)$ is supported on $(-1,1]$ since $-1$ cannot be an eigenvalue of $T(\pi/2)$ due to the Perron--Frobenius property.
\end{remark}

By construction, for any valid triple $(\mathbf X,i,i')$, we have
\begin{equation}
    \label{eq:cylop}
    \E_{\CYL{L}}[\mathbf X]
    =
    \lim_{M\to\infty}\P_{\T_{M,L}}[\mathbf X|\{\operatorname{balanced}\}]
    =\lim_{M\to\infty} \frac{\operatorname{Trace} (\frako_{\mathbf X}^{ii'} t(\pi/2)^{M-(i'-i)})}{\operatorname{Trace} t(\pi/2)^M}
    =v_0^\dagger\; \frakO_\mathbf X^{ii'} \;v_0.
\end{equation}
Indeed, the rightmost equality is obtained by observing that only the
largest eigenvalue of $t(\pi/2)$ contributes.

We conclude this subsection by observing that Lemma~\ref{lemma:Cylinder measure} is now straightforward.

\begin{proof}[Proof of Lemma~\ref{lemma:Cylinder measure}.]
    See Equation~\eqref{eq:cylop} above.
\end{proof}

\section{Spectral representation of general observables (Theorem~\ref{thm:spectral_representation_multi_point})}

\begin{proof}[Proof of Theorem~\ref{thm:spectral_representation_multi_point}]
    Consider the setting of the theorem,
    and fix $L$.
    Define the embeddings
    \begin{align}
        \calE^-&:\ \frakA^-\to\Omega,\quad\mathbf X\mapsto \frakO_{\mathbf X}^{(-i)0}\; v_0;\\
        \calE^+&:\ \frakA^+\to\Omega^\dagger,\quad\mathbf Y\mapsto v_0^\dagger\;\frakO_{\mathbf Y}^{0i},
    \end{align}
    where $i$ is chosen so large that $\mathbf X$ and $\mathbf Y$ are measurable
    with respect to the arrows in $E_{(-i)0}$ and $E_{0i}$ respectively.
    The definitions do not depend on $i$ since the composition rule implies that
    \begin{equation}
        v_0^\dagger\;\frakO_{\mathbf Y}^{0(i+1)}
        =
        v_0^\dagger T(\pi/2)\;\frakO_{\mathbf Y}^{0i}
        =
        v_0^\dagger\;\frakO_{\mathbf Y}^{0i}
    \end{equation}
    (the proof for $\calE^-$ is the same).
    Notice that $\calE^-$ and $\calE^+$ are linear maps that map real vector spaces
    to complex vector spaces.

\paragraph{Step 1: Definition of $\mu_{\mathbf X,\mathbf Y,L}$.}
   For $\mathbf X$ and $\mathbf Y$,   define the measure $\mu_{\mathbf X,\mathbf Y,L}$ on $[0,2)$ via
    \begin{equation}
        \label{eq:measure_definition}
        \mu_{\mathbf X,\mathbf Y,L}:=
        \sum_k
        \big(\calE^+(\mathbf Y)
        v_k\big)
        \big(
        v_k^\dagger
        \calE^-(\mathbf X)\big)
        \delta_{1-\Lambda_k(\pi/2)}.
    \end{equation}
  (Above, $\delta_x$ denotes the Dirac measure at the point $x$.)

  \paragraph{Step 2: Proof of (i).}
  Using Equation~\eqref{eq:cylop} and the composition rule in the first equality, horizontal shift invariance in the second and the definition of the embeddings in the third, we get
    \begin{align}
        \E_{\CYL{L}}[\mathbf X\cdot \tau_{(n,0)}(\mathbf Y)]
        &=
        v_0^\dagger\;\frakO_{\tau_{(n,0)}(\mathbf Y)}^{n(i+n)}\;T(\pi/2)^n\;\frakO_{\mathbf X}^{(-i)0}\;v_0
        \\&=
        v_0^\dagger\;\frakO_{\mathbf Y}^{0i}\;T(\pi/2)^n\;\frakO_{\mathbf X}^{(-i)0}\;v_0
        \\&=
        \calE^+(\mathbf Y) T(\pi/2)^n \calE^-(\mathbf X).\label{eq:7.2.1}
    \end{align}
   Using~\eqref{eq:measure_definition}, this implies the desired formula
    \begin{equation}\label{eq:preCS}
        \E_{\CYL{L}}[\mathbf X\cdot \tau_{(n,0)}(\mathbf Y)]
        =
        \int (1-a)^n\diff \mu_{\mathbf X,\mathbf Y,L}(a).
    \end{equation}

 \paragraph{Step 3: Proof of (ii).}
 For $\mathbf X$, reflection symmetry
    and the definition of the embeddings imply
    \begin{equation}
        \calE^+(\mathbf X^\dagger)
        =
        v_0^\dagger\; \frakO_{\mathbf X^\dagger}^{0i}
        =
        v_0^\dagger\; (\frakO_{\mathbf X}^{-i0})^\dagger
        =
        (\frakO_{\mathbf X}^{-i0}\; v_0)^\dagger
        =
        (\calE^-(\mathbf X))^\dagger.\label{eq:7.2.2}
    \end{equation}
    As a consequence, $\mu_{\mathbf X,\mathbf X^\dagger,L}$ is a positive measure since for every $k$,
    \begin{equation}
     \big(\calE^+(\mathbf X^\dagger)
        v_k\big)
        \big(
        v_k^\dagger
        \calE^-(\mathbf X)\big)=\big(\calE^-(\mathbf X)^\dagger
        v_k\big)
        \big(
        v_k^\dagger
        \calE^-(\mathbf X)\big)=\|\calE^-(\mathbf X)^\dagger
        v_k\|^2\ge0.
    \end{equation}
    It remains to prove the Cauchy-Schwarz inequality (Equation~\eqref{eq:Cauchy--Schwarz}).
Using \eqref{eq:preCS} (for $n=0$) and the definition of the total variation, we obtain
     \begin{equation}
        \label{eq:preCS2}
      |\E_{\CYL{L}}[\mathbf X\mathbf Y]|= |\mu_{\mathbf X,\mathbf Y,L}[[0,2)]|\le \|\mu_{\mathbf X,\mathbf Y,L}\|\leq \sum_k
        \Big|\big(\calE^+(\mathbf Y)
        v_k\big)
        \big(
        v_k^\dagger
        \calE^-(\mathbf X)\big)\Big|.
     \end{equation}
   Applying the Cauchy-Schwarz inequality to the sum on the right-hand side gives
       \begin{equation}
    \Big( \sum_k
        \big|\big(\calE^+(\mathbf Y)
        v_k\big)
        \big(
        v_k^\dagger
        \calE^-(\mathbf X)\big)\big|\Big)^2
        \le  \Big( \sum_k
        \|
        \calE^+(\mathbf Y)v_k\|^2\Big) \Big( \sum_k
       \|
        v_k^\dagger
        \calE^-(\mathbf X)\|^2\Big).   \end{equation}
   Using \eqref{eq:preCS} (for $n=0$) in the other direction together with the positivity of $\mu_{\mathbf X,\mathbf X^\dagger,L}$ gives
    \begin{equation}
     \sum_k
        \|
        v_k^\dagger
        \calE^-(\mathbf X)\|^2=\|\mu_{\mathbf X,\mathbf X^\dagger,L}\|=\int \diff\mu_{\mathbf X,\mathbf X^\dagger,L}(a)=   \E_{\CYL{L}}[\mathbf X\mathbf X^\dagger].
     \end{equation}
   Applying the same argument, we similarly obtain
  \begin{equation}
    \label{eq:blabla}
  \sum_k
       \|
        \calE^+(\mathbf Y)v_k\|^2
    =
        \E_{\CYL{L}}[\mathbf Y^\dagger\mathbf Y]. \end{equation}
Combining~\eqref{eq:preCS2}--\eqref{eq:blabla} yields the Cauchy-Schwarz inequality, completing the proof.
\end{proof}

\section{Spectral representation for the two-point function (Theorem~\ref{thm:spectral_representation_four_point})}

Before diving into the proof of Theorem~\ref{thm:spectral_representation_four_point}, let us start by introducing two more operators -- the vertical-arrow and up-shift operators -- and discuss their basic properties.
To measure arrow-arrow correlations between vertical arrows, we introduce some new operators encoding the direction of the vertical arrow between positions $j$ and $j+1$.
For $j\in(\Z/L\Z)$, set
\(
    s_j(\pi/2):=\frako_{\alpha_j}^{01}\),
    or equivalently
\begin{align}
    e_{\kappa'}^\dagger s_j(\pi/2) e_\kappa &:= Z_{\kappa\kappa'}^{01}
    \E_{\kappa\kappa'}^{01}[\alpha_j]
    \\
    &{\color{white}:}=
    \sum_{\alpha\in\{\pm1\}^L}\alpha_j \cdot
    \true{\text{ice rule for $(\kappa,\alpha,\kappa')$}}
    \cdot
    \c^{\#\{\text{$\c$-vertices in $(\kappa,\alpha,\kappa')$}\}}.
\end{align}
Write $S(\pi/2):=\frakO_{\alpha_0}^{01}=s_0(\pi/2)/\lambda_0(\pi/2)$.

\begin{remark} For Step~3, it suffices to compute the arrow-arrow correlation between two vertical arrows.
    We shall see that the general case follows from the addition rule
    and a mixing property. We therefore work exclusively with the operator that measures vertical arrows and avoid introducing operators for horizontal arrows altogether.
\end{remark}

Next, define the up-shift operator.
For any $\kappa\in\frakC$,
we let $\kappa^\uparrow\in\frakC$
denote the column configuration such that $(\kappa^\uparrow)_i=\kappa_{i-1}$.
Similarly, we define $\kappa^\downarrow\in\frakC$ such that
$(\kappa^{\downarrow})^{\uparrow}=\kappa$.
They are called the \emph{up} and \emph{down} shift respectively.
Define the \emph{up-shift operator} $T(0):\Omega\to\Omega$ by
\begin{equation}
    e_{\kappa'}^\dagger T(0)e_\kappa:=\true{\kappa' = \kappa^\uparrow}.
\end{equation}
By going back to the definition of $s_j(\pi/2)$, it is easy to see that
\begin{align}
    s_j(\pi/2) &= T(0)^js_0(\pi/2)T(0)^{-j};\\
    \frakO_{\alpha_j}^{01} &=\frac{s_j(\pi/2)}{\lambda_0(\pi/2)}=T(0)^jS(\pi/2)T(0)^{-j}.
\end{align}
Let us now collect some basic properties of these operators.

\begin{lemma}[The shift operator]
The basis $(v_k)_k$ introduced in Lemma~\ref{lem:properties transfer matrix} may be chosen such that it diagonalises $T(0)$ as well. If $\Lambda_k(0)$ denotes the eigenvalue associated with $v_k$, then the $\Lambda_k(0)$ are $L$-th roots of unity, and $\Lambda_0(0)=1$.\end{lemma}

\begin{proof}
    The operator $T(0)$ is normal since $T(0)$ commutes with $T(0)^\dagger=T(0)^{-1}$.
    Moreover, $T(0)$ and $T(\pi/2)$ commute thanks to shift-invariance of the six-vertex model.
    Therefore $T(0)$ and $T(\pi/2)$ are co-diagonalisable. We suppose without loss of generality that $T(0)$ was already diagonal in the basis $(v_k)_k$. This can be done without breaking the properties of $(v_k)_k$ from Lemma~\ref{lem:properties transfer matrix}. Indeed, in Lemma~\ref{lem:properties transfer matrix}
    we simply chose an arbitrary orthonormal basis diagonalising $T(\pi/2)$,
    and then ordered it, but the ordering does not play a role here (in particular because the eigenspace of the $\Lambda_0(\pi/2)$ is one dimensional).
    % \commentKarol{$\Lambda_0(0)$ can be a    degenerate eigenvalue of $T(0)$ since the latter is not Perron Frobenius}
    % \noticePiet{You're right, it should have been $\pi/2$. Updated}

    Since $T(0)^L=\operatorname{Identity}$,
    its eigenvalues are $L$-th roots of unity.
    Finally, $v_0$ is the Perron-Frobenius eigenvector of $T(\pi/2)$,
    and we imposed that all its entries in the basis $(e_\kappa)_\kappa$ are strictly positive.
    Since $T(0)$ has non-negative coefficients (it is a permutation matrix) in the basis $(e_\kappa)_\kappa$,
    this means that the entries of $T(0)v_0$ are also positive.
    Since $T(0)v_0=\Lambda_0(0)v_0$, it forces $\Lambda_0(0)=1$.
\end{proof}

\begin{lemma}[The edge measurement operator]
    The operator $S(\pi/2)$ is anti-Hermitian,
    and $v_0^\dagger S(\pi/2)v_0=0$.
\end{lemma}

\begin{proof}
    Focus on the first property.
    It is straightforward to see that $s_0(\pi/2)$ is anti-Hermitian,
    by going back to its definition in the basis $(e_\kappa)_{\kappa}$.
    Indeed, the definitions imply that
    \begin{align}
        e_{\kappa'}^\dagger s_0(\pi/2) e_\kappa
        &=
        \sum_{\alpha\in\{\pm1\}^L}\alpha_j \cdot
        \true{\text{ice rule for $(\kappa,\alpha,\kappa')$}}
        \cdot
        \c^{\#\{\text{$\c$-vertices in $(\kappa,\alpha,\kappa')$}\}}
        \\&=
        \sum_{\alpha\in\{\pm1\}^L}\alpha_j \cdot
        \true{\text{ice rule for $(\kappa',-\alpha,\kappa)$}}
        \cdot
        \c^{\#\{\text{$\c$-vertices in $(\kappa',-\alpha,\kappa)$}\}}
        \\
        &=-e_{\kappa}^\dagger s_0(\pi/2) e_{\kappa'}.
    \end{align}
    For the second property, notice that Equation~\eqref{eq:cylop} yields
    \begin{equation}\label{eq:flip symmetry}
        v_0^\dagger S(\pi/2)v_0
        =
        v_0^\dagger\;\frakO_{\alpha_0}^{01}\;v_0
        =
        \E_{\CYL{L}}[\text{the orientation of a fixed vertical edge}]=0,
    \end{equation}
    where the expectation is zero since the measure is invariant
    under a global arrow flip.
\end{proof}

\begin{proof}[Proof of Theorem~\ref{thm:spectral_representation_four_point}]
For any even integer $L$, define the measure
\begin{equation}\label{eq:def muL}
    \mu_L:=
    \sum_{k>0}
        \frac{|v_k^\dagger S(\pi/2) v_0|^2}{(1-\Lambda_k(\pi/2))^2}
        \delta_{(1-\Lambda_k(\pi/2),-\i\log\Lambda_k(0))}.
\end{equation}

The support of $\mu_L$ is contained in $(0,2)\times[-\pi,\pi)$, since $|\Lambda_k(\pi/2)|<1$ for every $k>0$.
This inequality also ensures that the denominator $1-\Lambda_k(\pi/2)$ is nonzero for every $k>0$, thereby validating the definition.

The proof now proceeds in three steps.
First, we prove Equation~\eqref{eq:thm:spectral_representation_four_point} under the constraint that $y_1=y_2=0$. Then, we extend it to the generality of Theorem~\ref{thm:spectral_representation_four_point}.
Finally, we conclude the proof.

\paragraph{Step 1: Equation~\eqref{eq:thm:spectral_representation_four_point} under the constraint that $y_1=y_2=0$.}
In this step, we consider $\u $ where $u_1$ and $u_1'$ are on the same horizontal line,
and that the same holds true for $u_2$ and $u_2'$.
In that case, $\Phi_{\CYL{L},2}(\u)$ can be written as a sum of
two-point functions for  $\u $ with $(u_1,u_1')$ and $(u_2,u_2')$
horizontally adjacent.
The additivity property of Equation~\eqref{eq:Phi-additive} therefore reduces the problem to proving Equation~\eqref{eq:thm:spectral_representation_four_point} in this special case.

 We work out the left and right of Equation~\eqref{eq:thm:spectral_representation_four_point}. Consider $\u $.
When $u_1'-u_1=u_2'-u_2=(1,0)$, we are just calculating an arrow-arrow correlation
between vertical arrows. Recall the definition of $\u _2-\u _1'=(x'_1,y'_1)$,
and that $y(u)$ denotes the second coordinate of $u\in\R^2$.
For the left-hand side, we get
\begin{align}
    \Phi_{\CYL{L},2}(\u )
    &=
    \E_{\CYL{L}}[(h(u_1')-h(u_1))(h(u_2')-h(u_2))]
    \\&=
    v_0^\dagger \; \frakO_{\alpha_{y(u_2)}}^{01} \; T(\pi/2)^{x_1'} \; \frakO_{\alpha_{y(u_1)}}^{01} \; v_0
    \\&=
    v_0^\dagger T(0)^{y(u_2)}S(\pi/2)T(0)^{-y(u_2)} T(\pi/2)^{x_1'}T(0)^{y(u_1)}S(\pi/2)T(0)^{-y(u_1)}v_0
    \\&=
    v_0^\dagger S(\pi/2) T(\pi/2)^{x_1'}T(0)^{-y_1'}S(\pi/2)v_0.
\end{align}
In the last line we use that $T(\pi/2)$ and $T(0)$ commute.
By expanding $S(\pi/2)$ in the basis $(v_k)_k$,
and using the anti-Hermitian property,
we get
\begin{equation}
    \Phi_{\CYL{L},2}(\u ) =-
    \sum_{k>0} |v_k^\dagger S(\pi/2) v_0|^2 \Lambda_k(\pi/2)^{x_1'}\Lambda_k(0)^{-y_1'}.
\end{equation}
We omitted the $k=0$ term since $v_0^\dagger S(\pi/2) v_0=0$ due to flip symmetry \eqref{eq:flip symmetry}.

For the right hand side of Equation~\eqref{eq:thm:spectral_representation_four_point}
we get, using that $x_1=x_2=1$ and $y_1=y_2=0$,
\begin{multline}
\int   -a^2
        (1-a)^{x_1'}e^{-\i by_1'}
        \diffi\mu_L(a,b)
        \\
    =-
    \sum_{k>0}
        \frac{|v_k^\dagger S(\pi/2) v_0|^2}{(1-\Lambda_k(\pi/2))^2}
        (1-\Lambda_k(\pi/2))^2
        \Lambda_k(\pi/2)^{x_1'}\Lambda_k(0)^{-y_1'}.
\end{multline}
This matches the expression we found for $\Phi_{\CYL{L},2}(\u )$.

\paragraph{Step 2 : Equation~\eqref{eq:thm:spectral_representation_four_point} in the general case.}

We now relax the condition that each pair $(u_i,u_i')$ lies on a horizontal line.

We shall first relax the condition that $y_1=0$,
and then the condition that $y_2=0$.
In fact, the two proofs are the same, and we focus on the first step.

Fix $\u $ such that $y_1\neq 0$ and $y_2=0$.
The idea is to write $\Phi_{\CYL{L},2}(\u )$ as the sum of three terms,
where two terms fall under the umbrella of Step~4,
and where the third term is an error term that tends to zero
in a certain limit.

More precisely, we write the height difference $h(u_1')-h(u_1)$
as a sum of increments along the path $(u_1,u_1-ne_1,u_1'-ne_1,u_1')$, with $e_1=(1,0)$.
The first and third steps are covered by Step~4, thus giving
\begin{multline}
    \Phi_{\CYL{L},2}(\u )
    -
    \Phi_{\CYL{L},2}\big(u_1-ne_1,u_1'-ne_1,u_2,u_2'\big)
    \\=
    \int   \left((1-a)^{x_2}-1\right)
        (1-a)^{x_1'}e^{-\i by_1'}
        \left(1-(1-a)^{x_1}e^{-\i by_1}\right)\left(1-(1-a)^n\right)\diffi\mu_L(a,b).
\end{multline}
Now let $n$ tend to infinity.
Since $\mu_L$ is a finite sum of Dirac measures with $0<a<2$ almost surely,
the right-hand side tends to
\begin{equation}
    \int   \left((1-a)^{x_2}-1\right)
        (1-a)^{x_1'}e^{-\i by_1'}
        \left(1-(1-a)^{x_1}e^{-\i by_1}\right)\diffi\mu_L(a,b).
\end{equation}
It suffices to prove that as $n$ tends to infinity, the following error term vanishes:
\begin{equation}
    \Phi_{\CYL{L},2}\big(u_1-ne_1,u_1'-ne_1,u_2,u_2'\big)=
    \E_{\CYL{L}}\big[\big(h(u_1'-ne_1)-h(u_1-ne_1)\big)\big(h(u_2')-h(u_2)\big)\big].
\end{equation}
But this expectation may be written as
\begin{equation}
    v_0^\dagger\frakO_2 T(\pi/2)^n \frakO_1 v_0,
\end{equation}
where each $\frakO_i$ measures one of the two height differences.
As $n$ tends to infinity,
this tends to $v_0^\dagger\frakO_2 v_0v_0^\dagger \frakO_1 v_0$,
since $\Lambda_0(\pi/2)=1$ and all other eigenvalues have a modulus strictly smaller than $1$.
But $v_0^\dagger\frakO_1 v_0=v_0^\dagger \frakO_2 v_0=0$ as each factor
encodes the expectation of a single height difference, which is zero
by global flip symmetry.
This proves the case that $y_1\neq 0$ and $y_2=0$;
the general case is proved similarly.

\paragraph{Step 3: Conclusion}
    We constructed the measure $\mu_L$
    above,
    we observed that it is supported on $(0,2)\times[-\pi,\pi)$,  and we established Equation~\eqref{eq:thm:spectral_representation_four_point} for all the desired points $\u$.
    The fact that $\mu_L[\{|b|\in (0,2\pi/L)\}]=0$
    follows from the observation that the eigenvalues
    $e^{-\i b}$
    of $T(0)$ are $L$-th roots of unity.

    Finally, we want to prove that $\mu_L$ is invariant under the reflection $(a,b)\mapsto (a,-b)$.
%    The above reflection corresponds to applying a complex conjugation to the weighted eigenvalues
%        of the two operators.
%        Complex conjugation is an element of the Galois group of the field extension $\R\subset \C$.
%        Since all operators are real-valued in the basis $(e_\kappa)_\kappa$,
%        the Galois group preserves the weighted joint spectra of $T(\pi/2)$ and $T(0)$.
%        In particular, $\mu_L$ is invariant under the reflection $(a,b)\mapsto (a,-b)$.
     But this is immediate, as we may simply replace $\mu_L$ by its symmetrised
        version $(\mu_L+\bar\mu_L)/2$ owing to the real-valuedness of the correlation function,
        which is invariant under the reflection by construction and clearly still satisfies all the other properties of Theorem~\ref{thm:spectral_representation_four_point}.
\end{proof}

\begin{remark}
The above reflection $(a,b)\mapsto (a,-b)$ corresponds to applying a complex conjugation to the weighted eigenvalues
        of the two operators.
        Complex conjugation is an element of the Galois group of the field extension $\R\subset \C$.
        Since all operators are real-valued in the basis $(e_\kappa)_\kappa$,
        the Galois group preserves the weighted joint spectra of $T(\pi/2)$ and $T(0)$.
        In particular, the formula in \eqref{eq:def muL} is itself invariant under the reflection $(a,b)\mapsto (a,-b)$.
\end{remark}

% </input src="sections/PART_C/new.tex" root="." version="0.0.1">
% <input src="sections/PART_D/main.tex" root="." version="0.0.1">
\part{\INGREDREGULARITY: Regularity estimates and qualitative behaviour}
\label{part:qualitative}

The purpose of this part is to prove the following intermediate results:
\begin{itemize}
    \item Existence of the infinite-volume six-vertex measure (Theorem~\ref{thm:infinite_volume_6V}),
    \item The results stated in \INGREDREGULARITY\ (Theorems~\ref{thm:Regularity}, \ref{thm:mixing}, \ref{thm:intro:arm_exponents}, \ref{thm:intro:flip_domination}, and Corollary~\ref{cor:Regularity_cylinder}).
\end{itemize}
These results have already been applied in Part~\ref{part:proofs},
and we will apply them again in Part~\ref{part:glimpse} (where we derive the remaining missing ingredients).

The proofs in this part are based on the Fortuin--Kasteleyn--Ginibre (FKG)
inequality and the Russo--Seymour--Welsh (RSW) theory.
We already mentioned that several representations of the six-vertex model
satisfy this FKG inequality for $\c\geq 1$.
We found the \emph{spin representation} of the six-vertex model the most
convenient for formalising the proof
(see~\cite{Lis_2022_SpinsPercolationHeight,Lis_2021_DelocalizationSixVertexModel,GlazmanLammers_2025_DelocalisationContinuity2D}).
We expect that the proofs can also be written down in terms of the
FKG inequality for the absolute value of the height function (see~\cite{Duminil-CopinKarrilaManolescu_2024_DelocalizationHeightFunction}),
but we do not pursue this route here.

This part is organised as follows.
Sections~\ref{sec:RSW}--\ref{sec:perco} introduce the spin representation and its properties.
Although the results are more or less known (for the spin representation,
the results may be found, for example, in~\cite{GlazmanLammers_2025_DelocalisationContinuity2D}), we still state everything
precisely because we need some subtle variations of the known results,
as well as minor extensions. Everything is written in a self-contained fashion,
except for the circuit estimate, which was proved in~\cite{Duminil-CopinKarrilaManolescu_2024_DelocalizationHeightFunction}
and~\cite{GlazmanLammers_2025_DelocalisationContinuity2D}.

Once the language and standard results for the spin representation have been established,
we prove the desired intermediate results one by one in Sections~\ref{sec:full_plane_spins}--\ref{sec:mixing}.

% <input src="sections/PART_D/12_spins/main.tex" root="." version="0.0.1">
\section{Spin representation and RSW theory}
\label{sec:RSW}

% <input src="sections/PART_D/12_spins/01_intro.tex" root="." version="0.0.1">
\subsection{Motivation of the spin representation}
\label{introintrosectionspin}

Recall that $h:F(\Z^2)\to\Z$ denotes the height function representation of the six-vertex model,
which is a function differing by $\pm1$ on neighbouring faces and which assigns
an even number to the face whose south-west corner is $(0,0)$.
We may partition $F(\Z^2)$ into two so that
$h$ assigns even numbers to $F_\sleven$ (the \emph{even} faces) and odd numbers to $F_\slodd$
(the \emph{odd} faces).

The gradient of $h$ can be recovered from
the values of $h$ modulo $4$, that is, from the function
\begin{equation}
    h/4\Z:F(\Z^2)\to\Z/4\Z,\,x\mapsto h(x)+4\Z.
\end{equation}
Moreover, this height function modulo $4$ can be encoded in terms of a family
$(\sigma_\sleven,\sigma_\slodd)\in\{\pm\}^{F_\sleven(\Z^2)}\times\{\pm\}^{F_\slodd(\Z^2)}$ of
\emph{spins}.
Indeed, we may simply define $(\sigma_\sleven,\sigma_\slodd)$ as follows:
\begin{itemize}
    \item
    If $x$ is even, then
    \begin{equation}
        \sigma_\sleven(x):=\begin{cases}
            + &\text{if $h(x)\in 0+4\Z$,}
            \\
            - &\text{if $h(x)\in 2+4\Z$;}
        \end{cases}
    \end{equation}
    \item
    If $x$ is odd, then
    \begin{equation}
        \sigma_\slodd(x):=\begin{cases}
            + &\text{if $h(x)\in 1+4\Z$,}
            \\
            - &\text{if $h(x)\in 3+4\Z$.}
        \end{cases}
    \end{equation}
\end{itemize}
The spin representation of a height function (that is, the height function modulo $4$)
thus encodes the height function up to constant shifts by constant multiples of $4$.
Spins at even faces are called \emph{even spins}
and spins at odd faces are called \emph{odd spins}.

The spin representation turns out to be useful because of two reasons:
\begin{itemize}
    \item The even spins satisfy the Fortuin--Kasteleyn--Ginibre inequality,
    \item The even spins have a ``smallest'' and ``largest'' value,
    namely $-$ and $+$ respectively, setting it appart from the height function
    which is unbounded and has no smallest or largest value.
\end{itemize}

Although $F(\Z^2)$ is a set of faces, we often identify each face with its face centre.
This way we may view $F(\Z^2)$ as a vertex set embedded in $\R^2$,
and we may identify edges between faces with line segments between face centres.
Note that this coincides with the standard dual graph of $\Z^2$, but we avoid referring to it explicitly since several other graphs will appear later on.

% </input src="sections/PART_D/12_spins/01_intro.tex" root="." version="0.0.1">
% <input src="sections/PART_D/12_spins/02_definition.tex" root="." version="0.0.1">
\subsection{Formal definition of the spin representation}

\begin{definition}[Even and odd sublattices]
We shall identify each face in $F(\Z^2)$ with its face centre.
The set of even faces $F_\sleven$ is endowed with an edge set $E_\sleven$
such that the four neighbours of a face at $(i,j)$ are given by $(i\pm1,j\pm1)$.
We shall also simply write $F_\sleven$ for the graph $(F_\sleven,E_\sleven)$.
Similar definitions apply to the odd sublattice $F_\slodd$.
See Figure~\ref{fig:lattices}.
\end{definition}

\begin{figure}[h]
    \begin{subfigure}{0.5\textwidth}
        \begin{center}
        \includegraphics{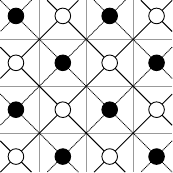}
        \end{center}
        \caption{The even and odd lattices}
    \end{subfigure}%
    \begin{subfigure}{0.5\textwidth}
        \begin{center}
        \includegraphics{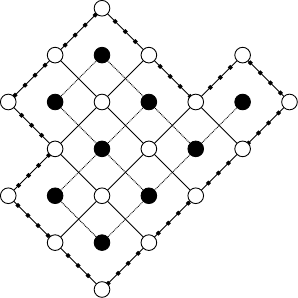}
        \end{center}
        \caption{An even domain}
    \end{subfigure}

    \begin{subfigure}{0.5\textwidth}
        \begin{center}
        \includegraphics{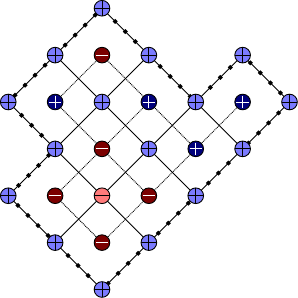}
        \end{center}
        \caption{A consistent spin configuration}
    \end{subfigure}%
    \begin{subfigure}{0.5\textwidth}
        \begin{center}
        \includegraphics{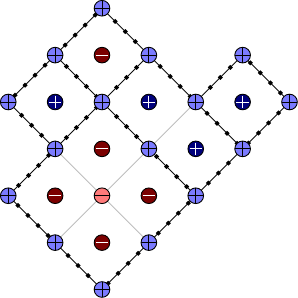}
        \end{center}
        \caption{A positive probability triple $(\sigma_\sleven,\omega,\sigma_\slodd)$}
    \end{subfigure}
    \caption{}
    \label{fig:lattices}
\end{figure}

\begin{definition}[Even domain]
    \label{def:even_domain}
In the context of percolation theory, we shall identify each edge
$\{x,y\}\subset\R^2$ with the straight closed line segment connecting the vertices $x$
and $y$.
Let $\partial\calD$ denote a finite self-avoiding circuit through the even sublattice
$F_\sleven$.
We let $F_\sleven(\calD)$ denote the subgraph of $F_\sleven$ consisting
of the vertices and edges which are entirely contained in the closure of the set of
points surrounded by $\partial\calD$ (in particular, $\partial\calD\subset F_\sleven(\calD)$).
Its edge set is denoted $E_\sleven(\calD)$ and its vertex set simply $F_\sleven(\calD)$.
The subgraph $F_\slodd(\calD)\subset F_\slodd$ is defined similarly,
except that obviously $\partial\calD$ cannot be a subgraph of $F_\slodd(\calD)$.
The triple $\calD:=(\partial\calD,F_\sleven(\calD),F_\slodd(\calD))$ is called
an \emph{even domain}, see Figure~\ref{fig:lattices}. We also write $F(\calD)$ for the subgraph
of the graph $F(\Z^2)$
(where faces sharing an edge are neighbours)
induced by the vertex set $F_\sleven(\calD)\cup F_\slodd(\calD)$.
We shall not define or use odd domains.
\end{definition}

\begin{definition}[Consistent spin configurations]
    Let $\calD$ denote an even domain.
    A \emph{spin configuration}  on $\calD$ is a pair
    \begin{equation}(\sigma_\sleven,\sigma_\slodd)\in\{\pm\}^{F_\sleven(\calD)}\times\{\pm\}^{F_\slodd(\calD)}.\end{equation}
    We call such a spin configuration \emph{consistent} if for any edge $uv\in E_\slodd(\calD)$
    with dual  edge $xy=uv^*\in E_\sleven(\calD)$ (here dual means intersecting it in its middle),
    we have
    \begin{equation}
        \sigma_\slodd(u)=\sigma_\slodd(v)
        \qquad
        \text{or}
        \qquad
        \sigma_\sleven(x)=\sigma_\sleven(y)
        \qquad\text{(or both)}.
    \end{equation}
    In this case we write $\sigma_\sleven\perp\sigma_\slodd$;
    see Figure~\ref{fig:lattices} for a consistent spin configuration.
\end{definition}

The ice rule and the consistency rule are two sides of the same coin:
they are the necessary and sufficient conditions for the existence of the associated height function.

\begin{definition}[Agreement edges]
    We also introduce an associated set of \emph{agreement edges}
    \begin{equation}
        A(\sigma_\sleven):=\{xy\in E_\sleven(\calD):\sigma_\sleven(x)=\sigma_\sleven(y)\}.
    \end{equation}
    We introduce the same notation for $\sigma_\slodd$.
    Thus, two spin configurations are consistent if and only if
    the \emph{complements} of the agreement edges do not cross each other.
\end{definition}

In the definitions below, we use the symbol $\propto$ to denote ``proportional to;'' The reader should bear in mind that a normalization constant is implicitly present to ensure that the measures are probability measures.

We would like to define a probability measure $\muSpin_{\calD}^+$ in which the random pair $(\sigma_\sleven,\sigma_\slodd)$
follows the distribution
\begin{equation}
    \label{eq:spinmpredef}
    \muSpin_{\calD}^+[(\sigma_\sleven,\sigma_\slodd)]
    \propto
    \true{\sigma_\sleven|_{\partial\calD}\equiv+}
    \cdot
    \true{\sigma_\sleven\perp\sigma_\slodd}
    \cdot \c^{\# A(\sigma_\sleven)} \cdot \c^{\# A(\sigma_\slodd)}.
\end{equation}
We shall prove that this is just a transformation of the six-vertex model
with fixed boundary conditions
at parameters $\a=\b=1$
and $\c\in[1,2]$.

We interpret the above formula as follows:
$\sigma_\sleven$ and $\sigma_\slodd$ are independent ferromagnetic Ising models
(on dual graphs),
conditioned on the event that their domain walls do not intersect.
The Ising model $\sigma_\sleven$ comes with fixed boundary conditions,
while $\sigma_\slodd$ comes with free boundary conditions.

After conditioning on $\sigma_\sleven$,
the distribution of $\sigma_\slodd$ may be interpreted as an Ising model,
except that some of its coupling constants are set to infinity
(due to the indicator function $\true{\sigma_\sleven\perp\sigma_\slodd}$).
This conditional Ising model has an FK--Ising coupling.
Rather than taking Equation~\eqref{eq:spinmpredef} as a definition,
we shall directly define the probability measure $\muSpin_{\calD}^+$
on a larger space, which also incorporates this FK--Ising coupling.
In what follows, the associated percolation configuration will be denoted by $\omega$.

\begin{definition}[Spin measure]
    \label{definition:spinmeasure}
    Consider an even domain $\calD$.
    Let $\Omega=\Omega_\calD$ denote the sample space
    \begin{equation}
        \Omega:=
        \{\pm\}^{F_\sleven(\calD)}\times\{0,1\}^{E_\sleven(\calD)}\times\{\pm\}^{F_\slodd(\calD)}
        .
    \end{equation}
    A typical element is denoted $(\sigma_\sleven,\omega,\sigma_\slodd)$.
    We often identify $\omega$ with the set $\{xy\in E_\sleven(\calD):\omega(xy)=1\}$,
    and use the standard percolation terminology.
    Below, for an edge $xy\in E_\sleven$, let $xy^*$ be the unique edge in $E_\slodd$ sharing the same middle.
    Define the probability measure $\muSpin_{\calD}^+$ on $(\sigma_\sleven,\omega,\sigma_\slodd)\in\Omega$ by
    \begin{align}
        \label{eq:spinmeasurefirstdef}
        \muSpin_{\calD}^+[(\sigma_\sleven,\omega,\sigma_\slodd)]
        \propto{}
        &\true{\sigma_\sleven|_{\partial\calD}\equiv+}\cdot\true{\partial\calD\subset\omega}
        \cdot\true{\omega\subset A(\sigma_\sleven)}
        \\
        &\qquad
        \cdot\true{\text{$xy\in\omega$ or $xy^*\in A(\sigma_\slodd)$ (or both) for any $xy\in E_\sleven(\calD)$}}
        \\
        &\qquad
        \cdot \c^{\#A(\sigma_\sleven)}
        \cdot \left(\tfrac1\c\right)^{\#\omega}
        \cdot\left({1-\tfrac1\c}\right)^{\#(A(\sigma_\sleven)\setminus\omega)}.
    \end{align}
    See Figure~\ref{fig:lattices} for a sample from this measure.
\end{definition}

\begin{lemma}
    \label{lemma:spinmpredef}
    The definition of $\muSpin_{\calD}^+$ is consistent with Equation~\eqref{eq:spinmpredef}.
\end{lemma}

\begin{proof}
    For fixed $(\sigma_\sleven,\sigma_\slodd)$ one recovers Equation~\eqref{eq:spinmpredef} by summing over $\omega$.
    \commentHugo{What is the link between the $(\sigma_\sleven,\sigma_\slodd)$ and Ashkin-Teller? HDC: the comment can be ignored for the first version.}
\end{proof}

\subsection{Basic properties of the spin representation}

Recall that an edge $xy\subset\R^2$ is identified with the straight line segment
from $x$ to $y$.
For example, we view $\omega$ as a (closed) subset of $\R^2$;
each edge is represented by a closed line segment between the two face centres.

\begin{lemma}[Flip symmetry]
    \label{lemma:oddspinsflip}
    Consider the measure $\muSpin_{\calD}^+$ conditional on $(\sigma_\sleven,\omega)$.
    Then the distribution of $\sigma_\slodd$ is given by flipping a fair coin for each
    bounded connected component of $\R^2\setminus\omega$.
\end{lemma}

\begin{proof}
    It is immediate from the definition of $\muSpin_{\calD}^+$ that conditionally on
    $(\sigma_\sleven,\omega)$, the distribution of $\sigma_\slodd$
    is uniform in the set of configurations $\{\pm\}^{F_\slodd(\calD)}$
    satisfying
    \begin{equation}
        \label{eq:condition_for_sigma_odd}
    \text{$xy^*\in A(\sigma_\slodd)$ for any $xy\in E_\sleven(\calD)\setminus\omega$},
    \end{equation}
    see also Figure~\ref{fig:lattices}.
    This leads to the distribution stated in the lemma.
\end{proof}

We now consider the marginal law of the pair $(\sigma_\sleven,\omega)$.
This marginal is particularly useful, as we often disregard the odd-spin configuration $\sigma_\slodd$.
For a set $A\subset\R^2$, let $f(A)$ be the number of connected components of $\R^2\setminus A$.

\begin{lemma}[$(\sigma_\sleven,\omega)$ marginal]
    \label{lemma:spin_marginal_weight}
   For every $(\sigma_\sleven,\omega)\in\{\pm\}^{F_\sleven(\calD)}\times\{0,1\}^{E_\sleven(\calD)}$,
    \begin{multline}
        \muSpin_{\calD}^+[(\sigma_\sleven,\omega)]
        \propto
        \\
        \true{\sigma_\sleven|_{\partial\calD}\equiv+}\true{\partial\calD\subset\omega}
      \true{\omega\subset A(\sigma_\sleven)} 2^{f(\omega)}
      \c^{\#A(\sigma_\sleven)-\#\omega}
       \left({1-\tfrac1\c}\right)^{\#(A(\sigma_\sleven)\setminus\omega)}.
    \end{multline}
\end{lemma}

\begin{proof}
    The proof is related to the previous lemma.
    For fixed $(\sigma_\sleven,\omega)$,
    one needs to count the number of configurations $\sigma_\slodd$
    which satisfy the condition in Equation~\eqref{eq:condition_for_sigma_odd}.
    This number if simply given by
$  2^{f(\omega)-1}$ (the $-1$ is due to the fact that the unique unbounded connected component is not involved).    This leads to the expression in the lemma.
\end{proof}

We now state a simple combinatorial observation.
To understand the law of $(\sigma_\sleven,\omega)$,
it is clearly important to evaluate $f(\omega)$, that is, to count the number of connected components of $\R^2\setminus\omega$.
Since $\omega\subset A(\sigma_\sleven)$, each edge
in $\omega$ connects two spins with the same sign.
Thus, we may write $A(\sigma_\sleven)=A^-(\sigma_\sleven)\cup A^+(\sigma_\sleven)$ and $\omega=\omega^-\cup\omega^+$,
where:
\begin{itemize}
    \item $A^-(\sigma_\sleven)$ and $\omega^-$ connect faces with $\sigma_\sleven$-spin $-$,
    \item $A^+(\sigma_\sleven)$ and $\omega^+$ connect faces with $\sigma_\sleven$-spin $+$,
    \item No face is incident to an edge of $A^-(\sigma_\sleven)$ and an edge of $A^+(\sigma_\sleven)$,
    \item No face is incident to an edge of $\omega^-$ and an edge of $\omega^+$.
\end{itemize}
In particular, we observe that the following formula holds:
\begin{equation}
    \label{eq:connected_components_formula}
  f(\omega)    =
    -1+
    \sum_{\#\in\{+,-\}}
   f(\omega^\#).
\end{equation}
We shall also write $\bar\omega$ for the
ordered pair of two percolation configurations $(\omega^+_\sleven,\omega^-_\sleven)$.
We would like to rewrite the weight in Lemma~\ref{lemma:spin_marginal_weight} in terms of these new objects.
The  (long) expression in the following lemma
shows that $\omega^+_\sleven$ and $\omega^-_\sleven$ only interact
via $\sigma_\sleven$, that is, they are independent after conditioning
on $\sigma_\sleven$.

\begin{lemma}
    \label{lemma:spin_marginal_weight_2}
    For any $(\sigma_\sleven,\bar\omega)\in\{\pm\}^{F_\sleven(\calD)}\times\{0,1\}^{E_\sleven(\calD)}\times\{0,1\}^{E_\sleven(\calD)}$,
    we have
    \begin{align}
        \muSpin_{\calD}^+[(\sigma_\sleven,\bar\omega)]
        \propto{}
        & \c^{\#A(\sigma_\sleven)}
        \\&\times
        \true{\partial\calD\subset \omega^+_\sleven}
        \true{\omega^+_\sleven\subset A^+(\sigma_\sleven)}2^{f(\omega^+)}\left(\tfrac1\c\right)^{\#\omega^+}
        \cdot\left({1-\tfrac1\c}\right)^{\#(A^+(\sigma_\sleven)\setminus\omega^+)}
        \\
        &\times        \true{\omega^-_\sleven\subset A^-(\sigma_\sleven)}
      2^{f(\omega^-)}
      \left(\tfrac1\c\right)^{\#\omega^-}
        \cdot\left({1-\tfrac1\c}\right)^{\#(A^-(\sigma_\sleven)\setminus\omega^-)}.
    \end{align}
\end{lemma}

\begin{proof}
    This follows from Equation~\eqref{eq:connected_components_formula}
    and straightforward manipulations.
\end{proof}

In Equation~\eqref{eq:connected_components_formula}, the simple topology
of the plane $\R^2$ plays an important role;
the equation does not immediately generalise to, for example, the torus.
When working with the torus, we must therefore slightly modify our setup
(see Section~\ref{section:cylinder} and Lemma~\ref{lemma:realFKGcylinder}).

Until now we defined all our measures with $+$ boundary conditions,
but they can equally be defined with $-$ boundary conditions;
we write $\muSpin_{\calD}^-$ for this measure.

\subsection{Six-vertex height function with fixed boundary conditions}

The spin representation is directly related to the six-vertex model. Consider an even domain $\calD$ and $(\sigma_\sleven,\omega,\sigma_\slodd)\sim\muSpin_{\calD}^+$.
Given $(\sigma_\sleven,\sigma_\slodd)$, one recovers the gradient
of the height function via the definition of the spins
at the beginning of this section (Page~\pageref{introintrosectionspin}).
More precisely, for adjacent faces $uv$
with $u\in F_\sleven$ and $v\in F_\slodd$, then
\begin{equation}
    \label{eq:relationheightsspins}
    h(v)-h(u)=\sigma_\sleven(u)\sigma_\slodd(v).
\end{equation}
\commentHugo{This sounds so much like bosonization. HDC: the comment can be ignored for a first version}
This defines the gradient of $h$ on $F(\calD)$
in the measure $\muSpin_{\calD}^+$.
The gradient is turned into a non-gradient function by imposing that
it equals $0$ on $\partial\calD$.

The function $h$ so defined is a \emph{height function} on $F(\calD)$:
an integer-valued function $h:F(\calD)\to\Z$
which differs by exactly $\pm1$ on adjacent faces
and which, as such, preserves the parity of each face.
By Lemma~\ref{lemma:spinmpredef} (Equation~\eqref{eq:spinmpredef}),
its law is given by:
\begin{equation}
    \mu_{\calD}^+[h]
    \propto
    \true{h|_{\partial\calD}\equiv 0}\cdot \c^{\# A(h)},
\end{equation}
where $A(h)\subset E_\sleven(\calD)\cup E_\slodd(\calD)$ denotes the \emph{agreement diagonals}.
The factor $\c^{\# A(h)}$ is consistent with the definition of the six-vertex model
(Equation~\eqref{eq:sixvertexweights}) since $\a=\b=1$
and since $\c$-type vertices induce one more agreement diagonal
than $\a$-type or $\b$-type vertices.

\begin{definition}[Level lines]
    From now on, for any $k\in2\Z$, write
    \begin{align}
        A^k&=A^k(h):=\{uv\in E_\sleven(\calD):h(u)=h(v)=k\};
        \\
        \omega^k&:=\omega\cap A^k.
    \end{align}
    The percolation $\omega^k\subset E_\sleven(\calD)$ is called the \emph{level line} of height $k$.
\end{definition}

Notice that
\begin{equation}
    \omega^+=\bigcup_{k\in 4\Z}\omega^k;
    \qquad
    \omega^-=\bigcup_{k\in 4\Z+2}\omega^k.
\end{equation}
Notice that $(\omega^k)_{k\in2\Z}$ does not just partition
$\omega$:
this partition also has the property that each connected component of
$\omega$ is contained in one $\omega^k$.
If a connected component $\alpha$ of $\omega$ is a subset of $\omega^k$,
then $h(x)=k$ for any $x$ incident to $\alpha$, and we simply say that $\alpha$
has height $k$.

Before, we saw that the odd spins can only change sign if they are separated
by $\omega$-edges (see for example Lemma~\ref{lemma:oddspinsflip}).
This immediately implies an intermediate value theorem,
which also motivates the terminology of \emph{level lines}.

\begin{lemma}[Intermediate value theorem]
    Consider the measure $\muSpin_{\calD}^+$ in some even domain $\calD$.
    Fix a target height $k\in 2\Z$
    as well as two faces $u,v\in F(\calD)$
    and two heights $a,b\in\Z$ with $a<k<b$.
    Then almost surely, the following statement holds true:
    if $h(u)=a$ and $h(v)=b$
    and if $\gamma:u\to v$ is any continuous path in $\R^2$ from $u$ to $v$,
    then $\gamma$ hits $\omega^k$.
\end{lemma}

\subsection{Markov property of the spin representation}

The spin representation satisfies a Markov property along even domains, a feature that will be fundamental to the analysis below. We present this property in the current section.

If $\calD$ and $\calB$ are two even domains, then we write
$\calD\subset\calB$, and say that $\calD$ is \emph{contained} in $\calB$,
whenever $E_\sleven(\calD)\subset E_\sleven(\calB)$.
Consider two even domains $\calD\subset\calB$.
Recall the definition of the sample space $\Omega_\calD$
(Definition~\ref{definition:spinmeasure}),
and write $\pi_\calD:\Omega_\calB\to\Omega_\calD$ for the natural projection map
(which simply erases the values of the spins and edges not relevant to $\Omega_\calD$).
Write $\pi_\calD^c$ for the complementary projection map,
so that $\pi_\calD\times\pi_\calD^c$ is the identity map on $\Omega_\calD$.

\begin{lemma}[Markov property]
    \label{lemma:realMarkov}
    Consider two even domains $\calD\subset\calB$.
    For $\#,\flat\in\{\pm\}$ such  that $\muSpin_{\calB}^\#[\{\partial\calD\subset\omega^\flat\}]>0$,
    \begin{itemize}
        \item The law $\muSpin_{\calB}^\#[\blank|\{\partial\calD\subset\omega^\flat\}]$ of $\pi_\calD(\sigma_\sleven,\bar\omega,\sigma_\slodd)$ is the same as the law $\mu_{\calD}^\flat$ of $(\sigma_\sleven,\bar\omega,\sigma_\slodd)$,
        \item The random variables $\pi_\calD$ and $\pi_\calD^c$ are independent in $\muSpin_{\calB}^\#[\blank|\{\partial\calD\subset\omega^\flat\}]$.
    \end{itemize}
\end{lemma}

\begin{proof}
    Take the expression in Lemma~\ref{lemma:spin_marginal_weight_2}
    and insert an extra indicator for the conditional event.
    It is then straightforward to work out that the weight factorises over $\calD$ and its complement,
    as desired.
\end{proof}

Since the distribution of $\sigma_\slodd$ conditional on $(\sigma_\sleven,\bar\omega)$
is very simple (Lemma~\ref{lemma:oddspinsflip}),
it makes sense to focus the analysis entirely on $(\sigma_\sleven,\bar\omega)$.
From now on, we shall write
\begin{equation}
    \Omega_\calD^\sleven
    :=
    \{\pm\}^{F_\sleven(\calD)}\times\{0,1\}^{E_\sleven(\calD)}.
\end{equation}
For any $\calD\subset\calB$,
we write $\pi_{\calD}^\sleven:\Omega_\calB^\sleven\to\Omega_\calD^\sleven$
for the associated projection map,
and $\pi_{\calD}^{c,\sleven}$ for the natural complementary projection map
such that $\pi_{\calD}^\sleven\times \pi_{\calD}^{c,\sleven}=\pi_\calB^\sleven$.
Finally, we shall often drop the subscript $\sleven$ from
$\sigma_\sleven$, $\omega$, $\bar\omega$, and $\omega^\pm_\sleven$
for brevity.

We want to prove one more Markov property, for so-called polar domains.
A \emph{polar domain} is an even domain $\calD$
together with a partition $\partial\calD$ into two segments
labelled $\partial^+\calD$ and $\partial^-\calD$.
More precisely, we impose that the vertices of these segments
partition the vertices on $\partial\calD$,
and that the edges of the segments partition $E(\partial\calD)$ except that the
two edges connecting the endpoints do not belong to any part
(see Figure~\ref{fig:polar_domain}).

\begin{figure}[h!]
    \begin{center}
        \includegraphics{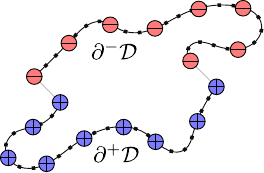}
    \end{center}
    \caption{A polar domain.}
    \label{fig:polar_domain}
\end{figure}

\begin{lemma}[Markov property for polar domains]
    \label{lem:markov_polar}
    Consider a polar domain $(\calD,\partial^+\calD,\partial^-\calD)$
    as well as another even domain $\calB\supset\calD$.
    Consider the event
    \begin{equation}
        \calE:=\{\text{$\partial^+\calD\subset\omega^+$ and $\partial^-\calD\subset\omega^-$}\}.
    \end{equation}
    Let $\#\in\{\pm\}$.
    If $\muSpin_{\calB}^\#[\calE]>0$,
    then in the conditional measure $\nu:=\muSpin_{\calB}^\#[\blank|\calE]$,
    there is a Markov property over $\partial\calD$
    for $(\sigma_\sleven,\omega)$. More precisely,
     $\pi_\calD^\sleven$ and $\pi_\calD^{c,\sleven}$ are independent in $\nu$.
\end{lemma}

\begin{proof}
    On the event $E$, the values of $\sigma_\sleven$ on $\partial\calD$
    are known. Thus, the only factor
    in the weight
    of Lemma~\ref{lemma:spin_marginal_weight_2}
    that may make
    $\pi_\calD^\sleven$ and $\pi_\calD^{c,\sleven}$ interact,
    is the factor
    \begin{equation}
        \label{eq:factorisable_weight}
        \prod_{\flat\in\pm}
        2^{f(\omega^\flat)}.
    \end{equation}
    Notice that on the event $E$, we have
    \begin{align}
       f(\omega^\flat)
        =
        1&+\#\text{conn.~comp.~of $\R^2\setminus\omega^\flat$ which are entirely surrounded by $\partial\calD$}
        \\
        &+
        \#\text{conn.~comp.~of $\R^2\setminus\omega^\flat$ which are entirely outside $\partial\calD$}
        .
    \end{align}
    The first term on the right is $\pi_\calD^\sleven$-measurable,
    and the second term is $\pi_\calD^{c,\sleven}$-measurable,
    proving the desired factorisation and independence.
\end{proof}

% </input src="sections/PART_D/12_spins/02_definition.tex" root="." version="0.0.1">
% <input src="sections/PART_D/12_spins/03b_FKG.tex" root="." version="0.0.1">
\section{FKG inequality of the spin representation}

We now state and prove the Fortuin-Kasteleyn-Ginibre (FKG) inequality. We also discuss several of its immediate consequences.

\begin{definition}[Increasing functions and the FKG property]
    A random variable is called \emph{$\sleven$-increasing}
    (or simply \emph{increasing})
    if it may be written as an increasing function of
    the triple $(\sigma_\sleven,\omega^+,-\omega^-)$.
    Notice the minus sign in the last entry;
    this means that $\omega^-_\sleven$-open edges are lower in the partial order on such triples.
    A random variable $X$ is called \emph{$\sleven$-decreasing} (or simply \emph{decreasing})
    whenever $-X$ is $\sleven$-increasing.
    An event is called $\sleven$-increasing or $\sleven$-decreasing
    if its indicator function is $\sleven$-increasing
    or $\sleven$-decreasing respectively.

    A probability measure $\mu$ is said to have \emph{$\sleven$-FKG}
    (or simply \emph{FKG})
    whenever
    \begin{equation}\operatorname{Cov}_\mu[X,Y]:=\mu[XY]- \mu[X]\mu[Y]\geq 0\end{equation}
    for any bounded $\sleven$-increasing random variables $X$
    and $Y$ which are measurable  in terms of finitely many spins and edges.
\end{definition}

The following statement captures the essence of this subsection.

\begin{lemma}
    \label{lemma:simpleFKG}
    The measures $\muSpin_{\calD}^+$
    and $\muSpin_{\calD}^-$
    have $\sleven$-FKG for any even domain $\calD$.
\end{lemma}

This lemma was proved in~\cite{GlazmanLammers_2025_DelocalisationContinuity2D},
but we shall state and prove a more general version of it.
More precisely, we shall define \emph{lattice events}, which are events $\calE$
with the property that conditioning on $\calE$ preserves FKG.
Many events of interest can be partitioned into lattice events,
which is very useful when applying the FKG inequality.

\begin{definition}[Lattice event]
A \emph{$\sleven$-lattice event} (or \emph{lattice event}) is an event of the form
\begin{equation}
    \{\omega|_Q=\zeta\}\cap\{\sigma_\sleven|_B=\xi\},
\end{equation}
where $B$ and $Q$ contain vertices and edges of $F_\sleven$ respectively,
and such that $B$ contains all endpoints of edges in $Q$.
Naturally, we require that $\zeta\in\{0,1\}^Q$
and $\xi\in\{\pm\}^B$.
Notice that if $Q=B=\emptyset$,
then $\calE$ is just the entire sample space.
\end{definition}

\begin{proposition}[FKG inequality]
    \label{proposition:realFKG}
    Consider an even domain $\calD$
    and a $\sleven$-lattice event $\calE$.
    Let $\#\in\{\pm\}$.
    If $\muSpin_{\calD}^\#[\calE]>0$,
    then the conditional probability measure $\muSpin_{\calD}^\#[\blank|\calE]$
    has $\sleven$-FKG.
\end{proposition}

    We closely follow~\cite[Theorem~2.8]{LammersOtt_2024_DelocalisationAbsolutevalueFKGSolidonsolid} and~\cite[Proposition~4.9]{GlazmanLammers_2025_DelocalisationContinuity2D}.
    The reader may choose to skip this technical proof on a first read.
    We focus on the case $\#=+$;
    the proof of the other case is identical.
   The proof of $\sleven$-FKG follows from the following lemma
    which is proved below.
    \begin{lemma}\label{lem:1} Fix a domain $\calD$ and a lattice event $\calE$ such that $\muSpin_{\calD}^+[\calE]>0$.
        Then, all of the following hold true in the conditional measure
    $\nu:=\muSpin_{\calD}^+[\blank|\calE]$.
     \begin{enumerate}
        \item The weights of $\sigma_\sleven$ satisfy the \emph{FKG lattice condition},
        which implies the FKG inequality for $\sigma_\sleven$~\cite{FortuinKasteleynGinibre_1971_CorrelationInequalitiesPartially}.
        \item Conditional on $\sigma_\sleven$,
        the percolations $\omega^+$ and $\omega^-$ are independent.
        \item Conditional on $\sigma_\sleven$,
        the law of $\omega^+$ satisfies the FKG inequality.
        \item The conditional law of $\omega^+$ is stochastically increasing in $\sigma_\sleven$.
        \item Conditional on $\sigma_\sleven$,
        the law of $\omega^-$ satisfies the FKG inequality.
        \item The conditional law of $\omega^-$ is stochastically decreasing in $\sigma_\sleven$.
    \end{enumerate}
\end{lemma}

Before proving the lemma, let us derive the FKG inequality.
    \begin{proof}[Proof of Proposition~\ref{proposition:realFKG}]
    As mentioned, we focus on the $+$ case. Fix $\calD$ and $\calE$ such that $\muSpin_{\calD}^+[\calE]>0$ and set $\nu=\muSpin_{\calD}^+[\blank|\calE]$.
   We apply the so-called \emph{tower property} for the FKG inequality.

    Lemma~\ref{lem:1}(ii)--(vi) implies that conditionally on $\sigma_\sleven$, the pair $(\omega^+,-\omega^-)$ satisfies the FKG inequality, and that the conditional law of the pair $(\omega^+,-\omega^-)$ is stochastically increasing in $\sigma_\sleven$.

    Now, let $X$ and $Y$ denote two bounded $\sleven$-increasing functions.
    Assert that
    \begin{equation}
        \nu[XY]= \nu[\nu[XY|\sigma_\sleven]]
        \geq
        \nu[\nu[X|\sigma_\sleven]\nu[Y|\sigma_\sleven]]
        \geq
        \nu[\nu[X|\sigma_\sleven]]\nu[\nu[Y|\sigma_\sleven]]
        =\nu[X]\nu[Y].
    \end{equation}
    This standard trick (see~\cite{LammersOtt_2024_DelocalisationAbsolutevalueFKGSolidonsolid})
    is proved as follows.
    The tower property implies the two equalities.
    The first inequality is the conditional FKG of the pair $(\omega^+,-\omega^-)$.
    Since the law of $(\omega^+,-\omega^-)$ is stochastically increasing in $\sigma_\sleven$,
    we see that $\nu[X|\sigma_\sleven]$ and $\nu[Y|\sigma_\sleven]$ are increasing functions
    of $\sigma_\sleven$.
    The second inequality then follows from the FKG inequality for $\sigma_\sleven$
    (Lemma~\ref{lem:1}(i)).\end{proof}

\begin{proof}[Proof of Lemma~\ref{lem:1}] We omit the subscript and write $\sigma=\sigma_\sleven$. To prove the lemma, we first  find an appropriate decomposition of $\nu$ and then derive the items one by one.

\paragraph{Step 1: Decomposition of $\nu$.}
    Since $\calE=\{\omega|_Q=\zeta\}\cap\{\sigma|_B=\tau\}$
    has positive probability,
    the edge set $Q$ may be written as the disjoint union
    \begin{equation}
        Q^+\cup Q^-\cup Q^\emptyset,
    \end{equation}
    where $Q^\emptyset=\{\zeta=0\}$
    and where $\tau$ is equal to $+1$ on the endpoints of $Q^+$
    and to $-1$ on the endpoints of $Q^-$.
    Without loss of generality, $\partial\calD\subset Q^+$.
    Inserting the indicator functions for the conditioning event
    in the expression of Lemma~\ref{lemma:spin_marginal_weight_2} yields
    \begin{align}
        \label{eq:bigexpressioninFKGproof}
        \nu[(\sigma,\bar\omega)]\propto{}&
        \true{\sigma|_B=\tau}
        \cdot
        \c^{\#A(\sigma)}
        \cdot
        2^{f(\omega^+)}Y_+(\sigma,\omega^+)
        \cdot
        2^{f(\omega^-)}Y_-(\sigma,\omega^-);
        \end{align}
        where
        \begin{align}
        \label{eq:bigexpressioninFKGproof_line2}
        \begin{split}
        Y_\pm(\sigma,\omega^\pm):={}&\true{Q^\pm\subset\omega^\pm}
            \true{\omega^\pm\cap Q^\emptyset=\emptyset}
        \true{\omega^\pm\subset A^\pm(\sigma)}
        \\
        &
        \qquad \cdot
        (\tfrac1\c)^{\#(\omega^\pm\cap(A^\pm(\sigma)\setminus Q))}
            (1-\tfrac1\c)^{\#((A^\pm(\sigma)\setminus Q)\setminus\omega^\pm)}.
        \end{split}
%            \\ \label{eq:bigexpressioninFKGproof_line3}
%        \begin{split}
%            Y_-(\sigma,\omega^-) :={}&\true{Q^-\subset\omega^-}
%            \true{\omega^-\cap Q^\emptyset=\emptyset}
%            \true{\omega^-\subset A^-(\sigma)}
%            \\
%            &\qquad\cdot
%             (\tfrac1\c)^{\#(\omega^-\cap(A^-(\sigma)\setminus Q))}
%            (1-\tfrac1\c)^{\#((A^-(\sigma)\setminus Q)\setminus\omega^-)}.
%        \end{split}
    \end{align}

    \paragraph{Step 2: Proof of (ii).}
    For fixed $\sigma$, the above weight may be written as a product
    of one factor depending only on $\omega^+$, and one depending
    only on $\omega^-$.
    This implies the desired independence.

\paragraph{Step 3: Proof of (i).}
Proving (i) is the most delicate part of the argument; once it is established, the remaining
claims follow easily.
To compute the weight of a spin configuration $\sigma$, we must sum over
$\omega^+$ and $\omega^-$ in
Equation~\eqref{eq:bigexpressioninFKGproof}.
Since in this expression only~$2^{f(\omega^+)}Y_+(\sigma,\omega^+)$ depends on $\omega^+$,
and only~$2^{f(\omega^-)}Y_-(\sigma,\omega^-)$ on $\omega^-$, the sums may be carried out separately.
This is almost the same for $\omega^+$ and $\omega^-$.
We first focus on $\omega^+$, and then explain how to adapt the argument for $\omega^-$.

The key idea is the following.
Conditional on~$\sigma$, the percolation configuration $\omega^+$ behaves like the dual
of a random-cluster model on $F_\slodd(\calD)$ with cluster weight $q=2$,
corresponding to an Ising model on the odd faces via the Edwards--Sokal coupling.
The law and partition function of both the random-cluster model and the Ising model
are well understood, and this correspondence allows us to verify the claims below.
For completeness, we give full detail.

Introduce the couplings
\begin{equation}
    \label{eq:definitionofacouplings}
a_{uv}^\pm(\sigma)
=
\begin{cases}
1    & uv^*\in Q^\pm,\\
1/\c & uv^*\in A^\pm(\sigma)\setminus Q,\\
0    & \text{otherwise}.
\end{cases}
\end{equation}
and the Ising model (on $F_\slodd(\calD)$) partition function $\mathcal{Z}_{\mathrm{Ising}}(a)$ with couplings $a$ defined by
\begin{equation}
\mathcal{Z}_{\mathrm{Ising}}(a)
=
\sum_{\tilde\sigma\in\{\pm1\}^{F_\slodd(\calD)}}
\prod_{uv\in E_\slodd(\calD)}
(a_{uv})^{\true{\tilde\sigma(u)\neq \tilde\sigma(v)}}.
\end{equation}
Our objective is to compute $\sum_{\omega^+}2^{f(\omega^+)}Y_+(\sigma,\omega^+)$.
First, rewrite
\begin{equation}
    \label{eq:standardESrewriting}
    2^{f(\omega^+)}=2\quad\sum_{\mathclap{\tilde\sigma\in\{\pm1\}^{F_\slodd(\calD)}}}\quad
    \true{\tilde\sigma\perp\omega^+},
\end{equation}
where $\tilde\sigma\perp\omega^+$ means that $\tilde\sigma$ is constant on each connected component of $\R^2\setminus\omega^+$.
The prefactor two on the right compensates for the unbounded face counted in $f(\omega^+)$.
Inserting the previous formula and exchanging the sums yields
\begin{equation}
\sum_{\omega^+}2^{f(\omega^+)}Y_+(\sigma,\omega^+)
=
2\sum_{\tilde\sigma\in\{\pm1\}^{F_\slodd(\calD)}}
\sum_{\omega^+,\,\tilde\sigma\perp\omega^+}
Y_+(\sigma,\omega^+).
\end{equation}
Although the expression for $Y_+$ is lengthy,
the inner sum (over $\omega^+\perp\tilde\sigma$) is easy to compute.
In fact, the indicators and the requirement $\omega^+\perp\tilde\sigma$ simply tell us that some edges must be open or closed.
The sum over the remaining edges can be performed independently.
By carrying out this computation, carefully bookkeeping the conditions on $\omega^+$,
one obtains
\begin{equation}
    \label{eq:sumoveromegaplus}
    \sum_{\omega^+}2^{f(\omega^+)}Y_+(\sigma,\omega^+)
    =
    2\mathcal{Z}_{\mathrm{Ising}}(a^+(\sigma)),
\end{equation}

Similarly, one obtains
\begin{equation}
\sum_{\omega^-}2^{f(\omega^-)}Y_-(\sigma,\omega^-)
=
\mathcal{Z}_{\mathrm{Ising}}(a^-(\sigma))
\end{equation}
(contrary to Equation~\eqref{eq:sumoveromegaplus}, we do not need the prefactor two,
as the unbounded face of $\R^2\setminus\omega^-$ intersects $F_\slodd(\calD)$ and therefore
its sign is already accounted for).
Putting the two expressions together yields
\begin{equation}
    \nu[\{\sigma\}]
\propto
\true{\sigma|_{\partial\calD}\equiv +}
\cdot
\true{\sigma|_B=\tau}
\cdot
\c^{\#A(\sigma)}
\cdot
\mathcal{Z}_{\mathrm{Ising}}(a^+(\sigma))
\cdot
\mathcal{Z}_{\mathrm{Ising}}(a^-(\sigma)).
\end{equation}
We now verify that each factor satisfies the FKG lattice condition:
\begin{equation}
    \label{eq:FKGlatticeconditionforweights}
    g(\sigma\vee\sigma')g(\sigma\wedge\sigma')
\ge
g(\sigma)g(\sigma'),
\end{equation}
where $\sigma \vee \sigma'$ and $\sigma \wedge \sigma'$ denote the pointwise maximum and minimum of $\sigma$ and $\sigma'$.

The first two factors trivially satisfy the condition as they are constant (recall that we are interested in configurations with positive $\nu=\nu_{\calD}^+[\cdot|\calE]$).
The third one is classical: the map $\sigma\mapsto\c^{\#A(\sigma)}$ corresponds to
ferromagnetic Ising interactions and it is straightforward to check the desired inequality.
We now handle $\calZ_{\operatorname{Ising}}(a^\pm(\sigma))$.
We focus on $+$, the case $-$ being similar.
The definition of $a^+$ (Equation~\eqref{eq:definitionofacouplings}) and the inequality $\c\ge1$ imply that
\begin{equation}
    \label{eq:FKGlatticeconditionforacouplings}
    a^+_{uv}(\sigma\vee\sigma')\geq a^+_{uv}(\sigma)\vee a^+_{uv}(\sigma');
    \qquad
    a^+_{uv}(\sigma\wedge\sigma')= a^+_{uv}(\sigma)\wedge a^+_{uv}(\sigma').
\end{equation}
Suppose now for a second that the map
\begin{equation}
    \label{eq:FKGlatticeconditionforacouplings2}
    a\mapsto\mathcal{Z}_{\mathrm{Ising}}(a)
\end{equation}
satisfies the FKG lattice condition over $a\in[0,1]^{E_\slodd(\calD)}$.
Since $\calZ_{\operatorname{Ising}}(a)$ is increasing in $a$,
Equation~\eqref{eq:FKGlatticeconditionforacouplings} and the FKG lattice condition for~\eqref{eq:FKGlatticeconditionforacouplings2}
imply the desired Equation~\eqref{eq:FKGlatticeconditionforweights} for $g(\sigma)=\mathcal{Z}_{\mathrm{Ising}}(a^+(\sigma))$.
The FKG lattice condition for Equation~\eqref{eq:FKGlatticeconditionforacouplings2} is classical
(after renormalising the partition function in a way that does not affect the FKG lattice condition),
and may be found in~\cite[Equations~(35), (36), (38)]{Frohlich_1982_TrivialityLphd4Theories}, \cite[Proposition A.1]{Chayes_1998_DiscontinuitySpinWaveStiffness},
\cite[Lemma~6.1 and Equation~(7)]{LammersOtt_2024_DelocalisationAbsolutevalueFKGSolidonsolid}, or~\cite{GlazmanLammers_2025_DelocalisationContinuity2D}.

We have established that all factors satisfy the FKG lattice condition,
and therefore the same holds true for their product $\nu[\{\sigma\}]$.
By~\cite{FortuinKasteleynGinibre_1971_CorrelationInequalitiesPartially}, this implies that $\sigma$ satisfies the FKG
property under~$\nu$. This completes the proof of (i).

    \paragraph{Proof of (iii)--(vi).}
    Let us start with (iii) and (v). Conditionally on $\sigma$,
    the percolation $\omega^+$ is the dual of a random-cluster model
    with cluster weight $q=2$,
    which is well-known to satisfy the FKG inequality.
    The same holds true for $\omega^-$.

  We now prove (iv); (vi) being derived similarly.
    Conditionally on $\sigma$,
    the coupling constants of the random-cluster model
    are encoded in $a^+(\sigma)$.
    Notice that $a^+(\sigma)$ (the inverse of the coupling strengths) is \emph{increasing}
    in $\sigma$.
    This means that law of $\omega^+$, which is the dual of the random-cluster model,
    is stochastically \emph{increasing} in $\sigma$.
    This proves (iv) and concludes the proof.
\end{proof}

We now state a classical consequence of the monotonicity properties established above, enabling to ``push'' domains ``away'', and to compare boundary conditions. It is used to circumvent the lack of independence in the model. Similar statements can be found in the theory of Ising and random-cluster models. This result will play a central role in the probabilistic analysis of the spin configuration that follows.

\begin{corollary}[Monotonicity in domains and boundary conditions]
    \label{cor:Monotonicity in domains and boundary conditions}
    Consider two even domains $\calD\subset\calB$
    as well as a bounded $\sleven$-increasing random variable $X$.
    Then
    \begin{equation}
        \muSpin_{\calD}^-[X]
        \leq
        \muSpin_{\calB}^-[X]
        \leq
        \muSpin_{\calB}^+[X]
        \leq
        \muSpin_{\calD}^+[X].
    \end{equation}
\end{corollary}

\begin{proof}
    Let $\calM$ be an even domain containing $\calB$ which is so big that $\partial\calB$
    does not intersect $\partial\calM$.
    Then the two events $\{\partial\calB\subset\omega^\pm\}$ have a positive probability
    in
    $\muSpin_{\calM}^+$.
    We claim that
    \begin{equation}
        \muSpin_{\calB}^+[X]
        =\muSpin_{\calM}^+[X|\partial\calB\subset\omega^+]
        \geq\muSpin_{\calM}^+[X].
    \end{equation}
    Indeed, the equality is the Markov property, and the inequality is the FKG inequality.
    This proves the claim.
    Similarly, we get
    \begin{equation}
        \muSpin_{\calB}^-[X]
        =\muSpin_{\calM}^+[X|\partial\calB\subset\omega^-]
        \leq\muSpin_{\calM}^+[X],
    \end{equation}
    leading to
    \begin{equation}
        \muSpin_{\calB}^-[X]\leq\muSpin_{\calB}^+[X].
    \end{equation}
    The other inequalities in the statement of the lemma are similar; for example,
    on the right side, we obtain
    \begin{equation}
        \muSpin_{\calB}^+[X]
        \leq
        \muSpin_{\calB}^+[X|\partial\calD\subset\omega^+]
        =
        \muSpin_{\calD}^+[X].
    \end{equation}
    This proves the lemma.
\end{proof}

% </input src="sections/PART_D/12_spins/03b_FKG.tex" root="." version="0.0.1">
% <input src="sections/PART_D/12_spins/04_percolation.tex" root="." version="0.0.1">

\section{Percolation estimates for the spin configuration}\label{sec:perco}

From now on, we shall consistently write
    \begin{alignat}{2}
            &\symInt{n}:=[-n,n];
        \qquad
        &&\symRect{n}{m}:=[-n,n]\times[-m,m];
        \label{Definition box}
    \\
    \label{eq:prelimpercostuff}
            &\bbox{n}:=\symRect{n}{n};
        \qquad
        &&\bbox{n}(x):=\bbox{n}+x;
    \\
        &\ann{R}{r}:=\bbox{R}\setminus\bbox{r};
        \qquad
        &&\ann{R}{r}(x):=\ann{R}{r}+x.
    \end{alignat}

    A \emph{rectangle} is a subset of $\R^2$ of the form
    $R=\symRect{a}{b}+x$.
    Its four sides (closed line segments which are subsets of $\partial R$)
    are denoted $\rectright{R}$, $\recttop{R}$, $\rectleft{R}$,
    and $\rectbottom{R}$ in the obvious way.

\subsection{Percolation events associated with the spin representation}

Recall that we view $\omega^+$ and $\omega^-$ as random subsets of $\R^2$
(each edge is viewed as the closed line segment between its two endpoints).
Our objective is to understand the random geometry of $\bar\omega=(\omega^+,\omega^-)$.
In this section we define some useful percolation events:
first ``simple'' events (defined in terms of either $\omega^+$ or $\omega^-$),
then ``alternating'' events (defined in terms of both $\omega^+$ and $\omega^-$).

\begin{definition}[Simple percolation events]
Define
\begin{align}
    \connect{A}{B}{D}
    :=\{S\subset\R^2:\text{$S\cap D$ contains a path from $A$ to $B$}\};
\end{align}
see Figure~\ref{fig:genericevent}.
%For example, the event
%\begin{equation}
%    \{\omega^+\in\connect{A}{B}{D}\}
%    =
%    \{\text{$\omega^+\cap D$ contains a path from $A$ to $B$}\}
%\end{equation}
%is an $\sleven$-increasing event.
Define the \emph{horizontal} and \emph{vertical crossings} of rectangles to be the sets
\begin{align}
    \HORI{R}&:=\connect{\rectleft{R}}{\rectright{R}}{R};\\
    \VERTI{R}&:=\connect{\recttop{R}}{\rectbottom{R}}{R}.
\end{align}
If $A\subset\R^2$ is a topological annulus, define the \emph{circuit} and \emph{arm} events
\begin{align}
    \CIRCUIT{A}&{}:=\{S\subset\R^2:\text{$S\cap A$ contains a non-contractible circuit  in $\mathbb R^2\setminus A$}\};
    \\
    \ARM{A}&{}:=\{S\subset\R^2:\text{$S\cap A$ has a path connecting the boundary components of $A$}\}.
\end{align}
See Figure~\ref{fig:percoevents} for an illustration of all four events we just defined.

\end{definition}

\begin{figure}
    \begin{center}
    \includegraphics{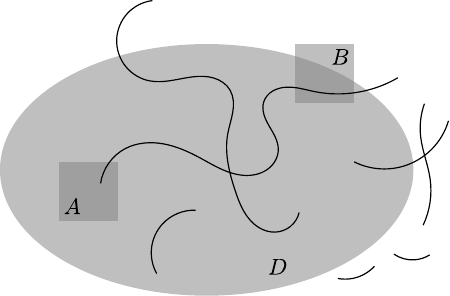}
    \end{center}
    \caption{An element of the generic percolation event $\connect{A}{B}{D}$}
    \label{fig:genericevent}
\end{figure}

\begin{figure}
    \begin{center}
    \includegraphics{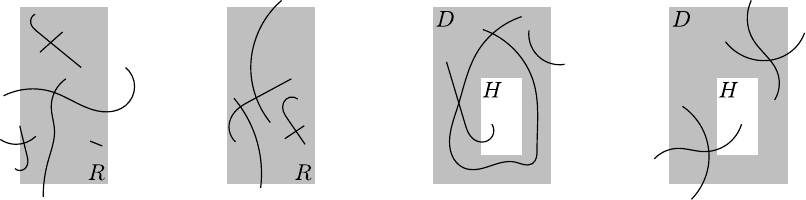}
    \end{center}
    \caption{Elements of $\HORI{R}$, $\VERTI{R}$, $\CIRCUIT{D\setminus H}$, and $\ARM{D\setminus H}$}
    \label{fig:percoevents}
\end{figure}

\begin{remark}[Combining circuit events]
    \label{remark:combinecircuits}
Circuit events are versatile as they can be combined to create (horizontal or vertical)
crossings (see Figure~\ref{fig:annulus_to_hor}); for example, for $k\geq 1$, we have
\begin{equation}
    \bigcap_{i=0}^{k-1} \CIRCUIT{\ann{2}{1}((2i,0))}
    \subset
    \HORI{[-1,2k-1]\times[1,2]}.
\end{equation}
They can also be combined to create more complicated percolation events.
For example,
for $r\in\Z_{\geq 1}$, we may find a subset
$Z\subset\R^2$ of cardinal $\# Z=4r$ such that
\begin{equation}
    \bigcap_{z\in Z}\CIRCUIT{\ann{2}{1}(z)}
    \subset
    \CIRCUIT{\ann{r+1}{r}}.
\end{equation}
It is often quite straightforward (but technically tedious) to construct a set $Z$
whose cardinal is optimal up to a constant factor.
In those cases, rather than giving a precise definition of $Z$,
we leave the choice to the reader, and refer to this remark instead.
\end{remark}

\begin{figure}
    \begin{center}
    \includegraphics{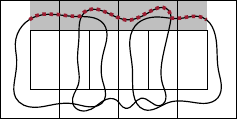}
    \end{center}
    \caption{Jointly the three circuits imply the crossing of the rectangle.}
    \label{fig:annulus_to_hor}
\end{figure}

\begin{definition}[Alternating percolation events]
\label{def:altpercoevents}
For the following definitions, let $\frakR$ denote the set
\begin{equation}\{(S_+,S_-):\text{$S_+,S_-\subset\R^2$ and $S_+\cap S_-=\emptyset$}\}.\end{equation}
\commentKarol{$S_{\pm}$ correspond to paths $\omega_{\pm}$ if I remember correctly. Maybe this could be said here?}
We now introduce our events; $R$ is a rectangle, and $A$ is a topological annulus.
\begin{itemize}
    \item $\ALTHORI{2k}{R}$ consists of the elements $(S_+,S_-)\in\frakR$
    such that
    $\rectleft{R}$ contains $2k$ distinct vertices
    \begin{equation}
        s^+_1,s^-_1,s^+_2,s^-_2,\dots,s^+_k,s^-_k
    \end{equation}
    appearing in descending order, such that
    each $s_i^\#$ is path-connected to $\rectright{R}$
    via a path in $S_\#\cap R$.

    \item $\ALTVERTI{2k}{R}$ is defined identically:
    it consists of the elements $(S_+,S_-)\in\frakR$
    such that
    $\recttop{R}$ contains $2k$ distinct vertices
    \(
        s^+_1,s^-_1,s^+_2,s^-_2,\dots,s^+_k,s^-_k
    \)
    appearing from left to right, such that
    each $s_i^\#$ is path-connected to $\rectbottom{R}$
    via a path in $S_\#\cap R$.
    \item $\ALTCIRCUIT{2k}{A}$ contains the elements $(S_+,S_-)\in\frakR$
    such that we may find $2k$ disjoint circuits of
    $A$,
    ordered from outside to inside,
    and where the circuits are alternately contained in $S_+$
    and $S_-$ with the first circuit belonging to $S_+$.
    \item $\ALTARM{2k}{A}$ is defined similarly;
    in this case, the arms are circularly ordered, but not ordered;
    there is no notion of topmost, leftmost, or outermost crossing or circuit.
\end{itemize}
    See Figure~\ref{fig:percoevents2} for an illustration.
\medbreak
We will use these definitions for different sets. We therefore add the following notation.

For any random subset $\alpha$ of $\R^2$, write
$\pHORI{\alpha}{R}:=\{\alpha\in\HORI{R}\}$, and similarly for other events.
If $\alpha$ is a random subset of $F(\Z^2)$,
then we think of $\alpha$ as being a subset of $\R^2$ defined by the union of all line segments connecting
centres of nearest-neighbour faces belonging to $\alpha\subset F(\Z^2)$.

    Suppose now that $\bar\alpha=(\alpha^+,\alpha^-)$ is a random
            pair of percolations.
    Finally, for each of the four events defined above,
    we introduce two more notations.
    Like above, introduce
    \begin{align}
                \pALTHORI{\bar\alpha}{2k}{R}&:=\{\bar\alpha\in\ALTHORI{2k}{R}\},\\
            \COUNTHORI{\bar\alpha}{R}&:=
            \max\{2k\in 2\Z_{\geq 0}:\bar\alpha\in\ALTHORI{2k}{R}\}
        .\end{align}
    These notations naturally adapt to the other three events.
\end{definition}

\begin{figure}[t]
    \begin{center}
    \includegraphics{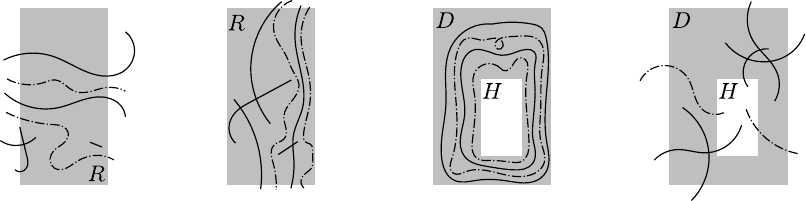}
    \end{center}
    \caption{Elements of $\ALTHORI{4}{R}$, $\ALTVERTI{4}{R}$, $\ALTCIRCUIT{4}{D\setminus H}$, and $\ALTARM{4}{D\setminus H}$.
    Full lines depict $\omega^+$; dotted lines $\omega^-$.}
    \label{fig:percoevents2}
\end{figure}

% </input src="sections/PART_D/12_spins/04_percolation.tex" root="." version="0.0.1">
% <input src="sections/PART_D/12_spins/05_renormalisation_input.tex" root="." version="0.0.1">
\subsection{Circuit estimate for the spin configuration}
\label{subsection:generalinput}

Percolation theory has seen remarkable progress over the past sixty years. For planar Bernoulli percolation, it was shown in the 1980s \cite{SeymourWelsh_1978_PercolationProbabilitiesSquare,Russo_1978_NotePercolation} (see also  \cite{Kohler-SchindlerTassion_2025_IntroductionRussoSeymourWelshTheory} and references therein) that at criticality, rectangles with a fixed aspect ratio have crossing probabilities -- that is, the probability that the percolation configuration contains a connected component crossing the rectangle -- that remain uniformly bounded away from both 0 and 1 as the size of the rectangle tends to infinity. This reflects the fact that connected components in the critical regime qualitatively exhibit scale-free behaviour and possess fractal-like geometric properties.

Originally developed for Bernoulli percolation, the Russo--Seymour--Welsh (RSW) theory has become indispensable in the analysis of critical phenomena. Over the past fifteen years, the theory has been significantly extended to encompass many dependent percolation models \cite{Duminil-CopinHonglerNolin_2011_ConnectionProbabilitiesRSWtype,Duminil-CopinSidoraviciusTassion_2017_ContinuityPhaseTransition,Duminil-CopinTassion_2015_RSWBoxcrossingProperty,Duminil-CopinTassionTeixeira_2018_BoxcrossingPropertyCritical,Duminil-CopinManolescuTassion_2021_PlanarRandomclusterModel,Duminil-CopinTassion_2019_RenormalizationCrossingProbabilities,Lammers_2023_DichotomyTheoryHeight,Kohler-SchindlerTassion_2023_CrossingProbabilitiesPlanar}. In this broader context, particular attention must be paid to the dependence between the configuration inside a given rectangle and its exterior. To be applicable, crossing estimates must therefore be uniform with respect to boundary conditions.

When studying the emergence of large connected components, an especially convenient geometric setting is that of an annulus rather than a rectangle. In this setting, one asks whether the percolation configuration contains a path that remains within the annulus and encircles its inner boundary. We adopt this annular framework throughout the present section.

The subsection states the key RSW-type input for
our percolation-type arguments. The theorem below yields that $\omega^-$ satisfies RSW-type estimates, even when boundary conditions are least favourable
(that is, $+$ boundary conditions).

%Recall that for any $x\in\R^2$ and $r\in(0,\infty)$,
%we defined
%$\bbox{r}(x)=[-r,r]^2+x$.
For any rectangle $R$,
we shall write $\muSpin_{R}^+$ for $\muSpin_{\calD}^+$,
where $\calD$ is the \emph{largest even domain}
whose face centres are all contained in $R$.
\begin{theorem}[Circuit estimate]
    \label{thm:renorminput}
   There exists a constant $\ccircuit>0$ (independent of $\c\in[1,2]$) such that
    for any $r\in[4,\infty)$ and $x\in\R^2$,
    \begin{equation}
        \muSpin_{\bbox{2r}(x)}^+[\pCIRCUIT{\omega^-}{\ann{2r}{r}(x)}]\geq\ccircuit.
    \end{equation}
    The same inequality remains true with $+$ and $-$ interchanged.
\end{theorem}

The result tells us that $\omega^-$-circuits have a uniformly positive
probability of appearing, even with the \emph{worst possible} boundary conditions (in the sense of increasing functions).
The uniformity in boundary conditions renders the result very flexible.

The theorem was first proved in~\cite[Theorem~1.4]{Duminil-CopinKarrilaManolescu_2024_DelocalizationHeightFunction}.
The article uses an input coming from the Bethe Ansatz to derive certain crossing
estimates,
then uses the FKG inequality (for the absolute value of the height function) to turn these crossing estimates into
circuit estimates.

A second proof not relying on the Bethe Ansatz was given in~\cite{GlazmanLammers_2025_DelocalisationContinuity2D},
via a \emph{renormalisation inequality}.
This technique was first used in~\cite{Duminil-CopinSidoraviciusTassion_2017_ContinuityPhaseTransition}
in the context of the random-cluster model.
The renormalisation inequality essentially asserts that if the theorem is \emph{false},
then the circuit probability decays exponentially fast in the radius $r$,
which in turn implies that the model is localised (contradicting the known delocalisation).

We now turn to the proof.
    While Theorem~\ref{thm:renorminput} is essentially identical to
    \cite[Theorem~1.4]{Duminil-CopinKarrilaManolescu_2024_DelocalizationHeightFunction};
    the two theorems are stated in slightly different settings.
    We will now ``translate''~\cite[Theorem~1.4]{Duminil-CopinKarrilaManolescu_2024_DelocalizationHeightFunction} into Theorem~\ref{thm:renorminput}.
    The reader may choose to skip this technical proof on a first read. We start with a lemma.

  Recall that the event $\pALTCIRCUIT{\bar\omega}{20}{\ann{2n}{n}}$ means that the annulus contains $20$ circuits which
    (from outside to inside) alternately belong to $\omega^+$, $\omega^-$, $\omega^+$, etc.
    \begin{lemma}\label{lem:8.1}
    There exist constants $c>0$ and $N\in\Z_{\ge0}$  (both independent of $\c\in[1,2]$) such that for every $n\ge N$,
    \begin{equation}
        \muSpin_{\bbox{(2n+2)}}^+[\pALTCIRCUIT{\bar\omega}{20}{\ann{2n}{n}}]\geq c.
    \end{equation}

    \end{lemma}

    \begin{proof}
        The first translation problem stems from the fact that boundary conditions
        are enforced differently in~\cite{Duminil-CopinKarrilaManolescu_2024_DelocalizationHeightFunction}.
        In~\cite{Duminil-CopinKarrilaManolescu_2024_DelocalizationHeightFunction}, boundary conditions are imposed
        on the heights of an $F(\Z^2)$-path of \emph{adjacent} faces,
        as opposed to an $F_\sleven$-path of \emph{diagonally adjacent even faces}
        which is what we do here.

        Fix $n\in\Z_{\geq 0}$
        and let $\calD^n$ denote the largest even domain whose face centres
        are contained in $\bbox{(2n+2)}$.
        Let $\gamma^n$ denote the self-avoiding $F(\calD^n)$-circuit
        which surrounds the largest area (this path alternately
        visits even and odd faces).
        It is straightforward to see that, as a subset of $\R^2$,
        $\bbox{2n}$ is surrounded by $\gamma^n$.

        Since each face visited by $\gamma^n$ lies on $\partial\calD^n$ or is adjacent to it,
        we have
        \begin{equation}
            \muSpin_{\bbox{(2n+2)}}^+[\{|h|_{\gamma^n}|\leq 1\}]=1.
        \end{equation}
        This is good news, because the conditional measure
        $\muSpin_{\bbox{(2n+2)}}^+[\blank|\,h|_{\gamma^n}]$
        has boundary conditions like in~\cite{Duminil-CopinKarrilaManolescu_2024_DelocalizationHeightFunction}.
        We may now directly apply~\cite[Theorem~1.4]{Duminil-CopinKarrilaManolescu_2024_DelocalizationHeightFunction}
        with $\ell=1$ and $k=100$,
        to find a constant $c>0$ such that  for sufficiently large $n$,
                \begin{equation}
            \muSpin_{\bbox{(2n+2)}}^+\left[
                \pCIRCUIT{\{h\geq 100\}}{\ann{2n}{n}}
            \middle|h|_{\gamma^n}\right]\geq c.
        \end{equation}
        In particular,
                       \begin{equation}
            \muSpin_{\bbox{(2n+2)}}^+[
                \pCIRCUIT{\{h\geq 100\}}{\ann{2n}{n}}
           ]\geq c.
        \end{equation}
        Since $\ann{2n}{n}$ contains a
        circuit of $\{h\leq 10\}$ almost surely,
        the intermediate value theorem
        asserts that we may find $\omega^k$-circuits at heights $k=12,14,\ldots,98$
        in the annulus.
        This implies the claim since the circuits alternately belong to $\omega^+$ and $\omega^-$.
    \end{proof}

\begin{proof}[Proof of Theorem~\ref{thm:renorminput}]
    Let $N$ and $c$ be provided by Lemma~\ref{lem:8.1}.
    We distinguish two cases: either $r\leq N$
    or $r> N$.
    \paragraph{Step 1: the case $r\leq N$.}
        In any measure $\mu_{\calD}^+$,
        every edge not incident to $\partial\calD$ is $\omega^-$-open with
        some uniformly positive probability $\eta>0$ independent of $\calD$
        and the chosen edge.
        Thus, using the FKG inequality for $\omega^-$, a uniform lower bound is given by $\eta^M$ where $M$ is
        the supremum over the minimal circuit lengths within annuli with $r\leq N$.
  \paragraph{Step 2: the case $r>N$.}
        Without loss of generality, $x\in [-2,2]^2$.
        Let $n=\lceil r\rceil$.
        Then
        \begin{equation}
            \pALTCIRCUIT{\bar\omega}{20}{\ann{2n}{n}}
            \subset
            \pALTCIRCUIT{\bar\omega}{2}{\ann{2r}{r}(x)}.
        \end{equation}
        Thus, the claim implies that
        \begin{equation}
            \muSpin_{\bbox{(2n+2)}}^+[\pALTCIRCUIT{\bar\omega}{2}{\ann{2r}{r}(x)}]
            \geq
            c.
        \end{equation}
        Let $\gamma$ denote the largest self-avoiding $\omega^+$-circuit in the annulus above
        (or $\gamma:=\emptyset$ if such a circuit does not exist),
        and let $\calD_\gamma\subset \bbox{2r}(x)$ denote the corresponding even domain
        (so that $\partial\calD_\gamma=\gamma$).
        By the tower property and the Markov property,
        we have
        \begin{multline}
            \muSpin_{\bbox{(2n+2)}}^+[\pALTCIRCUIT{\bar\omega}{2}{\ann{2r}{r}(x)}]
            \\
            =
            \int_{\{\gamma\neq\emptyset\}} \muSpin_{\calD_\gamma}^+[\pCIRCUIT{\omega^-}{\ann{2r}{r}(x)}] \diffi\muSpin_{\bbox{(2n+2)}}^+[\gamma]
            \geq c.
        \end{multline}
        Since $\muSpin_{\calD}^+[\pCIRCUIT{\omega^-}{\ann{2r}{r}(x)}]$ is increasing
        in $\calD$ and $\calD_\gamma\subset \bbox{2r}(x)$ almost surely,
        we get from Corollary~\ref{cor:Monotonicity in domains and boundary conditions} that
        \begin{equation}
            \muSpin_{\bbox{2r}(x)}^+[\pCIRCUIT{\omega^-}{\ann{2r}{r}(x)}]
            \geq c.
        \end{equation}
   This concludes the proof of the second step. Taken together, the two steps establish the theorem with $\ccircuit:=\eta^M\wedge c$.
\end{proof}

We record an easy consequence of the previous theorem together with the FKG inequality. This statement will prove convenient later on.

\begin{corollary}[Circuit estimate]
    \label{cor:renorminputcor}
    For any $r\in[4,\infty)$
    and $x\in\R^2$,
    and for any even domain $\calD$ such that $\partial\calD$
    surrounds $\bbox{2r}(x)$,
    we have
    \begin{equation}
        \muSpin_{\calD}^+[\pCIRCUIT{\omega^-}{\ann{2r}{r}(x)}|\calE]\geq\ccircuit
    \end{equation}
    for any event $\calE$ that is measurable in terms of the
    even spins and edges which do not intersect $\bbox{2r}(x)$.
    The same inequality remains true with $\omega^-$ replaced by $\omega^+$.
\end{corollary}

\begin{proof}
    The event $\calE$ may be written as a partition of finitely many $\sleven$-lattice
    events which are measurable in terms of the even spins and edges
    which do not intersect $\bbox{2r}(x)$.
    Without loss of generality, we may assume that $\calE$ itself is of this type.
    Let $\gamma\subset E_\sleven$ denote the boundary of the largest even domain in $\bbox{2r}(x)$.
    By the FKG inequality (Proposition~\ref{proposition:realFKG})
    for $\muSpin_{\calD}^+[\blank|\calE]$, we get
    \begin{align}
        \muSpin_{\calD}^+[\pCIRCUIT{\omega^-}{\ann{2r}{r}(x)}|\calE]
        &\geq
        \muSpin_{\calD}^+[\pCIRCUIT{\omega^-}{\ann{2r}{r}(x)}|\calE\cap\{\gamma\subset\omega^+\}]
        \\
        &=
        \muSpin_{\bbox{2r}(x)}^+[\pCIRCUIT{\omega^-}{\ann{2r}{r}(x)}]\\
        &>\ccircuit.
    \end{align}
    The equality is just the Markov property;
    the inequality on the right is Theorem~\ref{thm:renorminput}.
\end{proof}

% </input src="sections/PART_D/12_spins/05_renormalisation_input.tex" root="." version="0.0.1">
% <input src="sections/PART_D/12_spins/08_full_plane.tex" root="." version="0.0.1">
\section{Full-plane spin representation (incl.~Theorem~\ref{thm:intro:flip_domination})}
\label{sec:full_plane_spins}

The previous circuit estimate enables us to define a full plane analogue of the spin representation.

\subsection{Definition in the full plane}

We define an infinite volume version of our spin representation. We also show that the properties obtained in finite volume extend to the infinite volume setting.
Introduce the set
 \begin{equation}
    \Omega_{\Z^2}:=\{\pm\}^{F_\sleven}\times\{0,1\}^{E_\sleven}\times\{\pm\}^{F_\slodd}
\end{equation}
\begin{theorem}[Full-plane limit]
    \label{thm:infinite_volume_6V_spins}
    There exists a unique probability measure $\mu_{\Z^2}$
    on $\Omega_{\Z^2}$
   such that for any random variable $X$ measurable  in terms of finitely many spins and edges,
    \begin{equation}
        \lim_{\calD\nearrow\Z^2}
        \mu_{\calD}^+[X]
        =
        \lim_{\calD\nearrow\Z^2}
        \mu_{\calD}^-[X]
        =
        \mu_{\Z^2}[X].
    \end{equation}
    Moreover, $\mu_{\Z^2}$ satisfies the following properties:
    \begin{enumerate}
    \item $\R^2\setminus\omega$ contains $\mu_{\Z^2}$-almost surely
    no unbounded connected components.
        \item Conditional spin flip property for $\sigma^\slodd$ (analogous to Lemma~\ref{lemma:oddspinsflip}),
        \item Markov property (analogous to Lemma~\ref{lemma:realMarkov}),
        \item Markov property for polar domains (analogous to Lemma~\ref{lem:markov_polar}),
        \item FKG inequality (analogous to Proposition~\ref{proposition:realFKG}),
        \item Circuit estimate (analogous to Corollary~\ref{cor:renorminputcor}).
    \end{enumerate}
\end{theorem}

\begin{remark}
    We have not yet proved well-definedness of $\P_{\Z^2}$ (Theorem~\ref{thm:infinite_volume_6V}).
    This is done later, in Section~\ref{section:cylinder},
    where we also prove that the law of the gradient of $h$ in $\mu_{\Z^2}$ is precisely $\P_{\Z^2}$.
    This motivates our interest in the measure $\mu_{\Z^2}$.
\end{remark}

Before proving Theorem~\ref{thm:infinite_volume_6V_spins}, we introduce a lemma.
    \begin{lemma}
        The measures $\mu_{\calD}^+$ and $\mu_{\calD}^-$
        restricted to even spins and edges
        converge to the same limit,
        which we denote $\mu_{\Z^2,\sleven}$.
    \end{lemma}
    \begin{proof}
       Corollary~\ref{cor:Monotonicity in domains and boundary conditions} implies that for any $\sleven$-increasing positive random variable $X$ which is measurable  in terms of finitely many spins and edges,
        the limits
        \begin{equation}
            \ell^-:=\lim_{\calD\nearrow\Z^2}
            \mu_{\calD}^-[X];
            \qquad
            \ell^+:=\lim_{\calD\nearrow\Z^2}
            \mu_{\calD}^+[X]
        \end{equation}
        are well-defined (respectively as increasing and decreasing sequences) and satisfy $\ell^-\le \ell^+$.
                It therefore suffices to prove the other inequality $\ell^+\le \ell^-$. This will be done by showing that even under $\mu_{\calD}^+[X]$, there is a probability tending to 1 as $\calD$ tends to $\Z^2$ (this uses Corollary~\ref{cor:renorminputcor}) that there is a circuit in $\omega^-$ surrounding the vertices that serve to measure $X$.

    More formally, fix $r\geq 4$ so large that
    $X$ is measurable in terms of the spins and edges in $\bbox{r-2}$.
    Define the event $\calA_{n,m}:=\pCIRCUIT{\omega^-}{\bbox{2^mr}\setminus\bbox{2^nr}}$.
    Fix $m\in\Z_{\geq 0}$.
    If $\calD$ is an even domain whose perimeter surrounds $\bbox{2^mr}$,
    then
    Corollary~\ref{cor:renorminputcor} implies that
    \begin{equation}
        \mu_{\calD}^+[\calA_{n,n+1}|(\calA_{n+1,m})^c]
        \geq
        \ccircuit
    \end{equation}
    for any $n=0,1,\ldots,m-1$.
    By induction, this yields
    \begin{equation}
        \mu_{\calD}^+[\calA_{0,m}]\geq 1- (1-\ccircuit)^m.
    \end{equation}
    The right-hand side tends to $1$ as $m\to\infty$.
    Let $\gamma^m$ denotes the largest self-avoiding $\omega^-$-circuit
    contributing to $\calA_{0,m}$
    (if the event occurs), and set $\gamma^m:=\emptyset$ otherwise.
    Then
    \begin{equation}
        \mu_{\calD}^+[\indi{\calA_{0,m}}\cdot X]
        =
        \int_{\{\gamma^m\neq\emptyset\}} \mu_{\calD_{\gamma^m}}^-[X] \diffi\mu_{\calD}^+[\gamma^m]
        \leq \ell^-
    \end{equation}
    where $\calD_{\gamma^m}$ is the even domain whose boundary is $\gamma^m$.
    The identity is the Markov property, and the inequality
    is again Corollary~\ref{cor:Monotonicity in domains and boundary conditions}.
    If we first take $\calD\nearrow\Z^2$ and then $m\to\infty$,
    the left-hand side converges to $\ell^+$,
    proving the desired inequality.
    This finishes the proof of the lemma.
    \end{proof}

\begin{proof}[Proof of Theorem~\ref{thm:infinite_volume_6V_spins}]
    By reasoning as for the proof of the previous lemma,
    we observe that Corollary~\ref{cor:renorminputcor}
    implies that $\mu_{\Z^2,\sleven}$-almost surely all connected components of $\R^2\setminus\omega$ are bounded.
    Lemma~\ref{lemma:oddspinsflip} therefore extends to the full-plane limit,
    and $\mu_{\Z^2}$ may simply be obtained from $\mu_{\Z^2,\sleven}$ by
    flipping coins for the odd spins $\sigma^\slodd$
    in each connected component of $\R^2\setminus\omega$.
    The properties stated in the result now immediately follow (by passing to the limit) from the properties of $\mu_{\calD}^+$ and $\mu_{\calD}^-$.
\end{proof}

% </input src="sections/PART_D/12_spins/08_full_plane.tex" root="." version="0.0.1">

% </input src="sections/PART_D/12_spins/main.tex" root="." version="0.0.1">
% <input src="sections/PART_D/14_flip_dom.tex" root="." version="0.0.1">
\subsection{Flip domination (Theorem~\ref{thm:intro:flip_domination})}
\label{sec:flip_domination}

\begin{figure}[b]
    \centering
    \includegraphics{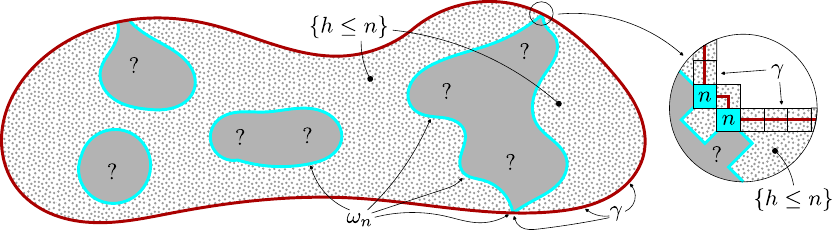}
    \caption{Flip domination: perfect flip symmetry around the height $n$ in the unexplored region; heights at most $n$ in the explored region.}
    \label{fig:flip_dom}
\end{figure}

Now that the infinite-volume measure is defined, we derive the flip domination property \emph{en passant}.
(In fact, we only formally prove that the law of $h$ in $\mu_{\Z^2}$ and $\P_{\Z^2}$ is the same
in Lemma~\ref{lemma:TorusLimits}.)

\begin{proof}[Proof of Theorem~\ref{thm:intro:flip_domination}]
    The proof is illustrated by Figure~\ref{fig:flip_dom}.
    Fix the $F(\Z^2)$-circuit $\gamma$, and define $F_\gamma$ as in the statement of the theorem.
    Fix $n\in 2\Z$.
    It suffices to prove that, for any even domain $\calD$ containing $\gamma$,
    and for any positive probability event $\calE
    \subset \{h|_\gamma \leq n\}$ that is measurable in terms of $h|_{F(\calD)\setminus F_\gamma}$,
    the height function $2n-h_{F_\gamma}$ stochastically dominates $h|_{F_\gamma}$ under $\mu_{\calD}^+[\blank|\calD]$.

    The idea is to simply explore the outermost $\omega_n$-loops within $\gamma$.
    Such loops may intersect $\gamma$; this is not a problem.
    Conditional on this exploration:
    \begin{itemize}
        \item The unrevealed heights (surrounded by $\omega_n$-loop) are flip-symmetric around $n$,
        \item The revealed heights are at most $n$.
    \end{itemize}
    This implies the desired stochastic domination.
\end{proof}

% </input src="sections/PART_D/14_flip_dom.tex" root="." version="0.0.1">
% <input src="sections/PART_D/15_cylinder.tex" root="." version="0.0.1">
\section{Torus and cylinder spin representation (Theorem~\ref{thm:infinite_volume_6V})}
\label{section:cylinder}

This section adapts the analysis developed above to the settings of the torus and the cylinder. While no essential difficulties arise, special care is required in defining the corresponding measures so as to preserve the FKG property. For completeness, we present the relevant results below.

\subsection{Definitions on the torus}

We want to replace fixed boundary conditions outside some even domain $\calD$
by periodic boundary conditions on the cylinder $\T_{M,L}$ defined in the introduction.
The graphs $F(\T_{M,L})$,
$F_\sleven(\T_{M,L})$,
and $F_\slodd(\T_{M,L})$ are defined in the obvious way,
in analogy with their definitions for $\calD$
in Section~\ref{sec:RSW}.

\begin{definition}[Torus spin measure]
    \label{definition:spinmeasureTORUS}
    For any torus $\T_{M,L}$,
    let $\Omega=\Omega_{\T_{M,L}}$ denote the sample space
    \begin{equation}
        \Omega:=
        \{\pm\}^{F_\sleven(\T_{M,L})}\times\{0,1\}^{E_\sleven(\T_{M,L})}\times\{\pm\}^{F_\slodd(\T_{M,L})}
        .
    \end{equation}
    A typical element is denoted by $(\sigma_\sleven,\omega,\sigma_\slodd)$.
    We often identify $\omega$ with the set $\{\omega=1\}\subset E_\sleven(\T_{M,L})$,
    and use the standard percolation terminology.
   Introduce a new, special event
    \begin{equation}
        \calE_8
        :=\{\text{the height gain of $h$ along any closed $F(\T_{M,L})$-circuit, belongs to $8\Z$}\}.
    \end{equation}
 Define the probability measure $\muSpin_{\T_{M,L}}$ on $(\sigma_\sleven,\omega,\sigma_\slodd)\in\Omega$ by
    \begin{align}
        \muSpin_{\T_{M,L}}[(\sigma_\sleven,\omega,\sigma_\slodd)]
        \propto{}
        &\true{(\sigma_\sleven,\sigma_\slodd)\in \calE_8}
        \cdot\true{\omega\subset A(\sigma_\sleven)}
        \\
        &\qquad
        \cdot\true{\text{$xy\in\omega$ or $xy^*\in A(\sigma_\slodd)$ (or both) for any $xy\in E_\sleven(\T_{M,L})$}}
        \\
        &\qquad
        \cdot \c^{\#A(\sigma_\sleven)}
        \cdot \left(\tfrac1\c\right)^{\#\omega}
        \cdot\left({1-\tfrac1\c}\right)^{\#(A(\sigma_\sleven)\setminus\omega)}.
    \end{align}
    \end{definition}

\begin{remark}
    Compare Definition~\ref{definition:spinmeasureTORUS} with Definition~\ref{definition:spinmeasure}.
    The only difference is that we dropped the first two indicators in Equation~\eqref{eq:spinmeasurefirstdef}
    (related to the boundary conditions on $\partial\calD$)
    and inserted an indicator for the event $\calE_8$ (which constrains the height gain on
    loops which wind nontrivially around the torus).
\end{remark}

\begin{remark}
  Samples from $\muSpin_{\T_{M,L}}$ are naturally interpreted as six-vertex configurations
    on $\T_{M,L}$, but the induced law is \emph{not} given by the measure $\P_{\T_{M,L}}$ defined in the introduction.
    The reasons are as follows.
   In $\P_{\T_{M,L}}$, with positive probability,
        the arrow configuration induces a height gain in $2\Z\setminus 4\Z$
        along some nontrivial loops.
        For the spin representation to be well-defined,
        the height gain along nontrivial loops must lie in $4\Z$, but in the previous definition we further constrain  the height gain
        to belong to $8\Z$ (via the event $\calE_8$). This is done to make the FKG inequality work
    (this is discussed below in further detail).
\end{remark}

\subsection{Markov property and the FKG inequality on the torus}

Let us first state a Markov property.
The maps $\pi_\calD$ and $\pi_\calD^c$ are defined the same as before.
The proof is identical to the original proof (Lemma~\ref{lemma:realMarkov}).

\begin{lemma}[Torus Markov property]
    \label{lemma:realMarkovTORUS}
    Consider a torus $\T_{M,L}$ and an even domain $\calD$
    such that (as a subset of $\T_{M,L}$) $\partial\calD$
    is a contractible self-avoiding circuit in $F_\sleven(\T_{M,L})$.
    For $\#\in\{\pm\}$,
    \begin{itemize}
        \item The laws $\muSpin_{\T_{M,L}}[\blank|\{\partial\calD\subset\omega^\#\}]$ of $\pi_\calD(\sigma_\sleven,\bar\omega,\sigma_\slodd)$ and $\mu_{\calD}^\#$ of $(\sigma_\sleven,\bar\omega,\sigma_\slodd)$ are the same,
        \item The random variables $\pi_\calD$ and $\pi_\calD^c$ are independent in $\muSpin_{\T_{M,L}}[\blank|\{\partial\calD\subset\omega^\#\}]$.
    \end{itemize}
\end{lemma}

We also mention an FKG inequality.

\begin{lemma}[Torus FKG inequality]
    \label{lemma:realFKGcylinder}
    Consider a torus $\T_{M,L}$
    and an $\sleven$-lattice event $\calE$.
    If $\muSpin_{\T_{M,L}}[\calE]>0$,
    then the conditional probability measure $\muSpin_{\T_{M,L}}[\blank|\calE]$
    has $\sleven$-FKG.
\end{lemma}

\begin{proof}
    The proof is entirely the same as the original proof (Proposition~\ref{proposition:realFKG}).
    There is, however, one critical point where we should pay attention.
    The decomposition into two independent Ising models is a bit
    more complicated due to the nontrivial topology.
    One may work out, however, that the decomposition
    in Equation~\eqref{eq:bigexpressioninFKGproof} remains valid on the torus
    for the measure $\muSpin_{\T_{M,L}}$ defined above
    (one should only replace the ambient space $\R^2$
    by $(\R/L\Z)\times (\R/M\Z)$ when counting the number of connected components).
    Crucially, Equation~\eqref{eq:bigexpressioninFKGproof} remains valid thanks
    to the event $\calE_8$ in Definition~\ref{definition:spinmeasureTORUS},
    and would fail otherwise. We refer to~\cite[Section~5]{GlazmanLammers_2025_DelocalisationContinuity2D} for details.
\end{proof}

\subsection{Circuit estimates on the torus}

With the Markov property and the FKG inequality,
we can essentially bring back our measures to the finite domain setup discussed
before. Let us now describe how to do this.

\begin{lemma}[Torus circuit estimate]
    \label{lemma:Torus circuit estimate}
    Fix $M,L$.
    Then, for any $4\le r<(M\wedge L)/4$
    and $x\in\R^2$,
    \begin{equation}
        \muSpin_{\T_{M,L}}[\pCIRCUIT{\omega^-}{\ann{2r}{r}(x)}|\calE]\geq\ccircuit
    \end{equation}
    for any event $\calE$ that is measurable in terms of the
    even spins and edges which do not intersect $\bbox{2r}(x)$.
    The same inequality remains true with $\omega^-$ replaced by $\omega^+$.
\end{lemma}

\begin{proof}
    The proof is identical to the proof of Corollary~\ref{cor:renorminputcor}.
\end{proof}

\subsection{Definition and main properties on the cylinder}

Recall the definitions of $\CYL{L}$ and $\P_{\CYL{L}}$.
The graphs $F(\CYL{L})$,
$F_\sleven(\CYL{L})$,
and $F_\slodd(\CYL{L})$ are defined in the obvious way,
in analogy with their definitions for $\calD$ and $\T_{M,L}$.

\begin{lemma}[Torus limits]
    \label{lemma:TorusLimits}
    All of the following statements hold true.
    \begin{enumerate}
        \item \textbf{Full-plane limit.} For any  random variable $X$ which is measurable  in terms of finitely many spins and edges,
        \begin{equation}
            \lim_{M,L\to\infty} \muSpin_{\T_{M,L}}[X] = \mu_{\Z^2}[X],
        \end{equation}
        where $\mu_{\Z^2}$ is the measure defined in Section~\ref{sec:RSW} (Theorem~\ref{thm:infinite_volume_6V_spins}).
        \item \textbf{Cylinder limit.} For any $L\in2\Z_{\geq 1}$,
        there exists a probability measure
        $\muSpin_{\CYL{L}}$ on the sample space
        \begin{equation}
            \Omega_{\CYL{L}}:=\{\pm\}^{F_\sleven(\CYL{L})}\times\{0,1\}^{E_\sleven(\CYL{L})}\times\{\pm\}^{F_\slodd(\CYL{L})}
        \end{equation}
        such that for any random variable $X$ which is measurable  in terms of finitely many spins and edges,
        \begin{equation}
            \lim_{M\to\infty} \muSpin_{\T_{M,L}}[X] = \muSpin_{\CYL{L}}[X].
        \end{equation}
        Moreover, all of the following are true:
        \begin{itemize}
            \item The Markov property (Lemma~\ref{lemma:realMarkovTORUS}) passes to the limit,
            \item The FKG inequality (Lemma~\ref{lemma:realFKGcylinder}) passes to the limit,
            \item The circuit estimate (Lemma~\ref{lemma:Torus circuit estimate}) passes to the limit,
            \item The conditional spin flip property for $\sigma^\slodd$ (Lemma~\ref{lemma:oddspinsflip}) holds true,
            \item As a measure on six-vertex configurations,
            $\muSpin_{\CYL{L}}$ equals $\P_{\CYL{L}}$.
        \end{itemize}
        \item \textbf{Full-plane as a limit of cylinder measures.}
        We have
        \begin{equation}
            \lim_{L\to\infty} \muSpin_{\CYL{L}}[X] = \mu_{\Z^2}[X]
        \end{equation}
        for any random variable $X$ which is measurable  in terms of finitely many spins and edges,
        due to the first item.
        If $X$ is also measurable in terms of $h$,
        then we furthermore have
        \begin{equation}
            \lim_{L\to\infty} \muSpin_{\CYL{L}}[X] = \mu_{\Z^2}[X]
            =
            \E_{\Z^2}[X].
        \end{equation}
        In particular, Theorem~\ref{thm:infinite_volume_6V} holds true.
    \end{enumerate}
\end{lemma}

\begin{proof}
    The proof of the first item is identical to the proof of Theorem~\ref{thm:infinite_volume_6V_spins}:
    one simply uses the circuit estimate (Lemma~\ref{lemma:Torus circuit estimate})
    and the Markov property (Lemma~\ref{lemma:realMarkovTORUS}) to compare the measure
    $\muSpin_{\T_{M,L}}$ with the measure $\muSpin_{\calD}^+$ for a large enough even domain $\calD$.

    The third items clearly follows from the first two since the limit in the first item
    may be taken in any order.
    We therefore focus on the second item.

    Recall from the introduction that $\{\operatorname{balanced}\}$
    denotes the event that a configuration is \emph{balanced},
    meaning that in each column of horizontal arrows,
    half of the arrows point to the left and the other half to the right.
    This event constrains the global topology, and equals the event that
    the height gain along any closed $F(\CYL{L})$-circuit
    that winds in the vertical direction, but not the horizontal direction,
    equals $0$.
    The limit
    \begin{equation}
        \lim_{M\to\infty} \muSpin_{\T_{M,L}}[\blank|\{\operatorname{balanced}\}]
    \end{equation}
    is well-defined and equal to $\P_{\CYL{L}}$ since we are essentially
    dealing with a recurrent Markov chain.
    We must therefore prove that $\{\operatorname{balanced}\}$ has high probability
    in the $M\to\infty$ limit.

     We now make one observation.
    If for some $n\in\Z$, the slice $[n,n+1]\times (\R/L\Z)$
            contains a nontrivial $\omega$-path, then the gradient of
            the height function is zero along that path and therefore
            the event $\{\operatorname{balanced}\}$ occurs. Write $S_n$ for this event.

     As discussed before, it is easy to find a uniform lower bound $\eta>0$ on the probability that an edge is $\omega^+$-open,
            even if we condition on the states of all edges not incident to that edge.
            \commentKarol{I could prove this in the unconstrained case, but not in the fully general constrained one.
I could deal with farther constraints though. I did not find the proof before in the paper. Could one include it, or, if this is really trivial, then I'd be happy to learn the argument HDC: I think this is very general but maybe I am missing something.}
        This implies that for any $k=0,\ldots, M/2-1$, we have
        \begin{equation}
            \muSpin_{\T_{M,L}}[S_{2k}|(S_0\cup S_2\cup S_4\cup\cdots \cup S_{2k-2})^c]\geq \eta^{L},
        \end{equation}
        and therefore
        \begin{equation}
            \muSpin_{\T_{M,L}}[\{\operatorname{balanced}\}]\geq 1-(1-\eta^{L})^{M/2}.
        \end{equation}
        The lower bounds clearly tends to $1$ as $M$ tends to infinity.

        The first three properties obviously pass to the limit,
        and the last property was already proved above.
        The conditional spin flip property for $\sigma^\slodd$
        is not obvious (it is not true on the torus due to the event $\calE_8$),
        but it is recovered in the limit
        thanks to the Markov chain structure of the six-vertex model on the cylinder.
    \end{proof}

% </input src="sections/PART_D/15_cylinder.tex" root="." version="0.0.1">
% <input src="sections/PART_D/16_loglin.tex" root="." version="0.0.1">
\section{Bounds on crossing counts (Theorem~\ref{thm:intro:arm_exponents})}
\label{sec:loglon}
\newcommand\clin{c_{\mathrm{lin}}}

The purpose of this section is to establish a log-linear bound on the number of alternating rectangle crossings (see the statement below).
We also prove the log-quadratic bound on arm events (Theorem~\ref{thm:intro:arm_exponents} in the introduction).

\subsection{The rectangle crossings case}

We start with a (far from optimal) bound on crossing counts.

Recall that $\COUNTHORI{\bar\omega}{R}$ counts the number of alternating horizontal crossings of the rectangle $R$.
\begin{lemma}[Log-linear bound on rectangle crossing counts]
    \label{lemma:loglin}
   There exists a constant $\clin\in(0,1)$ (independent of $\c\in[1,2]$) with the following property.
   For any $n\in \Z_{\geq 0}$ and any rectangle $R:=\llbracket w\times \rho w\rrbracket+u$,
    where $w\in\R_{\geq 1}$ and $\rho\in\R_{\geq 1}$,
    and $u\in\R^2$,
    \begin{equation}
        \mu_{\Z^2}[\{\COUNTHORI{\bar\omega}{R}\geq 2n\}|\calE]\leq (1-\clin^\rho)^n,
    \end{equation}
    where $\calE$ is any event that is measurable in terms of the even spins
    and edges at a $\ell^\infty$-distance at least $w/5$ of $R$.

Moreover, the same bounds hold with $\mu_{\Z^2}$ replaced by $\mu_{\CYL{L}}$, provided that
$2(\rho+\tfrac15) w \le L$, and with the understanding that $R$ is now regarded as a rectangle on the cylinder.
Analogous bounds also hold for vertical crossings of the rectangle, both in the full plane and on the cylinder.
\end{lemma}

\begin{remark} Requiring a macroscopic buffer zone between the rectangle and the spins or edges used to determine
$\calE$ is standard in this type of result. In fact, the conclusion may fail without such a buffer if
$\calE$ is allowed to depend on all edges and spins outside $R$.
\end{remark}

\begin{remark}\label{rmk:finite expectation rectangle} In the context of the lemma, summing over $n$ gives that
    \begin{equation}
        \mu_{\Z^2}[\COUNTHORI{\bar\omega}{R}|\calE]\leq 2\clin^{-\rho}.
    \end{equation}
    \end{remark}
For the proof of this lemma on rectangle crossings, we first require a strengthened version of the circuit estimate in which the annulus is allowed to be partially ``scarred'', provided that the boundary conditions induced on the remaining region are favorable.
This variation is illustrated by Figure~\ref{fig:approach_boundary}.

\begin{lemma}[Partial circuit estimate]
    \label{lem:circuitfavouredenviron}
    Fix $r\in[4,\infty)$ and $x\in\R^2$,
    and consider the following setup.
    \begin{itemize}
        \item Let $(\calD_i)_i$ denote a finite family of even domains.
        \item Let $\calB$ denote the set of edges in $\cup_i\partial\calD_i$ that are contained in $\bbox{2r}(x)$.
        \item Let $\calU:=\cup_i E_\sleven(\calD_i)$.
    \end{itemize}
    Then, for any event $\calE$ measurable in terms of the even spins and edges contained in $\calU$, and such that $ \muSpin_{\Z^2}[\calE\cap\{\calB\subset\omega^-\}]>0$, we have
       \begin{equation}
        \muSpin_{\Z^2}[\pCIRCUIT{\omega^-\cup\,\calU}{\ann{2r}{r}(x)}|\calE\cap\{\calB\subset\omega^-\}]
        \geq \ccircuit.
    \end{equation}
\end{lemma}

\begin{figure}
    \centering
        \includegraphics{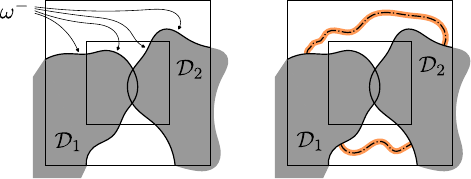}
    \caption{Lemma~\ref{lem:circuitfavouredenviron}.
   Suppose that we want to attach an  $\omega^-$-percolation
    to an $\omega^-$-open boundary segment,
    but we have conditioned on some ``messy'' event inside the unions of the domains $\calD_i$
    (the shaded area).
    The lemma tells us that we can still open part of the circuit which connects to the boundary.
    The proof uses the Markov property for polar domains to remove the conditioning on the adverse event.
    }
    \label{fig:approach_boundary}
\end{figure}
\begin{proof}
    Without loss of generality, $\calE$ is an $\sleven$-lattice event. Let $\gamma$ denote the smallest self-avoiding even circuit surrounding $\bbox{2r}(x)$.
    By using the FKG inequality, we may furthermore assume that $\calE$
    satisfies the following ``worst-case'' description:
    $\gamma\subset\calU$, and $\calE\subset \{\gamma\subset\omega^+\}$.
    We now claim that
    \begin{align}
       &\muSpin_{\Z^2}[\pCIRCUIT{\omega^-\cup\,\calU}{\ann{2r}{r}(x)}|\calE\cap\{\calB\subset\omega^-\}]
       \\
      &\qquad\qquad\qquad=
       \muSpin_{\Z^2}[\pCIRCUIT{\omega^-\cup\,\calU}{\ann{2r}{r}(x)}|\{\gamma\subset\omega^+\}\cap\{\calB\subset\omega^-\}]
       \\
       &\qquad\qquad\qquad\geq
        \muSpin_{\Z^2}[\pCIRCUIT{\omega^-}{\ann{2r}{r}(x)}|\{\gamma\subset\omega^+\}]\\
        &\qquad\qquad\qquad\geq \ccircuit.
    \end{align}
    The equality is the Markov property (Lemma~\ref{lemma:realMarkov}) and the Markov property for polar domains (Lemma~\ref{lem:markov_polar}).
    The first inequality is inclusion of events and FKG.
    The second inequality is the circuit estimate (Corollary~\ref{cor:renorminputcor}).
\end{proof}

    \begin{figure}
        \begin{center}
        \includegraphics{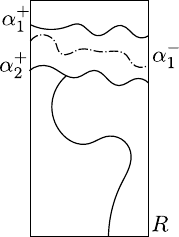}
        \end{center}
    \caption{When exploring the paths from top to bottom,
    each time we explore an $\omega^+$-crossing, there is a universally positive probability
    that this crossing connects to the bottom, blocking any further horizontal $\omega^+$-crossings from appearing.}
    \label{fig:linear_bound}
    \end{figure}

\begin{proof}[Proof of Lemma~\ref{lemma:loglin}]
    Define two sequences of paths, as follows:
    \begin{itemize}
        \item $\alpha^+_1$ is the highest horizontal $\omega^+$-crossing of $R$,
        \item $\alpha^-_i$ is the highest horizontal $\omega^-$-crossing of $R$ below $\alpha^+_i$, for $i=1,2,\ldots$,
        \item $\alpha^+_i$ is the highest horizontal $\omega^+$-crossing of $R$ below $\alpha^-_{i-1}$, for $i=2,3,\ldots$,
        \item If such a crossing does not exist, we set it (and all subsequent ones) equal to $\emptyset$.
    \end{itemize}
    Notice that $\{\COUNTHORI{\bar\omega}{R}\geq 2n\}=\{\alpha_{n}^-\neq\emptyset\}$.
    For Lemma~\ref{lemma:loglin}, it suffices to prove that for any $i$, we have
    \begin{equation}
        \muSpin_{\Z^2}[\{\alpha_i^-\neq\emptyset\}|\calE\cap\{\alpha_i^+\neq\emptyset\}]\leq 1-\clin^\rho
    \end{equation}
    for some fixed $\clin\in(0,1)$.

    We split into two cases, depending on the value of $w$.
    First suppose that $w\leq 1000$.
    Let $\eta>0$ denote a uniform constant such that the probability that an even spin is $+$ is at least $\eta$,
    even after conditioning on $\calE$ and all the other even spins.
    Then
    \begin{equation}
        \muSpin_{\Z^2}[\{\text{all even spins in $R$ below $\alpha^+_i$ are valued $+$}\}|\calE\cap\{\alpha_i^+\neq\emptyset\}]
        \geq(\eta^{1000000})^\rho.
    \end{equation}
    Since this event is disjoint from $\{\alpha_i^-\neq\emptyset\}$, any value $\clin\leq\eta^{1000000}$ will work.

    We are left with the (more interesting) case $w>1000$.
    By applying Lemma~\ref{lem:circuitfavouredenviron} at least $\lfloor 100\rho\rfloor $ times and applying the FKG inequality,
    we observe that
    with a $\muSpin_{\Z^2}[\blank|\calE\cap\{\alpha_i^+\neq\emptyset\}]$-probability of at least $(\ccircuit^{100})^{\rho}$,
    the path $\alpha^+_i$ is connected to
     $\rectbottom{R}$ within the rectangle (see Remark~\ref{remark:combinecircuits} and~Figure~\ref{fig:linear_bound}).
    In this case, it is impossible that the event $\{\alpha_i^-\neq\emptyset\}$ occurs.

    This proves that the value $\clin=\eta^{1000000}\wedge\ccircuit^{100}$ works.
\end{proof}

\subsection{The annulus arms case}

We now adapt the previous results to the context of arms crossing an annulus. For $r\in\R_{\geq 4}$, $R\in\R_{\geq 2r}$,
    and $x\in\R^2$,
    define the random variable
    \begin{equation}
        \label{eq:Kdef}
        K_{R,r,x}:=
        (1-\indi{\pALTCIRCUIT{\bar\omega}{2}{\ann{R}{r}(x)}}) \vee \COUNTARM{\bar\omega}{\ann{R-1}{r+1}(x)},
    \end{equation}
which, roughly speaking, counts alternating crossings from inside to outside in the annulus.     Notice that this random variable is measurable with respect to the even edges which
    are entirely contained in $\ann{R}{r}(x)$.

\begin{lemma}[Linear bound on arm exponents]
    \label{lemma:armexponent_for_spins_linear}
    There exists a constant $\carm'>0$ (independent of $\c\in[1,2]$) with the following property.
      For $r\in\R_{\geq 4}$, $R\in\R_{\geq 2r}$, $x\in\R^2$,
 and $n\in\Z_{\geq 0}$,
    \begin{equation}
        \label{eq:armexponentstargetLINEAR}
        \mu_{\Z^2}[\{K_{R,r,x}\geq n\}|\calE] \leq (r/R)^{\carm' n}
    \end{equation}
    for any $\calE$ with $ \mu_{\Z^2}[\calE]>0$ that is measurable with respect to the even spins and edges
    which do not intersect $\bbox{R}(x)$.

    The same bounds hold true if $\mu_{\Z^2}$ is replaced by $\mu_{\CYL{L}}$,
    provided that $L>2R$.
\end{lemma}

\begin{remark}\label{rmk:finite expectation annulus}
 In the context of the lemma, summing over $n$ gives that
    \begin{equation}\label{eq:finite expectation annulus}
        \mu_{\Z^2}[K_{R,r,x}|\calE] \leq \frac{(r/R)^{\carm'}}{1-2^{-\carm'}}.
    \end{equation}
\end{remark}

\begin{figure}
    \centering
    \includegraphics{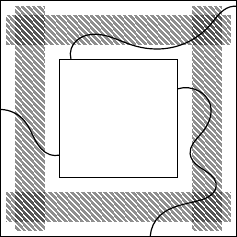}
    \caption{Illustration of the reduction from annulus arms to rectangle crossings.
    Each arm also crosses one of the four rectangles $R_1$, $R_2$, $R_3$, and $R_4$
    in the ``easy'' direction.}
    \label{fig:rectangle_to_annulus}
\end{figure}

\begin{proof}
    It suffices to consider the case that $R=2^mr$.
    We first prove this for $m=1$.

    Let us start by proving that with uniformly positive probability, $K_{R,r,x}=0$.
    Since the event $\pALTCIRCUIT{\bar\omega}{2}{\ann{R}{r}(x)}$ bars any arms from appearing,
    it suffices to show that this event occurs with uniformly positive probability.
    If $r\leq 100$ then we may simply choose two disjoint even circuits
    in $\ann{R}{r}(x)$, and lower bound the probability that they are open for $\omega^+$
    and $\omega^-$ (see the part of the proof of Lemma~\ref{lemma:loglin} where $w\leq 1000$).
    For $r>100$, we may combine $1000$ circuit estimates (Corollary~\ref{cor:renorminputcor})
    and the FKG inequality to show that $\pALTCIRCUIT{\bar\omega}{2}{\ann{R}{r}(x)}$ occurs
    with probability at least $\ccircuit^{1000}$
    (see also the part of the proof of Lemma~\ref{lemma:loglin} where $w> 1000$).

    To finish the proof for the case that $m=1$, it suffices to find constants $c>0$
    and $N\in 20\Z_{\geq 1000}$ such that
    for any $n\in\Z_{\geq 1}$, we have
    \begin{equation}
        \mu_{\Z^2}[\{K_{R,r,x}\geq N n\}|\calE] \leq e^{-cn}.
    \end{equation}
    Observe that any continuous path connecting $\bbox{r}(x)$ to $\partial\bbox{R}(x)$
    must necessarily traverse at least one of the following four rectangles in the ``easy'' direction:
    \begin{equation}
        R_1:=([5r/4,6r/4]\times [-7r/4,7r/4])+x
    \end{equation}
    and its images $R_2$, $R_3$, $R_4$,
    defined by the rotations by an angle $\frac{\pi}{2}$, $\pi$, and $\frac{3\pi}{2}$ around $x$.
    See Figure~\ref{fig:rectangle_to_annulus} for an illustration.
    In particular, we get
    \begin{equation}
        \{K_{R,r,x}\geq N n\} \subseteq
        \pALTHORI{\bar\omega}{\frac{Nn}{10}}{R_1}
        \cup
        \pALTVERTI{\bar\omega}{\frac{Nn}{10}}{R_2}
        \cup
        \pALTHORI{\bar\omega}{\frac{Nn}{10}}{R_3}
        \cup
        \pALTVERTI{\bar\omega}{\frac{Nn}{10}}{R_4}
    \end{equation}
    (the division by $10$ royally suffices; the safety margin compensates a few crossings that we may lose by how we set up the definitions).
    Lemma~\ref{lemma:loglin} then implies
    \begin{equation}
        \mu_{\Z^2}[\{K_{R,r,x}\geq N n\}|\calE] \leq 4 (1-\clin^{14})^{Nn/20}.
    \end{equation}
    It is then straightforward to find good values for $c$ and $N$.
    This concludes the case $m=1$.

    For the general case $m\geq 1$,
    we notice that
    \begin{equation}
        K_{2^m r,r,x} \leq \min_{i=0,\ldots,m-1} K_{2^{i+1} r,2^i r,x},
    \end{equation}
    and then use the $m=1$ case $m$ times at
    $m$ disjoint concentric annuli.
\end{proof}

% </input src="sections/PART_D/16_loglin.tex" root="." version="0.0.1">
% <input src="sections/PART_E/23_arms.tex" root="." version="0.0.1">
% \section{Quadratic bound on arm exponents }
% \label{sec:arm_exponents}

Theorem~\ref{thm:intro:arm_exponents} in the introduction is a direct corollary of the following stronger result.

The random variable $ K_{R,r,x}$ was defined in  equation~\eqref{eq:Kdef}.     The following lemma improves on Lemma~\ref{lemma:armexponent_for_spins_linear}.

\begin{theorem}[Quadratic bound on arm exponents]
    \label{thm:armexponent_for_spins_quadratic}
      For any $\c\in[1,2]$, there exists a constant $\carm\in(0,10^{-9})$  with the following property.
   For $r\in\R_{\geq 4}$, $R\in\R_{\geq 2r}$, $x\in\R^2$,
 and $n\in\Z_{\geq 0}$,
     \begin{equation}
    \label{eq:armexponentstargetQUADR}
        \mu_{\Z^2}[\{K_{R,r,x}\geq n\}|\calE] \leq (r/R)^{\carm n^2}
    \end{equation}
     for any $\calE$ that is measurable with respect to the even spins and edges
    which do not intersect $\bbox{R}(x)$.

    The same bounds hold true if $\mu_{\Z^2}$ is replaced by $\mu_{\CYL{L}}$,
    provided that $L>2R$.
\end{theorem}

\begin{figure}[ht]
    \centering
    \includegraphics{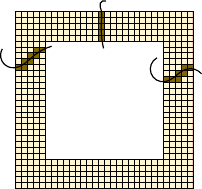}
    \caption{
        Each arm of the annulus $\ann{3r/2}{r}(x)$
        must touch a large number of small boxes,
        creating arm events around those smaller boxes.
    }
    \label{fig:quadratic_pigeonhole}
\end{figure}

\begin{proof}
    Just like in the proof of Lemma~\ref{lemma:armexponent_for_spins_linear},
    it suffices to consider the case that $R=2r$.
    We shall also suppose that $x=(0,0)$;
    this makes no difference to the proof.

    Recall the constant of $\carm'$ from Lemma~\ref{lemma:armexponent_for_spins_linear}
    on the linear bound on arm exponents.
    Since Lemma~\ref{lemma:armexponent_for_spins_linear} already handles Equation~\eqref{eq:armexponentstargetQUADR}
    for small values of $n$,
    it suffices to restrict ourselves to the case that $n=100\cdot K\cdot 2^k$
    for integers $k\geq 1$,
    where $K$ is a constant to be fixed later.

   Let $\bar r:=2^{-k-2}\cdot r$.
    The annulus $\ann{3r/2}{r}(x)$ can be tiled by exactly
    $5\cdot 2^{2k+2}$ squares of side length $2\bar r$.
    Write $S$ for the set of centres of these squares.
    Now for any percolation $\omega\subset\R^2$ which
    belongs to $\ARM{\ann{3r/2}{r}}$,
    we may find at least $2^k$ elements $u_1,\ldots,u_{2^k}$ in $S$ such that
    $\omega\in\ARM{\ann{2\bar r}{\bar r}(u_i)}$ for each $i=1,\ldots,2^k$;
    see Figure~\ref{fig:quadratic_pigeonhole}.
    By a pigeonhole argument,
    we get
    \begin{equation}
        K_{2r,r,x}\geq n
        \implies
        \sum_{u\in S} K_{2\bar r,\bar r,u} \geq 25\cdot K\cdot 2^{2k}.
    \end{equation}
    We must prove that the event on the right has a $\mu_{\Z^2}[\blank|\calE]$-probability of at most $2^{-\carm n^2}$.

    We call a set $S'\subset S$ \emph{separating} if the boxes $(\bbox{2\bar r}(u))_{u\in S'}$ are disjoint.
    We shall use the following input from
    Lemma~\ref{lemma:armexponent_for_spins_linear}:
    there exists some fixed constant $p>0$ such that,
    for any separating $S'\subset S$,
    the random variables $(K_{2\bar r,\bar r,u})_{u\in S'}$
    in $\mu_{\Z^2}[\blank|\calE]$
    are stochastically dominated by an i.i.d.\ family of random variables
    having a geometric random variable of parameter $p$.

    Write $\calS$ for some partition of $S$ into $25$ separating subsets.
      The pigeonhole principle implies that
        \begin{equation}
            \mu\Big[\Big\{\sum_{u\in S} K_{2\bar r,\bar r,u} \geq 25\cdot K\cdot 2^{2k}
            \Big\}\Big|\calE\Big]
            \leq
            25
            \max_{S'\in\calS}
            \mu\Big[\Big\{\sum_{u\in S'} K_{2\bar r,\bar r,u} \geq K\cdot 2^{2k}\Big\}
            \Big|\calE\Big].
        \end{equation}

        The probability on the right may be bounded by the probability that
        the sum of $5\cdot 2^{2k+2}$ i.i.d.\ geometric random variables of parameter $p>0$
        is at least $K\cdot 2^{2k}$.
        By standard large deviation estimates, this probability decays like $e^{-c 2^{2k}}$ for some constant $c>0$,
        provided that $K$ is some sufficiently large fixed constant (depending only on $p$).
        Since $n$ is of order $2^k$,
        this gives the desired quadratic bound.
\end{proof}

% </input src="sections/PART_E/23_arms.tex" root="." version="0.0.1">
% <input src="sections/PART_D/17_level_line_tree.tex" root="." version="0.0.1">

\section{Level line tree and branching function}
\label{sec:level line tree}

\subsection{Heuristic of the level line tree}

Before formally describing the level line tree,
we first give an analogy with the Gaussian free field,
and then proceed with some preliminary remarks on the combinatorial
structure important for the definition.

\paragraph{Discussion of the coupling of the GFF with $\operatorname{CLE}(4)$.}
The Gaussian free field in a disk with zero boundary conditions
has a natural coupling with $\operatorname{CLE}(4)$ \cite{BerestyckiPowell_2025_GaussianFreeField}.
Conditional on the $\operatorname{CLE}(4)$ loops,
we orient them counterclockwise or clockwise (independently and with equal probability).
The GFF is then morally equal to some constant $\lambda$ times the net winding of the loops around each point.
In this picture, the conditional variance between two points equals $\lambda^2$ times the number of loops
surrounding both points.

We can also think of the $\operatorname{CLE}(4)$ loops as a rooted tree:
the root is the whole disk,
the other tree vertices are simply connected subsets of the disk whose boundary is a $\operatorname{CLE}(4)$ loop,
and each tree vertex points to its parent, the smallest tree vertex strictly containing it.
Each point in the disk may be associated with the set of tree vertices containing it,
which may be interpreted as a tree path starting at the root and going downwards
(it may be finite or infinite, typically it is infinite).
In this formalism, the number of loops (and therefore the conditional covariance)
may be expressed in terms of the depth of the lowest common ancestor
of the tree paths of the two points.

In practice, we will be interested in the expectation of products of height differences.
The height difference between two points of the disk is naturally expressed
in terms of a tree path between them.
This tree path may be finite or infinite on both ends (for typical points, it is a bi-infinite paths).
The natural path $x\to y$ is the concatenation of the tree path from $x$ to the lowest common ancestor of $x$ and $y$,
and the tree path from that ancestor to $y$.
The conditional expectation of the product of $k$ height differences is then equal to
some function of the way that the $k$ tree paths intersect each other.

\paragraph{Notes on the combinatorial structure of the level line tree.}
We now put ourselves in the context of the measure
$\muSpin_{\calD}^+$, which means that the boundary height is equal to $0$.
Suppose for a second that we explore the outermost $\omega^-_\sleven$-circuits.
Then all of the following hold true:
\begin{itemize}
    \item On each $\omega^-_\sleven$-circuit, the height is equal to $+2$ or $-2$, depending
    on the value of the odd spins outside the circuit,
    \item The odd spins on each connected component of $\R^2\setminus \omega$ are fair coin flips
    (Lemma~\ref{lemma:oddspinsflip}),
    \item The Markov property (Lemma~\ref{lemma:realMarkov}) implies a renewal property:
 within each circuit $\partial\calB$, the conditional law of the spins and edges
    within $\calB$ is given by $\muSpin_{\calB}^-$ (which is nothing more than $\muSpin_{\calB}^+$
    with all even spins flipped).
\end{itemize}
This enables the definition of a tree structure bearing some resemblance to the tree defined above for the GFF.
At the same time, there are some differences.
\begin{itemize}
\item In the discrete the tree is finite, which makes things easier.
    \item The $\pm2$-labels of the outermost $\omega^-_\sleven$-circuits are \emph{not} independent,
    because it may happen that two such circuits are surrounded by the same connected component of odd spins
    in Lemma~\ref{lemma:oddspinsflip}, in which case they have the same label.
    \item There is a notion of odd and even circuits: starting from $\muSpin_{\calD}^+$,
    one explores the outermost $\omega^-_\sleven$-circuits,
    then, within each circuit, the outermost $\omega^+_\sleven$-circuits, et cetera.
    This parity issue is unrelated to the $\pm2$-labels
    (more precisely, the parity issue is related to the parity of the even spins,
    while the $\pm2$-labels relate to odd spins).
    \item The precise height at a face is not determined by the tree structure alone;
    sometimes we require small corrections, for example when the face itself is odd (and we need a correction of the form
    $\pm1$).
\end{itemize}
These issues slightly complicate the definition of the tree structure
(see the comments following the definition).
Nevertheless,
we can still apply the same logic as for the tree in the GFF case to bound correlations, as we will see for instance in the next section.

\begin{figure}[h]
    \centering
    \includegraphics{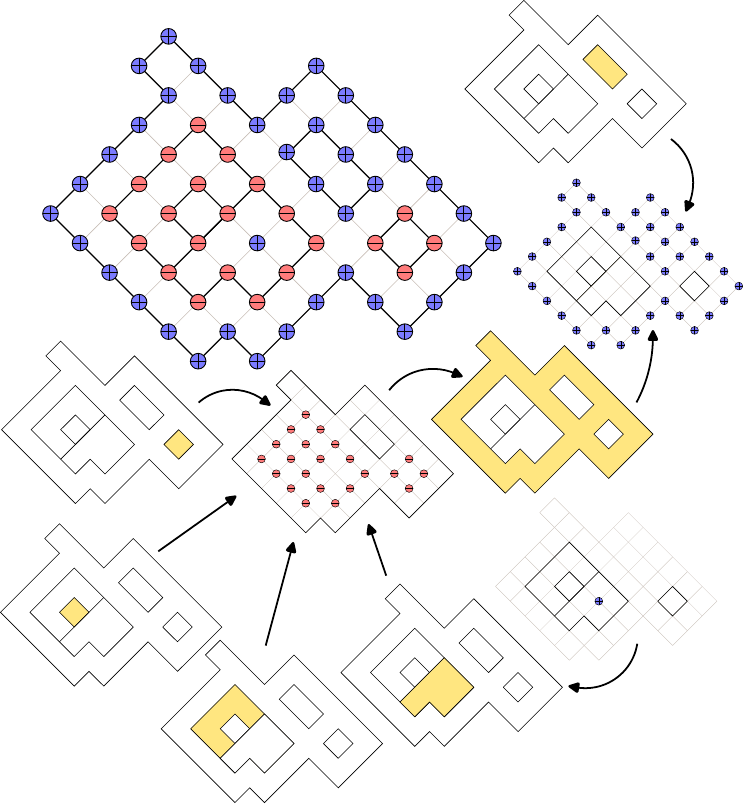}
    \caption{The directed level line tree $\calX$ associated with a configuration $(\sigma_\sleven,\bar\omega)$.}
    \label{fig:level_line_tree}
\end{figure}

\subsection{Formal definition of the level line tree}

\begin{definition}[Level line tree]
    \label{def:level_line_tree}
Consider a sample $(\sigma_\sleven,\bar\omega,\sigma_\slodd)$
from the measure $\muSpin_{\calD}^+$ for some even domain $\calD$.
The associated \emph{level line tree} $\calX=(V_\sleven(\calX),V_\slodd(\calX),E(\calX))$ is a
rooted directed tree, defined as follows.
\begin{itemize}
    \item
    The \emph{even} vertices $V_\sleven(\calX)$ are
    defined as follows.
    \begin{itemize}
    \item
        First, it has a bipartition $V_\sleven(\calX)=V_{\sleven}^+(\calX)\cup V_{\sleven}^-(\calX)$.
        \item $V_{\sleven}^+(\calX)$ is the partition of $\{\sigma_\sleven=+\}$
        such that two faces are in the same member of the partition if and only if they belong to
        the same connected component of $\R^2\setminus\omega^-_\sleven$.
        \item $V_{\sleven}^-(\calX)$ is the partition of $\{\sigma_\sleven=-\}$
        such that two faces are in the same member if and only if they belong to
        the same connected component of $\R^2\setminus\omega^+_\sleven$.
        \item The root of the tree is the member of $V_{\sleven}^+(\calX)$ containing $\partial\calD$.
    \end{itemize}
    \item The \emph{odd} vertices $V_\slodd(\calX)$ of the tree are formed
    by the bounded connected components of $\R^2\setminus\omega$.
% THE BELOW IS NOW REPLACED BY \sigma_\slodd(X), WHICH IS INTUITIVE AND DOESN'T ADD NEW DEFINITIONS.
%    Each vertex $v_\slodd$ has a value $\ell_\slodd(v_\slodd)=\pm$, depending on the values of the spins
%    $\sigma_\slodd$ on that connected component.
%    (For the avoidance of doubt, when conditioning on $\calX$, these labels are not taken into account in the sigma-algebra.)
    \item Each vertex (other than the root) has exactly one outgoing edge:
    \begin{itemize}
        \item Each \emph{odd} vertex $v_\slodd$ points towards the \emph{even} vertex $v_\sleven$
        containing the outer boundary of the connected component $v_\slodd\subset \R^2\setminus\omega$,        \item Each \emph{even} vertex $v_\sleven$ labelled
        $\pm$ points toward the odd vertex $v_\slodd$ such that:
        \begin{itemize}
            \item $v_\sleven$ and $v_\slodd$ lie in the same connected component $C$ of
            $\R^2\setminus \omega^{\mp}$,
            \item $v_\slodd$ is the unique such odd vertex pointing towards $V_{\sleven}^{\mp}(\calX)$
            (in fact, the other odd vertices in the same connected component $C$ point back to $v_\sleven$).
        \end{itemize}
    \end{itemize}
\end{itemize}
For each vertex $v$ other than the root, we let $p(v)$ denote the parent vertex.
The \emph{depth} of a vertex is defined as its distance to the root.
Notice that vertices in $V_{\sleven}^+(\calX)$, $V_{\sleven}^-(\calX)$,
and $V_\slodd(\calX)$ are at depth $4\Z$, $4\Z+2$, and $2\Z+1$ respectively.
Notice also that each \emph{odd} vertex has \emph{at most} one child pointing towards it.
\end{definition}

\begin{remark}
The odd spins are also integrated into the tree as odd vertices.
        This facilitates the definition of the height function at every face of the graph.
\end{remark}

\begin{remark} The definition of the even vertices is different from the above informal sketch.
        One surprising aspects lies in the fact that two $\omega^+$-circuits can
        surround each other, and still belong to the same even vertex of the tree.
        This is natural because no height difference is realised when no $\omega^-$-circuits
        separates the two.

        Ultimately, the definition is set up in such a way that the following lemma works.
\end{remark}

The lemma below says that conditionally on $\calX$,
the law of $h(v)$ is given by
the sum of $k$
independent $\pm2$-valued coin flips
if the depth of $v$ is $2k$,
and by the sum of  $k$ independent
 $\pm2$-valued coins and one $\pm1$-valued coin
    if the depth of $v$ is $2k+1$.

\begin{lemma}[Basic properties of the level line tree]
    \label{lemma:Basic properties of the level line tree}
    Let $\calD$ denote an even domain, and consider $\mu_{\calD}^+$.
    Then, all of the following hold true.
    \begin{enumerate}
        \item $\calX$ is measurable in terms of $(\sigma_\sleven,\bar\omega)$,
        \item Conditional on $\calX$, an independent fair coin flip $\sigma_\slodd(X)$ is attached to each element $X\in V_\slodd(\calX)$, so that $\sigma_\slodd(x)=\sigma_\slodd(X)$ for every $x\in X$,
        \item $h$ is almost surely constant on each vertex of $\calX$,
        \item Conditional on $\calX$, the height function $h$ has the following law:
        \begin{itemize}
            \item $h(r)=0$ on the root vertex $r$,
            \item $(h(v_\slodd)-h(p(v_\slodd)))_{v_\slodd\in V_\slodd(\calX)}$ has the law of independent fair $\pm1$-valued coin flips,
            \item $h(v_\sleven)-h(p(v_\sleven))=h(p(v_\sleven))-h(p(p(v_\sleven)))$ for any even non-root vertex $v_\sleven$.
        \end{itemize}
    \end{enumerate}
     Here, we recall that $p(v)$ denotes the parent vertex, as introduced in Definition \ref{def:level_line_tree}.
\end{lemma}

\begin{proof}
    While this lemma is important, its proof follows straightforwardly from the definitions.
    Property~(i) is immediate as the definition only involves $(\sigma_\sleven,\bar\omega)$.
    For Property~(ii), observe that the odd vertices are precisely the connected
    components of $\R^2\setminus\omega$, so that Lemma~\ref{lemma:oddspinsflip} applies.
    Properties~(iii) and~(iv) follow from the relation between spins and the height
    function, detailed in Equation~\eqref{eq:relationheightsspins}.
\end{proof}

\begin{definition}[Tree path]
    \label{def:tree_path}
    Let $\calD$ denote an even domain,
    and consider the level line tree $\calX$ in $\mu_{\calD}^+$.
    For any $u,v\in F(\calD)$, we define the \emph{tree path} $p^{uv}$ as the unique path in $\calX$ starting at the vertex containing $u$ and ending at the vertex containing $v$.
    It is viewed as a set of vertices of $\calX$.
\end{definition}

\subsection{Branching function}

We now introduce a convenient tool to analyse the covariance structure of the height function.
\begin{definition}[Branching function]
\label{def:branching_function}
    Let $\calD$ denote an even domain.
    Consider the level line tree $\calX$ in $\mu_{\calD}^+$.
    Let $p^{u\partial\calD}$ be the unique $\calX$-path from $u$ to $\partial\calD$.
    The \emph{branching function} is the random $\calX$-measurable function
    \begin{equation}
        \psi=\psi_\calD:F(\Z^2)\times F(\Z^2)\to\Z,\, (u,v)\mapsto
        \begin{cases}
            |p^{u\partial\calD}\cap p^{v\partial\calD}|-1
            &\text{if $u,v\in F(\calD)$,}
            \\
            0&\text{otherwise.}
        \end{cases}
    \end{equation}
    Its diagonal is denoted
    \begin{equation}
        \psi^*=\psi^*_\calD:F(\Z^2)\to\Z,\, u\mapsto \psi_\calD(u,u)
        =
        \begin{cases}
            \text{the $\calX$-depth of $u$} &\text{if $u\in F(\calD)$},\\
            0 &\text{otherwise}.
        \end{cases}
    \end{equation}
\end{definition}

\begin{lemma}[Basic properties of the branching function]
    \label{lem:branching_function_basic_properties}
    The branching function $\psi$ and its diagonal $\psi^*$ satisfy the following properties.
    \begin{enumerate}
    \item $\psi^*$ is a graph homomorphism from $F(\calD)$ to $\Z$ that equals $0$ on $\partial \calD$.
    \item $\psi^*$ preserves the parity of the faces.
    \item $\psi^*$ is equal to the maximum of $h$ over all possible realisations of $h$ given $\calX$.
    \item $\psi$ encodes the conditional covariance of each pair of faces as follows:
    \begin{equation}
        \label{eq:covariance_branching_function}
        \mu_{\calD}^+[h(u)h(v)|\calX]
        =
        \begin{cases}
            4k &\text{if $\psi(u,v)=2k$,}\\
            4k+1 &\text{if $\psi(u,v)=2k+1$ and $\psi^*(u)=\psi^*(v)=2k+1$,}\\
            4k+2 &\text{if $\psi(u,v)=2k+1$ and $\psi^*(u)\vee\psi^*(v)>2k+1$.}
        \end{cases}
    \end{equation}
\end{enumerate}
\end{lemma}

\begin{proof}
    The first two follow from the definition.
    Conditionally on $\calX$,
    the height function is obtained by sampling the labels $\sigma_\slodd(X)$ for every $X\in V_\slodd(\calX)$ and attributing the spin to each odd faces
    (see the comment following the proof of Lemma~\ref{lemma:Basic properties of the level line tree}).
    It is easy to see that if the labels $\sigma_\slodd(X)$ are chosen so that they maximise the height function,
    then we obtain $h=\psi^*$.

    For the fourth property, we make the link with the GFF picture sketched in the beginning of this section.
    Each odd vertex is labelled $\pm$ by flipping independent fair coins.
    Each $\omega^+$/$\omega^-$-circuit is associated with a height gain of $\pm2$,
    and so the conditional covariance of $h(u)$ and $h(v)$ is (roughly speaking)
    equal to $4$ times the number of alternating $\omega^+$/$\omega^-$-circuits surrounding both $u$ and $v$.
    We must be a bit more careful at odd heights, because the total height contains an additional $\pm1$-contribution.
    To make the distinction between the second and third case in Equation~\eqref{eq:covariance_branching_function},
    observe that since no odd $\calX$-vertex has more than one child,
    $\psi(u,v)=2k+1$ implies that either $u$ or $v$ is at depth $2k+1$ in the tree.
\end{proof}

\begin{definition}[Maximal domains]
    \label{def:maximal_domains}
    Let $D\subset\R^2$ denote any simply connected set,
    and consider an even spin configuration $(\sigma_\sleven,\bar\omega)$.
    Let $\calM^+(D)$ denote the set of maximal even domains $\calD$ subject to the following two conditions:
    \begin{equation}
        \partial\calD\subset D
        \qquad\text{and}\qquad
        \partial\calD\subset \omega^+.
    \end{equation}
    The set $\calM^-(\calD)$ is defined similarly by replacing $\omega^+$ with $\omega^-$.

    We also let $\partial\calM^\pm(D)\subset\R^2$ denote the union of $\partial\calD\subset\R^2$ over all $\calD\in\calM^\pm(D)$.
\end{definition}

\begin{lemma}[Recursion relation for the branching function]
    \label{lem:covariance_recursion}
    Consider the independent coupling of all measures $(\mu_{\calB}^+)_{\calB}$
    over all even domains.
    More precisely, define the probability measure $\P:=\prod_{\calB}\mu_{\calB}^+$,
    where the product is over all even domains.
    Let $\calD$ denote a fixed even domain.
    Then, in the probability measure $\P$,
    $ \psi_\calD$ and $ (\psi_\calD \wedge 2) + \sum_{\calB\in\calM^-(\calD)} \psi_{\calB}$ have the same distribution.\end{lemma}

\begin{proof}
    Suppose that we simply explore the domains $\calM^-(\calD)$ and everything that happens
    outside the union $\cup\calM^-(\calD)$ of those domains.
    Conditionally on $\calM^-(\calD)$,
    the configuration inside $\cup\calM^-(\calD)$
    is given by the independent product
    \begin{equation}
        \label{eq:covariance_recursion_independent_product}
        \prod_{\calB\in\calM^-(\calD)} \mu_{\calB}^-
    \end{equation}
    (this follows from the spatial Markov property).
    Notice that by definition of $\calM^-(\calD)$ and by the independent flip symmetry in each smaller domain,
    we get
    \begin{equation}
        \psi_\calD(u,v)\leq 2
    \end{equation}
    for any $u,v$ unless $u$ and $v$ belong to the same domain $\calB\in\calM^-(\calD)$.
    Moreover, if $u$ and $v$ belong to the same domain $\calB\in\calM^-(\calD)$,
    then \begin{equation}
        \psi_\calD(u,v)=2+\psi_\calB(u,v).
    \end{equation}
    This yields the lemma via Equation~\eqref{eq:covariance_recursion_independent_product}.
\end{proof}

We conclude by gathering two useful properties related to the branching function.
\begin{lemma}[Monotonicity for the branching function]
    \label{lem:covariance_mono}
    The following holds true.
    \begin{enumerate}
        \item \textbf{FKG inequality.} The law of $\psi$ satisfies the FKG inequality in any measure $\mu_{\calD}^+$,
        \item \textbf{Monotonicity in domains.} The law of $\psi$ in $\mu_{\calD}^+$ is stochastically increasing in the domain $\calD$.
    \end{enumerate}
\end{lemma}

\begin{proof}
    We prove that $\psi\wedge 2k$ satisfies both properties for all $k\geq 1$ by inducting on $k$.
    The base case $k=1$ is easy since $\psi\wedge 2$ is a decreasing function of the triple
    $(\sigma_\sleven,\omega^+,-\omega^-)$,
    for which both properties are known by $\sleven$-FKG
    (see Lemmas~\ref{lemma:simpleFKG} and~\ref{cor:Monotonicity in domains and boundary conditions}).

    By similar reasoning, the law of $\cup\calM^-(\calD)$ in $\mu_{\calD}^+$ is also stochastically increasing in $\calD$.

    We now start the induction step:
    suppose that $\psi\wedge 2k$ satisfies both properties.
    Then the law of $\sum_{\calB\in\calM^-(\calD)} (\psi_{\calB}\wedge 2k)$
    is increasing in $\cup\calM^-(\calD)$,
    and satisfies the FKG inequality.
    By the tower property (see for instance in the proof of Proposition~\ref{proposition:realFKG}) for the FKG inequality with respect to conditioning on $\cup\calM^-(\calD)$,
    this implies that
    \begin{equation}
        (\psi_\calD \wedge 2) + \sum_{\calB\in\calM^-(\calD)} (\psi_{\calB}\wedge 2k)=\psi_\calD \wedge (2k+2).
    \end{equation}
    satisfies both properties as well.
    This concludes the induction step and the proof.
\end{proof}

% </input src="sections/PART_D/17_level_line_tree.tex" root="." version="0.0.1">
% <input src="sections/PART_D/18_regularity.tex" root="." version="0.0.1">

\section{Regularity estimate in full plane (Theorem~\ref{thm:Regularity})}
\label{sections:proof:Theorem(regularity)}

We are now in a position to establish the regularity estimates stated in Theorem~\ref{thm:Regularity}.
It is proved in three steps:
first, we express the correlation function $\Phi_k$ in terms of
the level line tree (Subsection~\ref{sec:21.1}), second we introduce a way to relate the geometry of the level line tree to arm events (Subsections~\ref{sec:21.2} and \ref{sec:21.3}),
and third we use this relation to bound the correlation function (Subsection~\ref{sec:21.4}).

% The goal of this section is to prove the regularity estimate.
% To bound the expectation of products of height differences,
% we express them in terms of tree paths,
% and then bound the way that these tree paths intersect each other.
% We will usually bound the expectation somewhat brutally,
% by using an expression in terms of pairwise path intersections (see e.g.~Equation~\eqref{eq:treebound} below),
% and then bounding each pairwise intersection using arm events in disjoint annuli.
% The disjointness of the annuli is important, because it means that the arm exponent estimate
% (the motor block of the proof)
% factorises over the annuli.

\subsection{Bounding the $k$-point correlations in terms of the level line tree}\label{sec:21.1}

Recall Definition~\ref{def:tree_path}.
We first prove the following lemma.

\begin{lemma}\label{lem:h3}
  Fix $k\in2\Z_{\geq 1}$ and
    $\u \in (F(\Z^2))^{2k}$,
    and let $\calD$ denote an even domain containing all faces in $\u$.
    Then
 \begin{equation}\label{eq:h30}
        \Big|\mu_{\calD}^+\Big[\prod_{i=1}^{k} (h(u_i')-h(u_i))\Big]\Big|
        \leq
        2^k
        \sum_\pi
        \mu_{\calD}^+\Big[\prod_{ij\in\pi}
            I(u_iu_i',u_ju_j')\Big],
    \end{equation}
    where the sum runs over pairings of $\{1,\ldots,k\}$, and
   where $ I(vv',zz'):=|p^{vv'}\cap p^{zz'}\cap V_\slodd(\calX)|$.
   Here,  $p^{ z z'}$ refers to the unique $\calX$-path from $z$ to $z'$.
    \end{lemma}

\begin{proof}
    The idea is to bound the conditional expectation of the product of the height differences, given $\calX$.
    This conditional expectation may be bounded as follows.
     The height difference $h(u_i')-h(u_i)$
        may be written
        \begin{equation}
            h(u_i')-h(u_i)=\sum_{n=0}^{|p^{u_iu_i'}|-1}
            h(p^{u_iu_i'}_{n+1})-h(p^{u_iu_i'}_n)
        \end{equation}
        where $|\cdot|$ denotes the length of the path.

        We decompose the left-hand side of \eqref{eq:h30}
        by conditioning on $\calX$ and writing each height difference as in the previous displayed equation. We then expand the sum with respect to the product. At the end, we obtain a sum of products of (integer) powers of terms of the form $h(p^{u_iu_i'}_{n+1})-h(p^{u_iu_i'}_n)$.
        Each term in the sum contributes $-1$, $0$, or $1$, so that it suffices to
         upper bound the number of terms with a
        nonzero contribution.

        Since increments of $h$ on edges of $\calX$
        only interact on edges incident to the same odd vertex,
        we get, if $I(vv',zz'):=|p^{vv'}\cap p^{zz'}\cap V_\slodd(\calX)|$,
        \begin{equation}
            \label{eq:treebound}
            \textstyle
            \left|\mu_{\calD}^+\left[\prod_{i=1}^{k} (h(u_i')-h(u_i))\middle|\calX\right]\right|
            \leq \sum_\pi\prod_{ij\in\pi}
            4I(u_iu_i',u_ju_j'),
        \end{equation}
        where $\pi$ runs over pairings of $\{1,\ldots,k\}$
        and paths are viewed as subsets of $V(\calX)$.

    Clearly $|\pi|=k/2$.
    The tower property yields
    \begin{equation}  \textstyle
        \left|\mu_{\calD}^+\left[\prod_{i=1}^{k} (h(u_i')-h(u_i))\right]\right|
        \leq
        2^k
        \sum_\pi
        \mu_{\calD}^+\left[\prod_{ij\in\pi}
            I(u_iu_i',u_ju_j')\right],
    \end{equation}
    where the $I(u_iu_i',u_ju_j')$ are viewed as random variables.
    \end{proof}

\subsection{Key input for the regularity estimate}
\label{sec:21.2}

We now want to bound the right hand side of Equation~\eqref{eq:h30}.
While the appropriate bound requires a bit of geometrical analysis,
the kea idea is fairly simple and explained in the following lemma.

\begin{lemma}
    \label{lemma:simple_bounds_on_I}
    Let $u,u',v,v'\in\Z^2$ denote four points contained in some finite domain $\calD$.
    Then all of the following hold true.
    \begin{enumerate}
        \item Suppose that $\{u,u'\}$
        is contained in one connected component of $\R^2\setminus \ann{R}{r}(x)$
        and $\{v,v'\}$ in the other connected component,
        where $r\geq 4$, $R\geq 2r$, and $x\in\R^2$.
        Then
        \begin{equation}
            I(uu',vv')\leq 4K_{R,r,x},
        \end{equation}
        where $K_{R,r,x}$ counts alternating arms in the annulus
        as defined in
        Equation~\eqref{eq:Kdef}.
        \item If $u$ and $u'$ are neighbours (or $v$ and $v'$), then
        \begin{equation}
            I(uu',vv')\leq 4.
        \end{equation}
    \end{enumerate}
\end{lemma}

\begin{proof}
    The second part is easy: if $u\sim u'$,
    then $I(\blank)\leq |p^{uu'}\cap V_\slodd(\calX)|\leq 4$.
    For the first part, notice that as we walk along the path
    $p^{uu'}\cap p^{vv'}$,
    we discover alternating $\omega$-circuits, which each have the property
    that they separate $u$ from $u'$ and $v$ from $v'$.
    In particular, each such circuit creates an arm in the annulus.
    The factor $4$ royally suffices.
    This implies the desired result.
\end{proof}

To bound the right-hand side of Equation~\eqref{eq:h30},
we would like to proceed as follows:
if we can find an annulus for each pair $ij\in\pi$ such that the $|\pi|=k/2$ annuli are disjoint,
then we can simply apply the previous lemma and Remark~\ref{rmk:finite expectation annulus}
to bound the expectation of the product of the arm counts.

For the general case (when the scale separation between some of the pairs is small), we must do more work.
To reduce to the case with good scale separation,
we shall further decompose each variable $I(\blank)$.
This random variable clearly satisfies the triangular inequality
$I(vv'',zz')\leq I(vv',zz')+I(v'v'',zz')$.
In the next subsection,
we decompose the right hand side of Equation~\eqref{eq:h30},
to the point that we can use Lemma~\ref{lemma:simple_bounds_on_I} in combination
with Lemma~\ref{lemma:armexponent_for_spins_linear}.

\subsection{Organizing the points in a suitable fashion}\label{sec:21.3}

The purpose of the following lemma is roughly as follows: if the points $a,a',b,b'$ do not have good scale separation,
then we may find a path from $a$ to $a'$ such that we can apply one of the two scenarios of Lemma~\ref{lemma:simple_bounds_on_I}
to each step of the path (against $\{b,b'\}$).
The shortest appropriate path is called an \emph{optimal path}.

\begin{lemma}
    \label{lemma:optimal_path_decomposition}
    Recall the discrete scale separation function $S_{\R^2}'$
    in Equation~\eqref{eq:scale_separation_functions_discrete}.
    Then for each $k\in2\Z_{\geq 1}$,
    there exists a constant $N_k\in\R_{>0}$ with the following properties.
    For any two pairs of points $\{a,a'\}$
    and $\{b,b'\}$ in $\Z^2$ with $|a'-a|\leq |b'-b|$,
    there exists a path $p=(p(n))_{0\leq n\leq \ell}\subset\Z^2$ of length at most $N_k\cdot S_{\R^2}'(\{a,a'\},\{b,b'\})$
    from $a$ to $a'$ such that for each $0\leq n\leq \ell-1$,
    one of the following two holds true:
    \begin{itemize}
        \item $\gamma(n)$ and $\gamma(n+1)$ are neighbours,
        \item $S_{\R^2}(\{\gamma(n),\gamma(n+1)\},\{b,b'\})\geq 20k^2$.
    \end{itemize}

    The shortest such path is called \emph{optimal}.
\end{lemma}

    The lower bound $20k^2$ on the scale separation may appear arbitrary at this point.
    Its choice is motivated by the following problem.
    Broadly speaking, we want that the annuli do not overlap in order to apply Lemma~\ref{lemma:armexponent_for_spins_linear}.
    If the scale separation is large enough, then we may extract appropriate \emph{disjoint} annuli.
    This is proved in the following ``shrinking annuli'' lemma which one can skip in a first reading.

    \begin{lemma}[Shrinking annuli lemma]\label{lem:shrinking}
    Fix $N\in\Z_{\geq 2}$ and a family
    $(R_i,r_i,x_i)_{i=1,\ldots,N}\subset (0,\infty)\times(0,\infty)\times\R^2$
    satisfying $R_i/r_i\geq e^{40N^2}$.
    Then, we may find some $(R_i',r_i')_{i=1,\ldots,N}\subset(0,\infty)\times(0,\infty)$
    such that:
    \begin{enumerate}
        \item The radii satisfy $r_i\leq r_i'\leq R_i'\leq R_i$,
        \item The radii satisfy $R_i'/r_i'= \sqrt[4N^2]{R_i/r_i}\geq e^{10}$ for each $i$,
        \item The annuli $(\ann{R_i'}{r_i'}(x_i))_{i=1,\ldots,N}$ are pairwise disjoint.
    \end{enumerate}
\end{lemma}

\begin{proof}
    Set $Q=2N^2$
    and $\rho_i:=\sqrt[2Q]{R_i/r_i}=\sqrt[4N^2]{R_i/r_i}\geq e^{10}$.
    Let $\P$ denote the uniform probability measure on
    the random element $s\in\{0,\ldots,Q-1\}^N$.
    Define the random annuli
    \begin{equation}
        A_i:=\ann{R_i'}{r_i'}(x_i);
        \qquad
        R_i':=R_i'(s_i):=r_i\cdot\rho_i^{2s_i+2};
        \qquad
        r_i':=r_i'(s_i):=r_i\cdot\rho_i^{2s_i+1}
    \end{equation}
    in this probability space.
    These radii clearly satisfy the first two properties;
    it suffices to
    prove that the annuli are disjoint with positive $\P$-probability,
    a fact which would follow from $\P[\{A_i\cap A_j\neq\emptyset\}]\leq 2/N^2$
    for any distinct $i$ and $j$.

    Let us turn to the proof of this fact. Fix $i\ne j$ and
    let $O\subset \{0,\ldots, Q-1\}^2$
    denote the set of pairs $(s_i,s_j)$ which lead to overlapping annuli.
    Fix $(a,b)\in O$; we then claim that
    \begin{equation}\label{eq:ab}
           O\cap  \{(a',b'):a'>a,\,b'\leq b\}=\emptyset\qquad\text{or}\qquad O\cap\{(a',b'):a'\leq a,\,b'> b\}=\emptyset.
        \end{equation}
  Indeed, without loss of generality (by swapping $i$ and $j$ if necessary),
    we may assume that
    \(
        R_i'(a)\geq R_j'(b)\).
   One can then see that for any $a'>a$
    and $b'\leq b$,
    the annulus $\ann{R_i'(a')}{r_i'(a')}(x_i)$
    is disjoint from $\bbox{R_j'(b)}(x_j)$,
    hence from $\ann{R_j'(b')}{r_j'(b')}(x_j)$.

With \eqref{eq:ab} in hand, we get that  $|O|\leq 4Q$.
    This implies the desired bound $\P[\{A_i\cap A_j\neq\emptyset\}]\leq 4Q/Q^2=2/N^2$, and concludes the proof of the lemma.
\end{proof}

\subsection{Proof of the regularity estimate}\label{sec:21.4}

We are now ready to dive into the proof of Theorem~\ref{thm:Regularity}.

\begin{proof}[Proof of Theorem~\ref{thm:Regularity}]
 Recall that the law of $h$ is the same in $\mu_{\Z^2}$ and $\E_{\Z^2}$ (Lemma~\ref{lemma:TorusLimits}). We therefore only need to bound the correlations in an even domain with $+$ boundary conditions, and then to let the domain go to $\Z^2$ to obtain our result. From now on, we fix a large enough even domain $\calD$ containing all the points of $\u$.

   By Lemma~\ref{lem:h3}, it suffices to bound for every $\pi$,
  \begin{equation}\label{eq:h30_second_time}
        \mu_{\calD}^+\Big[\prod_{ij\in\pi}
            I(u_iu_i',u_ju_j')\Big].
    \end{equation}
    Below, we shall view $\pi$ in \eqref{eq:h30} as a set of \emph{ordered} pairs $ij=(i,j)$,
    where we order each pair such that $|u_i'-u_i|\leq |u_j'-u_j|$.
    For each $ij\in\pi$,
    let $q_{ij}$ denote the optimal path from $u_i$ to $u_i'$
    (relative to the pair $\{u_j,u_j'\}$).
    Write $|q_{ij}|$ for the length of this path
    (this quantity is bounded via Lemma~\ref{lemma:optimal_path_decomposition}). Partition the pairs in $\pi$ into two sets, as follows.
    \begin{itemize}
        \item The set $\pi_g$ is defined as the set ``good'' pairs, that is, the pairs $ij$ which satisfy $S_{\R^2}(\{u_i,u_i'\},\{u_j,u_j'\})\geq 20k^2$.
        This means that automatically, $|q_{ij}|=1$ for any $ij\in\pi_g$.
        \item The remaining pairs are ``bad''; the set of bad pairs is denoted $\pi_b$.
        For $ij\in\pi_b$, it is still possible that $|q_{ij}|=1$ for $ij\in\pi_b$, namely when $u_i$ and $u_i'$ are neighbours.
    \end{itemize}

    Define $\calI(\pi):=\prod_{ij\in\pi}\{0,\ldots,|q_{ij}|-1\}$;
    the path decomposition leads to
    \begin{equation}
         \mu_{\calD}^+\Big[\prod_{ij\in\pi}
            I(u_iu_i',u_ju_j')\Big]\le
        \sum_{\n \in\calI(\pi)}
        \mu_{\calD}^+\left[\prod_{ij\in\pi}
        I(q_{ij}(\n _{ij})q_{ij}(\n _{ij}+1),u_ju_j')\right].
    \end{equation}

    \begin{figure}
        \centering
        \includegraphics{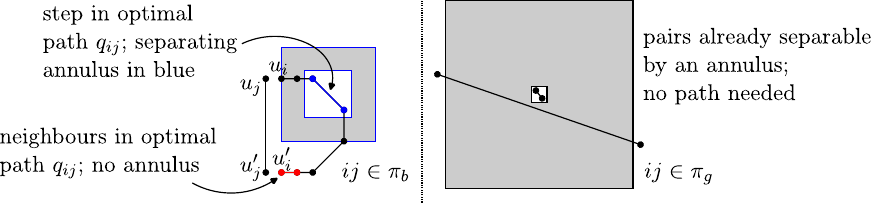}
        \caption{Left: An optimal path $q_{ij}$ from $u_i$ to $u_i'$
        for a pair $ij\in\pi_b$.
        Right: If $ij\in\pi_g$, then no optimal path is needed.
        }
        \label{fig:optimal_path}
    \end{figure}

    Next, we are going to bound the random variables $I(\blank)$
    using Lemma~\ref{lemma:simple_bounds_on_I}.
    We distinguish three cases (see Figure~\ref{fig:optimal_path}).
    \begin{enumerate}
        \item If $ij\in\pi_b$ and $q_{ij}(\n _{ij})$ and $q_{ij}(\n _{ij}+1)$ are neighbours,
        then $I(\blank)\leq 4$ deterministically.
        \item If $ij\in\pi_b$ and the two faces are not neighbours,
        then there exists an annulus $A_{ij\n }:=\ann{R_{ij\n }}{r_{ij\n }}(x_{ij\n })$
        such that $\{q_{ij}(\n _{ij}),q_{ij}(\n _{ij}+1)\}$ is contained in one connected component
        of $\R^2\setminus A_{ij\n }$, and $\{u_j,u_j'\}$ in the other connected component,
        and such that $R_{ij\n }/r_{ij\n }\geq e^{10k^2}$.
        In that case, Lemma~\ref{lemma:simple_bounds_on_I} gives
        \begin{equation}
            \label{eq:boundonIntersections}
            I(\blank)\leq 4K_{R_{ij\n },r_{ij\n },x_{ij\n }}.
        \end{equation}
        \item If $ij\in\pi_g$, then we may argue as for the previous case, except that we may choose the annulus such that
        $R_{ij\n }/r_{ij\n }\geq e^{S_{\R^2}(\{u_i,u_i'\},\{u_j,u_j'\})/2}\geq e^{10k^2}$.
    \end{enumerate}
    Writing $\pi_{b,\n }\subset \pi_b$ for the pairs in Case~(ii),
    we get
     \begin{equation}
        \mu_{\calD}^+\Big[\prod_{ij\in\pi}
            I(u_iu_i',u_ju_j')\Big]\le4^k
        \sum_{\n \in\calI(\pi)}
        \mu_{\calD}^+\left[\prod_{ij\in\pi_g\cup\pi_{b,\n }}
        K_{R_{ij\n },r_{ij\n },x_{ij\n }}\right].
    \end{equation}
        The right-hand side is easy to upper bound if the annuli do not overlap (using Remark~\ref{rmk:finite expectation annulus}),
    but the disjointness is not guaranteed by our construction. We therefore apply the shrinking annuli lemma (Lemma~\ref{lem:shrinking}) with $N=k/2$ to find a family of non-overlapping annuli     $A_{ij\n }':=\ann{R_{ij\n }'}{r_{ij\n }'}(x_{ij\n })\subset A_{ij\n }$ (with $ij\in\pi_g\cup\pi_{b,\n }$) with radii $r_{ij}\geq 4$ and $R_{ij\n }'/r_{ij\n }'=\sqrt[k^2]{R_{ij\n }/r_{ij\n }}\geq e^{10}$.

Thanks to the inclusion of the annuli, the previous estimate turns into
         \begin{equation}
        \mu_{\calD}^+\Big[\prod_{ij\in\pi}
            I(u_iu_i',u_ju_j')\Big]
        \leq
        4^k
        \sum_{\n \in\calI(\pi)}
        \mu_{\calD}^+\left[\prod_{ij\in\pi_g\cup\pi_{b,\n }}
        K_{R_{ij\n }',r_{ij\n }',x_{ij\n }}\right].
    \end{equation}
    The expectation on the right
    may now be bounded as follows:
    first we bound the expectation of $K_{(\blank)}$
    for the annulus with the largest outer radius, then the \emph{conditional}
    expectation of the $K_{(\blank)}$ with the next-largest outer radius,
    et cetera.
    Lemma~\ref{lemma:armexponent_for_spins_linear} (more precisely Remark~\ref{rmk:finite expectation annulus}) asserts that the conditional expectation of
    each $K_{R_{ij\n }',r_{ij\n }',x_{ij\n }}$
    is bounded by $(R_{ij\n }'/r_{ij\n }')^{-\carm}/(1-2^{-\carm})$ in this procedure.
    We therefore get
    \begin{equation}
       \mu_{\calD}^+\Big[\prod_{ij\in\pi}
            I(u_iu_i',u_ju_j')\Big]        \leq
        \frac{4^k}{(1-2^{-\carm})^{k/2}}
        \sum_{\n \in\calI(\pi)}
        \prod_{ij\in\pi_g\cup\pi_{b,\n }}
        (R_{ij\n }'/r_{ij\n }')^{-\carm}.
    \end{equation}
    Using the lower bounds on $R_{ij\n }'/r_{ij\n }'$ provided by  Lemma~\ref{lem:shrinking} and Lemma~\ref{lemma:optimal_path_decomposition}
    (which bounds $|q_{ij}|$ and therefore $|\calI(\pi)|$),
    we get the  bound
    \begin{equation}
     \mu_{\calD}^+\Big[\prod_{ij\in\pi}
            I(u_iu_i',u_ju_j')\Big]        \leq
        C_k\prod_{ij} \begin{cases}
            e^{-(\carm/2k^2) S_{\R^2}(\{u_i,u_i'\},\{u_j,u_j'\})} & \text{if $ij\in\pi_g$,} \\
            1\vee -S_{\R^2}'(\{u_i,u_i'\},\{u_j,u_j'\}) & \text{if $ij\in\pi_b$.}
        \end{cases}
    \end{equation}
   As mentioned in the preamble of the proof, this implies the claim, as one can let $\calD$ to $\Z^2$ and sum over every $\pi$ to get a bound on $|\Phi_k(\u )|$.
\end{proof}

% </input src="sections/PART_D/18_regularity.tex" root="." version="0.0.1">
% <input src="sections/PART_D/19_cylinder_bounds.tex" root="." version="0.0.1">

\section{Regularity estimate (cylinder, Corollary~\ref{cor:Regularity_cylinder})}

The proof is quite straightforward relative to the proof of Theorem~\ref{thm:Regularity}.
We can essentially work as for the full-plane case,
except when the cylinder is very thin relative to the distance between the points
(see the picture on the right in Figure~\ref{fig:cylinder_large}).
In that case, we must slightly modify our proof.
This modification is very natural: rather than counting alternating arms in an annulus,
we count alternating arms going through a thin subcylinder.
This leads to an even better bound than the one we need.

\begin{proof}[Proof of Equation~\eqref{eq1:cor:Regularity_cylinder} in Corollary~\ref{cor:Regularity_cylinder}]
    We first prove that there is some constant $C$ such that
    \begin{equation}
            |\Phi_{\CYL{L},2}((0,0),(k,0),(2k,0),(3k,0))|\leq C
    \end{equation}
    for any $k$ and $L$.
    Fix $N=10^9$.
    For simplicity we split in three cases.

    \paragraph{Small values for $k$ ($k\leq N$).}
    If $k\leq N$, then $\Phi_{\CYL{L},2}((0,0),(k,0),(2k,0),(3k,0))\leq N^2$.

    \paragraph{Values for $k$
    that are large, but smaller than the cylinder
    ($ N<k\leq L/40$).}
    This is the ``full-plane case''.
    Set $u:=(k/2,0)$, $r=5k/8$, and $R=2r=5k/4$.
    By arguing as in the proof of Theorem~\ref{thm:Regularity}
    (Section~\ref{sections:proof:Theorem(regularity)}),
    we see that
    \begin{equation}
        |\Phi_{\CYL{L},2}((0,0),(k,0),(2k,0),(3k,0))|
        \leq
        8\mu_{\CYL{L}}[K_{R,r,u}],
    \end{equation}
   where $K_{R,r,u}$ is defined in~\eqref{eq:Kdef}; see Figure~\ref{fig:cylinder_large}.
    This expectation is uniformly bounded by Remark~\ref{rmk:finite expectation annulus}.

        \begin{figure}
            \centering
            \includegraphics{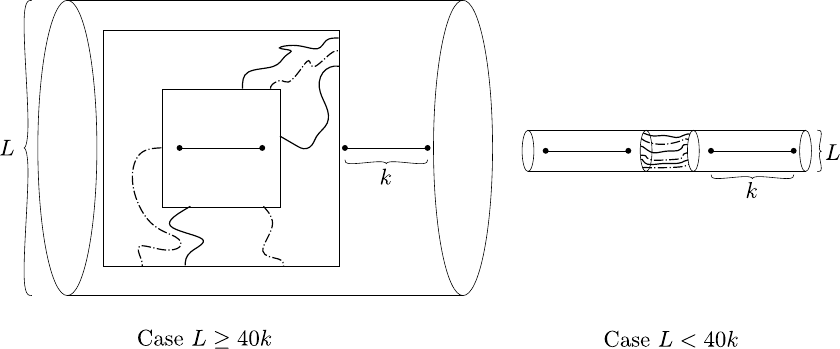}
            \caption{Regularity estimates for the cylinder}
            \label{fig:cylinder_large}
        \end{figure}

    \noticePiet{Made the Figure~22, Right larger}
    \paragraph{Values where $k$ is on the scale of the cylinder ($k> N \vee (L/40)$).}
    This is the ``cylinder case''.
    To illustrate why this case is different, suppose for a moment that $L$ is in fact much smaller than $k$
    (for example, $L\approx k/100$).
    The problem that arises is that the annulus considered in the previous case wraps around the cylinder many times,
    and therefore the estimate that we used before does not make sense.
    To circumvent this problem, we are going to replace the annulus by a different shape,
    which leverages the topology of the cylinder.
    This leads to an \emph{even better} upper bound, even though a constant bound suffices for our purposes.

    Define the subcylinder $A=[\frac54k,\frac74k]\times (\R/L\Z)]\subset\CYL{L}$,
    and let $K$ denote the $2\Z_{\geq 0}$-valued random variable defined to be maximal subject to
    $A$ having $K$ alternating $\bar\omega$ crossings from left to right (see Figure~\ref{fig:cylinder_large}).
    Notice that, just like in the definition of $\pALTARM{\bar\omega}{K}{A}$,
    there is not really a highest or leftmost crossing, due to the topology of $A$.

    By arguing as before,
    we see that
    \begin{equation}
        |\Phi_{\CYL{L},2}((0,0),(k,0),(2k,0),(3k,0))|
        \leq
        8(1+\mu_{\CYL{L}}[K]).
    \end{equation}
    Importantly, the aspect ratio $L/ \frac12k$ of $A$ is upper bounded thanks
    to our assumption.
    We may therefore argue as before to see that $K$ has exponentially decaying tails,
    uniformly in the choice of $k$ and $L$ (subject to the bound on the aspect ratio).
    This yields the desired uniform bound on $\mu_{\CYL{L}}[K]$, which is also uniformly bounded by Remark~\ref{rmk:finite expectation annulus}.
\end{proof}

\begin{remark} Although not necessary, in this last case it is straightforward to see that the upper bound on the correlation function
    decays \emph{exponentially fast} in $k/L$.
    This is consistent with an intuition coming from the transfer matrix perspective:
    the eigenvalues of $T(0)$ are $L$-th roots of unity,
    and therefore they cannot be too close to one (without being equal to one).
    This is obvious because $T(0)^L$ is the identity operator.
    Such a one-line proof does not exist for the eigenvalues of the Hermitian matrix $T(\pi/2)$,
    but it can be proved that there are no eigenvalues in
    the range $(1-\epsilon/L,1)$ by RSW-type arguments.
\end{remark}

\begin{proof}[Proof of Equation~\eqref{eq2:cor:Regularity_cylinder} in Corollary~\ref{cor:Regularity_cylinder}]
    We aim to prove that for any $L\in 2\Z_{\geq 1}$ and for any integers $0<\ell\leq L/2$
        and $k\geq \ell/8$,
        \begin{equation}
            |\Phi_{\CYL{L},2}((0,0),(0,\ell),(k,0),(k,\ell))|\leq C(\ell/k)^c.
        \end{equation}
Again, we divide into three cases.

    \paragraph{Values of $k\leq 10^6\ell$.}
    It suffices to bound the correlation function by a uniform constant.
    If $\ell\leq 1000$ then this is trivial.
    Suppose that $\ell>1000$.
    Let
    \begin{align}
        K&:=\COUNTARM{\bar\omega}{A};\\        A&:=
        \left(\symInt{\ell/10}\times [-\tfrac\ell3,\tfrac{4\ell}{3}]\right)\setminus\left(\symInt{\ell/20}\times [-\tfrac\ell{6},\tfrac{7\ell}{6}]\right).
    \end{align}
    Then, the correlation function is bounded by $8(1+\mu_{\CYL{L}}[K])$.
    It is easy to prove that $\mu_{\CYL{L}}[K]$
    is uniformly bounded, by arguing as in the proof of Lemma~\ref{lemma:armexponent_for_spins_linear}.

    \paragraph{Values of $k>10^6\ell$ satisfying    $\sqrt{\ell k}\le L$.}
    This is the ``full-plane case''.
    Fix $r:=4\ell \geq 4$ and $R:=\frac1{10}\sqrt{\ell k}$.
    Recall the definition of $K_{r,R,(0,\ell/2)}$ from the statement of Lemma~\ref{lemma:armexponent_for_spins_linear} (roughly speaking, it counts the number of alternating crossings from inside to outside in the annulus).
    The level line tree picture can be developed for the cylinder:
    without going into the details, it is not difficult to see that the two-point
    correlation function is bounded by the number of alternating circuits separating the two points
    in each pair (Lemma~\ref{lemma:simple_bounds_on_I}),
    which in turn is bounded by (see the picture on the left in Figure~\ref{fig:cylinder_large})

    \begin{equation}
        8\mu_{\CYL{L}}[K_{r,R,(0,\ell/2)}]
        \leq
        \frac{(r/R)^{\carm}}{1-2^{-\carm}}
        \propto
        (\ell/k)^{\carm/2},
    \end{equation}
    which is a bound of the desired form.

    \paragraph{Values of $k>10^6\ell$ satisfying
    $\sqrt{\ell k}>L$.}
    This is the ``cylinder case''.
    First observe that
    \(k/L\geq \sqrt{k/\ell}\geq 1000\).
    Define the subcylinders
    \begin{align}
        U_0 &:= [L,2L]\times (\R/L\Z);\\
        U_1 &:= [3L,k-L]\times (\R/L\Z).
    \end{align}
    Write $K$ for the $2\Z_{\geq 0}$-valued random variable defined to be maximal subject to
    \(\pALTARM{\bar\omega}{K}{U_0}\)
    occurring.
    Write $\calE:=\pCIRCUIT{\omega}{U_1}$.
    Then, the correlation function is bounded by
    \begin{equation}
        8\mu_{\CYL{L}}[\indi{\calE^c}(1+K)].
    \end{equation}
    Yet, the conditional expectation $\muSpin_{\CYL{L}}[K|\calE^c]$ is uniformly bounded, and the probability $\muSpin_{\CYL{L}}[\calE^c]$ tends to zero exponentially fast in $k/L$.
    Thus, we get a bound of the form
    \begin{equation}
        |\Phi_{\CYL{L},2}((0,0),(0,\ell),(k,0),(k,\ell))|\leq C e^{-c k/L}\leq C e^{-c \sqrt{k/\ell}}.
    \end{equation}
    The stretch-exponential decay is even stronger than the desired polynomial decay.
\end{proof}

% </input src="sections/PART_D/19_cylinder_bounds.tex" root="." version="0.0.1">
% <input src="sections/PART_D/20_mixing.tex" root="." version="0.0.1">

\section{Mixing estimate (Theorem~\ref{thm:mixing})}

\label{sec:mixing}

This section is split into two subsections.
The first subsection contains the main proof of the mixing estimate.
The second subsection analyses in further detail the covariance
structure of the height function,
which is used in the last step of the main proof.
The ideas on the covariance structure are also used in
the last section (Section~\ref{sec:ldp_split}).

\subsection{Main part of the proof}

\begin{proof}[Proof of Theorem~\ref{thm:mixing}]
    Fix $\c\in[1,2]$, $k\in 2\Z_{\geq 3}$,
    and $\epsilon:=1/10^9$ throughout this section.
    Consider $\u\in\calD_k$.

 \paragraph{Step 1: Conveniently positioning the points $\u$.}
 We first argue that without loss of generality, $\u $ satisfies the following conditions (cf.~Figure~\ref{fig:mixing-2}):
    \begin{itemize}
        \item $u_2=(0,0)$, $u_2'=(-1,0)$, $u_1'=(1,0)$,
        \item $|u_i-u_j|\geq 1/\epsilon^2$ for any distinct $i,j\geq 2$,
        \item $|u_i'-u_i|\leq \epsilon^2$ for any $i\geq 3$.
    \end{itemize}
Indeed, by translating the system, we may assume that $u_2=(0,0)$.
    By decomposing each pair $(u_i,u_i')$ into paths of tiny steps (like in the proof of Proposition~\ref{proposition:harmonicity}) and using
    additivity of the correlation functions,
    we may assume that $\inf_{i\neq j}|u_j-u_i|/\sup_i |u_i'-u_i|\geq \epsilon^5$.
    We may then rescale the system such that the second and third properties are satisfied.
    Finally, the statement we are trying to prove does not depend on the position of
    $u_1'$ and $u_2'$ by the additivity property and the regularity bound, and therefore we may choose them as described in the claim.

Recall that $u_1$ is the variable of interest,
and we would like to let $u_1$ tend to $(0,0)=u_2$.

\paragraph{Step 2: Splitting the height increments.}
Consider $|u_1|<\epsilon$ and $\delta<\epsilon^2$,
and write $\w:=(u_3,u_3',\ldots)\in\calD_{k-2}$.
Write $\bar u_i\in F(\Z^2)$ for the face corresponding to $u_i/\delta$,
and define $\bar u_i'$, and $\bar\w$ similarly.
It now suffices to find a universal constant $C<\infty$
(depending only on $k$)
such that $|(\star)|\leq C$, where
\begin{equation}
    \label{eq:mixing_bound}
 (\star):=|\Phi_k(\bar u_1,\bar u_1',\bar u_2,\bar u_2',\bar\w)-\Phi_2(\bar u_1,\bar u_1',\bar u_2,\bar u_2')\Phi_{k-2}(\bar\w)|.
\end{equation}

\begin{figure}
    \centering
    \includegraphics{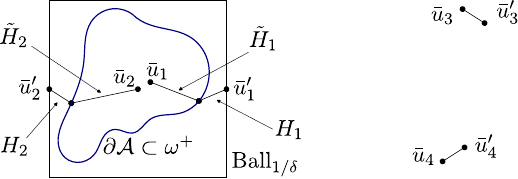}
    \caption{The points are positioned such that the pairs $(\bar u_i,\bar u_i')$
    for $i\geq 3$ have a relatively small diameter and are far from each other and from the point $(0,0)$.
    For the proof of the mixing estimate, it is convenient to write the height difference $h(\bar u_i)-h(\bar u_i')$
    as $H_i+\tilde H_i$
    for $i=1,2$, illustrated by the figure.}
    \label{fig:mixing-2}
\end{figure}

We rephrase our correlation functions in terms of the spin measure $\mu_{\Z^2}$.
The key step in the proof is to consider the random domain $\calA$
defined as the unique maximal even domain such that $0=\bar u_2\in F(\calA)$ and $\partial\calA\subset \omega^+\cap \bbox{(1/\delta)}$.
We set $\calA=\emptyset$ and $\partial\calA=\{\bar u_2\}$ when such an even domain does not exist.
Notice that almost surely $h$ is constant on $\partial\calA$;
write $h(\partial\calA)$ for this constant value.
Define the following random variables (cf.~Figure~\ref{fig:mixing-2}):
\begin{alignat*}{3}
     &P\ :=\textstyle\prod_{i=3}^k (h(\bar u_i') - h(\bar u_i)) ;\qquad&&a\ :=\
        \begin{cases}
            h(\partial\calA) & \text{if $F(\calA)\ni\bar u_1$,} \\
            h(\bar u_1) & \text{otherwise;}
        \end{cases}\\
     &H_1\ :=\ a - h(\bar u_1') ;\qquad
    &&\tilde H_1\ :=\ h(\bar u_1) - a ;\qquad
     \\[1ex]
    &H_2\ :=\ h(\partial\calA) - h(\bar u_2'); \qquad
    &&\tilde H_2\ :=\ h(\bar u_2) - h(\partial\calA).
\end{alignat*}
\commentHugo{I find the choice of notation weird... and isn't there something strange as well with the sign $a-h(\bar u_1')$, since in the product they are rather $h(\bar u_i')-h(\bar u_i)$? Of course there are two of them, but I would rather change $H_i$ for the term with $a-h(\bar u_i)$, and $H'_i$ for the term with $h(\bar u_i')-a$. What do you think? HDC: not urgent}
Then,
\begin{align}
    \textstyle\Phi_{k}(\bar u_1,\bar u_1',\bar u_2,\bar u_2',\bar\w) &=
    \mu_{\Z^2}[P(H_1+\tilde H_1)(H_2+\tilde H_2)]
    ;
    \\
    \textstyle\Phi_{2}(\bar u_1,\bar u_1',\bar u_2,\bar u_2')
    &=
    \mu_{\Z^2}[(H_1+\tilde H_1)(H_2+\tilde H_2)]
    ;\\
   \Phi_{k-2}(\bar\w)
    &=
    \mu_{\Z^2}[P];
    \\
    (\star)&=\Cov_{\mu_{\Z^2}}[P,(H_1+\tilde H_1)(H_2+\tilde H_2)].
\end{align}
\paragraph{Step 3: Rewriting $(\star)$ as a finite sum of bounded terms.}
Conditionally on $\partial\calA$,
the Markov property over $\calA$ applies.
This means that the tuples $(P,H_1,H_2)$
and
$(\tilde H_1,\tilde H_2)$ are independent.
Moreover, we know that the second tuple has zero mean due to flip symmetry (Lemma~\ref{lemma:oddspinsflip}).
Thus, we get
\begin{equation}
(\star)=
\Cov_{\mu_{\Z^2}}[P,H_1H_2 + \tilde H_1 \tilde H_2]
=
\Cov_{\mu_{\Z^2}}[P,H_1H_2]
+
\Cov_{\mu_{\Z^2}}[P,\tilde H_1 \tilde H_2].
\end{equation}
Since $P$ and $\tilde H_1\tilde H_2$ are independent conditionally on $\calA$,
the second term on the right-hand side is bounded by
\begin{equation}
    \big|\Cov_{\mu_{\Z^2}}\big[
        \mu_{\Z^2}[P|\calA],\mu_{\Z^2}[\tilde H_1 \tilde H_2|\calA]
    \big]\big|
    \leq
    \sqrt{
        \mu_{\Z^2}[\mu_{\Z^2}[P|\calA]^2]
        \Var_{\mu_{\Z^2}}[\mu_{\Z^2}[\tilde H_1 \tilde H_2|\calA]]
    }.
\end{equation}
Combining yields
\begin{equation}
    \label{eq:targetformixing}
|(\star)|
\leq
|\Cov_{\mu_{\Z^2}}[P,H_1H_2]| +
\sqrt{
        \mu_{\Z^2}[\mu_{\Z^2}[P|\calA]^2]
        \Var_{\mu_{\Z^2}}[\mu_{\Z^2}[\tilde H_1 \tilde H_2|\calA]]
}.
\end{equation}
To get Equation~\eqref{eq:mixing_bound},
it suffices to bound each of the three probabilistic terms appearing on the right in~\eqref{eq:targetformixing} by a universal constant
(depending only on $k$). This is neither short nor straightforward.
So the three terms are handled in Lemmata~\ref{lem:Bound1},  \ref{lem:Bound3}, and \ref{lem:covariance_bounded_variance} below.
\end{proof}

\subsection{Proofs of the lemmata}

\begin{lemma}\label{lem:Bound1}
In the context of the proof of Theorem~\ref{thm:mixing} (in particular Equation~\eqref{eq:targetformixing}),
there exists a constant $C<\infty$ (depending only on $k$) such that
\begin{equation}
 \mu_{\Z^2}[\mu_{\Z^2}[P|\calA]^2]\le C.
\end{equation}
\end{lemma}

\begin{proof}
    Since $\calA$ is $\omega^+$-measurable,
    we get
    \begin{equation}
         \mu_{\Z^2}[\mu_{\Z^2}[P|\calA]^2]
         \leq
         \mu_{\Z^2}[\mu_{\Z^2}[P|\sigma_\sleven,\bar\omega]^2]
         .
    \end{equation}
    Recall the definition of the level line tree from Section~\ref{sec:level line tree}.
We shall argue as in Section~\ref{sections:proof:Theorem(regularity)}
    on the regularity estimate.
    Let $\calD$ be an extremely large domain.
    In Section~\ref{sections:proof:Theorem(regularity)}, we argued in Equation~\eqref{eq:treebound}
    that
    \begin{equation}
        |\mu_{\calD}^+[
            P
        |\calX]|
        \leq \sum_\pi\prod_{ij\in\pi}
            4I(\bar u_i \bar u_i', \bar u_j \bar u_j'),
    \end{equation}
    where $\pi$ runs over the pairings of $\{3,4,\ldots,k\}$.
    In Section~\ref{sections:proof:Theorem(regularity)} our ultimate
    goal was to bound the first moment of the right-hand side,
    but now we want to bound its second moment.

    This follows straightforwardly from Lemma~\ref{lemma:armexponent_for_spins_linear}.
    Indeed, later in Section~\ref{sections:proof:Theorem(regularity)} (Equation~\eqref{eq:boundonIntersections}),
    we bounded each intersection count by an arm count around $\bar u_i$:
    \begin{equation}
        I(\bar u_i \bar u_i', \bar u_j \bar u_j')
        \leq 4K_{(1/\delta),(\epsilon/\delta),\bar u_i}.
    \end{equation}
    Letting $\calD$ converge to $\Z^2$ yields
    \begin{equation}
        \left|\mu_{\Z^2}[
            P
        |\sigma_\sleven,\bar\omega]\right|
        \leq \sum_\pi\prod_{ij\in\pi}
            16 K_{(1/\delta),(\epsilon/\delta),\bar u_i}
        .
    \end{equation}
    But the annuli corresponding to the $\bar u_i$ do not overlap,
    and therefore we may use our (exponential) bounds on the tail of $K_{(\blank)}$
    from Lemma~\ref{lemma:armexponent_for_spins_linear} (which bounds the second moment of each $K_{(1/\delta),(\epsilon/\delta),\bar u_i}$)
    to see that the second moment of
    the random variable $\mu_{\Z^2}[
            P
        |\sigma_\sleven,\bar\omega]$ is universally bounded (with a bound
        depending on $k$ and $\c$ only).
    \end{proof}

Let us turn to Lemma~\ref{lem:Bound3}, which is slightly more involved but relies on similar ideas.

\begin{lemma}\label{lem:Bound3}
In the same context (of Equation~\eqref{eq:targetformixing}),
there exists a constant $C<\infty$ (depending only on $k$) such that
\begin{equation}
|\Cov_{\mu_{\Z^2}}[P,H_1H_2]|\le  C.
\end{equation}
\end{lemma}

Before diving into its proof, let us state a convenient intermediary result.

\begin{lemma}\label{lem:Bound2}
In the same context (of Equation~\eqref{eq:targetformixing}), there exists a constant $C<\infty$ such that
\begin{equation}
 \mu_{\Z^2}[\mu_{\Z^2}[H_1H_2|\sigma_\sleven,\bar\omega]^2]\le C.
\end{equation}
\end{lemma}
\begin{proof}
    Since $\partial\calA\subset\omega^+$, the set $\partial\calA$ is contained in a single $\calX$-vertex.
    Identify $\partial\calA$ with this $\calX$-vertex.
    It therefore makes sense to consider the unique $\calX$-path from $\bar u_2'$ to $\partial\calA$,
    which we denote by $p^{\bar u_2'\partial A}$.

    Let $\calD$ denote an extremely large domain.
    By arguing as in Section~\ref{sections:proof:Theorem(regularity)}
     (Equation~\eqref{eq:treebound}),
        we observe that
        \begin{equation}
            \label{eq:bound2mixing}
            |\mu_{\calD}^+[
                H_1 H_2
            |\calX]|
            \leq 4|p^{\bar u_2'\partial\calA}\cap p^{\bar u_1'\bar u_1}\cap V_\slodd(\calX)|.
        \end{equation}
    Notice that $p^{\bar u_2'\partial\calA}$ is a truncated version of the $\calX$-path
    $p^{\bar u_2'\bar u_2}$, which will work in our favour.

    The right-hand side of the previous display is a random variable,
    and our objective is to bound its second moment with a universal constant.

    Define the following geometric objects (illustrated in Figure~\ref{fig:mixing-1}):
    \begin{equation}
        S:=\bbox{\tfrac1\delta};
        \qquad
        A:=\ann{\tfrac4\delta}{\tfrac1\delta};
        \qquad
        R^1:=[\tfrac1{2\delta},\tfrac1\delta]\times[-\tfrac4\delta,\tfrac4\delta],
    \end{equation}
    and $R^2$, $R^3$, $R^4$ the rotations of $R^1$ by the angles $\pi/2$, $\pi$, and $3\pi/2$ around $(0,0)$.

\begin{figure}
    \centering
    \includegraphics{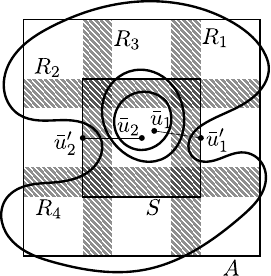}
    \caption{For Lemma~\ref{lem:Bound2}, each loop which surrounds exactly one point of each pair,
    must either be contained in the inner boundary of the annulus (the small square),
    or cross one of the four rectangles in the easy
    direction, or realise an arm event for the annulus.
    The figure contains an example for each of these three cases.}
    \label{fig:mixing-1}
\end{figure}

    It can be verified
    (by analysing Figure~\ref{fig:mixing-1})
    that any self-avoiding loop $\gamma\subset\R^2$
    surrounding exactly one point of the pair $\{\bar u_1,\bar u_1'\}$ and exactly one of $\{\bar u_2,\bar u_2'\}$ satisfies one of the following three properties:
    \begin{enumerate}[label=\arabic*.]
        \item $\gamma\subset S$,
        \item $\gamma$ crosses $R^1$ or $R^3$ horizontally, or $R^2$ or $R^4$ vertically,
        \item $\gamma$ realises an arm event for $A$.
    \end{enumerate}

    To upper bound Equation~\eqref{eq:bound2mixing},
    we observe that each $\omega$-connected component
    in $p^{\bar u_2'\partial\calA}\cap p^{\bar u_1'\bar u_1}\cap V_\sleven(\calX)$
    must contribute to at least one of the three cases above.
    In fact, by definition of $\calA$, at most two such components can be contained in $S$.
    Thus, we get
    \begin{equation}
        |p^{\bar u_2'\partial\calA}\cap p^{\bar u_1'\bar u_1}\cap V_\slodd(\calX)|
        \leq Q,
        \end{equation}
  where
  \begin{equation}
       Q :=
        24+
        \left(\sum_{i=1,3}\COUNTHORI{\bar\omega}{R^i}\right)
        +
        \left(\sum_{i=2,4}\COUNTVERTI{\bar\omega}{R^i}\right)
        +
        \COUNTARM{\bar\omega}{A}.
    \end{equation}
    The term $24$ compensates for the potential components in $S$ (at most two),
    and a few other components that we may loose because of boundary effects when counting
    crossings
    (for example, in our definitions we always imposed that we start counting at the highest $\omega^+$-crossing, thus missing out on a potential
    higher $\omega^-$-crossing).

    Putting our bounds together and letting $\calD$ tend to $\Z^2$ yields
    \begin{equation}
        \left|\mu_{\Z^2}[
                H_1 H_2
            |\sigma_\sleven,\omega]\right|
            \leq 4Q.
    \end{equation}
    It suffices to prove that $4Q$ has a uniformly bounded second moment.
    This follows from Lemma~\ref{lemma:loglin}
    (for the tails of the number of rectangle crossings)
    and Lemma~\ref{lemma:armexponent_for_spins_linear}
    (for the tails of the number of arms).
\end{proof}

We are now in a position to prove Lemma~\ref{lem:Bound3}.

\begin{proof}[Proof of Lemma~\ref{lem:Bound3}]
    Lemmata~\ref{lem:Bound1} and \ref{lem:Bound2} imply that the expectations of the conditional expectations $P$ and
    $H_1 H_2$ are uniformly bounded (since the second moments are). The product of the expectations is therefore bounded and
    it suffices to bound the expectation of the product of the two conditional expectations.

    By arguing as before,
    we get
    \begin{equation}
        \left|\mu_{\calD}^+[
            P
            H_1 H_2
        |\calX]\right|
        \leq
        \sum_{\pi}\prod_{ij\in\pi} 4I(\bar u_i \bar u_i', \bar u_j \bar u_j'),
    \end{equation}
    where we abusively write $\bar u_2 := \partial\calA$.
    The sum runs over pairings of $\{1,\ldots,k\}$.
    It suffices to bound the term corresponding to each pairing $\pi$ separately.

 Suppose first that $\{1,2\}\not\in\pi$.
        Then, each pair may be ordered such that $i\neq 1,2$ for any $ij\in\pi$. This assumption enables us to
     bound each factor as in Lemma~\ref{lem:Bound1}, yielding
        \begin{equation}
            \lim_{\calD\nearrow\Z^2}\mu_{\calD}^+[\prod_{ij\in\pi}4I(\bar u_i \bar u_i', \bar u_j \bar u_j')]
            \leq
            \mu_{\Z^2}[\prod_{ij\in\pi}16 K_{(1/\delta),(\epsilon/\delta),\bar u_i}].
        \end{equation}
        This leads to the desired uniform upper bound via Lemma~\ref{lemma:armexponent_for_spins_linear}.

Suppose now that $\{1,2\}\in\pi$.
        Then, by a reasoning similar to the previous proof, we get
        \begin{equation}
            \lim_{\calD\nearrow\Z^2}\mu_{\calD}^+[\prod_{ij\in\pi}4I(\bar u_i \bar u_i', \bar u_j \bar u_j')]
            \leq \mu_{\Z^2}[4Q\prod_{ij\in\pi\setminus\{\{1,2\}\}}16 K_{(1/\delta),(\epsilon/\delta),\bar u_i}],
        \end{equation}
        which is uniformly bounded by applying again our bounds for arms and crossings.
\end{proof}

It remains to bound the last term in Equation~\eqref{eq:targetformixing}.
This proof is quite different.

% <input src="sections/PART_D/20b_cov.tex" root="." version="0.0.1">
\begin{lemma}
    \label{lem:covariance_bounded_variance}
    In the same context (of Equation~\eqref{eq:targetformixing}),
    there exists some universal constant $C>0$ (depending only on $\c$ and $k$)
    such that
    \begin{equation}\label{eq:h40}
        |
        \Var_{\mu_{\Z^2}}[\mu_{\Z^2}[\tilde H_1 \tilde H_2|\calA]]|\leq C.
    \end{equation}
\end{lemma}

\begin{proof}
    It follows immediately from the definitions and from
    the Markov property over $\calA$ that
    \begin{equation}
        \mu_{\Z^2}[\tilde H_1 \tilde H_2|\calA]
        =
        X_{\bar u_1,\bar u_2}(\calA);
     \end{equation}
     where
     \begin{equation}
        X_{v_1,v_2}(\calA)
        :=
        \begin{cases}
            \mu_{\calA}^+[h(v_1)h(v_2)] &\text{if $F(\calA)\ni v_1,v_2$,}\\
            0 &\text{otherwise.}
        \end{cases}
    \end{equation}
Write $X:=X_{v_1,v_2}$.
   Note that $X$ is an increasing function of the domain.
    Indeed, $X(\calD)$ can be expressed as the expectation of an increasing function of the branching function $\psi$
    (see Lemma~\ref{lem:branching_function_basic_properties}),
    and the branching function is stochastically increasing in the choice of
    the domain (Lemma~\ref{lem:covariance_mono}).

    To prove the lemma, we are going to prove the following stronger statement.
    There exists a constant $C<\infty$ (depending only on $\c$)
    with the following properties.
    Let $B\subset\R^2$ denote any bounded simply connected set,
    and fix any two faces $v_1,v_2\in F(\Z^2)$
    (say $v_2$ is the face at $(0,0)$ without loss of generality).
    Let $\calB$ denote the largest even domain such that
    $\partial\calB\subset B$ and such that $v_2\in F(\calB)$
    (if $v_2$ exists), and set $\calB=\emptyset$ otherwise.
    Then, we claim that
    \begin{equation}
        \label{eq:variance_bound_to_prove}
        \Var_{\mu_{\Z^2}}[X_{v_1,v_2}(\calB)]\le C.
    \end{equation}
The monotonicity in the domain gives that \eqref{eq:variance_bound_to_prove} implies \eqref{eq:h40}.
 We are now going to prove Equation~\eqref{eq:variance_bound_to_prove}.

Let $\calD_n$ be the largest even domain contained in $\bbox{e^n}$.
    We claim that the increasing function
    \begin{equation}
        q:\Z_{\geq 1}\to\R,\,n\mapsto X(\calD_n)
    \end{equation}
    is Lipschitz with a uniform Lipschitz constant.
    Since $q_n\leq 8^n$,
    it suffices to prove that the increments
    $q_{n+1}-q_n$ are bounded (uniformly in all input data) for all $n\geq 100$.

    Consider the measure $\mu_{\calD_{n+1}}^+$.
    Let $\calB$ denote the largest even domain containing $v_2$
    and such that $\partial\calB\subset\omega^-$.
    By the circuit estimate, there exists some uniform probability $p>0$ such that,
    with probability at least $p$,
    the even domain $\calD_{n}$ contains an $\omega^+$-circuit which
    surrounds $(0,0)$ and is connected to $\partial\calD_{n+1}$.
    By inclusion, the event $\{\calB\subset\calD_n\}$ has probability at least $p$ as well.
    By recursion (Lemma~\ref{lem:covariance_recursion}) and monotonicity (Lemma~\ref{lem:covariance_mono}),
    we get
    \begin{equation}
        q_{n+1}\leq 4+\mu_{\calD_{n+1}}^+[X(\calB)]
        \leq
        4 + p X(\calD_n) + (1-p)X(\calD_{n+1})
        =
        4 + p q_n + (1-p) q_{n+1}.
    \end{equation}
    This implies the desired uniform Lipschitz bound $q_{n+1}-q_n\leq 4/p$.

    Finally, define
    \begin{equation}
        \operatorname{Scale}(\calB):=0\vee\sup\{n\in\Z_{\geq 1}:\calD_n\subset\calB\}.
    \end{equation}
    To finish the proof, it suffices to prove that:
    \begin{itemize}
        \item The variance of $\operatorname{Scale}(\calB)$ is uniformly bounded, and
        \item $|X(\calB)-q_{\operatorname{Scale}(\calB)}|$ is uniformly bounded.
    \end{itemize}

    Lemma~\ref{lemma:armexponent_for_spins_linear} implies that the second moment of $\operatorname{Scale}(B)-\operatorname{Scale}(\calB)$
    is uniformly bounded (by exploring from the outside towards the inside, we find the desired $\omega^+$-circuit with a uniformly positive probability at each scale, independently of the past).
     This implies that the variance of $\operatorname{Scale}(\calB)$ is also uniformly bounded.

    It suffices to prove the second statement. Since $X$ is an increasing function, it suffices to uniformly upper bound
    $X(\calB)-X(\calD_{\operatorname{Scale}(\calB)})$.
    By arguing as for the Lipschitz property above, we may find some uniform $p'>0$ such that
    \begin{equation}
        \mu_{\calB}^+\left[\left\{\parbox{20em}{there is some $\omega^-\cap\calD_{\operatorname{Scale}(\calB)}$-circuit around $v_2$ that is also connected to $\partial\calB$}\right\}\right]\geq p'.
    \end{equation}
    Analogously to what we did above, we deduce that
    \begin{equation}
        X(\calB) \leq 4 + p' X(\calD_{\operatorname{Scale}(\calB)}) + (1-p')X(\calB)
        \qquad\text{and}\qquad
        X(\calB)-X(\calD_{\operatorname{Scale}(\calB)})\leq 4/p'.
    \end{equation}
    This concludes the proof.
\end{proof}

% </input src="sections/PART_D/20b_cov.tex" root="." version="0.0.1">

% </input src="sections/PART_D/20_mixing.tex" root="." version="0.0.1">

% </input src="sections/PART_D/main.tex" root="." version="0.0.1">
% <input src="sections/PART_E/main.tex" root="." version="0.0.1">
\part{\INGREDSCALE: Glimpse of scale invariance}\label{part:glimpse}

Theorem~\ref{thm:computation_deriv_f} is proved in Section~\ref{sec:computation_deriv_f};
Theorem~\ref{thm:glimpse_scale_invariance} in Sections~\ref{sec:overview_glimpse}--\ref{sec:ldp_split}.

% <input src="sections/PART_E/21_computation.tex" root="." version="0.0.1">
\section{Free energy second derivative (Theorem~\ref{thm:computation_deriv_f})}
\label{sec:computation_deriv_f}

It was proved in \cite[Theorem~2]{Duminil-CopinKozlowskiKrachun_2022_SixVertexModelsFree} that the free
energy $f$ is twice differentiable at $0$ whenever  $\c \in [0,2]$,
with the value of the derivative depending on the parameter $\c$.
Moreover, in that article, the second derivative $f''(0)$ is characterised, but not explicitly computed.
We give the explicit computation below.
Theorem~\ref{thm:computation_deriv_f} is an immediate consequence of Proposition~\ref{prop:f''Q} and Lemma~\ref{lem:WienerHopf} below.

The statements in this section use another parametrisation of the six-vertex
model, namely the one given in Equation~\eqref{eq:parametrization}.
Recall that this amounts to $\Delta=-\cos\zeta$.

In this section, we use the convention that the Fourier transform of any $F\in L^1(\R)$, denoted $\widehat F$, where $\widehat F:\R\to\C$ is defined by $t\mapsto\int_\R e^{-\i t x}F(x)\diff x$.

\subsection{Reduction to a Wiener--Hopf equation}

The proposition below is essentially in~\cite{Duminil-CopinKozlowskiKrachun_2022_SixVertexModelsFree},
except that it is not written there as such.
Below, we shall describe how to derive the proposition from the (intermediate)
results stated in~\cite{Duminil-CopinKozlowskiKrachun_2022_SixVertexModelsFree}.

\begin{proposition}[\cite{Duminil-CopinKozlowskiKrachun_2022_SixVertexModelsFree}]
    \label{prop:f''Q}
    For any fixed $\zeta\in[0,2\pi/3]$,
    let $T:\R \to \R$ be the unique $L^1(\R)$-solution of
    \begin{align}\label{eq:WienerHopf}
    	T(x) -\int_0^\infty  R(x-y)T(y)dy = \mathfrak{e}(x),
    \end{align}
    where $\frake$ and $R$ are given by Table~\ref{tab:key_quantities}
    (in particular, $R$ is defined via its Fourier transform).
    Then
    \begin{align}\label{eq:f''Q}
    f''(0) = -  2 \frac{\displaystyle \int_0^\infty \frac{\frake(x)}{\frake(0)} T(x)\diff x}{\displaystyle\left(\frac{ \pi }{ \pi - \zeta } \int_{0}^{\infty} T(x)\diff x\right)^2}
            \qquad\text{where}\qquad
        \frac{\frake(x)}{\frake(0)}=\begin{cases}
            e^{-\frac{\pi}{\zeta}x} &\text{if $\zeta>0$,}\\
            e^{-\pi x} &\text{if $\zeta=0$.}
        \end{cases}
    \end{align}
\end{proposition}

\begin{table}
\centering
\begin{tabular}{@{}lll@{}}
\toprule
 & $\zeta = 0$ & $\zeta \in (0,\frac23\pi]$ \\
 \toprule
$\widehat R(t)$ &
$\frac{e^{-\frac{|t|}{2}}}{2\cosh(\frac{t}{2})}$ &
$\frac{\sinh((\pi - 2 \zeta) \tfrac{t}{2})}{2 \sinh((\pi - \zeta) \tfrac{t}{2}) \cosh(\zeta \tfrac{t}{2})}$ \\\midrule
$1-\widehat R(t)$ &
$\frac{e^{\frac{|t|}{2}}}{2\cosh(\frac{t}{2})}$ &
$\frac{\sinh(\pi \tfrac{t}{2})}{2\sinh((\pi-\zeta)\tfrac{t}{2})\cosh(\zeta \tfrac{t}{2})} $ \\
\midrule
$\mathfrak{e}(x)$ &
$e^{-\pi x} \indi{\mathbb{R}_{\geq 0}}(x)$ &
$\frac{1}{\zeta} e^{- \frac{\pi}{\zeta}x} \indi{\mathbb{R}_{\geq 0}}(x)$ \\
\midrule
$\widehat{\mathfrak{e}}(t)$ &
$\frac{1}{\pi + i t}$ &
$\frac{1}{\pi + i \zeta t}$ \\
\midrule
$\alpha(t)$, $t \in \mathbb{H}^+$ &
$
\left(\frac 1e \frac{-\i t}{2\pi}\right)^{\frac{-\i t}{2\pi}}
\frac{{\sqrt{2\pi}}}{\Gamma(\frac12 - \frac{\i t}{2\pi})}
$ &
$
        ( 1-\tfrac{\zeta}{\pi} )^{ -\i (1-\frac{\zeta}{\pi}) \frac{t}{2}} \cdot ( \tfrac{\zeta}{\pi} )^{ -\i \frac{  \zeta}{\pi} \frac{t}{2} }
        \frac{\Gamma(1-\i \frac{t}2)}{\Gamma(1-\i(1-\frac\zeta\pi)\frac{t}2)}
        \frac{\sqrt{2(\pi-\zeta)}}{\Gamma(\tfrac12-\i\tfrac\zeta\pi\tfrac{t}2)}$ \\
\midrule
$\alpha_+(0)^2$ &
$2=2\frac{\pi-\zeta}\pi$ &
$2\frac{\pi-\zeta}{\pi}$
         \\
\bottomrule
\end{tabular}
\caption{The summary of key quantities for the cases $\zeta = 0$ and $\zeta > 0$. The letter $\Gamma$ refers to the Euler gamma function.}
\label{tab:key_quantities}
\end{table}

\begin{remark}
    Basic properties of the integral equation were discussed in~\cite[Propositions~25 and~27]{Duminil-CopinKozlowskiKrachun_2022_SixVertexModelsFree}.
    Note that the integral kernel $R$ has constant sign and integrates to $\widehat R(0)\in[-1/2,1/2]$. By Young's convolution inequality,
    the integral equation~\eqref{eq:WienerHopf}
    is contractive and therefore has a unique $L^1(\R)$-solution
    which may be written $T=\sum_{k=0}^\infty \frakR^k(\frake)$
    where $\frakR(\fraka):=R* (\indi{\R_{\geq 0}}\fraka)$.

    Moreover, if $\zeta>0$, then the integral kernel $R$ has exponentially decaying tails,
    and so does the solution $T$.
    The case $\zeta=0$ is similar;  the integral kernel $R$ and $T$ have tails of order $O(x^{-2})$.

     We stress that to solve  \eqref{eq:WienerHopf}, one first restricts it to $\mathbb{R}_{\geq 0}$, solves it on $L^1(\R_{\geq 0})$ and then
    uses the relation $T(x) =\int_0^\infty  R(x-y)T(y)dy $ to extend $T$ to $\mathbb{R}_{< 0}$.
\end{remark}

\begin{proof}
%
%    \cite{Duminil-CopinKozlowskiKrachun_2022_SixVertexModelsFree}. For the precise statement, see \cite[eqn.~(8)]{Duminil-CopinKozlowskiKrachun_2022_SixVertexModelsFree}
%with $\theta = \pi/2$ and $n = (1+  \alpha) N/2$.
%
%By diving into the proof of \cite[Thm.~2]{Duminil-CopinKozlowskiKrachun_2022_SixVertexModelsFree} one may obtain the following expression for $f''(0)$ in terms of the integral of some implicit function $T$.
%
%The proof is fully contained in \cite{Duminil-CopinKozlowskiKrachun_2022_SixVertexModelsFree}, but may be difficult to piece together. We describe below the necessary steps.
%For the existence and uniqueness of the function $T$, see \cite{Duminil-CopinKozlowskiKrachun_2022_SixVertexModelsFree} and references therein.

First, \cite[eqn.~(60)]{Duminil-CopinKozlowskiKrachun_2022_SixVertexModelsFree} defines a function $\delta f$ which it relates to $f''(0)$ by
\begin{align}\label{eq:f''Q1}
	f''(0) = - \lim_{\alpha \searrow 0} \frac{2}{\alpha^2} \cdot \delta f\big(Q\big(\tfrac{1-\alpha}{2}\big)\big).
\end{align}
The function $Q$ is defined in \cite[eqn.~(16)]{Duminil-CopinKozlowskiKrachun_2022_SixVertexModelsFree}, but is not important to us.
The asymptotics of $f\big(Q\big(\frac{1-\alpha}2\big)\big)$ as $\alpha \searrow 0$ are then computed in \cite[Sec.~7.2]{Duminil-CopinKozlowskiKrachun_2022_SixVertexModelsFree} for $0\leq c <2$ and \cite[Sec.~7.3]{Duminil-CopinKozlowskiKrachun_2022_SixVertexModelsFree} for $c = 2$.
Combining them with \eqref{eq:f''Q1} we obtain
\begin{align}\label{eq:f''Q2}
f''(0) =
-  \frac{2}{C_\Delta^2}{ \int_0^\infty  \frac{\frake(x)}{\frake(0)} T(x)dx},
\end{align}
with $C_\Delta$ defined at the start of \cite[Sec.~7]{Duminil-CopinKozlowskiKrachun_2022_SixVertexModelsFree}.
The value of $C_\Delta$  is computed in~\cite[Prop. 25]{Duminil-CopinKozlowskiKrachun_2022_SixVertexModelsFree} as
\begin{align}
C_\Delta =  \frac{ \pi }{ \pi - \zeta } \int_{0}^{+\infty} T(\lambda)d \lambda.
\end{align}
Inserting this into \eqref{eq:f''Q2}, we obtain \eqref{eq:f''Q}. % The line of thought, with minor details changing, is similar at $\zeta=0$.
\end{proof}

The purpose of the next sections is to compute the ratio of the integrals in Equation~\eqref{eq:f''Q}, or equivalently to derive the following result. Together with the previous proposition,
it implies Theorem~\ref{thm:computation_deriv_f}.

\begin{lemma}\label{lem:WienerHopf}
	For any $\zeta\in[0,2\pi/3]$ (corresponding to $\c \in [1,2]$), we have
	\begin{align}\label{eq:WienerHopf_solution}
	 \frac{\displaystyle \int_0^\infty \frac{\frake(x)}{\frake(0)} T(x)\diff x}{\displaystyle\left(\int_{0}^{\infty} T(x)\diff  x\right)^2} = \frac{\pi^2}{4 ( \pi - \zeta)}.
	\end{align}
\end{lemma}

We proceed in several steps. First, we rephrase the Wiener--Hopf equation in Fourier space. Then, we express $T$ in terms of the solutions of a well-chosen Riemann-Hilbert problem. This new expression enables an explicit calculation of the ratio in Equation~\eqref{eq:f''Q}.

\subsection{Fourier transform of the Wiener--Hopf equation}

Equation~\eqref{eq:WienerHopf} is an integral equation
that can be solved explicitly via the Wiener--Hopf method \cite{HopfWienerMethodWienerHopfEqns}, see \cite[7.4.1]{AblowitzFokas_2003_ComplexVariablesIntroduction}
for a modern exposition.
Let $T=T_\uparrow-T_\downarrow$
where $T_\uparrow:=\indi{\R_{\geq 0}}T$ and $T_\downarrow:=-\indi{\R_{<0}}T$.
The Fourier transforms of $T_\uparrow$ and $T_\downarrow$
then extend to holomorphic functions on the lower and upper half plane, respectively.
Equation~\eqref{eq:WienerHopf} may then be written in Fourier space,
leading to a Riemann--Hilbert problem for the corresponding holomorphic extensions.
This Riemann--Hilbert problem can be solved uniquely and explicitly,
and the solution for $T$ is given by the inverse Fourier transform.
Once this is done, it is straightforward to calculate the ratio of the two integrals in \eqref{eq:f''Q}.

We now implement this strategy. Let $\H^\pm\subset\C$ denote the open upper and lower half plane.
Let $\calO(\C\setminus\R)$ denote the set of holomorphic functions $\varphi$ on $\C\setminus\R$,
such that $\varphi|_{\H^+}$ has a continuous extension to $\bar \H^+$,
and such that $\varphi|_{\H^-}$ has a continuous extension to $\bar \H^-$.
For any such $\varphi\in\calO(\C\setminus\R)$,
we define the continuous functions $\varphi_+$ and $\varphi_-$ on $\R$ via
\begin{equation}
    \varphi_\pm(t):=\lim_{\lambda\to 0^+}\varphi(t\pm \i\lambda).
\end{equation}
In the context of our Wiener--Hopf equation,
define
\begin{equation}
    \tilde T:\C\setminus\R,\,t\mapsto\begin{cases}
        \tilde T_\uparrow(t):=\int_\R e^{-\i t x}T_\uparrow(x)\diff x &\text{if $t\in\H^-$,}\\
        \tilde T_\downarrow(t):=\int_\R e^{-\i t x}T_\downarrow(x)\diff x &\text{if $t\in\H^+$.}
    \end{cases}
\end{equation}
Notice that $\tilde T\in\calO(\C\setminus\R)$ by the dominated convergence theorem,
 $\lim_{|t|\to\infty}\tilde T(t)=0$ by integration by parts,
 and $\tilde T_-=\widehat T_\uparrow$ and $\tilde T_+=\widehat T_\downarrow$ by the definition of the Fourier transform.

In Fourier space, Equation~\eqref{eq:WienerHopf}
is written as
\begin{equation}
    \label{eq:Wiener_Hopf_equation_1}
    (1-\widehat R)\widehat T_\uparrow-\widehat T_\downarrow =
    (1-\widehat R)\tilde T_- - \tilde T_+
    = \widehat \frake.
\end{equation}
We shall see below that this equation admits a unique solution $\tilde T\in\calO(\C\setminus\R)$ that tends to $0$
as $|t|\to\infty$.
We explicitly solve it via two Riemann--Hilbert problems:
we first solve a variant of the equation without the driving term $\frake$,
and then we use our first solution to solve a second Riemann--Hilbert problem
which incorporates the driving term.

\subsection{Two Riemann--Hilbert problems}

Branch cuts will start to play a role.
We view the Gamma function $\Gamma$ involved in the definition of $\alpha$ as a holomorphic function on the right half plane $\{t\in\C:\Re(t)>0\}$,
 and we view the function $t\mapsto t^t$ as a holomorphic function on the set $\C\setminus \R_{\leq 0}$.
Recall that $t\mapsto t^t$ extends continuously to the point $t=0$, where it takes the value $1$.

 Define $\alpha\in\calO(\C\setminus\R)$ as follows:
    \begin{equation}
        \alpha(t):=\begin{cases}
            \text{the value given by Table~\ref{tab:key_quantities}} &\text{if $t\in\H^+$,}\\
            1/\alpha(-t) &\text{if $t\in\H^-$.}
        \end{cases}
    \end{equation}

\begin{lemma}[First Riemann-Hilbert problem]
    \label{lem:first_Riemann_Hilbert_problem}
    The function $\alpha$ satisfies the following properties:
    \begin{enumerate}
        \item The functions $\alpha$ and $\alpha_\pm$ do not vanish,
        \item The function $\alpha$ satisfies $\lim_{|t|\to\infty}\alpha(t)=1$,
        \item We have $\alpha_-/\alpha_+=1-\widehat R$.
    \end{enumerate}
\end{lemma}

\begin{proof}
    We view $\Gamma$ as a holomorphic function that is non-vanishing on the right half plane,
    and that Stirling's approximation gives an estimate up to a factor of order $1+o(1)$
    as $|t|\to\infty$.
    The first two properties follow by basic manipulations.

    Finally, we must show that for any $t\in\R$,
    we have
    \begin{equation}\frac{\alpha_-(t)}{\alpha_+(t)}=\frac{1}{\alpha_+(-t)\alpha_+(t)}=(1-\widehat R)(t).\end{equation}
    This is an elementary consequence of the well-known identities
    \begin{equation}
        \Gamma(1+\i \lambda)\Gamma(1-\i \lambda)=\frac{\pi \lambda}{\sinh(\pi \lambda)};
        \qquad
        \Gamma(\tfrac12+\i \lambda)\Gamma(\tfrac12-\i \lambda)=\frac{\pi}{\cosh(\pi \lambda)},
    \end{equation}
    which are valid for any $\lambda\in\mathbb{C} \setminus \i \mathbb{Z}$.
    \commentPiet{I need to check this later}
\end{proof}

Lemma~\ref{lem:first_Riemann_Hilbert_problem} enables to rewrite \eqref{eq:Wiener_Hopf_equation_1} as
\begin{equation}
    \label{eq:Wiener_Hopf_equation_2}
%        \alpha_- \widehat T_\uparrow - \alpha_+ \widehat T_\downarrow
%        =
        \alpha_-\tilde T_- - \alpha_+ \tilde T_+
        =
        \alpha_+ \widehat\frake.
\end{equation}

\begin{lemma}
    \label{lemma:second_Riemann_Hilbert_problem}
    There exists a unique solution $G\in\calO(\C\setminus\R)$
    to the equation
    \begin{align}\label{eq:second_Riemann_Hilbert_problem}
        G_- - G_+ = \alpha_+ \widehat\frake
    \end{align}
    with $\lim_{|t|\to\infty}G(t)=0$. In particular, one has  $\tilde T=G/\alpha $ and $\widehat T_\uparrow(t)=\alpha_+(-t)G_-(t)$.
\end{lemma}

\begin{proof}
    Equation~\eqref{eq:Wiener_Hopf_equation_2} implies that $G=\alpha \tilde T$
    is a solution to Equation~\eqref{eq:second_Riemann_Hilbert_problem}.
    This solution also satisfies $\widehat T_\uparrow(t)=\tilde T_-(t)=G_-(t)/\alpha_-(t)=\alpha_+(-t)G_-(t)$.

    It suffices to prove uniqueness of the solution $G$.
    Let $G'\in\calO(\C\setminus\R)$ denote another solution.
        Then $D:=G-G'\in\calO(\C\setminus\R)$ satisfies
        \begin{equation}
            D_- - D_+ = 0
            \qquad\text{and}\qquad
            \lim_{|t|\to\infty}D(t)=0.
        \end{equation}
        The first relation implies that $D$ extends continuously to $\C$ and
        by Morera's theorem this extension is an entire function.
        The limit at infinity being 0, we get $D = 0$ and $G = G'$ by the second relation.
\end{proof}

\begin{lemma}\label{lem:second_Riemann_Hilbert_problem_solution}
    We have $\widehat T_\uparrow=\tilde T_-$.
   The functions
    $\widehat T_\uparrow=\tilde T_-$, and  $\tilde T|_{\H^-}$
    take the explicit form
    \begin{equation}
        \widehat T_\uparrow(t)
        =
        \alpha_+(-t)\alpha(t_\zeta)\widehat\frake(t);
        \qquad
        \tilde T_\uparrow|_{\H^-}(t)
        =
        \alpha(-t)\alpha(t_\zeta)\widehat\frake(t).
    \end{equation}
\end{lemma}

\begin{proof}
   The unique solution $G\in\calO(\C\setminus\R)$ of Equation~\eqref{eq:second_Riemann_Hilbert_problem} is given by
    \begin{align}\label{eq:second_Riemann_Hilbert_problem_solution}
        G(t):=
        \big(\alpha(t_\zeta)-\alpha(t)\indi{\H^+}(t)\big)\widehat\frake(t)
        \qquad
        \text{where}
        \qquad
        t_\zeta:=\begin{cases}
            \i\pi/\zeta &\text{if $\zeta>0$,}\\
            \i\pi &\text{if $\zeta=0$,}
        \end{cases}
    \end{align}
    is the unique pole of $\widehat\frake$ in $\C$.
    Indeed, the function $G$ defined in \eqref{eq:second_Riemann_Hilbert_problem_solution} is clearly holomorphic and has the desired limit at infinity.
    The jump condition is straightforwardly verified.

    The expressions for $\widehat T_\uparrow$ and $\tilde T_\uparrow|_{\H^-}$
    now follow from the previous lemma.
\end{proof}

\subsection{Proof of Lemma~\ref{lem:WienerHopf}}

We are now in a position to prove Lemma~\ref{lem:WienerHopf}.
\begin{proof}[Proof of Lemma~\ref{lem:WienerHopf}]
Recall that $\tilde T|_{\H^-}$ is a holomorphic function
    that extends continuously to the real line, where it equals $\widehat T_\uparrow$. As a consequence, using Lemma~\ref{lem:second_Riemann_Hilbert_problem_solution} in the third equality gives
    \begin{equation}
        \int_0^\infty T(x)\diff x
        =
        \int_\R T_\uparrow(x)\diff x
        =
        \widehat T_\uparrow(0)=\alpha_+(0)\alpha(t_\zeta)\widehat\frake(0)
        =
        \tfrac1\pi\alpha_+(0)\alpha(t_\zeta).
    \end{equation}

The lemma therefore follows from the value of $\alpha_+(0)$ given in Table~\ref{tab:key_quantities} and the identity
\begin{equation}\label{eq:ah}
\int_0^\infty \frake(x) T(x) \diff x =
        \frac{|t_\zeta|}{2\pi^2} \alpha(t_\zeta)^2.
        \end{equation}

    To prove the latter, observe first that $T_\uparrow$ and $\frake$ belong to $L^1(\R)\cap L^2(\R)$.
   Indeed, we already know that $T\in L^1(\R)$.
    Since $T_\uparrow$ is also bounded, we get $T_\uparrow\in L^2(\R)$.
    For $\frake$ the claim is obvious.

   Plancherel's theorem can therefore be used to obtain
    \begin{equation}
        \int_0^\infty \frake(x) T(x) \diff x
        =
        \int_\R \frake(x) T_\uparrow(x) \diff x
        =
        \frac1{2\pi}\int_\R \widehat\frake(-t)\widehat T_\uparrow(t) \diff t.
    \end{equation}
    Now, $\tilde T|_{\H^-}$ is holomorphic and extends continuously to $\R$, where it equals $\widehat T_\uparrow$. Cauchy's theorem implies that for some small
    $\epsilon>0$,
    \begin{equation}
        \int_0^\infty \frake(x) T(x) \diff x = \frac1{2\pi}\int_{\R-\epsilon\i} \widehat\frake(-t)\tilde T(t) \diff t.
    \end{equation}
    (We use that the integrand is  $O(1/t^2)$ as $|t|\to\infty$.)
    Cauchy's integral formula (again using the $O(1/t^2)$ decay) gives
        \begin{equation}
        \int_0^\infty \frake(x) T(x) \diff x = \frac{|t_\zeta|}{\pi} \tilde T(-t_\zeta)
        = \frac{|t_\zeta|}{\pi} \alpha(t_\zeta)^2\widehat\frake(-t_\zeta)
        =
        \frac{|t_\zeta|}{2\pi^2} \alpha(t_\zeta)^2.
    \end{equation}
    This is exactly \eqref{eq:ah}, so that the proof is completed.
\end{proof}

% </input src="sections/PART_E/21_computation.tex" root="." version="0.0.1">
% <input src="sections/PART_E/22_GFF_LDP.tex" root="." version="0.0.1">
\section{Proof of Theorem~\ref{thm:glimpse_scale_invariance}}
\label{sec:overview_glimpse}

This section establishes Theorem~\ref{thm:glimpse_scale_invariance}.
The proof relies on \INGREDREGULARITY\ (Part~\ref{part:qualitative}).
In particular, we use our bound on arm exponents (Theorem~\ref{thm:intro:arm_exponents})
and
flip domination (Theorem~\ref{thm:intro:flip_domination}).
Finally, Proposition~\ref{prop:split_into_ridge_events} plays an important
role -- it is stated in the current section, but its proof is deferred to Section~\ref{sec:ldp_split}.

The proof is structured as follows.
We first state a preparatory identity, linking the free energy
to a large deviations principle.
Then, we present an intuitive version of our strategy, which has some clear flaws.
This guides our rigorous proof, which consists of three steps.
The first step entails Proposition~\ref{prop:split_into_ridge_events},
which is thus deferred to Section~\ref{sec:ldp_split}.

\subsection{Preparatory identity}

Let us first introduce the notion of \emph{constant even height circuit} of $h$. It is a circuit $\gamma$ of edges in $E_\sleven$ with the property that the height difference between the endpoints is equal to zero.  By definition, all faces corresponding to endpoints in $\gamma$ have the same height, which we denote by $h(\gamma)$. We view each edge of the circuit as a line segment in $\R^2$ connecting the centres of the two corresponding faces;
so that the circuit itself can be seen as a random subset of $\R^2$ in this way.

Consider a \emph{topological annulus}, that is, a subset $A\subset\R^2$
that is bounded and topologically equivalent to an annulus.
For any topological annulus $A$ and any $k>0$, define
\begin{equation}
    \pCIRCUIT{+k}{A} :=\left\{\parbox{26em}{$A$ contains two even height circuits $\gamma_0$, $\gamma_+$ of $h$
    with nontrivial winding in $A$,
    such that $\gamma_0$ is the outermost even height circuit included in $A$,
    and such that $h(\gamma_+)-h(\gamma_0)\geq k$}\right\}.
\end{equation}
Notice that this event is measurable in terms of the gradient of $h$,
that is, it has a well-defined $\P_{\Z^2}$-probability.
We shall start by giving an alternative way of characterising $f''(0)$, where $f$ is the free energy.

\begin{proposition}[Variational principle]
    \label{prop:ZtoR_globalNEW}
        For any $\alpha\in(0,1)$, we have
        \begin{equation}
            \label{prop:ZtoR_global:char_for_sigma}
            \lim_{\rho\to\infty}\lim_{L\to\infty}
            \frac{1}{4\rho L^2}\log
            \P_{\Z^2}[\pCIRCUIT{+\alpha L}{
                A_{\rho,L}
            }]=f(\alpha)-f(0),
        \end{equation}
        where $A_{\rho,L}:=\symRect{(\rho+1) L}{L}\setminus\symRect{\rho L}{0}$, while $\symRect{a}{b}$is as defined in Equation~\eqref{Definition box}.
\end{proposition}

Proposition~\ref{prop:ZtoR_globalNEW} can be understood as a \emph{variational principle}.
It tells us that the likelihood of height function deviations of order $L$
in a domain of size $L$ can be expressed in terms of the free energy functional
(which provides the entropy of the system at different slopes).
For height functions, such a principle was first established in
the work of Cohn, Kenyon, and Propp in the context of the domino tiling model~\cite{CohnKenyonPropp_2001_VariationalPrincipleDomino}
(the domino tiling model is integrable, but the proof of the variational principle does not use the integrable structure).
All the essential ingredients for the proof of Proposition~\ref{prop:ZtoR_globalNEW} are already present
in~\cite{CohnKenyonPropp_2001_VariationalPrincipleDomino}.
The variational principle was later proved in a general finite-range setting (including the six-vertex model)
in a work of Sheffield~\cite{Sheffield_2005_RandomSurfaces}.
Finally, we mention~\cite{LammersTassy_2024_MacroscopicBehaviorLipschitz}, which also establishes the variational principle
for height functions which are potentially infinite-range (also including the six-vertex model).
A proof of Proposition~\ref{prop:ZtoR_globalNEW} may thus be found in any of the references~\cite{CohnKenyonPropp_2001_VariationalPrincipleDomino,Sheffield_2005_RandomSurfaces,LammersTassy_2024_MacroscopicBehaviorLipschitz};
we do not reproduce it here.

\subsection{Naive strategy and its problems}
To explain our strategy for establishing Theorem~\ref{thm:glimpse_scale_invariance},
we first ``prove'' the result by using two assumptions that are sensible,
but which are not exactly true or, at least, which we cannot rigorously establish at this stage. Below, $\approx$ means that the events have the same probability up to the relevant precision.

Assume first that
\begin{equation}
       \P_{\Z^2}[\pCIRCUIT{+k}{
                A_{\rho,L}}]
        \approx
       \P_{\Z^2}\left[\left\{\parbox{22em}{the average of $h$ on the line segment $\symRect{\rho L}{0}$
        minus
        the average of $h$ on $\partial\symRect{(\rho+1)L}{L}$
        is at least $k$}\right\}\right].
    \end{equation}

If this assumption were true, then the GFF convergence (a hypothesis in Theorem~\ref{thm:glimpse_scale_invariance}) and a GFF calculation that can be made precise quite easily (we shall do this later in full detail) would imply
that for fixed $k$, we would have
\begin{multline}
    \lim_{L\to\infty}
            \P_{\Z^2}[\pCIRCUIT{+k}{
                A_{\rho,L}
    }]
    \\\approx
    \lim_{L\to\infty}
    \P_{\Z^2}\left[\left\{\parbox{22em}{the average of $h$ on the line segment $\symRect{\rho L}{0}$
        minus
        the average of $h$ on $\partial\symRect{(\rho+1)L}{L}$
        is at least $k$}\right\}\right]
    \approx e^{-\frac{(4\rho)k^2}{2\sigma^2}}.
    \label{eq:simplegffcalc}
\end{multline}

Let us now make a second assumption, which is even more optimistic: suppose that the previous formula is true even when
    we set $k=\alpha L$ rather than keeping $k$ fixed.
    Such an assumption is a big stretch, since GFF convergence covers events with a probability
    of order $1$, while setting $k=\alpha L$ means passing to events in the large deviation regime.

Such an assumption would allow us to combine Equations~\eqref{prop:ZtoR_global:char_for_sigma} and~\eqref{eq:simplegffcalc}
with $k=\alpha L$
to get
\begin{equation}
    f(\alpha)-f(0)
    \approx
    -\frac{\alpha^2}{2\sigma^2}.
\end{equation}
Since $f$ is twice differentiable at $\alpha=0$, this
is equivalent to Theorem~\ref{thm:glimpse_scale_invariance}.

In the remainder of this section, we describe a rigorous proof of Theorem~\ref{thm:glimpse_scale_invariance}.
It does not really establish the previous two assumptions,
but it is inspired by the same ideas.
Remark that the equality $\sigma^2=-1/f''(0)$ may be viewed as the combination of two inequalities;
unfortunately, we will have to treat the two inequalities separately in each step.

\subsection{Step~1. Splitting the large deviation event into ``independent''  events}

This step circumvents the problems with the second assumption.
We essentially split up the large deviation event which has a probability of order $e^{-cL^2}$
into $O(L^2)$ ``independent'' events which have a probability of order $e^{-c}$.

\begin{definition}[Straightened annulus]
    \label{def:straightened_annulus}
    An \emph{$\eta$-straightened path} is a path in $\R^2$ which is a union of line segments of the square lattice
    graph $\eta\Z^2$.
    A \emph{$(\rho,\eta)$-straightened annulus} is a topological annulus $A\subset\symInt{\rho+1}\times\R$
    such that its inner and outer boundaries are $\eta$-straightened paths,
    and such that the inner boundary crosses the vertical strip $[-\rho,\rho]\times\R$.
\end{definition}

\begin{proposition}[Splitting]
    \label{prop:split_into_ridge_events}
    Fix $\c\in[1,2]$.
    \begin{enumerate}
        \item \textbf{Straight bound.}
        For fixed $\rho,k,N\in1000\Z_{\geq 1}$ with $N\geq 2k$,
        \begin{equation}
        \log\P_{\Z^2}[\pCIRCUIT{+k}{
                A_{\rho,N}
            }]
            \leq  4\rho N^2\Big(f(\tfrac{k-24}{N})-f(0)\Big).
        \end{equation}
        In particular,
        \begin{equation}
            \limsup_{k\to\infty}
            \limsup_{N\to\infty}
            \tfrac{1}{4\rho k^2}\log
            \P_{\Z^2}[\pCIRCUIT{+k}{
                A_{\rho,N}
            }]
            \leq \frac12f''(0).
        \end{equation}

        \item \textbf{Curly bound.}
        For fixed $\rho\in2\Z_{\geq 1}$ and $\epsilon\in(0,1/100)$, there exist constants $\eta>0$ and $C,r=r(\rho,\epsilon)<\infty$
        such that, for sufficiently large $k\in 2\Z_{\geq 1}$
        and for sufficiently large $N\in 2\Z_{\geq 1}$
        (depending on $k$),
        we have
            \begin{multline}
                    \lim_{n\to\infty}
            \frac1n\log
           \Big( \max_{(A_i)\in\frakA(r,\rho,\eta,\epsilon,n)}
            \prod_i
            \P_{\Z^2}[\pCIRCUIT{+2\lceil(1-\epsilon)k\rceil}{NA_i}]\Big)
            \\
            \geq
             4(\rho+1)N^2\left(
                f(\tfrac{2(k+4)}{N})-f(0)
                \right)
                +
                200k^2\log\ccircuit -C,
        \end{multline}
        where $\frakA(r,\rho,\eta,\epsilon,n)$ is the set of all families of $\lceil(1-\epsilon)n\rceil$ disjoint $(\rho,\eta)$-straightened annuli $A_i$ for $1\leq i\leq \lceil(1-\epsilon)n\rceil$
        of diameter at most $2r=2r(\rho,\epsilon)$ that are contained in $ \symRect{(\rho+1)}{n}$.

             In particular,
                     \begin{multline}
                        \label{eq:curly_bound_second_deriv}
                        \liminf_{k\to\infty}
                        \liminf_{N\to\infty}
                    \lim_{n\to\infty}
            \frac1{4(\rho+1)nk^2}\log
         \Big(   \max_{(A_i)\in\frakA(r,\rho,\eta,\epsilon,n)}
            \prod_i
            \P_{\Z^2}[\pCIRCUIT{+2\lceil(1-\epsilon)k/2\rceil}{NA_i}]\Big)
            \\
            \geq
              \frac12 f''(0)
                +
                \frac{100}{\rho+1}\log\ccircuit.
        \end{multline}
           \end{enumerate}
\end{proposition}

The proof of Proposition~\ref{prop:split_into_ridge_events} is deferred to Section~\ref{sec:ldp_split}.
We now turn to the second part of the proof, that corresponds to a variation of the first assumption we made.

\subsection{Step~2. GFF counterparts of circuit events}

First, notice that our first assumption is not yet unambiguously defined,
since the notion of ``average'' requires the introduction of a probability measure,
which we did not explicitly do.
It turns out that, from the perspective of the GFF, there is a unique natural choice of
probability measures.
Let us describe this choice first.
In this subsection, we shall work in the generality of an arbitrary topological annulus $A\subset\R^2$.

For any topological annulus $A$, let $\partial_{\operatorname{ext}} A$  denote its exterior boundary,
and let $\partial_{\operatorname{int}} A$ denote its interior boundary.
Introduce the function
\begin{equation}
    H_A:\R^2\to[0,1],\,x\mapsto\P[\{\text{a Brownian motion started at $x$ hits $\partial_{\operatorname{int}} A$ before $\partial_{\operatorname{ext}} A$}\}].
\end{equation}
This is the unique bounded harmonic extension of
\commentKarol{I would suggest to provide a reference. In the end, I could figure out a proof, but mine was not that direct and
the result was new to me}
\begin{equation}
  \partial A\to\R,\,x\mapsto \begin{cases}
    1 &\text{if $x\in\partial_{\operatorname{int}} A$,}\\
    0 &\text{if $x\in\partial_{\operatorname{ext}} A$,}
  \end{cases}
\end{equation}
to $\R^2$.
Let $\nu_A$ denote the ``Laplacian measure'' associated to $A$:
it is defined as $\nu_A:=-\Delta H_A$,
and may also be characterised as the unique finite signed measure supported on $\partial A$
such that for any smooth compactly supported function $F:\R^2\to\R$,
we have \begin{equation}
    \label{eq:char_nuA}
    \int F(x)\diff \nu_A(x)=\int \nabla F(x)\cdot\nabla H_A(x)\diff x.
\end{equation}
Decompose $\nu_A:=\nu_A^+ -\nu_A^-$ where both terms on the right are (positive) finite measures.
By definition, $\nu_A^+$ is supported on $\partial_{\operatorname{int}} A$,
and $\nu_A^-$ is supported on $\partial_{\operatorname{ext}} A$.
Moreover, Equation~\eqref{eq:char_nuA} with $F=H_A$ shows that
\begin{equation}
    \nu_A^{+}(\partial_{\operatorname{int}} A)
    =\nu_A^{-}(\partial_{\operatorname{ext}} A)
    =\int |\nabla H_A(x)|^2\diff x=:\operatorname{Dirichlet}(H_A).
\end{equation}
Define
\begin{equation}
    \varphi_A:=\frac{1}{\operatorname{Dirichlet}(H_A)}\nu_A.
\end{equation}
This measure is normalised in the sense that it decomposes as the difference $\varphi_A^+ -\varphi_A^-$ of two probability measures
with disjoint support.
We are now ready to state the main result of this subsection.
Recall that a {\em convergence sequence} is a sequence $(\delta_n)_n$
tending to zero such that $h^{(\delta_n)}$ converges to a multiple of the GFF.

\begin{proposition}[Formal version of the first assumption]
    \label{prop:mollification_of_ridge_eventsSIMPLER}
   For $(\delta_n)_n$  a convergence sequence and $A$ a topological annulus,
    \begin{equation}
        \label{eq:formal_assumption_I}
            \lim_{k\to\infty}
            \lim_{n\to\infty}
            \frac1{k^2}
            \log
            \frac{
            \P_{\Z^2}[\{
                \langle h,\varphi_{A/\delta_n}\rangle
                \geq k
            \}]
            }{\P_{\Z^2}[\pCIRCUIT{+k}{
                A/\delta_n
            }]}
            = 0.
    \end{equation}
\end{proposition}

Here and below,  $\Gamma$ denotes a normalised GFF  on $\R^2$ (in the sense of Section~\ref{Sec:2.3}).
To prove the previous proposition, we first need some very basic information on the behaviour of the random variable
$\langle \Gamma,\varphi_A\rangle$.

\begin{lemma}[GFF analysis of $\langle\Gamma,\varphi_A\rangle$]
    \label{eq:lemma:ldp_gff_equivalence_useful}
       For every  topological annulus $A$,
    \begin{enumerate}
        \item The random variable $\langle\Gamma,\varphi_A\rangle$ has the distribution $\calN(0,1/\operatorname{Dirichlet}(H_A))$,
        \item We have
   \begin{equation}
   \lim_{k\to\infty}
            \frac1{k^2}
            \log
            \P[\{
                \langle \Gamma,\varphi_A\rangle
                \geq k
            \}]
            = -\frac12\operatorname{Dirichlet}(H_A),\end{equation}
        \item The Gaussian process $\Gamma$ may be decomposed as the sum
        \begin{equation}
            \Gamma = \Gamma^A + \langle\Gamma,\varphi_A\rangle H_A,
        \end{equation}
        where $\Gamma^A$ is a Gaussian process that is independent of $\langle\Gamma,\varphi_A\rangle$.
    \end{enumerate}
\end{lemma}

\begin{proof}
    Fix any finite Dirichlet energy generalised test function $\mu$.
    Since the Green function $G_{\R^2}$ is the inverse of the negative-Laplacian,
    we get
    \begin{align}
        \operatorname{Dirichlet}(H_A)\Cov[\langle\Gamma,\varphi_A\rangle,\langle\Gamma,\mu\rangle]
        &=
        \Cov[\langle\Gamma,\nu_A\rangle,\langle\Gamma,\mu\rangle]
        \\
        &=
        \int G_{\R^2}(x,y)\diffi \nu_A(x)\diffi\mu(y)
        \\
        &=
        \int \diffi\mu(y) ((-\Delta)^{-1}(-\Delta) H_A)(y)
        \\
        &=
        \int H_A(y)\diffi\mu(y).
    \end{align}
    In particular, we get
    \begin{align}
        \Cov[\langle\Gamma,\varphi_A\rangle,\langle\Gamma,\varphi_A\rangle]
        &\stackrel{\phantom{\eqref{eq:char_nuA}}}=
        \frac1{\operatorname{Dirichlet}(H_A)^2}
        \int H_A(y)\diffi\nu_A(y)\\
        &\stackrel{\eqref{eq:char_nuA}}=\frac1{\operatorname{Dirichlet}(H_A)^2}\int |\nabla H_A(y)|^2dy
        \\
        &\stackrel{\phantom{\eqref{eq:char_nuA}}}=
        \frac1{\operatorname{Dirichlet}(H_A)};
    \label{eq:cov_varphiA}
    \end{align}
        and similarly
        \begin{align}
              \Cov[\langle\Gamma,\varphi_A\rangle,\langle\Gamma,\delta_x-\varphi_A^-\rangle]
        &\stackrel{\phantom{\eqref{eq:char_nuA}}}=
        \frac1{\operatorname{Dirichlet}(H_A)}
                \int H_A(y)\diffi(\delta_x-\varphi_A^-)(y)\\
                &\stackrel{\phantom{\eqref{eq:char_nuA}}}=\frac{H_A(x)}{\operatorname{Dirichlet}(H_A)}.
                \label{eq:cov_varphiA_delta}
    \end{align}
    Item (i) follows from \eqref{eq:cov_varphiA}.
    Item (ii) follows from item (i).
    Let us now focus on (iii).
    Since $\Gamma$ is a centred Gaussian process,
    it must clearly have a decomposition of the form
    \(
        \Gamma = \Gamma^A + \langle\Gamma,\varphi_A\rangle F
    \).
    It suffices to show that $F=H_A$, which is done via~\eqref{eq:cov_varphiA} and~\eqref{eq:cov_varphiA_delta}.
\end{proof}

The next lemma is a direct corollary of the previous lemma.

\begin{lemma}
    \label{lemma:ldp_gff_equivalence_boring}
   For a convergent sequence $(\delta_n)_n$ and a topological annulus $A$,
        \begin{align}
        \label{eq:boringequation}
            \lim_{k\to\infty}
            \lim_{n\to\infty}
            \frac1{k^2}
            \log
            \P_{\Z^2}[\{
                \langle h,\varphi_{A/\delta_n}\rangle
                \geq k
            \}]
            &=
            \lim_{k\to\infty}
            \frac1{k^2}
            \log
            \P[\{
                \langle \sigma\Gamma,\varphi_A\rangle
                \geq k
            \}]\\
            &= -\frac1{2\sigma^2}\operatorname{Dirichlet}(H_A),
    \end{align}
    where $\sigma^2$ is the variance of the limiting GFF along $(\delta_n)_n$.
\end{lemma}

The heavy lifting is now done in the following lemma. Below, a subset of $F(\Z^2)$ is seen as a subset of $\R^2$ by considering the union of the line segments joining the middles of adjacent faces.

\begin{lemma}
    \label{lemma:circuit_event_given_average}
    Fix $\c\in[1,2]$.
  For any convergence sequence $(\delta_n)_n$, any topological annulus $A$
    and any $\epsilon>0$, we may find some constant $\alpha>0$ such that the following holds
    true.
    Let $h'$ denote the representative of $h$ such that $\varphi_{A/\delta_n}^-(h')\in[0,2)$
    (this depends implicitly on $n$).
    Suppose that $x\in A$ and $r>0$ are chosen such that $\bbox{r}(x)\subset A$.
    Then,
       \begin{equation}
        \lim_{k\to\infty}
        \lim_{n\to\infty}
        \P_{\Z^2}\left[
            \left\{\parbox{19em}{$\{ H_A(x)-\epsilon\leq \frac{h'}{k}\leq H_A(x)+\epsilon\}$
            contains \\
           a nontrivial circuit in $\ann{r}{\alpha r}(x)/\delta_n$}\right\}
        \middle|
        \{
            \langle h,\varphi_{A/\delta_n}\rangle
            \geq k
        \}
        \right]
        = 1.
    \end{equation}
\end{lemma}

\begin{figure}
    \centering
    \includegraphics{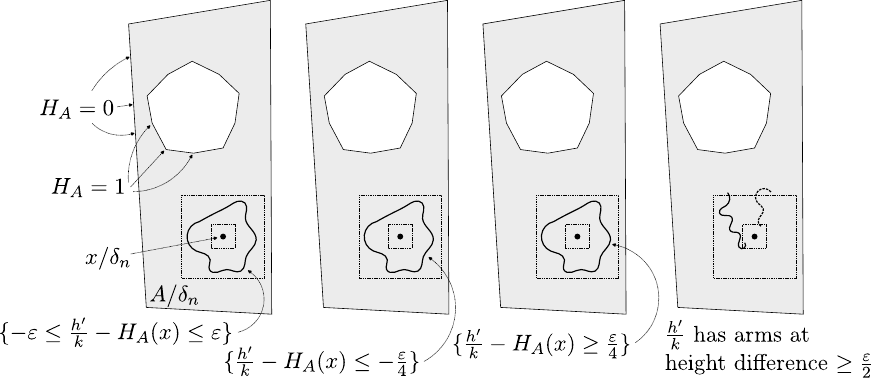}
    \caption{By topological considerations,
    one of the following three events must occur
    for the scaled height function $(h')^{(\delta_n)}/k$:
    the event in the statement of Lemma~\ref{lemma:circuit_event_given_average} (left picture),
    the event $E_-\cup E_+$ in the proof (middle pictures),
    or the event $E_*$ in the proof (right picture).}
    \label{fig:annulus_distribution_micro}
\end{figure}

We start with an informal explanation of the statement of this lemma.
The height function $h^{(\delta_n)}$ converges to a multiple of the GFF
in the sense of finite-dimensional distributions.
In particular, the distribution of $(h')^{(\delta_n)}/k$ in the conditional measure
$\P_{\Z^2}[\blank|\{\langle h,\varphi_{A/\delta_n}\rangle\geq k\}]$
converges to $H_A+\sigma\Gamma^A/k$ as $n$ tends to infinity
(Lemma~\ref{eq:lemma:ldp_gff_equivalence_useful}).
This implies that in the double limit, $(h')^{(\delta_n)}/k$ converges to the deterministic function $H_A$,
\emph{but only in the sense of finite-dimensional distributions}.

The lemma provides a link between this distributional convergence,
and the appearance of circuits in the level sets of $(h')^{(\delta_n)}/k$ (which are defined on the microscopic level).
More precisely, it says that with high probability, we may find a microscopic circuit around any point $x\in A$
along which the height of $(h')^{(\delta_n)}/k$ is close to $H_A(x)$
(Figure~\ref{fig:annulus_distribution_micro}).

\begin{proof}
    Define the following three events:
    \begin{align}
        E_-:&=
        \left\{\parbox{17em}{$\{ \frac{h'}{k}\leq H_A(x)-\frac\epsilon4\}$
            contains a nontrivial circuit in $\ann{r}{\alpha r}(x)/\delta_n$}\right\};
        \\
        E_+:&=
        \left\{\parbox{17em}{$\{ \frac{h'}{k}\geq H_A(x)+\frac\epsilon4\}$
            contains a nontrivial circuit in $\ann{r}{\alpha r}(x)/\delta_n$}\right\};
        \\
        E_*:&=
        \left\{
            \parbox{24em}{we may find some $a\in2\Z$
            such that $\{h+a\leq 0\}\subset F(\Z^2)$ and $\{h+a\geq \frac{\epsilon}{2} k\}$ both
            contain a path connecting the two boundary components of $\ann{r}{\alpha r}(x)/\delta_n$}
            \right\}.
    \end{align}
    We claim that if the event in the statement of the lemma does \emph{not} occur,
    then the event $E_-\cup E_+\cup E_*$ must occur.
    The claim is true for topological reasons (Figure~\ref{fig:annulus_distribution_micro}).
    Indeed,
    if the event in the statement does not occur,
    then one of
    \begin{equation}
        \label{eq:three_events22221324}
        \{h'/k-H_A(x) \geq 7\epsilon/8\}
        \quad\text{or}\quad
        \{h'/k-H_A(x) \leq -7\epsilon/8\}
    \end{equation}
        must contain an arm of the annulus.
        If the event $E_-\cup E_+$ does not occur, then both
    \begin{equation}
        \label{eq:three_events22221324_bis}
        \{h'/k-H_A(x) \geq -3\epsilon/8\}
        \quad\text{and}\quad
        \{h'/k-H_A(x) \leq 3\epsilon/8\}
    \end{equation}
    must contain an arm of the annulus.
    But if one of~\eqref{eq:three_events22221324} and both
    of~\eqref{eq:three_events22221324_bis} contain an arm, then $E_*$ occurs.
    This proves the claim.

    By the claim, it suffices to prove that the events $E_-$, $E_+$, and $E_*$ each occur with a low conditional probability in the limit.
    We first handle $E_*$ by fixing a suitable value for $\alpha$.
Note that the bound on arm exponents (Theorem~\ref{thm:intro:arm_exponents}) implies that
     \begin{equation}
   \mathbb P_{\Z^2}[E_*]\le
         \alpha^{c \epsilon^2k^2/4}.
         \end{equation}
     If $\alpha$ is sufficiently small, then Lemma~\ref{lemma:ldp_gff_equivalence_boring} implies
     that this probability is negligible compared to the probability of the conditioning event.
     Fix such an $\alpha>0$ from now on.

 It suffices to prove that $E_-$ occurs with a conditional probability tending to zero
    (the case of $E_+$ is similar).
    Let $\pi_n$ denote the uniform probability measure on a circle of radius $\alpha/\delta_n$
    centred at $x/\delta_n$.
    By flip domination (Theorem~\ref{thm:intro:flip_domination}), we get
    \begin{equation}
        \P_{\Z^2}\left[
        \left\{
            \frac{\langle h,\pi_n-\varphi^-_{A/\delta_n}\rangle}{k}
            \leq H_A(x)-\tfrac\epsilon8
        \right\}
        \middle|E_-\cap\{
            \langle h,\varphi_{A/\delta_n}\rangle
            \geq k
        \}\right]
        \geq \frac12.
    \end{equation}
    But without $E_-$ as conditioning event,
    Lemma~\ref{eq:lemma:ldp_gff_equivalence_useful} tells us
    that
    \begin{equation}
        \lim_{k\to\infty}
        \lim_{n\to\infty}
        \P_{\Z^2}\left[
        \left\{
            \frac{\langle h,\pi_n-\varphi^-_{A/\delta_n}\rangle}{k}
            \leq H_A(x)-\tfrac\epsilon8
        \right\}
        \middle|\{
            \langle h,\varphi_{A/\delta_n}\rangle
            \geq k
        \}\right]
        =0.
    \end{equation}
    This implies that the conditional probability of $E_-$ tends to zero.
\end{proof}

\begin{proof}[Proof of Proposition~\ref{prop:mollification_of_ridge_eventsSIMPLER}]
 Consider a convergence sequence $(\delta_n)$.
 Let $\sigma^2$ be the variance of the limiting GFF.
    We view Equation~\eqref{eq:formal_assumption_I} as the combination of two inequalities,
    which are treated somewhat differently.
    More precisely, it suffices to prove that, for fixed $\epsilon>0$,
    \begin{align}
        \label{eq:formal_assumption_I_first_half}
     \liminf_{k\to\infty}
        \liminf_{n\to\infty} \frac{
        \P_{\Z^2}[\{
                \langle h,\varphi_{A/\delta_n}\rangle
                \geq k
            \}]}{\P_{\Z^2}[\pCIRCUIT{+k+4}{A/\delta_n}]}
            &\geq \tfrac14;
            \\
            \label{eq:formal_assumption_I_second_half}
                  \liminf_{k\to\infty}
        \liminf_{n\to\infty} \frac{
                    \P_{\Z^2}[
                    \pCIRCUIT{+k}{A/\delta_n}]}{\P_{\Z^2}[
                    \{
                \langle h,\varphi_{A/\delta_n}\rangle
                \geq (1+4\epsilon)k
            \}]}
            &\geq \tfrac14.
    \end{align}

    \paragraph{Proof of Equation~\eqref{eq:formal_assumption_I_first_half}.}
    In fact, we shall prove this inequality for fixed $k$ and $n$ using flip domination.
    Define  $\gamma_0$ to be the outermost even height circuit in $A/\delta_n$, and $\gamma_\ell$ to be the outermost even height circuit in $A/\delta_n$ with $h(\gamma_\ell)-h(\gamma_0)=\ell$.
    These circuits may or may not be well-defined (if they are not, we set them equal to $\emptyset$).
   Introduce the event
    \(
        C_\ell:=
        \{\gamma_\ell\neq\emptyset\}\).
    For $\ell\in2\Z\setminus\{0\}$, define the random variables
    \begin{align}
        X_0&:=
        h(\gamma_0)-\varphi_{A/\delta_n}^-(h);\\
        X_\ell&:=\varphi_{A/\delta_n}^+(h)-h(\gamma_\ell).
    \end{align}
    On $C_\ell$, we have that
    \begin{equation}
        \langle h,\varphi_{A/\delta_n}\rangle = X_0 +\ell+ X_\ell.
    \end{equation}
    Flip domination (Theorem~\ref{thm:intro:flip_domination}) implies that
    for any even $\ell\geq 4$,
        \begin{equation}
            \P[\{X_0\leq 0\}\cap C_\ell]
            \leq
            \P[\{X_0\leq 0\}\cap C_{-\ell+4}]
            =
            \P[\{X_0\geq 0\}\cap C_{\ell-4}],
        \end{equation}
    where the equality is just total flip symmetry.
    Since also $C_{\ell-4}\subset C_\ell$, we get
    \begin{equation}
        \P[\{X_0\geq 0\}\cap C_{\ell-4}] \geq \tfrac12 \P[ C_{\ell}].
    \end{equation}
    Using flip domination (Theorem~\ref{thm:intro:flip_domination}) again, we obtain
    \begin{equation}
        \P[\{X_{\ell-4}\geq 0\}|\{X_0\geq 0\}\cap C_{\ell-4}]\geq \tfrac12.
    \end{equation}
    Putting things together, we arrive at
    \begin{equation}
        \P[\{\langle h,\varphi_{A/\delta_n}\rangle \geq \ell -4\}]
        \geq
        \P[\{X_0\geq 0\}\cap C_{\ell-4} \cap \{X_{\ell-4}\geq 0\}]
        \geq \tfrac14\P[C_\ell].
    \end{equation}
    Setting $\ell=k+4$ yields Equation~\eqref{eq:formal_assumption_I_first_half}.

    \paragraph{Proof of Equation~\eqref{eq:formal_assumption_I_second_half}.}
    Define $h'$ as in Lemma~\ref{lemma:circuit_event_given_average}.
    Using Lemma~\ref{lemma:circuit_event_given_average}
    and a union bound (see also Remark~\ref{remark:combinecircuits}), we obtain
    \begin{equation}
        \lim_{k\to\infty}
        \lim_{n\to\infty}
        \P_{\Z^2}\left[
            \left\{\parbox{15.5em}{$\{\frac{h'}{k}\leq\epsilon\}$
            and $\{\frac{h'}{k}\geq 1+3\epsilon\}$
             contain a nontrivial circuit in $A/\delta_n$}\right\}
        \middle|
        \{
            \langle h,\varphi_{A/\delta_n}\rangle
            \geq (1+4\epsilon)k
        \}
        \right]
        = 1.
    \end{equation}
    By the intermediate value theorem,
    we deduce from this that
    \begin{equation}
        \lim_{k\to\infty}
        \lim_{n\to\infty}
        \P_{\Z^2}\left[
            \left\{\parbox{16em}{there are two even height circuits $\gamma_0,\gamma_+$ winding nontrivially in $A/\delta_n$
            with $h(\gamma_+)-h(\gamma_0)\geq (1+\epsilon)k$} \right\}
        \middle|
        \{
            \langle h,\varphi_{A/\delta_n}\rangle
            \geq (1+4\epsilon)k
        \}
        \right]
        = 1.
    \end{equation}
    Equation~\eqref{eq:formal_assumption_I_second_half} follows by combining this with \eqref{eq:formal_assumption_I_first_half}, Lemma~\ref{lemma:ldp_gff_equivalence_boring}, and the inequality
    \begin{equation}
        \label{eq:useful_flip_domination_consequence}
        \P[C_{\ell+2\ell'-4}] \geq \frac12 \P[C_\ell \cap C_{-\ell'}]
        \qquad\text{for any even $\ell\geq\ell'\geq 2$},
    \end{equation}
which can be obtained in the same way as in the proof of Equation~\eqref{eq:formal_assumption_I_first_half} above. This concludes the proof.
\end{proof}

\subsection{Proof of Theorem~\ref{thm:glimpse_scale_invariance}}

With everything in hand, the proof of the theorem will basically boil down to an analysis of the Dirichlet energy.

\begin{proof}[Proof of Theorem~\ref{thm:glimpse_scale_invariance}]
  Consider a convergence sequence $(\delta_n)_n$ and let $\sigma^2$ be the variance of the associated limiting GFF.

  \paragraph{Proof of $f''(0)\geq - 1/\sigma^2$.}
    First, the straight bound in Proposition~\ref{prop:split_into_ridge_events}
    combined with Proposition~\ref{prop:mollification_of_ridge_eventsSIMPLER}
    and Lemma~\ref{lemma:ldp_gff_equivalence_boring} yields
    \begin{equation}
        \frac12 f''(0)
        \geq -\frac{\operatorname{Dirichlet}(H_{A_{\rho,1}})}{4\rho}\frac1{2\sigma^2}.
    \end{equation}
    Introduce a new function
    \begin{equation}
        H'_\rho:\R^2\to[0,1],\,x\mapsto 0\vee (1-\operatorname{dist}(x,\symRect{\rho}{0}))
    \end{equation}
    which coincides with $H_{A_{\rho,1}}$ on $\partial A_\rho$.
   As harmonic functions minimise the Dirichlet energy,
    \begin{equation}
        \operatorname{Dirichlet}(H_{A_{\rho,1}})
        \leq \operatorname{Dirichlet}(H'_\rho)
        = 4\rho + C,
    \end{equation}where $C$ is a constant independent of $\rho$.
     Letting $\rho$ goes to infinity yields the result.

\paragraph{Proof of $f''(0)\leq -1/\sigma^2$.}
    Consider the curly bound in Proposition~\ref{prop:split_into_ridge_events}(ii).
    Importantly, for fixed $\rho$ and $\epsilon$,
    there are only finitely many \emph{shapes} of $(\rho,\eta)$-straightened annuli $A_i$
    of diameter at most $2r$
    (where we say that two topological annuli have the same shape if they differ by a translation).
    Since there are only finitely many shapes,
    we may apply Proposition~\ref{prop:mollification_of_ridge_eventsSIMPLER} and Lemma~\ref{lemma:ldp_gff_equivalence_boring}
    to each of these shapes to get
    \begin{equation}
                        \label{eq:curly_bound_second_derivXXX}
- \lim_{n\to\infty}
            \min_{(A_i)\in\frakA(r,\rho,\eta,\epsilon,n)}
            \frac1{4(\rho+1)n}
            \sum_{i}
            \frac{\operatorname{Dirichlet}(H_{A_i})}{2\sigma^2}
            \\
            \geq
              \frac12 f''(0)
                +
                \frac{100}{\rho+1}\log\ccircuit.
        \end{equation}
        Recall that $\frakA(r,\rho,\eta,\epsilon,n)$ is the set of all families of $K_n:=\lceil(1-\epsilon)n\rceil$ disjoint $(\rho,\eta)$-straightened annuli $A_i$
        of diameter at most $2r$ that are contained in $\symRect{(\rho+1)}{n}$.

        To deduce $f''(0)\leq -1/\sigma^2$ from~\eqref{eq:curly_bound_second_derivXXX},
        it suffices to demonstrate that
        \begin{equation}
            \label{eq:curly_bound_second_derivXXX_bis}
            \sum\limits_{i=1}^{K_n}
            \operatorname{Dirichlet}(H_{A_i})\geq 4\rho n (1-\epsilon)^2,
        \end{equation}
        since $\rho$ and $n$ can be taken large while $\epsilon$ can be taken small.
        Equation~\eqref{eq:curly_bound_second_derivXXX_bis} follows from a straightforward
        calculation which we now describe.

        Fix $(A_i)\in \frakA(r,\rho,\eta,\epsilon,n)$;
        we are now going to establish~\eqref{eq:curly_bound_second_derivXXX_bis}.
       Since the annuli are disjoint, we get
        \begin{equation}
           \sum_{i}\operatorname{Dirichlet}(H_{A_i}) =
           \operatorname{Dirichlet}\big({\textstyle\sum_{i}H_{A_i}}\big)
            \geq
            \int_{\symRect{\rho}{n}} \big|\nabla_2\big({\textstyle\sum_{i}H_{A_i}}\big)(z) \big|^2 \diffi z.
        \end{equation}
        For the avoidance of doubt: on the right, we only integrate the square of the \emph{vertical} derivative
        over a \emph{smaller} set than $\R^2$, hence the inequality.
        By writing $z=(x,y)$, we may write this out explicitly as
        \begin{equation}
            \sum\limits_{i}
            \operatorname{Dirichlet}(H_{A_i})
            \geq \int_{-\rho}^{\rho} \int_{-n}^{n} \big| \partial_{y}\big({\textstyle\sum_{i}H_{A_i}}\big)(a,b) \big|^2 \diffi b \diffi a.
        \end{equation}
        We now claim that for fixed $a\in[-\rho,\rho]$,
        we have
        \begin{equation}
            \label{eq:curly_bound_second_derivXXX_bis_bis}
            \int_{-n}^{n} \big| \partial_{y}\big({\textstyle\sum_{i}H_{A_i}}\big)(a,b) \big|^2 \diffi b
            \geq
            2n(1-\epsilon)^2,
        \end{equation}
        which clearly suffices for Equation~\eqref{eq:curly_bound_second_derivXXX_bis}.
        Notice that the total variation of $\sum_{i}H_{A_i}$ along the vertical line $\{a\}\times [-n,n]$ is at least $2K_n\geq 2(1-\epsilon)n$,
        since each annulus contributes at least $2$ to the total variation along this line.
        Equation~\eqref{eq:curly_bound_second_derivXXX_bis_bis} now follows from the lemma below.
\end{proof}

\begin{lemma}
    Let $g:[-n,n]\to\R$ denote any continuous piecewise smooth function
    whose total variation is at least $2(1-\epsilon)n$.
    Then $\int g'(x)^2\diff x\geq 2n(1-\epsilon)^2$.
\end{lemma}

\begin{proof}
    We may assume that $g(-n)=0$ and that $g$ is non-decreasing,
    by replacing it by $\tilde g(x):=\int_{-n}^x |g'(t)| \diff t$
    if necessary (this function has the same total variation and Dirichlet energy).
    This also implies that $g(n)\geq 2(1-\epsilon)n$.
    But given the values $g(-n)$ and $g(n)$,
    we know which function $\hat g$ minimises the one-dimensional Dirichlet energy:
    it is a one-dimensional harmonic function, that is, a linear interpolation.
    We therefore obtain
    \begin{equation}
        \int g'(x)^2\diff x\geq
        \int \hat g'(x)^2\diff x
        =
        \frac{g(n)^2}{2n}\geq 2n(1-\epsilon)^2.
    \end{equation}
    This finishes the proof.
\end{proof}

% </input src="sections/PART_E/22_GFF_LDP.tex" root="." version="0.0.1">
% <input src="sections/PART_E/24_LDP_SPLIT/main.tex" root="." version="0.0.1">
\section{Proof of the splitting (Proposition~\ref{prop:split_into_ridge_events})}
\label{sec:ldp_split}

We conclude this article by proving the two bounds in Proposition~\ref{prop:split_into_ridge_events}.
This final section consists of two subsections;
each one is dedicated to one of the two bounds.
The first bound is significantly easier to prove than the second.

This section builds on the following ideas:
the general spin representation in Section~\ref{sec:RSW},
the level line tree and
the covariance structure in Section~\ref{sec:level line tree},
and the combinatorial argument used in Section~\ref{sec:loglon} (proof of Theorem~\ref{thm:armexponent_for_spins_quadratic}).

% <input src="sections/PART_E/24_LDP_SPLIT/straight.tex" root="." version="0.0.1">
\subsection{Proof of the straight bound (Proposition~\ref{prop:split_into_ridge_events}(i))}

 The proof of Proposition~\ref{prop:split_into_ridge_events}(i) consists of four steps.
    First, we rewrite the quantity $\chi:=\P_{\Z^2}[\pCIRCUIT{+k}{
                A_{\rho,N}
    }]$ in terms of the branching function of the spin representation.
  Second, we use the FKG inequality to prove (roughly) that the probability of a circuit
        in an extremely wide topological annulus $A_{2n\rho,N}$  is at least $\chi^{2n}$.
     Third, we stack the annuli vertically to show that we can get many horizontal crossings
        in an extremely large rectangle with a probability lower bounded by some power of $\chi$.
      Finally, we relate the final quantity to the free energy $f(\alpha)$.

\begin{proof}[Proof of Proposition~\ref{prop:split_into_ridge_events}(i)]
   Fix $\rho$, $k$, and $N$.

    \paragraph{Step~1: Rewriting in terms of the branching function.}
    Recall that the law of $h$ is the same in $\P_{\Z^2}$ and $\mu_{\Z^2}$.
    We shall work solely in the measure $\mu_{\Z^2}$.
    By inclusion of events, we get
    \begin{equation}
        \chi:=\mu_{\Z^2}[\pCIRCUIT{+k}{
                A_{\rho,N}
        }]
        \leq
        \mu_{\Z^2}[\pALTCIRCUIT{\bar\omega}{k/2-4}{A_{\rho,N}}].
    \end{equation}
    Recall the definition of maximal domains (Definition~\ref{def:maximal_domains}).
    Let $\calD\in\calM^+(A_{\rho,N})$ denote the largest domain such that $\partial\calD$
    surrounds the interior boundary of $A_{\rho,N}$
    (and set $\calD:=\emptyset$ if such a domain does not exist).
    By the tower property and the Markov property,
    we obtain
    \begin{equation}
       \chi\leq \mu_{\Z^2}[\pALTCIRCUIT{\bar\omega}{k/2-4}{A_{\rho,N}}]
       =
       \mu_{\Z^2}\big[\mu_{\calD}^+[\pALTCIRCUIT{\bar\omega}{k/2-4}{A_{\rho,N}}]\true{\calD\ne \emptyset}\big].
    \end{equation}
    Now, recall the definition of the branching function and its diagonal $\psi^*$ (Definition~\ref{def:branching_function}).
    Roughly speaking, $\psi^*/2$ counts the number of alternating disjoint circuits around each point.
    Since $\pALTCIRCUIT{\bar\omega}{k/2-4}{A_{\rho,N}}\subset\pCIRCUIT{\{\psi^*\geq k-12\}}{A_{\rho,N}}$,
    we may use the previous bound and monotonicity in domains (Lemma~\ref{lem:covariance_mono}) to push $\calD$ away and get
    \begin{equation}
        \chi\leq
        \mu_{\symRect{(\rho+1)N}{N}}^+[\pCIRCUIT{\{\psi^*\geq k-12\}}{A_{\rho,N}}].
    \end{equation}

    \paragraph{Step~2: Estimating the probability of a circuit in an extremely wide annulus.}
    We claim that for any $n\in\Z_{\geq 1}$,
    \begin{equation}
        \label{eq:piece2_claim}
        \chi^{2n}\leq
        \mu_{\symRect{(2n\rho+1)N}{N}}^+[\pCIRCUIT{\{\psi^*\geq k-12\}}{A_{2n\rho,N}}].
    \end{equation}
    We call $2n$ the \emph{horizontal multiplicity}.
    To see that the claim is true,
    observe first that monotonicity in domains and the FKG inequality (Lemma~\ref{lem:covariance_mono}) imply
    \begin{align}
        \chi^{2n}&\leq
        \mu_{\symRect{(2n\rho+1)N}{N}}^+\Big[
            \bigcap_{i=1}^{2n}
            \pCIRCUIT{\{\psi^*\geq k-12\}}{A_{\rho,N}+((2i-2n-1) \rho N,0)}
        \Big]
        \\
        \label{eq:step2_fkg}
        &\leq
        \mu_{\symRect{(2n\rho+1)N}{N}}^+[\pCIRCUIT{\{\psi^*\geq k-12\}}{A_{2n\rho,N}}].
    \end{align}
    The second inequality is just inclusion of events (see Remark~\ref{remark:combinecircuits}).

    If we let $\calA\in\calM^-(\symRect{(2n\rho+1)N}{N})$  denote the largest domain containing the hole of the annulus,
    then boundary pushing yields
    \begin{equation}
        \label{eq:step2_fkg2}
        \chi^{2n}\leq
        \mu_{\Z^2}[\pCIRCUIT{\{\psi^*_\calA\geq k-14\}}{A_{2n\rho,N}}|\calE]
    \end{equation}
    for any event $\calE$ measurable with respect to the even edges outside $\symRect{(2n\rho+1)N}{N}$.
    By inclusion of events, we then conclude that
    \begin{equation}
        \chi^{2n}\leq
        \mu_{\Z^2}[\pALTHORI{\bar\omega}{k-20}{\symRect{2n\rho N}{N}}|\calE].
    \end{equation}

    \paragraph{Step~3: Stacking extremely wide annuli vertically.}
    Fix $\rho',n\in\Z_{\geq 1}$.
    Apply the previous step $2n$ times in $2n$ vertically stacked rectangles,
    which have horizontal multiplicity $2\rho'n$.
    This yields
    \begin{equation}
        \chi^{4\rho'n^2}\leq
        \mu_{\Z^2}[E_+\cap E_-],
    \end{equation}
    where
    \begin{equation}
        E_\pm:=
        \pALTHORI{\bar\omega}{n(k-22)}{\symInt{\rho\rho'2n N}\times \pm[0,2nN]}.
    \end{equation}

    \paragraph{Step~4: Relating our quantity to the free energy.}
    We first claim that
    \begin{equation}
        \mu_{\Z^2}[\pALTCIRCUIT{\bar\omega}{n(k-23)}{A_{\rho\rho',2nN}}| E_+\cap E_-]
        \geq c^{n^2N^2}
    \end{equation}
    for some small fixed constant $c>0$.
    We prove the claim via a straightforward \emph{finite energy} or \emph{surgery} argument.
    Indeed, condition on $E_+\cap E_-$,
    and reveal the necessary edges to verify that this event occurs.
    Then, any \emph{unrevealed} even edge still has a uniformly positive probability of being
    open (unless it connects two vertices with known opposite spins).
    By conditioning on the states of at most $100n^2N^2$ such edges,
    we may wire up the horizontal crossings in such a way that the event $\pALTCIRCUIT{\bar\omega}{n(k-23)}{A_{\rho\rho',2nN}}$ occurs.
    This proves the display above.

    As a consequence, we get
    \begin{equation}
        \mu_{\Z^2}[\pALTCIRCUIT{\bar\omega}{n(k-23)}{A_{\rho\rho',2nN}}]
        \geq c^{n^2N^2}\cdot \chi^{4\rho'n^2}.
    \end{equation}
    Finally, by flipping the coins (for each element of $V_\slodd(\calX)$, see Lemma~\ref{lemma:Basic properties of the level line tree}) to determine the odd spins,
    we find
    \begin{equation}
        \mu_{\Z^2}[\pCIRCUIT{+2n(k-24)}{A_{\rho\rho',2nN}}]
        \geq 2^{-nk}\cdot c^{n^2N^2}\cdot \chi^{4\rho'n^2}.
    \end{equation}
    By the variational principle (Proposition~\ref{prop:ZtoR_globalNEW}),
    sending first $n$ and then $\rho'$ tend to infinity,
    we get
    \begin{equation}
        f(\tfrac{k-24}{N})-f(0)\geq \frac1{4\rho N^2}\log\chi.
    \end{equation}
    This is the desired inequality.
\end{proof}

% </input src="sections/PART_E/24_LDP_SPLIT/straight.tex" root="." version="0.0.1">
% <input src="sections/PART_E/24_LDP_SPLIT/curly.tex" root="." version="0.0.1">
\subsection{Proof of the curly bound (Proposition~\ref{prop:split_into_ridge_events}(ii))}

We start the proof with the key step, which is called {\em ridge splitting} (Lemma~\ref{lemma:ridge_splitting} below).
After the key step, we connect one side of the inequality in Lemma~\ref{lemma:ridge_splitting} to the variational principle
(Proposition~\ref{prop:ZtoR_globalNEW}), see Lemma~\ref{lemma:curly_bound_LDP_event} below.
Finally, we perform three simplification steps to connect the other side to the quantity in Proposition~\ref{prop:split_into_ridge_events}.
The simplification steps are:
cutoff of the domain diameter (Lemma~\ref{lemma:curly_bound_regularity_step1}),
decoupling (Lemma~\ref{lemma:curly_bound_decoupling_step}),
and straightening of the annuli (Lemma~\ref{lemma:curly_bound_regularity_step3}).
The curly bound in Proposition~\ref{prop:split_into_ridge_events} is an immediate corollary of Lemma~\ref{lemma:curly_bound_regularity_step3}.

The next lemma follows immediately from twice-differentiability of the free energy obtained via the Bethe Ansatz (a result that requires exact integrability).
We want this section to be independent of such integrability,
and therefore we provide an alternative proof of this lemma that only relies on the circuit estimate.
    \begin{lemma}[Lower bound on the free energy via RSW theory]
        \label{lemma:free_energy_lower_bound}
        There exists a constant $c_{\rm fe}<\infty$ such that
        $f(\alpha)-f(0)\geq -c_{\rm fe}\alpha^2$ for any $\alpha<1$.
    \end{lemma}

    \begin{proof}
        Since $f(\alpha)$ is a convex function,
        it suffices to bound $f(\alpha)$
        for $\alpha\approx 0$.
        Recall Proposition~\ref{prop:ZtoR_globalNEW}.
        It is straightforward to see that
        the event $\pALTCIRCUIT{\bar\omega}{2\lceil\alpha L/4\rceil}{ A_{\rho,L}}$
        is contained in the intersection of
        $O(\rho L^2\alpha^2)$ circuit events,
        thus having a probability of at least $(\ccircuit)^{O(\rho L^2\alpha^2)}$.
        The circuits are combined in such a way
        that they form alternating $\omega^+$-loops
        and $\omega^-$-loops in the topological annulus (cf.~Remark~\ref{remark:combinecircuits}).
        Conditional on this event, the event
        $\pCIRCUIT{+\alpha L}{
                A_{\rho,L}
        }$ has a probability of at least $e^{-O(\alpha L)}$ since every circuit
        is oriented upwards with probability $1/2$.
        This lower bound leads to the desired bound in the lemma.
    \end{proof}

\subsubsection{Ridge splitting}

    In order to define more complicated events,
    it will be useful to introduce the following notion.

    \begin{definition}[Local level line forest]
        \label{def:local_level_line_forest}
        Recall Definition~\ref{def:level_line_tree} for the level line tree and
        Definition~\ref{def:maximal_domains} for maximal domains.
        Consider a fixed continuum domain $D\subset\R^2$.
        The \emph{local level line forest} is the family
        of level line trees
        \( \calL^D:=(\calX_\calD)_{\calD\in\calM^+(D)}\).
        This means that we first find the maximal domains $\calD$ in $D$
        with $\partial\calD\subset\omega^+$, and then construct the level line tree $\calX_\calD$ for each such domain $\calD$.

        We are interested in $\omega$-circuits at a certain \emph{forest depth},
        which means that they are contained in a certain vertex of some level line tree $\calX_\calD$ at that distance from the root of $\calX_\calD$.
    \end{definition}

   \begin{figure}
        \centering
        \includegraphics{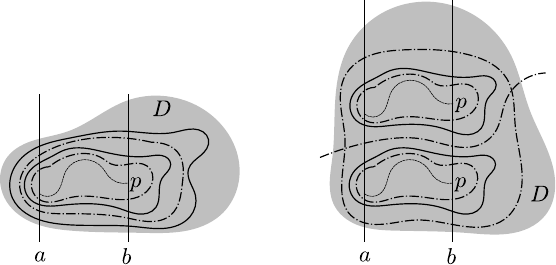}
        \caption{Left: the event $\pALTRIDGE{\bar\omega}{4}{D,a,b}$.
        Right: the event $\{\#\pALTRIDGE{\bar\omega}{2}{D,a,b}=2\}$.}
        \label{fig:percoRidge}
    \end{figure}

    We can use this notion to define \emph{ridge events}. The ridge event basically means that there is some outermost $\omega^+$-circuit
        in which we may find alternating $\omega^+/\omega^-$-circuits
        such that the $k$-th circuit still surrounds some path crossing $[a,b]\times\R$.

    \begin{definition}[Ridge events]
        Let $D\subset\R^2$ denote a domain, fix $k\in2\Z_{\geq 1}$, and let $a\leq b$.
        Then, define
        \begin{equation}
            \pALTRIDGE{\bar\omega}{k}{D,a,b}
            :=
            \bigcup_{p} \pALTCIRCUIT{\bar\omega}{k}{D\setminus p}
            =\bigcup_p \left\{\parbox{12em}{some $\omega^-\cap D$-circuit at forest depth $2k-2$ surrounds $p$}\right\},
        \end{equation}
        where the union runs over all paths $p$ from the left of $[a,b]\times\R$ to the right
        (see Figure~\ref{fig:percoRidge}).

        We also define
        \begin{equation}
            \#\pALTRIDGE{\bar\omega}{k}{D,a,b}
            :=\#\left\{\calD\in\calM^+(D):\parbox{20em}{some $\omega^-\cap D$-circuit at depth $2k-2$ in $\calX_\calD$ surrounds some path $p$
            that crosses $[a,b]\times\R$}\right\};
        \end{equation}
        which simply counts the number of maximal domains in which the ridge event occurs.
    \end{definition}

    The key step in the proof of the curly bound is \emph{ridge splitting}
    (Lemma~\ref{lemma:ridge_splitting} and Figure~\ref{fig:splitsplit}).
    This lemma involves \emph{ridge events}.
    On one side in the comparison, we consider the event
    that the large rectangle contains many maximal domains containing several very wide nested $\omega^+/\omega^-$-circuits (right side in Figure~\ref{fig:splitsplit}).
    On the other side, we chop the large rectangle into smaller ones,
    and consider the event that the smaller rectangles contain many maximal domains
    containing several nested $\omega^+/\omega^-$-circuits (left side in Figure~\ref{fig:splitsplit}).

      \begin{figure}
        \centering
        \includegraphics{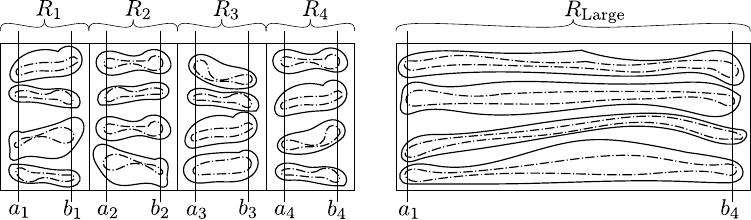}
        \caption{Left: the left event in Lemma~\ref{lemma:ridge_splitting}.
        Right: the right event in Lemma~\ref{lemma:ridge_splitting}.}
        \label{fig:splitsplit}
    \end{figure}

      For $\rho,k,N,\rho',n\in 2\Z_{\geq 1}$
        with $k,N/k\geq 100$, define the following rectangles:
        \begin{itemize}
            \item The rectangle with side lengths $4n\rho'(\rho+1)N$ and $4nN$ given by
            \begin{equation}
            \label{eq:RLARGEDEF}
            R_{\operatorname{Large}}:=\symRect{2n\rho'(\rho+1)N}{2nN},
        \end{equation}
            \item This rectangle may be sliced vertically into $2n\rho'$ rectangles of width $2(\rho+1)N$ and height $4nN$,
        which we shall call $R_i$ for $i=1,\ldots,2n\rho'$ (indexed from left to right),
            \item Write $R_i'\subset R_i$ for the rectangle of width $2\rho N$ and height $4nN$
        centred within $R_i$,
        \item         Write $a_i$ and $b_i$ for the $x$-coordinates of the left and right sides of $R_i'$ respectively.
        \end{itemize}
Define the event
\begin{equation}  \calE_{\rho,k,N,\rho',n}:=\left\{\textstyle\sum_{i=1}^{2n\rho'}\#\pALTRIDGE{\bar\omega}{k-4}{R_i,a_i,b_i}\geq 4n^2\rho'\right\}.
        \end{equation}

%   \begin{remark}
%        We shall eventually take limits in the parameters.
%        They are always taking in the following order (rightmost index first):
%        \begin{equation}
%            \rho\to\infty,\,\epsilon\to0,\,k\to\infty,\,N\to\infty,\,\rho'\to\infty,\,n\to\infty.
%        \end{equation}
%        The parameters $\rho'$ and $n$ have roughly the same role as in the proof of the straight bound.
%        \commentHugo{I don't think this remark is necessary... or at least not here as I don't see any limits in the near future...}
%        \noticePiet{I thought it may be useful to check that there are no circular dependencies between the parameters; we can also decide to delete.}
%    \end{remark}

    \begin{lemma}[Ridge splitting, cf.~Figure~\ref{fig:splitsplit}]
        \label{lemma:ridge_splitting}
        For $\rho,k,N,\rho',n\in 2\Z_{\geq 1}$
        with $k,N/k\geq 100$,
        \begin{equation}
            \label{eq:curly_key_step}
            \mu_{\Z^2}\left[\calE_{\rho,k,N,\rho',n}\right]
            \geq
            (\ccircuit)^{100 k^2n^2\rho'}
            \cdot
            \mu_{\Z^2}[\{\#\pALTRIDGE{\bar\omega}{k}{R_{\operatorname{Large}},a_1,b_{2n\rho'}}\geq 2n\}].
            \end{equation}
    \end{lemma}

        The idea of the proof is that if the event on the right occurs,
        then we can somehow locally rearrange our percolations $\bar\omega$
        to create the event on the left.
        We do so by essentially rewiring the configurations in the vertical slits
        between the slightly slimmer rectangles $R_i'$.
        The rewiring is done by first exploring the correct event (conditional on the event on the right, see Figure~\ref{fig:tough_exploration2}),
        then using the circuit estimate to obtain the appropriate wiring (see Figure~\ref{fig:tough_exploration2_2}).

        Before diving into the proof, let us first define an exploration process.
        This exploration process may succeed or fail,
        with $\calS$ denoting the event that it succeeds.
        This event contains the event on the right in Equation~\eqref{eq:curly_key_step}.
        Once the exploration process is done, we will prove that conditional on $\calS$,
        the event $\calE_{\rho,k,N,\rho',n}$ has a probability of at least $(\ccircuit)^{100 k^2n^2\rho'}$.
        This will imply the lemma.

        To facilitate the construction, define,
        for any $A,\omega\subset \R^2$ and any rectangle $R\subset\R^2$,
        the sets
        \begin{gather}
            \frakB(A,\omega,R):=\left\{u\in A\cap R:\parbox{16em}{$\{u\}$ and $\partial A$ do not intersect
            the same connected component of $R\setminus\omega$}\right\};
            \\
            \frakC(A,\omega,R):=\left\{D:\parbox{20em}{$D$ is a connected component of $\frakB$ intersecting both the left and right of $R$}\right\}.
        \end{gather}
        We think of $\frakB$ as the set of points \emph{blocked} (or \emph{shielded away}) from $\partial A$ by $\omega$ within $R$,
        and of $\frakC$ as the set of \emph{crossings} of such blocked points in $R$.
        See Figure~\ref{fig:frakBfrakC}.

        \begin{figure}[b]
        \centering
        \includegraphics{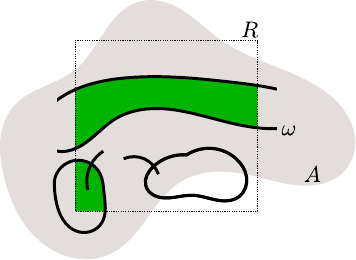}
        \caption{The green set depicts $\frakB$. Only the wide green component belongs to $\frakC$.}
        \label{fig:frakBfrakC}
        \end{figure}

        In the construction below,
        $R_{i,j}'\subset R_i$ is defined to be the rectangle of width $2(\rho+j/k) N$
        and height $4nN$ centred within $R_i$.

       \begin{remark}
       We think of $R_{i,j}'$ as ``interpolating'' between $R_i'$ and $R_i$,
        in the sense that       \begin{equation}
            R_i' = R_{i,0}' \subset R_{i,1}'\subset\cdots\subset R_{i,k-1}'\subset R_{i,k}'=R_i.
        \end{equation}\end{remark}

        In the setting of the lemma, the exploration process is represented by a tree $\calT=(V(\calT),E(\calT))$
        with root $R_{\operatorname{Large}}$.
        For each $x\in V(\calT)$ at distance $j$ from the root,
        let $\calN(x)$ denote the neighbours of $x$ at distance $j+1$ from the root
        (the letter $\calN$ refers to \emph{nesting} for reasons becoming clear shortly).
        The tree is constructed iteratively by explicitly constructing the function $\calN$ and the vertices at a distance $j$ from the root (let $V_j(\calT)$ denote this set)  as follows:
        \begin{itemize}
            \item $\calN(R_{\operatorname{Large}}):=\{\text{bounded connected components
            of $\R^2\setminus\partial\calM^+(R_{\operatorname{Large}})$}\}$.
            \item For any $A\in V_1(\calT)$, set
            $\calN(A):=\{\text{bounded connected components of $\R^2\setminus\partial\calM^-(A)$}\}$.
            \item For any $A\in V_2(\calT)$, set
            $\calN(A):=\cup_{i=1}^{2n\rho'}\frakC(A,\omega^+,R_{i,k-2}')$,
            \item For any $A\in V_3(\calT)$, set
            $\calN(A):=\cup_{i=1}^{2n\rho'}\frakC(A,\omega^-,R_{i,k-3}')$,
            \item Then, repeat the last two steps, alternating $+$ and $-$,
            until $V_k(\calT)$ has been defined.
        \end{itemize}
      With this definition, we are now in a position to prove the lemma.

        \begin{figure}
        \centering
        \includegraphics{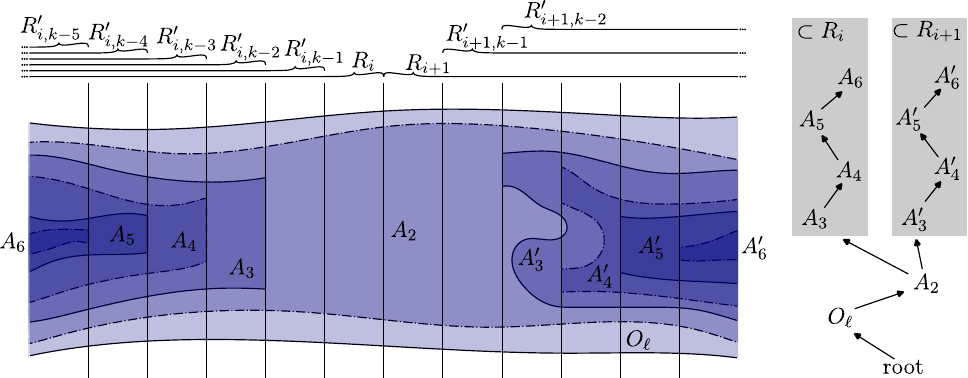}
        \caption{The nested sets in the exploration tree $\calT$ (assuming $\calS$ occurs).}
        \label{fig:tough_exploration2}
        \end{figure}

        \begin{proof}[Proof of Lemma~\ref{lemma:ridge_splitting}]
        Let us first make some remarks.
        Notice that the tree is \emph{nesting} in the sense that
        $A'\subset A$ for any $A'\in \calN(A)$.
        Notice also that any $A\in\cup_{j>2}V_j(\calT)$ is contained in some $R_i$.
        Let us write $(V_{j,i})_i$ for the partition of $V_j(\calT)$
        such that any $A\in V_{j,i}$ is contained in $R_i$.

      Let $\calS$ denote the event
        \begin{equation}
            \calS:=\left\{\parbox{34em}{there exist $2n$ distinct elements $(O_\ell)_{\ell=1,\ldots,2n}\subset V_1(\calT)$
            such that for every $1\leq \ell\leq 2n$ and $1\leq i\leq 2n\rho'$,
            the descendants of $O_\ell$ contain an element of $V_{k,i}$
            }
            \right\}.
        \end{equation}
        Figure~\ref{fig:tough_exploration2} contains a detailed illustration of this event,
        zooming in on a small rectangle and a single element $O_\ell$
        between two rectangles $R_i$ and $R_{i+1}$.
        It is tedious but straightforward to check that $\calS$ is included in the event on the right-hand side of Equation~\eqref{eq:curly_key_step}:
        \begin{equation}\calS\subset \{\#\pALTRIDGE{\bar\omega}{k}{R_{\operatorname{Large}},a_1,b_{2n\rho'}}\geq 2n\}.
        \end{equation}
        To finish the proof,
        introduce the event
        \begin{equation}
            \calC:=\left\{
                \parbox{28em}{any $A\in \cup_{2<j\leq k-2}V_j(\calT)$
                contains an $\omega^+$-circuit (if $j$ is odd)
                or an $\omega^-$-circuit (if $j$ is even)
                surrounding all elements of $\calN(A)$
                }
                \right\}.
        \end{equation}
        An impression of the event $\calC$ is given in Figure~\ref{fig:tough_exploration2_2}.
        In that figure, we already explored the tree $\calT$ (it is the same realisation as used for Figure~\ref{fig:tough_exploration2}).
        The dotted area represents the explored set,
        where we had to reveal states of the spins and edges to construct the tree.
        The event $\calC$ is the event that in the \emph{unexplored} set,
        the percolations $\omega^+$ and $\omega^-$ link up
        the already existing segments to form the desired circuits.
        In the figure, the circuits are formed of already explored segments
        (marked with blue) and unexplored segments (marked with red).

        \begin{figure}
        \centering
        \includegraphics{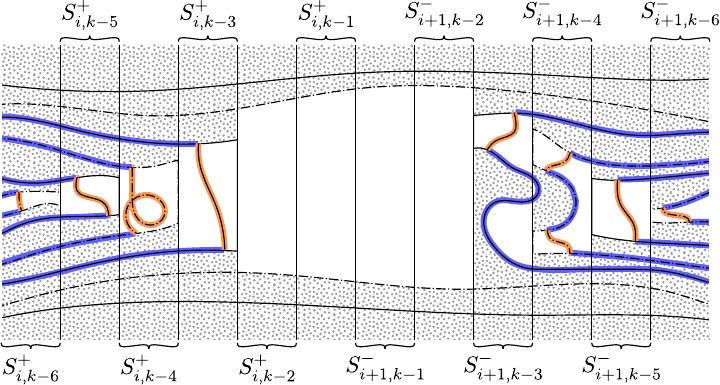}
        \caption{The explored set (dotted area) and the event $\calC$.}
        \label{fig:tough_exploration2_2}
        \end{figure}

        It is again tedious but straightforward to check that
        \begin{equation}
            \calC\cap\calS
            \subset
            \bigcap_{i=1}^{2n\rho'}\{\#\pALTRIDGE{\bar\omega}{k-4}{R_i,a_i,b_i}\geq 2n\}
            \subset
             \calE_{\rho,k,N,\rho',n}.
        \end{equation}
        Thus, to finish the proof of the claim,
        it suffices to prove that
        \begin{equation}
            \mu_{\Z^2}[\calC|\calS]\geq (\ccircuit)^{100 k^2n^2\rho'}.
        \end{equation}

        Let $\eta$ denote the set of even edges which have been revealed in the exploration process.
        For any $i=1,\ldots,2n\rho'$ and $j=0,\ldots,k-1$, the set $R_{i,j+1}'\setminus R_{i,j}'$
        consists of two vertical strips of width $N/k$ and height $4nN$.
        Write $S_{i,j}^-$ and $S_{i,j}^+$ for the left and right strips respectively
        (see Figure~\ref{fig:tough_exploration2_2}).
        Then
        \begin{equation}
            \calC\supset\calC':= \bigcap_{i=1}^{2n\rho'}\bigcap_{j=3}^{k-2}\bigcap_{\#=\pm}
            \begin{cases}
                \pVERTI{\omega^+\cup\eta}{S_{i,k-j}^\#} & \text{if $j$ is odd;}\\
                \pVERTI{\omega^-\cup\eta}{S_{i,k-j}^\#} & \text{if $j$ is even.}
            \end{cases}
        \end{equation}
        In Figure~\ref{fig:tough_exploration2_2}, this means, for example, that $S_{i,k-3}^+$
        is crossed vertically by the union of the explored (dotted) set with $\omega^+$ (red segment).
        It boils down to proving that
        \begin{equation}
        \mu_{\Z^2}[\calC'|\calS]\geq (\ccircuit)^{100 k^2n^2\rho'}.
        \end{equation}

        Notice that within each strip $S_{i,j}^\#$,
        the boundary conditions on the top and bottom of each unexplored domain
        are favourable for $\omega^+$ (if $j$ is odd) and $\omega^-$ (if $j$ is even).
        We may therefore apply the partial circuit estimate (Lemma~\ref{lem:circuitfavouredenviron}).
        The aspect ratio (the height divided by the width) of each strip is equal to $4nk$.
        The vertical crossing probability in each strip may thus be lower bounded by $(\ccircuit)^{25nk}$
        (where the factor $25$ royally suffices).
        Since there are $4n\rho'(k-4)$ such strips,
        we get the desired bound
        \begin{equation}
            \mu_{\Z^2}[\calC|\calS]\geq
            \mu_{\Z^2}[\calC'|\calS]
            \geq (\ccircuit)^{4n\rho'(k-4)\cdot 25nk}\geq (\ccircuit)^{100 k^2n^2\rho'}.
        \end{equation}
        This finishes the proof.
    \end{proof}

\subsubsection{Connecting to the free energy}

    To finish the proof of the curly bound, we must connect the quantities on either side in Lemma~\ref{lemma:ridge_splitting}
    to the quantities on either side in Proposition~\ref{prop:split_into_ridge_events}.
    We start by relating the probability of $\calE_{\rho,k,N,\rho',n}$ to the free energy,
    by analysing the right hand side in Lemma~\ref{lemma:ridge_splitting}.

    \begin{lemma}[Analysis of the large deviation event]
        \label{lemma:curly_bound_LDP_event}
        For $\rho,k,N\in 2\Z_{\geq 1}$
        with $k,N/k\geq 100$,
        \begin{equation}
        \lim_{\rho'\to\infty}\lim_{n\to\infty}
        \frac{1}{\rho'n^2}
        \log
        \mu_{\Z^2}[\calE_{\rho,k,N,\rho',n}]
        \geq
        16(\rho+1)N^2\left(
        f(\tfrac{2(k+4)}{N})-f(0)
        \right)
        +
        100k^2\log \ccircuit.
    \end{equation}
    \end{lemma}

    \begin{proof}
            We claim that
    \begin{align}
        \mu_{\Z^2}[\{\#\pALTRIDGE{\bar\omega}{k}{R_{\operatorname{Large}},a_1,b_{2n\rho'}}\geq 2n\}]
        &\geq
        \tilde c^{n^2N^2}
        \mu_{\Z^2}[\pALTCIRCUIT{\bar\omega}{2n(k+4)}{A_{\rho'(\rho+1),2nN}}]
        \\&\geq
        \tilde c^{n^2N^2}
        \P_{\Z^2}[\pCIRCUIT{+4n(k+4)+8}{A_{\rho'(\rho+1),2nN}}].
    \end{align}
    The second inequality is inclusion of events.
    The first inequality follows by a standard surgery argument (as in Step~4 in the proof for the straight bound).
    More precisely, conditional on the alternating crossing event,
    we may explore the horizontal rectangle crossings that it induces.
    Conditional on the exploration,
    we may now open $O(n^2N^2)$ $\omega^+/\omega^-$-edges in order
    to realise the ridge event.
    Since each edge is open with a uniformly positive probability,
    it is easy to find the desired universal constant $\tilde c$.

    By the variational principle (Proposition~\ref{prop:ZtoR_globalNEW}),
    \begin{multline}
        \lim_{\rho'\to\infty}\lim_{n\to\infty}
        \frac{1}{\rho'n^2}
        \log \P_{\Z^2}[\pCIRCUIT{+4n(k+4)+8}{A_{\rho'(\rho+1),2nN}}]
        \\
        =
        16(\rho+1)N^2\left(
        f(\tfrac{2(k+4)}{N})-f(0)
        \right).
    \end{multline}
    The previous lemma now yields the desired inequality.
    \end{proof}

\begin{remark}
By Lemma~\ref{lemma:free_energy_lower_bound}, we deduce that
    \begin{equation}
        \lim_{\rho'\to\infty}\lim_{n\to\infty}
        \frac{1}{\rho'n^2}
        \log
        \mu_{\Z^2}[\calE_{\rho,k,N,\rho',n}]
        \geq 100k^2(-c_{\rm fe}\rho + \log \ccircuit).
    \end{equation}
\end{remark}

\subsubsection{Simplification steps}
    To close the gap between the event $\calE_{\rho,k,N,\rho',n}$
    and the curly bound in Proposition~\ref{prop:split_into_ridge_events},
    we perform three simplification steps.
    \begin{itemize}
        \item \textbf{Diameter cutoff (Lemma~\ref{lemma:curly_bound_regularity_step1})}.
        First, we control the diameter of each domain: we impose that the outermost circuit of each ridge event has
        a diameter proportional to $N$.
        \item
        \textbf{Decoupling (Lemma~\ref{lemma:curly_bound_decoupling_step})}.
        Next, we split up the occurrence of many ridges in a single probability measure,
        into the occurrence of a single ridge in many independent probability measures.
        \item
        \textbf{Straightening (Lemma~\ref{lemma:curly_bound_regularity_step3})}.
        Finally, we impose that the ridge events occur in straightened domains (Definition~\ref{def:straightened_annulus}).
    \end{itemize}
    These steps are slightly technical.
    Their order is not important, and we do not exclude the possibility that some steps may be combined.

        Recall Lemma~\ref{lemma:ridge_splitting}.
        Split $R_{\operatorname{Large}}$ into $4n^2\rho'(\rho+1)$ squares of side length $2N$,
        and let $Q$ denote the set of centres of such squares.
        Recall the definition of the random variable $K_{2N,N,x}$ from Lemma~\ref{lemma:armexponent_for_spins_linear}
        (cf.\ Theorem~\ref{thm:armexponent_for_spins_quadratic}).
        We first prove the following lemma;
        its proof is similar to that of Theorem~\ref{thm:armexponent_for_spins_quadratic}.

    \begin{lemma}[Bound on arm events]
        \label{lemma:curly_bound_arm_events}
       Let $\rho,k,N,\rho',n\in 2\Z_{\geq 1}$
        with $k,N/k\geq 100$.
        Then for any $\epsilon,t>0$, we get
        \begin{equation}
            \frac{1}{\rho'n^2}\log\mu_{\Z^2}\Big[\Big\{\sum_{x\in Q} \Big\lfloor \frac{K_{2N,N,x}}{\epsilon k} \Big\rfloor \geq t |Q|\Big\}\Big]
            \leq
            4(\rho+1)(1+t(1-\carm \epsilon^2 k^2 / 9))\log 2.
        \end{equation}
    \end{lemma}

\begin{remark}\label{rmk:20}
        In particular, if $k\geq 9 /(\epsilon\sqrt{\carm})$, and if
        \begin{equation}
            t:=t_\epsilon:=\frac{1000}{\epsilon^2\carm}(c_{\rm fe}-\log\ccircuit),
        \end{equation}
     (recall that $c_{\rm fe}$ comes from Lemma~\ref{lemma:free_energy_lower_bound} and $\ccircuit$ from Theorem~\ref{thm:renorminput})  then the event $\{\sum_{x\in Q} \lfloor \frac{K_{2N,N,x}}{\epsilon k} \rfloor \geq t |Q|\}$ has a much smaller probability than $\calE_{\rho,k,N,\rho',n}$.
\end{remark}

    \begin{proof}
      The union bound gives that
        \begin{align}
            \mu_{\Z^2}\Big[\Big\{\sum_{x\in Q} \Big\lfloor \frac{K_{2N,N,x}}{\epsilon k} \Big\rfloor \geq t |Q|\Big\}\Big]
            &
            \leq \sum_{\substack{T:Q\to\Z_{\geq 0},\\\sum_xT_x=\lceil t|Q|\rceil}}
            \mu_{\Z^2}\Big[\Big\{\Big(\Big\lfloor \frac{K_{2N,N,x}}{\epsilon k}\Big\rfloor\Big)_x  \geq T\Big\}\Big]
            \end{align}
     Using a greedy algorithm to find a subset $Q'\subset Q$ such that
    the corresponding annuli do not overlap and such that $\sum_{x\in Q'}T(x)\geq T/9$, we obtain via Theorem~\ref{thm:armexponent_for_spins_quadratic} that for any $T$,
           \begin{align}
            \mu_{\Z^2}\Big[\Big\{\Big(\Big\lfloor \frac{K_{2N,N,x}}{\epsilon k}\Big\rfloor\Big)_x  \geq T\Big\}\Big]
            &
            \leq
            2^{-\carm \epsilon^2 k^2  t|Q| / 9}.\end{align}
            Bounding the number of functions $T$ by $2^{(1+t)|Q|}$ implies that
            \begin{align}
            \mu_{\Z^2}\Big[\Big\{\sum_{x\in Q} \Big\lfloor \frac{K_{2N,N,x}}{\epsilon k} \Big\rfloor \geq t |Q|\Big\}\Big]
            &\leq 2^{(1+t)|Q|-\carm \epsilon^2 k^2  t|Q| / 9}
            =2^{4n^2\rho'(\rho+1)(1+t(1-\carm \epsilon^2 k^2   / 9))}.
        \end{align}
      The result follows.
    \end{proof}

    We now use the quantitative bound in the previous lemma
    to analyse our event $\calE_{\rho,k,N,\rho',n}$.
    We first need some more definitions.
    Recall the definition of the level line forest from Definition~\ref{def:local_level_line_forest}.

    \begin{definition}[Local level lines forest with diameter cut-off]
        Consider a fixed continuum domain $D\subset\R^2$
        and a configuration $(\sigma_\sleven,\bar\omega,\sigma_\slodd)$.
        Recall that $\omega^\pm\cap D $ denotes the set of edges of $\omega^\pm$
        which (as line segments embedded in $\R^2$) are entirely contained in $D$.
        Let $T_r(\omega^\pm\cap D)$ denote the set of edges in $\omega^\pm\cap D $
        which belong to a connected components whose diameter is at most $2r$.
        Let $\calM^+_r(D)$ denote the set of even domains $\calD$
        which satisfy $\partial\calD\subset T_r(\omega^+\cap D)$
        and which are maximal subject to this condition.
        The \emph{local level line forest with diameter cut-off} is the family
        of level line trees
        \( \calL^{D,r}:=(\calX_\calD)_{\calD\in\calM^+_r(D)}\).
    \end{definition}

        \begin{definition}[Ridge events with diameter cut-off]
        The event $\pALTRIDGE{\bar\omega}{k,r}{D,a,b}$
        and the random variable $\#\pALTRIDGE{\bar\omega}{k,r}{D,a,b}$
        are defined exactly as before,
        except that they are defined with respect to the forest $\calL^{D,r}$
        instead of $\calL^D$.
    \end{definition}

       \begin{figure}
        \centering
        \includegraphics{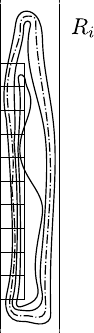}
        \caption{Diameter cutoff. If alternating circuits cover a lot of vertical distance,
        then this creates many arm events (see the squares on the left), violating the bound obtained in Lemma~\ref{lemma:curly_bound_arm_events}.}
        \label{fig:diameter_control}
    \end{figure}

 Let $\rho,k,N,\rho',n\in 2\Z_{\geq 1}$
    with $N\geq 100 k$.
    Recall the definition of $t_{\epsilon}$ from Remark~\ref{rmk:20}.
    For $r:=r(\rho,\epsilon):=\frac{1000\rho}{\epsilon} t_{\epsilon/8\rho}$, define the event
   \begin{equation}
            \calE^*_{\rho,\epsilon,k,N,\rho',n}:=\left\{\sum_{i=1}^{2n\rho'}\#\pALTRIDGE{\bar\omega}{2\lceil(1-\epsilon)k/2\rceil,rN}{R_i,a_i,b_i}\geq (1-\epsilon)4n^2\rho'\right\}.
    \end{equation}

    \begin{lemma}[Diameter cutoff]
        \label{lemma:curly_bound_regularity_step1}
       Let $\rho,k,N,\rho',n\in 2\Z_{\geq 1}$
        with $N\geq 100 k$.
       Fix $\epsilon\in(0,1/100)$ and suppose that $k\geq 72\rho /(\epsilon\sqrt{\carm})$.
        Then
        \begin{equation}
            \lim_{\rho'\to\infty}\lim_{n\to\infty}
            \frac{1}{\rho'n^2}
            \log
            \mu_{\Z^2}[\calE^*_{\rho,\epsilon,k,N,\rho',n}]
            \geq
                16(\rho+1)N^2\left(
                f(\tfrac{2(k+4)}{N})-f(0)
                \right)
                +
                100k^2\log \ccircuit.
        \end{equation}
    \end{lemma}

    \begin{proof}
        This proof is illustrated by Figure~\ref{fig:diameter_control}.
        By Lemma~\ref{lemma:curly_bound_arm_events}, it suffices to prove that
        \begin{equation}
            \calE_{\rho,k,N,\rho',n}
            \setminus \bigg\{\sum_{x\in Q} \lfloor \frac{K_{2N,N,x}}{(\epsilon/8\rho) k} \rfloor \geq t_{\epsilon/8\rho} |Q|\bigg\}
            \subset\calE^*_{\rho,\epsilon,k,N,\rho',n}.
        \end{equation}
        Define the auxiliary event
        \begin{equation}
            \calA:=\left\{
                \parbox{20em}{the forests $(\calL^{R_i})_i$ contain
        at most $2\epsilon n^2 \rho'$ trees
        that contain a vertex of depth $ 2\lfloor \epsilon k/4\rfloor$
        that has a diameter larger than $2rN$}
            \right\}.
        \end{equation}
        Suppose that $\calA$ does not occur.
        Then, every ``big'' tree (that is, a tree that contains a vertex of depth $ 2\lfloor \epsilon k/4\rfloor$
        that has a diameter larger than $2rN$)
        contributes at least $r/2$ to the sum
        in the arm event (of Lemma~\ref{lemma:curly_bound_arm_events}), so that the arm event certainly occurs.
        Therefore, it suffices to prove that
        \begin{equation}
            \calE_{\rho,k,N,\rho',n}\cap\calA
            \subset\calE^*_{\rho,\epsilon,k,N,\rho',n}.
        \end{equation}

        Yet, this is straightforward to see:
        if a tree in $(\calL^{R_i})_i$ contributes to the trees counted in $\calE_{\rho,k,N,\rho',n}$
        and does not belong to the trees counted in $\calA$,
        then it must necessarily contain a tree contributing to those counted in $\calE^*_{\rho,\epsilon,k,N,\rho',n}$.
    \end{proof}

    \begin{lemma}[Decoupling step]
        \label{lemma:curly_bound_decoupling_step}
        Fix $\rho$, $\epsilon$, $k$, and $N$ as in the previous lemma.
        Then
        \begin{multline}
            \lim_{n\to\infty}
            \frac1n\log
            \max_{(\calD_i)_{1\leq i\leq \lceil(1-\epsilon)n\rceil}}
            \prod_i
            \mu_{\Z^2}[\pALTRIDGE{\bar\omega}{2\lceil(1-2\epsilon)k/2\rceil}{\calD_i,-\rho N,\rho N}]
            \\
            \geq
             4(\rho+1)N^2\left(
                f(\tfrac{2(k+4)}{N})-f(0)
                \right)
                +
                100k^2\log \ccircuit.
        \end{multline}
        where the maximum runs over all families of disjoint even domains $\calD_i$
        which have a diameter of at most $2r_{\rho,\epsilon}N$,
        and which are contained in $ \symRect{(\rho+1)N}{nN}$.
    \end{lemma}

    \begin{proof}
        Consider the measure $\mu_{\Z^2}$.
        Let $(\calD_{ij})_{j=1,\ldots,|\calM^+_{rN}(R_i)|}$ denote the family of even domains in $\calM^+_{rN}(R_i)$
        (where the index $j$ is consistently chosen in a suitable way, for example via a dictionary order on $\R^2$).
        Recall that $(\calD_{ij})_j$ is a family of disjoint even domains
        of diameter at most $2rN$ contained in $R_i$.
        The idea of the proof is to show that:
        \begin{itemize}
            \item We do not lose too much probability by further conditioning on the exact number of
            even domains $\calD_{ij}$ where a ridge occurs,
            \item We do not loose too much probability by selecting on beforehand the domains $\calD_{ij}$
            where the ridges occur.
        \end{itemize}
        The proof is mostly technical, but we shall provide full detail.

        For any even domain $\calD$, let $\eta_\calD$ denote the set of even edges
        that are not contained in $\calD$.
        For each $ij$,
        define the events
        \begin{alignat}{2}
            &A_{ij}:=
            \pALTRIDGE{\bar\omega}{2\lceil(1-\epsilon)k/2\rceil,rN}{\calD_{ij},a_i,b_i};
            \qquad&&p_{ij}:=\mu_{\calD_{ij}}^+[A_{ij}];
            \\
            &B_{ij}:=\{\text{$R_i'$ is crossed vertically by $\omega^+\cup\eta_{\calD_{ij}}$}\};
            \qquad&&q_{ij}:=\mu_{\calD_{ij}}^+[B_{ij}].
        \end{alignat}
        Observe that $A_{ij}$ and $B_{ij}$ are disjoint and that
        \begin{equation}
            \calE^*_{\rho,\epsilon,k,N,\rho',n}
            =
            \{\textstyle\sum_{ij}\indi{A_{ij}}\geq (1-\epsilon)4n^2\rho'\}.
        \end{equation}
        Set $T:=\lceil (1-\epsilon)4n^2\rho'\rceil$ and
        define the events
        \begin{gather}
            \calA:=\{\textstyle\sum_{ij}\indi{A_{ij}}=T\};
            \\
            \calB:=\{\textstyle\sum_{ij}\indi{B_{ij}^c}=T\};
            \\
            \calG:=\{\text{the event $A_{ij}\cup B_{ij}$ occurs for every $ij$}\}.
        \end{gather}
        Our first aim is to prove that
        \begin{equation}
            \label{eq:A_G_conditional_bound}
            \mu_{\Z^2}[\calA\cap\calG|\calE^*_{\rho,\epsilon,k,N,\rho',n}]
            \geq (\ccircuit)^{100n^2\rho'}.
        \end{equation}
        We first claim that $\mu_{\Z^2}$-almost surely
        \begin{equation}
            \label{eq:A_G_conditional_bound_claim}
            \prod_{ij}q_{ij}=\mu_{\Z^2}\Big[\bigcap_{ij}B_{ij}\Big|(\calD_{ij})_{ij}\Big]\geq (\ccircuit)^{100n^2\rho'}.
        \end{equation}
        To prove the claim,
        we first use the tower property and the Markov property, and then the FKG inequality
        to obtain
        \begin{align}
            {\textstyle\prod_{ij}q_{ij}}            &=
                \Big({\bigotimes_{ij}\mu_{\calD_{ij}}^+}\Big)\Big[\bigcap_{ij}B_{ij}\Big]
            \\
            &=
            \mu_{\Z^2}\Big[\bigcap_{ij}B_{ij}\Big|(\calD_{ij})_{ij}\Big]
            \\
&=
            \mu_{\Z^2}\Big[\{\text{each $R_i$ is crossed vertically by $\omega^+$}\}\Big|\Big\{2R_{\operatorname{Large}}\cap\Big(\bigcap_{ij}\eta_{\calD_{ij}}\Big)\subset\omega^+\Big\}\Big]
            \\
           &\geq
            \mu_{\Z^2}[\{\text{each $R_i$ is crossed vertically by $\omega^+$}\}]
            \\
            &\geq (\ccircuit)^{100n^2\rho'}.
        \end{align}
        This proves the claim.
        It is now straightforward to prove
        Equation~\eqref{eq:A_G_conditional_bound}
        by running the following exploration process.
        We inspect $\calD_{ij}$ one by one, counting how often the event $A_{ij}$
        occurs.
        Let $Q$ denote this count.
        If $Q<T$,
        then we first ask if $A_{ij}$ occurs.
        If it does not occur, then we condition on the event that $B_{ij}$ occurs
        (which happens with a probability of at least $q_{ij}$ since $A_{ij}$ and $B_{ij}$
        are disjoint, and we know that $A_{ij}$ does not occur).
        If $Q=T$, then we simply condition
        on the event that $B_{ij}$ occurs.
        The product of the probabilities of the additional conditioning events in this algorithm,
        is clearly lower bounded by $\prod_{ij}q_{ij}$.
        The desired lower bound then follows by Equation~\eqref{eq:A_G_conditional_bound_claim}.
        Combining with the previous lemma, this yields the asymptotic bound
        \begin{equation}
            \label{eq:asympttttt}
            \mu_{\Z^2}[\calA\cap\calG]
            \geq \exp\Big[(n^2\rho')\left(16(\rho+1)N^2\left(
                f(\tfrac{2(k+4)}{N})-f(0)
                \right)
                +
                400k^2\log \ccircuit+o(1)\right)\Big].
        \end{equation}

        Now, notice that $\calA\cap\calG=\calA\cap\calB$.
        Conditional on $\calB$,
        let $(\calD'_\ell)_{\ell=1,\ldots,T}\subset (\calD_{ij})_{ij}$
        denote the set of even domains where $B_{ij}$ does not occur.
        By the tower property and the Markov property,
        we get
        \begin{equation}
            \mu_{\Z^2}[\calA\cap\calB]
            =\int_\calB
            \prod_{\ell=1}^T
            \mu_{\calD'_\ell}^+[A_{\ell}|B_{\ell}^c]
            \diffi\mu_{\Z^2}[(\calD_{\ell}')_{\ell}]
            \leq
            \max_{(\calD_\ell)_{\ell}}\prod_{\ell=1}^T
            \mu_{\calD_\ell}^+[A_{\ell}|B_{\ell}^c],
        \end{equation}
        where the maximum runs over all families of $T$ disjoint even domains $\calD_\ell$
        of diameter at most $2rN$ such that each $\calD_\ell$ is contained in some $R_i$.

      Set
        \begin{equation}
            B_\calD:=\{\text{$\symInt{\rho N}\times\R$ is crossed vertically by $\omega^+\cup\eta_{\calD}$}\}.
        \end{equation}
        By stacking the rectangles $(R_i)_i$ vertically,
        we get the following bound:
        \begin{equation}
            \label{eq:stacking_argument}
            \max_{(\calD_\ell)_{1\leq\ell\leq\lceil (1-\epsilon)4n^2\rho')\rceil}}\prod_{\ell}
            \mu_{\calD_\ell}^+[\pALTRIDGE{\bar\omega}{2\lceil(1-\epsilon)k/2\rceil,rN}{\calD_{\ell},-\rho N,\rho N}|B_\calD^c]
            \geq \mu_{\Z^2}[\calA\cap\calG],
        \end{equation}
        where the maximum runs over tuples of disjoint even domains $\calD_\ell$
        which are
        of diameter at most $2rN$ and which are contained in $R_{\operatorname{Stacked}}:=\symRect{(\rho+1)N}{4n^2\rho'N}$.

      Following the same arguments as those employed to go from Equation~\eqref{eq:step2_fkg}
        to Equation~\eqref{eq:step2_fkg2} in the proof of the straight bound,   we obtain that for every $\calD$ contained in $R_{\operatorname{Stacked}}$,
        \begin{equation}
            \label{eq:conditional_to_unconditional}
            \mu_{\calD}^+[\pALTRIDGE{\bar\omega}{k+2,rN}{\calD,-\rho N,\rho N}|B_\calD^c]
            \leq
            \mu_{\Z^2}[\pALTRIDGE{\bar\omega}{k}{\calD,-\rho N,\rho N}].
        \end{equation}
        Together with Equations~\eqref{eq:asympttttt} and~\eqref{eq:stacking_argument}, this inequality finishes the proof.
    \end{proof}

    Recall the notions of \emph{straightened paths} and \emph{straightened annuli}
    from Definition~\ref{def:straightened_annulus}.

    \begin{lemma}[Straightening step]
        \label{lemma:curly_bound_regularity_step3}
        Fix $\rho\in 2\Z_{\geq 1}$ and $\epsilon\in(0,1/100)$.
        Then, we may find constants $\eta\in(0,1/100)$ and $r,C<\infty$ such that the following holds true.
        Fix $k,N\in 2\Z_{\geq 1}$ with $k\geq 72\rho/(\epsilon\sqrt{\carm})$ and $N\geq 100k/\eta$.
        Then,
            \begin{multline}
                    \lim_{n\to\infty}
            \frac1n\log
            \max_{(A_i)_{1\leq i\leq \lceil(1-4\epsilon)n\rceil}}
            \prod_i
            \P_{\Z^2}[\pCIRCUIT{+4\lceil(1-8\epsilon)k/2\rceil}{NA_i}]
            \\
            \geq
             4(\rho+1)N^2\left(
                f(\tfrac{2(k+4)}{N})-f(0)
                \right)
                +
                200k^2\log\ccircuit -C,
        \end{multline}
        where the maximum runs over all families of disjoint $(\rho,\eta)$-straightened annuli $A_i$
        of diameter at most $2r$ and
         contained in $ \symRect{(\rho+1)}{n}$.
\end{lemma}

    \begin{proof}
        Fix $\rho$, $\epsilon$, $r=r_{\rho,\epsilon}$, and $k$ as in the previous step.
        We start with a series of simple bounds.

     Fix $\eta>0$ very small, and suppose that $N\in2\Z_{\geq 1}$ satisfies
        $N>100/\eta$.
        Recall the definition of the random variable $K_{R,r',x}$ from Lemma~\ref{lemma:armexponent_for_spins_linear}
        (see also Theorem~\ref{thm:armexponent_for_spins_quadratic}).
        Then, for any $x\in\R^2$,
        \begin{equation}
            \mu_{\Z^2}[\{K_{N/2,2\eta N,x}\geq \epsilon k/10\}]
            \leq (4\eta)^{\carm \epsilon^2 k^2 / 100}.
        \end{equation}
 By a union bound over the points $x\in S:=(\eta N\Z)^2 \cap \symRect{2(\rho+1)N}{4r_{\rho,\epsilon}N}$,
        we deduce
        \begin{equation}
                        \mu_{\Z^2}[\{\max_{x\in S}K_{N/2,2\eta N,x}\geq \epsilon k/10\}]
                            \leq 1000\rho r_{\rho,\epsilon}\cdot (4\eta)^{\carm \epsilon^2 k^2 / 100-2}=:p_{\rho,\epsilon,\eta,k}.
        \end{equation}
     By inclusion of events, it is then easy to deduce that
        \begin{equation}
                                \mu_{\Z^2}[\{\max_{x\in  \symRect{(\rho+1)N}{2r_{\rho,\epsilon}N}}K_{N,\eta N,x}\geq \epsilon k/10\}]
                            \leq
                            p_{\rho,\epsilon,\eta,k}.
        \end{equation}

The probability that this event occurs at least $\epsilon n$ times in $n$ independent samples,
        is at most
        \(
            (2 (p_{\rho,\epsilon,\eta,k})^\epsilon)^n
        \). For sufficiently small $\eta=\eta_{\rho,\epsilon}$ (depending only on $\rho$ and $\epsilon$),
        this probability is much smaller than the probability of the event in the previous lemma (the decoupling step, Lemma~\ref{lemma:curly_bound_decoupling_step}).

 Now, consider the previous lemma.
        We view the product over the probabilities as a single probability
        of a cylinder event $\calR$ in the product measure.
        In this product measure, the probability
        \begin{equation}\calR\cap\{\text{in at least $(1-3\epsilon)n$ of the domains,  too many arms contribute to $K_{N,\eta N,x}$}\}\end{equation}
        has the same asymptotics as the event $\calR$ itself.
        By choosing the domains in which the arms occur,
        we get the bound
        \begin{align}
                    \lim_{n\to\infty}
            \tfrac1n\log
            \max_{(\calD_i)_{1\leq i\leq \lceil(1-3\epsilon)n\rceil}}
            &\prod_i
            \mu_{\Z^2}[\calE_{\calD_i}]
            \\
           & \geq
             4(\rho+1)N^2\left(
                f(\tfrac{2(k+4)}{N})-f(0)
                \right)
                +
                100k^2\log \ccircuit
                                -\log 2;\end{align}
              where
              \begin{equation}  \calE_\calD:=\pALTRIDGE{\bar\omega}{2\lceil(1-2\epsilon)k/2\rceil}{\calD_i,-\rho N,\rho N}\cap\{\max_{x\in\calD_i}K_{N,\eta N,x}\leq \epsilon k/10\}.
        \end{equation}
        (the extra $\log 2$ comes from the combinatorial choice of domains).

        Let us now study event $\calE_\calD$.
        It is easy to see that
        \begin{equation}
            \calE_\calD\subset
            \bigcup_{A} \pALTCIRCUIT{\bar\omega}{2\lceil(1-4\epsilon)k/2\rceil}{NA},
        \end{equation}
        where the union is over all $(\rho,\eta)$-straightened annuli $A$
        such that $NA$ is completely surrounded by $\partial\calD$.
        Notice that the number of such annuli is at most $2^{99\rho r_{\rho,\epsilon}/\eta_{\rho,\epsilon}^2}$
        (by simply upper bounding the number of $\eta\Z^2$-edges that may or may not belong to the boundary of the annulus).
        Thus, we get
                \begin{align}
                    \lim_{n\to\infty}
            \tfrac1n\log
            &\max_{(A_i)_{1\leq i\leq \lceil(1-3\epsilon)n\rceil}}
            \prod_i
            \mu_{\Z^2}[\pALTCIRCUIT{\bar\omega}{2\lceil(1-4\epsilon)k/2\rceil}{NA_i}]
            \\
            &\geq
             4(\rho+1)N^2\left(
                f(\tfrac{2(k+4)}{N})-f(0)
                \right)
                +
                100k^2\log \ccircuit-\frac{100\rho r_{\rho,\epsilon}}{\eta_{\rho,\epsilon}}\log 2,
        \end{align}
        where the maximum is over all families of disjoint $(\rho,\eta)$-straightened annuli $A_i$
        such that each annulus has a diameter of at most $2r_{\rho,\epsilon}$
        and remains contained in $\symRect{(\rho+1)}{2n}$.

        To finish the proof, notice simply that
        \begin{equation}
            \P_{\Z^2}[\pCIRCUIT{+4\lceil(1-8\epsilon)k/2\rceil}{NA_i}]
            \geq
            2^{-2\lceil(1-4\epsilon)k/2\rceil}
            \mu_{\Z^2}[\pALTCIRCUIT{\bar\omega}{2\lceil(1-4\epsilon)k/2\rceil}{NA_i}],
        \end{equation}
        by flipping the coins corresponding to each level line.
        This implies the desired bound.
\end{proof}

% </input src="sections/PART_E/24_LDP_SPLIT/curly.tex" root="." version="0.0.1">

% </input src="sections/PART_E/24_LDP_SPLIT/main.tex" root="." version="0.0.1">

% </input src="sections/PART_E/main.tex" root="." version="0.0.1">

\paragraph{Acknowledgements.}
 This project has received funding from the Swiss National Science Foundation and the NCCR SwissMAP.
 HDC acknowledges the support from the Simons collaboration on localization of waves.
 The work of KKK is supported by the ERC Project LDRAM: ERC-2019-ADG Project 884584.
 KKK acknowledges the support from CNRS and from the joint AND-DFG TSF24 project ANR-24-CE92-0033.
 PL acknowledges the support from the French National Research Agency (ANR), project number ANR-23-CPJ1-0150-01.
}
\newcommand\strictlygeq{>}

\addcontentsline{toc}{part}{References}

\printbibliography
\printtodolist

\end{document}